\documentclass[]{aa}  

\usepackage[fleqn]{amsmath}
\usepackage{amssymb}
\usepackage{textcomp}
\usepackage{graphicx}
\usepackage{txfonts}
\usepackage{natbib}
\usepackage{url}
\usepackage{multirow}
\usepackage{subfigure}
\usepackage[dvipsnames]{xcolor}
\usepackage{placeins}
\usepackage[version=3]{mhchem}
\usepackage{lscape}

\begin{document}

\title{Complex organic molecules in protoplanetary disks}

\author{Catherine Walsh\inst{\ref{inst1},\ref{inst2}} 
\and T. J. Millar\inst{\ref{inst2}} 
\and Hideko Nomura\inst{\ref{inst3},\ref{inst4},\ref{inst5}} 
\and Eric Herbst\inst{\ref{inst6},\ref{inst7}}
\and Susanna Widicus Weaver\inst{\ref{inst8}} 
\and Yuri Aikawa\inst{\ref{inst9}}
\and Jacob C. Laas\inst{\ref{inst8}}
\and Anton I. Vasyunin\inst{\ref{inst10},\ref{inst11}}}

\institute{Leiden Observatory, Leiden University, P.~O.~Box 9513, 2300 RA Leiden, The Netherlands \\
\email{cwalsh@strw.leidenuniv.nl} \label {inst1} 
\and Astrophysics Research Centre, School of Mathematics and Physics, 
Queen's University Belfast, University Road, Belfast, BT7 1NN, UK \label{inst2}
\and Department of Astronomy, Graduate School of Science, Kyoto University, Kyoto 606-8502, Japan \label{inst3}
\and National Astronomical Observatory of Japan, Osawa, Mitaka, Tokyo 181-8588, Japan \label{inst4}
\and Department of Earth and Planetary Sciences, Tokyo Institute of Technology, 2-12-1 Ookayama, Meguro-ku, Tokyo 152-8551, Japan \label{inst5}
\and Departments of Physics, Chemistry and Astronomy, The Ohio State University, Columbus, OH 43210, US \label{inst6}
\and Departments of Chemistry, Astronomy, and Physics, University of Virginia, Charlottesville, VA 22904, US \label{inst7}
\and Department of Chemistry, Emory University, Atlanta, GA 30322, US \label{inst8}
\and Department of Earth and Planetary Sciences, Kobe University, 1-1 Rokkodai-cho, Nada, Kobe 657-8501, Japan \label{inst9}
\and Department of Chemistry, University of Virginia, Charlottesville, VA 22904, US \label{inst10}
\and Visiting Scientist, Ural Federal University, Ekaterinburg, Russia \label{inst11}
}

\date{Recieved $\cdots$/Accepted $\cdots$}
 
\abstract
{Protoplanetary disks are vital objects in star and planet formation, possessing 
all the material, gas and dust, which may form a planetary system orbiting 
the new star.  
Small, simple molecules have traditionally been detected in protoplanetary disks; 
however, in the ALMA era, we expect the 
molecular inventory of protoplanetary disks to significantly increase.  }
{We investigate the synthesis of complex organic molecules 
(COMs) in protoplanetary disks to 
put constraints on the achievable chemical complexity and to predict 
species and transitions which may be observable with ALMA.  
}
{We have coupled a 2D steady-state physical model of a protoplanetary disk around a 
typical T Tauri star with a large gas-grain chemical network including COMs.  
We compare the resulting column densities with those derived from observations 
and perform ray-tracing calculations to predict line spectra.  
We compare the synthesised line intensities with current observations 
and determine those COMs which may be observable in 
nearby objects. 
We also compare the predicted grain-surface abundances with those 
derived from cometary comae observations.}
{We find COMs are efficiently formed in the disk midplane 
via grain-surface chemical 
reactions, reaching peak grain-surface fractional abundances $\sim$~10$^{-6}$--10$^{-4}$ 
that of the H nuclei number density.  
COMs formed on grain surfaces are returned to the 
gas phase via non-thermal desorption; however, 
gas-phase species reach lower fractional abundances than their grain-surface equivalents, 
$\sim$~10$^{-12}$--10$^{-7}$.  
Including the irradiation of grain mantle material helps build further complexity 
in the ice through the replenishment of grain-surface radicals which take part in 
further grain-surface reactions.  
There is reasonable agreement with several line transitions of \ce{H2CO} observed 
towards T~Tauri star-disk systems.  
There is poor agreement with \ce{HC3N} lines observed towards LkCa~15 and 
GO~Tau and we discuss possible explanations for these discrepancies.  
The synthesised line intensities for \ce{CH3OH} are consistent with upper limits 
determined towards all sources.   
Our models suggest \ce{CH3OH} should be readily observable 
in nearby protoplanetary disks with ALMA; however, detection of more complex species may prove challenging, even 
with ALMA `Full Science' capabilities.    
Our grain-surface abundances are consistent with those derived from cometary comae observations 
providing additional evidence for the hypothesis that comets (and other planetesimals) formed 
via the coagulation of icy grains in the Sun's natal disk.  
}
{}


\maketitle

\section{Introduction}
\label{introduction}

Protoplanetary disks are crucial objects in star formation. 
They dissipate excess angular momentum away from the protostellar system, 
facilitate the accretion of matter from the natal cloud onto the 
new star, and contain all the material, dust and gas, which 
will likely go on to form a surrounding planetary system 
\citep[for a review, see, e.g.,][]{williams11}.   
The study of the detailed chemistry of these objects has gained impetus in recent years 
driven by the impending completion of the Atacama Large Millimeter/Submillimeter Array (ALMA).   
ALMA, with its unprecedented sensitivity and spatial and spectral resolution,
will reveal, for the first time, the composition of protoplanetary 
disks on $\sim$~milliarcsecond scales, probing material $\lesssim$~10~AU of the 
parent star in objects relatively close to Earth ($\approx$~140~pc).    
This spatial resolution will be achievable using the most extended configuration 
(with maximum baseline, $B$~$\approx$~16~km) at its highest operational frequencies 
($\nu$~$>$~275~GHz).
This will allow the study of the detailed composition of the cold molecular material within
the `planet-forming' region of nearby disks, which will 
advance our understanding of the process of planetary system formation, 
and help answer questions regarding the morphology and composition 
of our own Solar System.  

The molecules observed in protoplanetary disks have thus far 
been restricted to 
small species and associated isotopologues due to their relatively 
high abundance and simple rotational spectra leading to observable line emission.  
The sources in which these molecules have been detected are also 
limited to a handful of nearby, and thus well-studied, objects.  
Molecules have been observed at both infrared (IR) and (sub)mm wavelengths with the IR 
emission originating from the inner warm/hot material 
(T~$\gtrsim$~300~K, R~$\lesssim$~10~AU) and the (sub)mm emission 
originating from the outer cold disk (T~$<$~300~K, R~$\gtrsim$~10~AU).  
The molecules detected at (sub)mm wavelengths 
include CO, \ce{HCO+}, CN, HCN, CS, \ce{N2H+}, SO and \ce{C2H} 
\citep[see, e.g.,][]{kastner97,dutrey97,vanzadelhoff01,thi04,fuente10,henning10}.  
Also detected are several isotopologues of the listed species, e.g., $^{13}$CO, C$^{18}$O, 
H$^{13}$CO$^{+}$, \ce{DCO+} and DCN \citep[see, e.g.,][]{vandishoeck03,thi04,qi08}.  
Several relatively complex molecules have also been observed:  
\ce{H2CO} \citep{dutrey97,aikawa03,thi04,oberg10,oberg11}, \ce{HC3N} \citep{chapillon12}, 
and $c$-\ce{C3H2} \citep{qi13b}.  
Line emission in the (sub)mm can be observed from the ground and such observations 
have historically been conducted
using single-dish telescopes, 
e.g., the JCMT (James Clerk Maxwell Telescope), 
the CSO (Caltech Submillimeter Observatory), 
the IRAM (Institut de Radioastronomie Millim\'{e}trique) 30~m telescope, 
and APEX (Atacama Pathfinder Experiment).   
More recently, several interferometers have been available, e.g., 
the SMA (Submillimeter Array), 
CARMA (Combined Array for Research in Millimeter-wave Astronomy), and 
PdBI (Plateau-de-Bure Interferometer).  
These latter facilities have enabled spatially-resolved mapping 
of very nearby objects including the archetypical protoplanetary disk, 
TW Hydrae, located at a distance of $\approx$~56~pc \citep{oberg10,oberg11,hughes11,qi11}.  
Due to its proximity, TW Hydrae was observed during ALMA Science 
Verification which utilised between six and nine antennae working in conjunction to 
map the line emission from this source \citep{oberg12,rosenfeld12}.  
Early results from ALMA also include the first detection
of the location of the CO snowline\footnotemark[1]~in the disk around HD~163296 
using \ce{DCO+} line emission \citep{mathews13}, 
and in the disk around TW~Hydrae using \ce{N2H+} line emission \citep{qi13c}. 

\footnotetext[1]{The CO snowline marks the transition zone in the disk midplane 
($T$~$\approx$~17~K) beyond which CO is depleted from the gas via freezeout 
onto dust grains.}

The launch of the {\em Herschel} Space Observatory allowed 
the first detection of ground-state transitions of 
ortho- and para-\ce{H2O} (at 557~GHz and 1113~GHz, respectively) 
in the disk of TW Hydrae \citep{hogerheijde11}.  
\citet{bergin13} also report the first detection of 
HD in TW~Hydrae using {\em Herschel}, allowing, for the first time, a direct 
determination of the disk mass without relying on analysis of 
dust thermal emission or CO rotational line emission.  
\citet{bergin13} determine a disk mass $<$~0.05~M$_\odot$ 
confirming that TW~Hydrae, although considered a rather old system 
($\sim$~10~Myr), contains sufficient material 
for the formation of a planetary system.
Also detected in the far-IR using {\em Herschel}, 
is the molecular ion, \ce{CH+}, in the disk of the Herbig~Be star, 
HD~100546 \citep{thi11}, and multiple lines of OH 
and warm \ce{H2O} have also been detected in numerous sources  
\citep[][]{fedele12,meeus12,rivieremarichalar12}.  

Most detections of line emission in the mid-IR have been conducted with 
the {\em Spitzer} Space Telescope and molecules observed include 
OH, \ce{H2O}, \ce{C2H2}, HCN, CO, and \ce{CO2}
\citep{lahuis06,carr08,salyk08,pontoppidan10,bast13}. 
Species detected at IR wavelengths are also 
limited to abundant, small, simple molecules 
with strong rovibrational transitions and/or vibrational modes, which are able to 
survive the high temperatures encountered in the inner disk.    
\citet{mandell12} also report the detection of near-IR 
emission lines of \ce{C2H2} and 
HCN, for the first time, using ground-based observatories
(CRIRES on the Very Large Telescope and NIRSPEC on the Keck II Telescope).  

The greatest chemical complexity (outside of our Solar System) is seen in 
massive star-forming regions towards the Galactic centre \citep[e.g., Sgr~B2(N),][]{turner91} and 
in objects called `hot cores' and `hot corinos', considered important stages in 
high-mass ($M_{\ast}$~$\gtrsim$~10~$M_{\odot}$) and low-mass 
($M_{\ast}$~$\lesssim$~10~$M_{\odot}$) star formation, respectively \citep[see, e.g., ][]{herbst09}. 
Hot cores are remnant, often clumpy, cloud material left over from the 
explosive process of high-mass star formation which is heated by the embedded massive star.  
They are warm ($T$~$\sim$~100~K), dense ($n$~$\gtrsim$~10$^{6}$~cm$^{-3}$), 
relatively large ($R$~$\sim$~0.1~pc) objects which are heavily shielded by dust from both the 
internal stellar radiation and the external 
interstellar radiation ($A_\mathrm{v}$~$\sim$~100~mag).  
Hot corinos, considered the equivalent early stage of low-mass star formation, 
possess similar densities and temperatures to hot cores, yet, are much less massive and 
smaller in spatial extent (typically, $R$~$\sim$~100~AU). 
The line emission from the hot corino arises from a very compact 
region on the order of 1" in size for a source at the distance of Taurus (140~pc).  
Hence, we are limited to studying a handful of nearby sources 
\citep[see, e.g.,][]{ceccarelli05}.
Nevertheless, hot corinos are certainly as chemically complex 
as their more massive counterparts (if not more so), attested by the detection of 
glycolaldehyde, \ce{HOCH2CHO}, in IRAS 16293+2422 during ALMA Science Verification  
\citep{jorgensen12}.

Hot cores and corinos are typified by the detection of rotational 
line emission from complex organic molecules 
(henceforth referred to as COMs), the formation of which remains one of the great puzzles in astrochemistry.  
The generally accepted mechanism is that simple ices formed on grain surfaces 
in the molecular cloud at 10~K, either via direct 
freezeout from the gas phase or via H-addition reactions on the grain 
(e.g., CO, \ce{H2O}, \ce{H2CO}, \ce{CH3OH}),  
undergo warming to $\approx$~30~K where they achieve 
sufficient mobility for grain-surface chemistry to occur via radical-radical
association to create more complex ice mantle species (e.g., \ce{HCOOCH3}).  
The grain-surface radicals necessary for further molecular synthesis 
are thought to be produced by dissociation via UV photons created 
by the interaction of cosmic rays with \ce{H2} molecules.  
Dissociation and/or ionisation via energetic
electrons, created along the impact track as a cosmic ray particle 
penetrates a dust grain, is an alternative scenario 
\citep[see, e.g.,][]{kaiser97}.
Further warming to $T$~$\gtrsim$~100~K allows the removal of these more complex species from the ice mantle 
via thermal desorption thus `seeding' the gas with gas-phase COMs.  
Typically, the observed rotational line emission is characterised by a gas temperature of 
$\gtrsim$~100~K with COMs observed at abundances 
$\sim$~10$^{-10}$ to $\sim$~10$^{-6}$ times that of the \ce{H2} number density 
\citep[see, e.g., ][]{herbst09}.  

Comparing the physical conditions in hot cores/corinos 
with those expected in the midplane and molecular regions of protoplanetary disks, 
it appears a similar chemical 
synthesis route to COMs may be possible; however, to date, targeted searches for gas-phase COMs 
in nearby protoplanetary disks have been unsuccessful \citep[see, e.g.,][]{thi04,oberg10,oberg11}.  
The possible reasons for this are severalfold: 
(i) gas-phase COMs are relatively abundant in disks;  
however, due to their more complex spectra and resulting weaker emission and the small intrinsic size of disks,
existing telescopes are not sufficiently sensitive to detect line emission from COMs on realistic integration 
time scales, (ii) gas-phase COMs are relatively abundant in disks;  
however, previous targeted searches have not selected the best candidate lines for detection with 
existing facilities, and (iii) gas-phase COMs achieve negligible abundances in disks.  
The latter reason may be related to the major difference between hot cores/corinos 
and disks: the presence of external UV and X-ray radiation.  
Certainly, observations using ALMA, with its superior sensitivity and spectral resolution, 
will elucidate which scenario is correct.     
The confirmation of the presence (or absence) of COMs in disks is of 
ultimate astrobiological importance; is it possible for prebiotic molecules 
to form in the disk and survive assimilation into planets and other 
objects such as comets and asteroids?  Looking at our own Solar System, it appears possible.  
Many relatively complex molecules have been observed in the comae of multiple comets: 
\ce{H2CO}, \ce{CH3OH}, HCOOH, \ce{HC3N}, \ce{CH3CN}, \ce{C2H6} 
\citep[see, e.g.,][and references therein]{mumma11}. 
The brightest comet in modern times, Hale-Bopp, displayed immense chemical complexity with 
additional detections of \ce{CH3CHO}, \ce{NH2CHO}, \ce{HCOOCH3}, 
and ethylene glycol, \ce{(CH2OH)2} \citep{crovisier04a,crovisier04b}.  
In addition, the simplest amino acid, glycine (\ce{NH2CH2COOH}), was identified in 
samples of cometary dust from comet 81P/Wild 2 returned by the Stardust mission \citep{elsila09}.  
The detection of gas-phase glycine  is considered one 
of the `holy grails' of prebiotic chemistry; however,  
thus far, searches for gas-phase glycine towards hot cores have been 
unsuccessful \citep[see, e.g.,][]{snyder05}.  

In \citet{walsh10} and \citet{walsh12}, henceforth referred to as WMN10 and WNMA12, 
we calculated the chemical composition of a protoplanetary disk using a gas-phase chemical 
network extracted from the UMIST Database for Astrochemistry 
(\citeauthor{woodall07} \citeyear{woodall07}\footnotemark[2]), termed `{\sc Rate}06', and 
the grain-surface chemical network from \citet{hasegawa92} and \citet{hasegawa93}.  
We included the accretion of gas-phase species onto dust grains and 
allowed the removal of grain mantle species via both thermal and non-thermal desorption.  
In WMN10, our aim was to study the effects of cosmic-ray-induced desorption, photodesorption, and 
X-ray desorption on the chemical structure of the disk, 
whereas, in WNMA12, we extended our investigations to cover the importance 
of photochemistry and X-ray ionisation on disk composition.   
In both works, we focussed our discussions on species detected in disks, 
the most complex of which, at that time, was formaldehyde, \ce{H2CO}.    

\footnotetext[2]{\url{http://www.udfa.net}}

{\sc Rate}06 includes several gas-phase COMs, including 
methanol (\ce{CH3OH}), 
formaldehyde (\ce{H2CO}), formic acid (HCOOH), methyl formate (\ce{HCOOCH3}), 
dimethyl ether (\ce{CH3OCH3}) and acetone (\ce{CH3COCH3}).  
These represent the most simple alcohol, aldehyde, carboxylic acid, 
ester, ether and ketone, respectively.  
The network also includes several larger members of these families, 
e.g., ethanol (\ce{C2H5OH}) and acetaldehyde (\ce{CH3CHO}).    
In WMN10 and WNMA12, we adopted the grain-surface 
network of \citet{hasegawa92} and \citet{hasegawa93} which 
includes the grain-surface synthesis of several of these   
more complex species.  
However, this network concentrates on simple atom-addition reactions, more likely to 
occur at the lower temperatures encountered in dark clouds.  
Hence, to date, the grain-surface chemistry that has been included 
is by no means comprehensive regarding the grain-surface synthesis of 
COMs.    

In this work, we study the efficiency of the synthesis of COMs in protoplanetary 
disks using a chemical network typically used for hot core 
and hot corino chemical models.  
In Sect.~\ref{protoplanetarydiskmodel}, we describe our 
protoplanetary disk model (Sect.~\ref{physicalmodel}) 
and chemical network (Sect.~\ref{chemicalmodel}).  
In Sects.~\ref{results} and \ref{discussion}, we present and discuss our results, 
respectively, and in Sect.~\ref{summary} we state our conclusions.  

\section{Protoplanetary Disk Model}
\label{protoplanetarydiskmodel}

\subsection{Physical Model}
\label{physicalmodel}

Our protoplanetary disk physical structure is calculated according to the methods outlined in 
\citet{nomura05} with the addition 
of X-ray heating as described in \citet{nomura07}.  
We model an axisymmetric disk in Keplerian rotation about a typical classical T Tauri star 
with mass, $M_\ast$~=~0.5~$M_\odot$, radius, $R_\ast$~=~2~$R_\odot$ and effective 
temperature, $T_\ast$~=~4000~K \citep[see, e.g.,][]{kenyon95}.  
The surface density distribution is determined by the
central star's mass and radius and assuming a constant
disk mass accretion rate, $\dot{M}$, \citep[see, e.g.,][]{pringle81} 
and we parameterise the kinematic viscosity, $\nu$, using the $\alpha$-disk 
model of \citet{shakura73}.  
We use a viscous parameter, $\alpha$~=~0.01, and a mass accretion rate, 
$\dot{M}$~=~$10^{-8}$~$M_\odot$~yr$^{-1}$, typical values for 
accretion disks around classical T~Tauri stars.  
We self-consistently solve the equation of hydrostatic equilibrium in the vertical direction 
and the local thermal balance between the heating and cooling of the gas to 
model the gas temperature, dust temperature, and density structure of the disk.  
The heating mechanisms included are grain photoelectric heating by UV photons 
and heating due to hydrogen ionisation by X-rays.  
We include gas-grain collisions and line transitions as cooling mechanisms.

The UV field in our disk model has two sources: the central star and the interstellar medium.  
The central star's UV radiation field has three components: black-body radiation 
at the star's effective temperature, hydrogenic bremsstrahlung emission, and strong 
Ly-$\alpha$ line emission.  
The latter two components are necessary for accurately modelling the excess 
UV emission often observed towards T~Tauri stars, which is thought to arise from an accretion shock 
as material from the disk impinges upon the stellar surface \citep[see, e.g.,][]{johnskrull00}.
For the UV extinction, we include absorption and scattering by dust grains.  
The combined UV spectrum originating from the T Tauri star 
is displayed in Fig.~C.1 in \citet{nomura05} and replicated in 
Fig.~1 in WNMA12.  The total UV luminosity is $L_\mathrm{UV}$~$\sim$~10$^{31}$~erg~s$^{-1}$.  

We model the X-ray spectrum of the T Tauri star by fitting the {\em XMM-Newton} 
spectrum observed towards the classical T Tauri star, TW Hydrae, with a 
two-temperature thin thermal plasma model \citep[see, e.g.,][MEKAL model]{liedahl95}.  
The best-fit parameters for the temperatures are $kT_1$~=~0.8~keV and 
$kT_2$~=~0.2~keV.  For the foreground interstellar hydrogen column density, we find 
$N$(\ce{H2})~=~2.7~$\times$~10$^{20}$~cm$^{-2}$. 
For the X-ray extinction, we include attenuation due to all 
elements and Compton scattering by hydrogen.  
The resulting X-ray spectrum is shown in Fig.~1 in 
\citet{nomura07} and is replicated in Fig.~1 in WNMA12.  
The total X-ray luminosity of the star is $L_\mathrm{X}$~$\sim$~10$^{30}$~erg~s$^{-1}$. 

We assume the dust and gas in the disk are well mixed, and 
we adopt a dust-grain size distribution which reproduces the extinction 
curve observed in dense clouds \citep{weingartner01}.  
The dust grains are assumed to consist of silicate and carbonaceous 
material, and water ice. 
The resulting wavelength-dependent dust absorption coefficient is shown in 
Fig.~D.1 in \citet{nomura05}.  
We acknowledge that this is a simplistic treatment of the dust-grain 
distribution in protoplanetary disks since it is thought that gravitational 
settling and grain coagulation (grain growth) will perturb the 
dust size and density distribution from that observed in dense clouds 
\citep[see, e.g.,][]{dullemond04,dullemond05,dalessio06}. 
To keep the chemical calculation computationally tractable, 
we adopt average values for the dust grain size and density that are 
consistent with the dust model adopted in the disk structure 
calculation \citep[see, e.g.,][]{bergin11}.

\begin{figure*}
\includegraphics[width=0.5\textwidth]{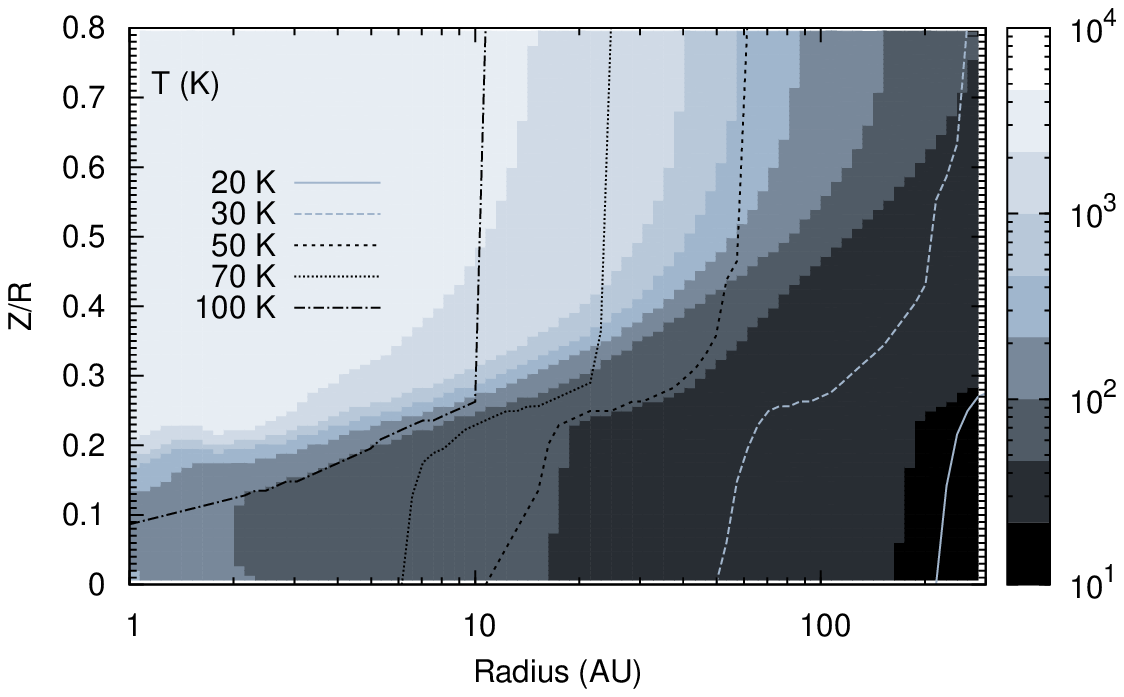}
\includegraphics[width=0.5\textwidth]{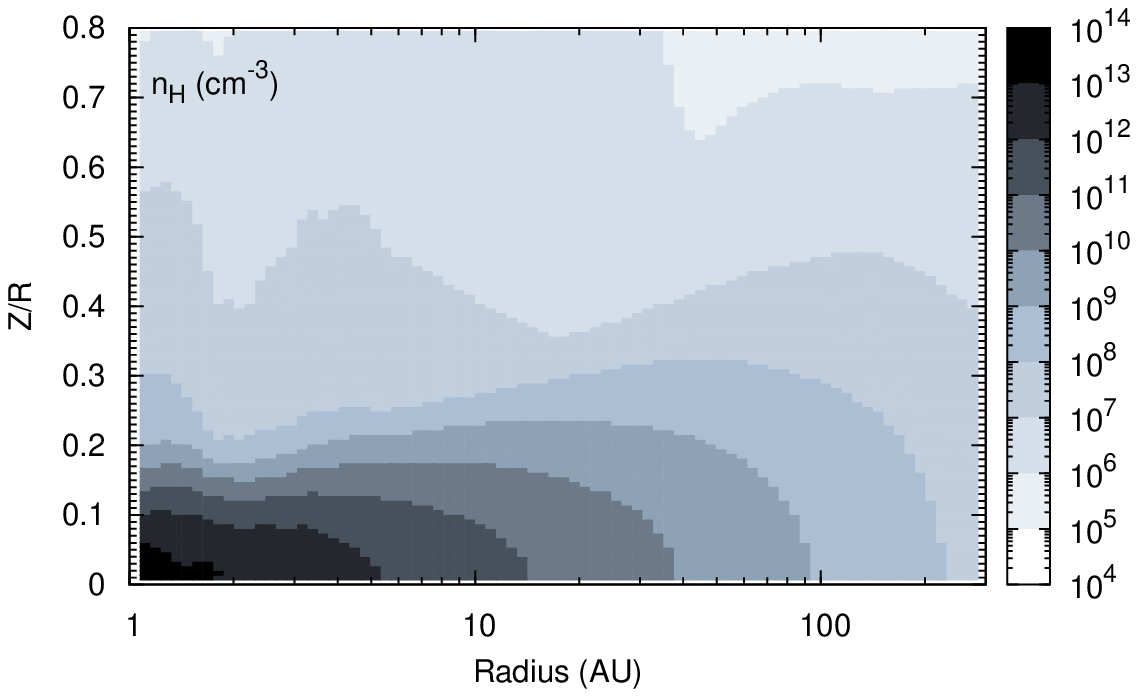}
\includegraphics[width=0.5\textwidth]{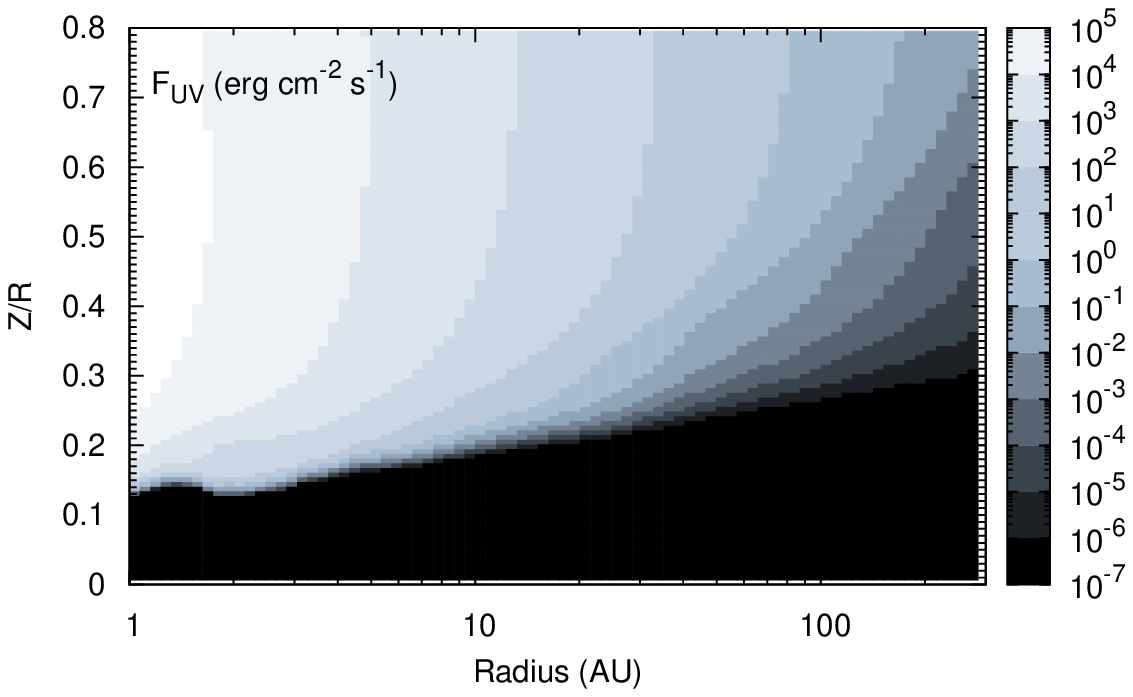}
\includegraphics[width=0.5\textwidth]{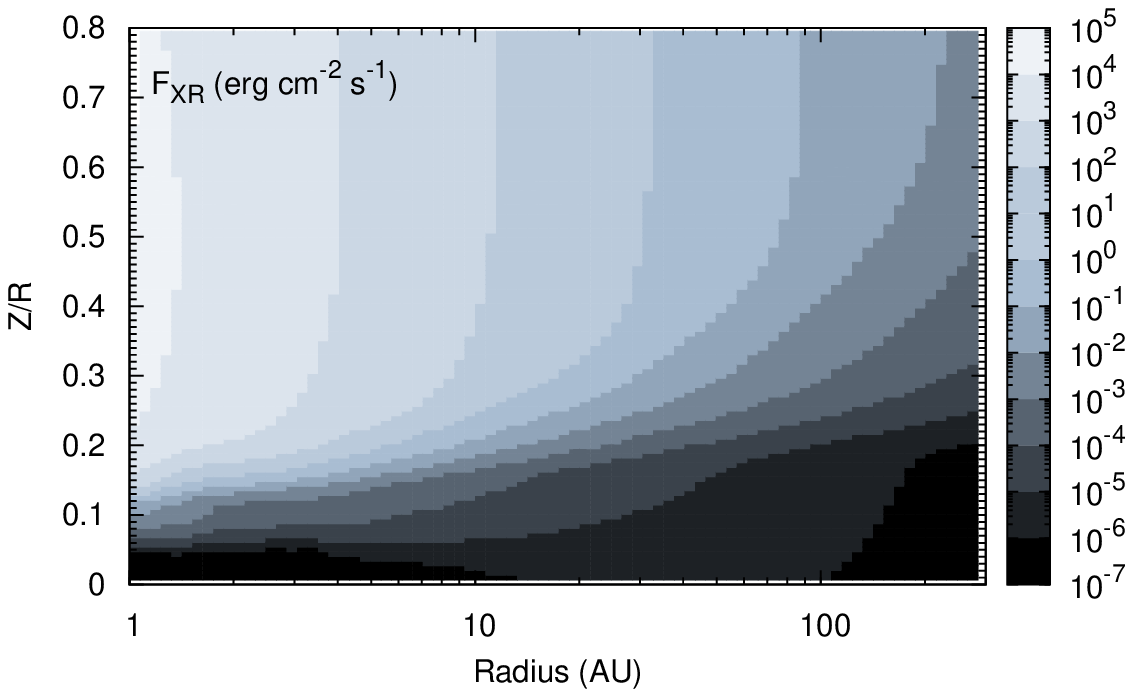}
\caption{Gas temperature (top left panel), H nuclei number density (top right panel), 
integrated UV flux (bottom left panel), and integrated X-ray flux (bottom right panel) 
as a function of disk radius, $R$, and disk height, $Z/R$. 
The contours in the top left panel represent the dust temperature.  
}
\label{figure1}
\end{figure*}

In Fig.~\ref{figure1}, to guide the discussion in the paper, 
we present the resulting gas temperature (top left panel), 
total H nuclei number density, $n_\ce{H}$ (top right panel), 
integrated UV flux (bottom left panel), and integrated X-ray flux (bottom right panel), 
as a function of disk radius, $R$, and height, $Z/R$.  
The contours in the top left panel represent the dust temperature.  
Here, we do not discuss the resulting disk structure in detail as this is covered in 
a series of previous publications \citep{walsh10,walsh12}.  
In WMN10, we also present the two-dimensional gas and dust temperatures and gas number 
density, with supporting material and comprehensive discussions 
in the Appendix of that paper.  In WNMA12, 
we display and discuss the incident UV and X-ray spectra and the 
resulting two-dimensional wavelength-integrated UV and X-ray fluxes.    

The physical conditions throughout the disk cover many different regimes 
that generally differ from the conditions typical of dark clouds and hot cores and corinos.  
The gas temperature ranges from $\approx$~17~K in the outer disk midplane 
to $>$~6000~K in the inner disk surface.  
The H nuclei number density in the dark disk midplane spans many orders of magnitude from 
$\sim$~10$^7$~cm$^{-3}$ in the outer disk (R~$\approx$~300~AU) to $\sim$~10$^{14}$~cm$^{-3}$ 
in the inner disk (R~$\sim$~1~AU). 
We also remind the reader that protoplanetary disks around T~Tauri stars are irradiated 
by X-rays in addition to cosmic rays and UV photons; hence, the ionisation rates and 
induced photodestruction rates in the molecular regions of protoplanetary disks 
can reach values much higher than those experienced in molecular clouds 
and hot cores and corinos (see Fig.~\ref{figure1}). 

\subsection{Chemical Model}
\label{chemicalmodel}

We include various processes in our chemical network: gas-phase two-body reactions, 
photoreactions, cosmic-ray and X-ray reactions, gas-grain interactions, 
grain-surface two-body reactions, grain-surface photoreactions, and grain-surface 
cosmic-ray-induced and X-ray induced photoreactions.  
The latter three processes allow the chemical processing of grain-surface 
material by the radiation field present in the disk.   
The gas-phase chemistry (including the photochemistry and cosmic-ray chemistry) 
is from \citet{laas11} which is based on
the OSU chemical network\footnotemark[3] 
and which includes new possible routes to the 
formation of methyl formate and its structural isomers.
To render the network suitable for protoplanetary disk chemistry, we have added 
reactions and rate coefficients 
applicable at higher temperatures \citep{harada10,harada12}.    
We include the direct X-ray ionisation of elements and calculate the X-ray ionisation rate 
throughout the disk using the prescription outlined in WNMA12.  
To simulate the X-ray chemistry, we duplicate the set of cosmic-ray-induced 
photoreactions and scale the rates by the X-ray ionisation rate.  
We also include the explicit calculation of the photochemical rates 
using the UV spectrum at each point in the disk for  
those species for which photodissociation and photoionisation cross sections exist 
(around 60 species, see \citeauthor{vandishoeck06} \citeyear{vandishoeck06}\footnotemark[4]).  
In the absence of cross-sectional data, we approximate the rate
by scaling the expected interstellar rate by the ratio of the 
integrated UV flux to that of the ISRF ($\approx$~1.6~$\times$~10$^{-3}$ erg~cm$^{-2}$~s$^{-1}$).  

\footnotetext[3]{\url{http://www.physics.ohio-state.edu/}$\sim$\url{eric/}} 
\footnotetext[4]{\url{http://www.strw.leidenuniv.nl/}$\sim$\url{ewine/photo}} 

For gas-grain interactions, we include accretion from the gas onto dust grains (i.e., freezeout) 
and desorption from grain surfaces back into the gas phase.  
For calculating the gas accretion and thermal desorption rates, we use the 
theory of \citet{hasegawa92} as in our previous work (see Equations~2,~3,~and~4 in WMN10).  
We assume, for simplicity, that the grains are negatively charged 
compact spheres with a radius, $a$~=~0.1~$\mu$m, and a constant fractional abundance 
(relative to the H nuclei number density) of $\sim$~10$^{-12}$, 
equivalent to a gas-to-dust mass ratio of $\sim$~100.  
We assume a sticking coefficient, $S$~$\sim$~1, for all species. 
We model the grain-surface formation of molecular hydrogen by assuming 
the rate of \ce{H2} formation equates to half the rate of arrival of H atoms 
on the grain surface and use a reduced sticking coefficient for atomic hydrogen ($S$~$\sim$~0.3).
We adopt binding energies from the work of \citet{laas11} 
which originate from \citet{garrod06} and \citet{garrod08} and references therein 
(also see Table~\ref{table1}).  
\citet{garrod06} assume the value for the molecule 
measured in water ice (which makes up the largest component of the ice mantle) 
and we maintain this convention \citep[see, e.g.,][]{collings04}. 
We include the dissociative 
recombination of cations on grain surfaces with the products returned to the gas phase.  
We adopt the same branching ratios as for the equivalent gas-phase  
dissociative recombination reaction.  
Note that in the densest regions of the disk (n~$\gtrsim$~10$^{10}$~cm$^{-3}$), the assumption of 
negatively charged grains becomes invalid and explicit grain-charging reactions should 
be included.  
\citet{umebayashi09} found that neutral grains dominate the grain 
population in the midplane in the inner disk ($R$~$<$~1.5~AU).  
In this region, we are likely overestimating the recombination of cations 
on grain surfaces where n(G$^{0}$)/n(G$^{-}$) is of the order of a factor of a few.  
Because we are concerned with the chemistry occurring in the 
outer disk (R~$\gg$~1~AU), the neglect of explicit grain charging will not 
affect the discussion and conclusions presented in this work.

In addition to thermal desorption, we include photodesorption, cosmic-ray-induced desorption (via heating), 
and reactive desorption.  
For the photodesorption rates, 
we use the most recent experimental values for the photodesorption yields 
\citep{oberg09a,oberg09b}.  
In this model, we include a `coverage' factor, $\theta_s$, accounting for recent experimental 
results which suggest that photodesorption occurs from the top few monolayers only \citep{bertin12}.   
For molecules which do not have constrained photodesorption yields, $Y_i$, we 
use the early experimental value determined for water ice by \citet{westley95} 
of 3~$\times$~$10^{-3}$ molecules photon$^{-1}$ (see Table~1 and Equation~1 in WNMA12). 
The photodesorption rate for species $i$, $k^\mathrm{pd}_{i}$, is thus given by
\begin{equation*}
k^\mathrm{pd}_{i} = F_\mathrm{UV} \, Y_i \, \sigma_d \, n_d \left ( \frac{n^s_i}{n^s_\mathrm{tot}}\right) \theta_s 
\qquad \mathrm{cm}^{-3}\mathrm{s}^{-1} 
\end{equation*} 
where $F_\mathrm{UV}$ (photons~cm$^{-2}$~s$^{-1}$) is the wavelength-integrated UV photon flux, 
$\sigma_d$ (cm$^{2}$) is the dust-grain cross section, $n_d$ is the number density 
of dust grains, 
$n^s_i$ (cm$^{-3}$) is the number density 
of species $i$ on the grain, and $n^s_\mathrm{tot}$ (cm$^{-3}$) is the total number density 
of grain surface species.  
The surface coverage factor, $\theta_s$ is given by
\begin{equation*}
\theta_s =
\begin{cases}
 1   &\quad \mathrm{for} \quad M \ge 2 \\
 M/2 &\quad \mathrm{for} \quad M < 2,
\end{cases}
\end{equation*}
where $M$ is the total number of monolayers per grain. 
We include photodesorption by both external and internal UV photons, the latter 
of which are produced by the interaction of cosmic rays with \ce{H2}.  
We adopt a value for the integrated cosmic-ray-induced UV photon flux equal to 
$\sim$~10$^{4}$~photons~cm$^{-2}$~s$^{-1}$ \citep{prasad83}.  
In regions where cosmic rays are attenuated, we scale the internal 
UV photon flux by the corresponding cosmic-ray ionisation rate.  
In WMN10, we investigated the influence of X-ray desorption on molecular abundances 
in protoplanetary disks and found it a potentially very powerful mechanism for returning 
grain-surface molecules to the gas phase.  
We used the theoretical framework 
of \citet{leger85}, \citet{najita01}, and \citet{dwek96} to estimate the X-ray absorption cross sections 
and desorption rates.  
However, the interaction of X-ray photons with ice is still not well understood and indeed, recent 
experiments suggest that the picture is somewhat complicated with soft X-rays inducing 
chemistry in the ice via the production of ionic fragments \citep[see, e.g.,][]{andrade10,jimenezescobar12}.  
In general, there is a significant lack of quantitative data on X-ray induced desorption 
of astrophysical ices, hence, we choose not to include this process explicitly  
in this work and instead, we treat X-ray photodesorption as we treat UV photodesorption.  
Note that we also allow X-rays to dissociate and ionise grain mantle material 
in line with that seen in experiments (see discussion below).

For the calculation of the cosmic-ray-induced thermal desorption rates 
we use the method of \citet{hasegawa93} (see Equation~5 in WMN10).  
Here, we also include the process of reactive desorption for the first time.    
We follow the method of \citet{garrod07} and assume, for each grain-surface 
reaction which leads to a single product, a proportion of the product will be released 
into the gas phase.  
This assumes a probability that, upon reaction, a proportion of the energy released goes into 
desorbing the molecule from the grain surface.  
Investigations into the efficacy of reactive desorption in dark cloud models 
by \citet{garrod07} constrain the value for the probability of desorption to 
$P_\mathrm{rd}$~=~0.01 and we adopt this value in our work. 
Recently, \citet{vasyunin13b} suggested reactive desorption from grain surfaces 
followed by radiative association in the gas phase as a potential mechanism for 
the production of several complex molecules recently detected in dark clouds 
and prestellar cores \citep{bacmann12,cernicharo12}.  
The species detected include the methoxy radical (\ce{CH3O}), ketene (\ce{CH2CO}), 
acetaldehyde (\ce{CH3CHO}), methyl formate (\ce{HCOOCH3}), and
dimethyl ether (\ce{CH3OCH3}).  
Certainly, the detection of gas-phase complex molecules in regions
with a temperature $\lesssim$~15~K was unexpected and further 
adds to the puzzle regarding the chemical origin of COMs.  
In addition, recent experiments have investigated grain-surface chemistry induced  
by the irradiation of a single monolayer of \ce{O2} ice by a beam of
D atoms \citep{dulieu13}.  
These experiments suggest that reactive desorption is particularly efficient 
for the reformation of doubly-deuterated water (\ce{D2O}) and \ce{O2} via the 
surface reactions, {\em s-}D + {\em s-}OD and {\em s-}O + {\em s-}O, respectively.  
\citet{dulieu13} find these reactions release \ce{D2O} and \ce{O2} 
into the gas phase with efficiencies, $>$~90~\% and $\approx$~60~\%, respectively
We discuss the sensitivity of our results to the assumed 
probability for reactive desorption in Appendix A.  

Our grain-surface network is also from \citet{laas11} which 
itself is derived from \citet{garrod06} and \citet{garrod08} with the  
grain-surface reaction rates calculated according to \citet{hasegawa92}.  
We assume a density of surface sites 
equal to $\approx$~1.5~$\times$~10$^{15}$~cm$^{-2}$ 
and for the barrier between surface sites, $E_b$~$\approx$~0.3~$E_D$, 
where $E_D$ is the binding (desorption) energy to the grain surface 
of the reactant of interest.  
We discuss the sensitivity of our results to the assumed 
diffusion barrier in in Appendix A.  
For light reactants i.e., H and \ce{H2}, the diffusion rate is replaced with the 
quantum tunnelling rate assuming a barrier thickness of 1~$\AA$ (10$^{-8}$~cm). 
Whether or not there is a quantum component in the diffusion rate 
of H atoms on grain surfaces remains a controversial topic.  
Analysis of experimental work on \ce{H2} formation on bare grain 
surfaces concluded there was no quantum diffusion component 
\citep{pirronello97,pirronello99,katz99}.
A reanalysis of this experimental work determined a quantum component is necessary to 
explain the rate of formation of HD observed in the experiments \citep{cazaux04}.  
More recent experiments on H atom diffusion on amorphous solid water 
(ASW) have proved inconclusive \citep[see, e.g.,][]{watanabe10}.  
Here, since the bulk of our ice mantle is composed of water ice, 
we choose the `optimistic' case and allow quantum tunnelling for H and \ce{H2}.

The dissociation and ionisation of grain-surface species via UV photons 
(originating both externally and internally via cosmic rays) and X-rays
are new processes, not included in our previous work.  
The importance of UV processing for building chemical complexity 
in interstellar ice analogues has been known for some time 
\citep[see, e.g.,][]{allamandola88}.  
An example of a grain-surface photoreaction is 
the photodissociation of grain-surface methanol, {\em s-}\ce{CH3OH}, into 
its constituent radicals, {\em s-}\ce{CH3} and {\em s-}\ce{OH}, which are then 
available on the grain to take part in further surface-association reactions.  
Note that grain-surface (ice) species are prefixed with `{\em s-}'. 
For these reactions, we assume the rate of the equivalent gas-phase reaction.  
This is supported by recent estimates of the grain-surface photodestruction
of pure methanol ice which also show that photochemistry occurs deep within the 
bulk ice \citep{oberg09c}.
The various reaction channels possible are estimated by assuming a 
grain-surface molecule will likely dissociate into its functional group components 
(as demonstrated above for methanol), i.e., allowing no rearrangement of the constituent atoms, 
and allowing reactions involving destruction of the weaker bonds to have 
higher branching ratios \citep{garrod08}.  
For each ionisation event, we assume that the cation 
dissociatively recombines with the electron, 
and the excess energy in the products lost via translation energy on the 
grain mantle, i.e., the products remain on the grain surface \citep[][]{garrod08}. 
We adopt the same branching ratios for the equivalent gas-phase dissociative 
recombination reaction.
For X-ray induced dissociation and ionisation of grain-surface species, we 
follow the same formulation as that adopted for the gas-phase X-ray reactions: 
we duplicate the set of cosmic-ray-induced photoreactions and scale the reaction 
rates by the X-ray ionisation rate calculated in the disk.  
In line with experiments, we allow X-ray and UV photodissociation and ionisation 
to occur throughout the ice mantle \citep[see, e.g.][]{oberg09c,andrade10,jimenezescobar12}.

In addition to cosmic-ray-induced photoreactions, the direct impact of 
cosmic ray particles, which, in the Galaxy, consist predominantly of protons 
and stripped nuclei  \citep[$\approx$~98\%, see, e.g.][]{simpson83} can penetrate 
the dust grain and induce cascades of up to 10$^{2}$ suprathermal atoms along the impact track.  
These atoms, in turn, ionise the molecular material within the ice mantle also creating 
energetic electrons ($\sim$~keV) which also dissociate ionise the ice mantle material 
\citep[see, e.g.,][]{kaiser97,bennett07,bennett11}.  
Here, we simulate this process in the ice by also adopting the direct 
cosmic-ray ionisation rates for the equivalent gas-phase reaction.

Our complete chemical network has $\approx$~9300 reactions involving $\approx$~800 
species and is one of the most complex chemical models of a protoplanetary disk 
constructed to date. 

\subsection{Molecular Line Emission}
\label{molecularlineemission}

In order to compare our model results with current observations
and select potential molecules and line transitions which may 
be observable in protoplanetary disks, we have
calculated the molecular line emission from the disk to determine
rotational line transition intensities for molecules of interest in this work. 

For simplicity, we assume the disk is `face-on', i.e., has an inclination of 
0\textdegree, and that local thermodynamic equilibrium (LTE) holds throughout.  
The former assumption allows us to quickly and efficiently calculate the 
line emission without worrying about geometrical effects 
due to disk inclination, whilst the latter assumption makes the calculation 
more computationally tractable and
allows us to compute an entire spectrum of a particular molecule in a single 
calculation.  
This is important for COMs, in particular, which typically have 
many energy levels and transitions and thus, relatively complex spectra.  
In order to perform a non-LTE calculation, we require accurate collisional rate 
coefficients which are only available for a handful of molecules 
(see, e.g., the Leiden Atomic and Molecular Database or LAMDA\footnotemark[5]).  
For molecules considered in this work, collisional data 
for \ce{H2CO}, \ce{CH3OH}, \ce{HC3N} and \ce{CH3CN} only are available.
We expect the disk conditions to depart from LTE mainly in the outer, colder,  
more diffuse regions of the disk.  
However, we also only expect our 
LTE disk-integrated line intensities to deviate from those calculated assuming 
non-LTE conditions by no more than a factor of a few \citep[see, e.g.,][]{pavlyuchenkov07}.  
Since this work is exploratory in nature, the line intensities we calculate are 
still very useful for determining which transitions in which molecules 
may be detectable using ALMA.  

\footnotetext[5]{\url{http://home.strw.leidenuniv.nl/}$\sim$\url{moldata/}}

There are numerous caveats when assuming LTE.  
The intense background thermal radiation 
in warm regions of the disk may radiatively pump particular transitions 
leading to weakly (or possibly, strongly) masing lines 
($T_\mathrm{rot}$~$>$~$T_{k}$).  
Formaldehyde (\ce{H2CO}) and methanol (\ce{CH3OH}) masers have been 
observed in the local interstellar 
medium and are commonly associated with massive star forming regions
\citep[see, e.g.,][]{elitzur92}.  
Indeed, there have been suggestions that methanol masers trace the 
protoplanetary disk material around massive embedded protostars 
\citep[see, e.g.,][]{norris98}.  
Certainly, the potential for observable maser emission from formaldehyde 
and methanol in disks 
around low-mass stars should be explored in future work.

The disk-integrated line flux density, $F_\nu$, is determined by integrating 
the solution of the radiative transfer equation 
in the vertical direction and summing over radial sections of the disk, i.e.,
\begin{equation}
F_\nu = \frac{1}{4\pi\,D^2}\int_{r_\mathrm{min}}^{r_\mathrm{max}}  
\int_{-z_\mathrm{max}(r)}^{+z_\mathrm{max}(r)} 2\pi \, r \, \bar{\eta}_\nu(r,z)\, \mathrm{d}z\,\mathrm{d}r
\end{equation}
where $D$ is the distance to source and 
$\bar{\eta}_\nu(r,z)$ is the emissivity at a grid point $(r,z)$ 
times the absorption in the upper disk, i.e.,
\begin{equation}
\bar{\eta}_\nu(r,z) = n_\mathrm{u}(r,z)\, A_\mathrm{ul} \, \phi_\nu \frac{h\nu}{4\pi}  
\exp{\left[ -\tau_\nu(r,z)\right ]}.
\end{equation}
Here, $n_\mathrm{u}$ is the abundance in the upper energy level of the transition, 
$A_\mathrm{ul}$ is the Einstein coefficient for spontaneous emission 
from the upper level, $u$, to the lower level, $l$, 
$\phi_\nu$ is the value of the line profile function 
at the frequency, $\nu$ (assumed to be Gaussian in shape),  
and $h$ is Planck's constant.   
The optical depth, $\tau_\nu(r,z)$, is 
\begin{equation}
\tau_\nu(r,z) = \int_{z}^{z_\mathrm{max}} \chi_\nu(r,z')\, \mathrm{d}z',
\end{equation}
where the absorption coefficient, $\chi_\nu$, is given by 
\begin{equation}
\chi_\nu = \rho\kappa_\nu + \left( n_\mathrm{l}B_\mathrm{lu} - n_\mathrm{u}B_\mathrm{ul} \right) \, 
\phi_\nu \frac{h\nu}{4\pi}.
\end{equation}
Here, $\rho$ is the dust mass density (g~cm$^{-3}$), 
$\kappa_\nu$ is the dust mass absorption coefficient (cm$^2$~g$^{-1}$), 
$n_\mathrm{l}$ is the abundance in the lower 
energy level of the transition and $B_\mathrm{lu}$ and $B_\mathrm{ul}$ 
are the Einstein coefficients for absorption and stimulated emission, 
respectively.  

Since we assume LTE holds throughout, our level populations are given
by the Boltzmann distribution, i.e., the population of an energy level, $i$, is 
determined using
\begin{equation}
\frac{n_i}{n} = \frac{g_i \exp{ \left( -E_{i}/T\right)}}{Z(T)}
\end{equation}
where $n_i$ is the number density in level $i$, $n$ is the total number density 
of the molecule ($\sum_i n_i$), $g_i$ is the degeneracy of 
the level, $E_i$ is the energy (in units of K) and 
$T$ is the gas temperature.  
We explicitly calculate the rotational partition function, $Z_\mathrm{rot}(T)$, 
by summing over 
populated energy levels in colder regions, i.e., 
$Z_\mathrm{rot}(T) = \sum_i g_i \exp{\left(-E_i/T\right)}$,  
and swapping to the high temperature approximation once the higher energy 
levels become sufficiently populated, that is, 
$Z_\mathrm{rot}(T) \approx \sqrt{\pi}(kT)^{3/2}/(\sigma \sqrt{ABC}) $, 
where, $k$ is Boltzmann's constant, $\sigma$ is the symmetry factor, i.e., 
the number of indistinguishable rotational orientations of the molecule, and 
$A$, $B$, and $C$ are the rotational constants in energy units.  
In the inner regions of the disk, the gas temperature is sufficiently 
high that molecules can become vibrationally excited.  
COMs, in particular, can be vibrationally excited at 
relatively moderate temperatures, $\approx$~300~K; hence, we also include 
the vibrational partition function in our determination of the total partition function, 
$Z(T) = Z_\mathrm{vib}(T)\times Z_\mathrm{rot}(T)$, where
\begin{equation}
Z_\mathrm{vib}(T) = \prod_i \frac{1}{1 - \exp{\left(-h\nu_i/kT\right)}}
\end{equation}
with $\nu_i$ representing the set of characteristic vibrational frequencies for 
each molecule.  
Typically, $i$~$\gg$~1 for complex molecules, e.g., methanol has 12 characteristic frequencies 
of vibration.  

We use the molecular rotational line lists from either the 
Cologne Database for Molecular Spectroscopy (CDMS, \citeauthor{muller05} \citeyear{muller05}\footnotemark[6])
or the Jet Propulsion Laboratory (JPL) molecular spectroscopic database 
(\citeauthor{pickett98} \citeyear{pickett98}\footnotemark[7]) and 
molecular constants (rotational constants and vibrational frequencies) 
from CDMS, JPL, and 
the NIST database\footnotemark[8]$^{,}$\footnotemark[9].  
For line lists from CDMS, we use internally generated Einstein A coefficients provided by the database.  
For JPL, we convert the listed line intensities, $S_{\nu}$, to transition probabilities 
using the partition function provided by the database at the reference temperature, 300~K  
\citep[see equation 9 in][]{pickett98}.  
We have benchmarked our partition functions with those provided by CDMS and JPL.  
Our results for \ce{H2CO} and \ce{CH3OH} at low temperatures ($T$~$\leq$~37.5~K) 
agree with those provided by the databases 
to $>$~3 significant figures.  
At higher temperatures, our results agree to within $\approx$~15\% 
of the database values over the temperature range applicable to our disk model  
($T$~$\lesssim$~150~K, see Fig.~\ref{figure1}).  

\footnotetext[6]{\url{http://www.astro.uni-koln.de/cdms/}}
\footnotetext[7]{\url{http://spec.jpl.nasa.gov/}}
\footnotetext[8]{\url{http://webbook.nist.gov/chemistry}}
\footnotetext[9]{\url{http://cccbdb.nist.gov}}

In our calculations, we assume a distance to source, $D$~=~140~pc, 
the distance to the Taurus molecular cloud complex 
where many well-studied protoplanetary disks are located. 
To convert line flux densities to sources at other distances, one 
must simply scale the values by the square of the ratio of the distances, e.g., for a source 
at 400~pc, for example, the Orion molecular cloud, the line flux density is reduced by a factor, 
(140/400)$^2$ = 0.123, whilst for a source at 56~pc, for example, TW~Hya, the line flux density 
is enhanced by a 
factor, (140/56)$^{2}$~=~6.25.     

\section{Results}
\label{results}
We determine the chemical structure of the disk by calculating the time-dependent chemical 
evolution at each grid point in our model.  
Firstly, we investigate the influence of different chemical processes 
by running a reduced grid at a fixed radius in the outer disk (R~=~305~AU).  
Secondly, we present results from our full disk model,  
i.e., by running the entire grid ($\approx$~10,000 points) as a series of single-point models.  
We calculate the chemical structure between a radius of $\approx$~1~AU and $\approx$~305~AU.  
We concentrate our results and discussions on the outer cold disk ($T$~$\lesssim$~100~K) where 
sufficient freezeout allows grain-surface synthesis to occur. 
We map the chemical structure of the disk at a time of 10$^6$~yr, 
the typical age of classical T Tauri stars.  

\subsection{Initial Abundances}
\label{initialabundances}

Our initial abundances are extracted from the results of a 
simple time-dependent dark cloud model with constant physical 
conditions ($n$~=~10$^{5}$~cm$^{-3}$, $T$~=~10~K, $A_v$~=~10~mag) at a time of 10$^5$~yr. 
We use the same chemical network for the generation of the initial abundances
as in our full disk model.  
Our initial elemental abundance ratios for 
H:He:O:C:N:S:Na:Mg:Si:Cl:Fe are
1:9.75(-2):1.76(-4):7.3(-5):2.14(-5):2(-8):3(-9):3(-9):3(-9):3(-9):3(-9) 
where $a$($b$) represents $a$~$\times$~10$^b$ \citep{graedel82}. 

The initial abundances for a selection of gas-phase and grain-surface (ice) 
molecules are shown in Table~\ref{table1} along with their corresponding 
grain-surface binding energies. 
The species are ordered by mass and we list the initial fractional abundances and 
binding energies of the complex 
molecules of interest in this work separately.  

The calculations begin with appreciable 
fractional abundances (relative to total H nuclei number density, 
$n_\mathrm{H}$~=~$n$(H)~+~2$n$(\ce{H2})) of relatively simple ices, such as, 
{\em s-}\ce{H2O} ($\approx$~1~$\times$~10$^{-4}$),
{\em s-}\ce{CO}  ($\approx$~3~$\times$~10$^{-5}$), 
{\em s-}\ce{CH4} ($\approx$~3~$\times$~10$^{-5}$), 
{\em s-}\ce{NH3} ($\approx$~1~$\times$~10$^{-5}$), and 
{\em s-}\ce{N2}  ($\approx$~4~$\times$~10$^{-6}$). 
Formaldehyde ({\em s-}\ce{H2CO}), methylamine ({\em s-}\ce{CH3NH2}), methanol ({\em s-}\ce{CH3OH}), 
formamide ({\em s-}\ce{NH2CHO}), and ethane 
({\em s-}\ce{C2H6}) also achieve relatively high fractional abundances on the grains 
($\gtrsim$~10$^{-8}$).   
These species are efficiently formed via atom-addition reactions at 10~K.  
{\em s-}\ce{C3H4} also reaches an appreciable fractional abundance ($\approx$~2~$\times$~10$^{-6}$).  
In this network we do not distinguish between isomers of \ce{C3H4}.   
Because propyne has the lower zero-point energy and 
also possesses a rotational spectrum, we choose to treat \ce{C3H4} 
as propyne (\ce{CH3CCH}) as opposed to allene (\ce{CH2CCH2}). 

Under dark cloud conditions, methanol and formaldehyde are formed on the grain 
via the sequential hydrogenation of CO ice,
\begin{equation*}
s\mbox{-}\ce{CO}   \xrightarrow{s\mbox{-}\ce{H}}
s\mbox{-}\ce{HCO}  \xrightarrow{s\mbox{-}\ce{H}}
s\mbox{-}\ce{H2CO} \xrightarrow{s\mbox{-}\ce{H}}
\genfrac{}{}{0pt}{}{s\mbox{-}\ce{CH3O}}{s\mbox{-}\ce{CH2OH}} \xrightarrow{s\mbox{-}\ce{H}}
s\mbox{-}\ce{CH3OH}.
\end{equation*}
Methylamine is formed via the hydrogenation of {\em s-}\ce{CH2NH}, 
\begin{equation*}  
s\mbox{-}\ce{CH2NH}\xrightarrow{s\mbox{-}\ce{H}}
\genfrac{}{}{0pt}{}{s\mbox{-}\ce{CH2NH2}}{s\mbox{-}\ce{CH3NH}}\xrightarrow{s\mbox{-}\ce{H}}
s\mbox{-}\ce{CH3NH2}
\end{equation*}
{\em s-}\ce{CH2NH} has multiple formation pathways originating from 
atom addition to small hydrocarbon radicals, 
\begin{equation*} 
s\mbox{-}\ce{CH2} \xrightarrow{s\mbox{-}\ce{N}}~s\mbox{-}\ce{H2CN}~\xrightarrow{s\mbox{-}\ce{H}} s\mbox{-}\ce{CH2NH} \\
\end{equation*}
and
\begin{equation*}
s\mbox{-}\ce{CH3} \xrightarrow{s\mbox{-}\ce{N}} s\mbox{-}\ce{CH2NH} .
\end{equation*}  
Propyne, {\em s-}\ce{CH3CCH}, forms via successive hydrogenation 
of {\em s-}\ce{C3} and {\em s-}\ce{C3H}, both of which form readily in the 
gas phase and subsequently freeze out onto the dust grains. 
Grain-surface formamide originates from {\em s-}OCN e.g., 
\begin{equation*}
s\mbox{-}\ce{OCN}\xrightarrow{s\mbox{-}\ce{H}} s\mbox{-}\ce{HNCO}\xrightarrow{s\mbox{-}\ce{H}}
s\mbox{-}\ce{NHCHO}\xrightarrow{s\mbox{-}\ce{H}} s\mbox{-}\ce{NH2CHO}. 
\end{equation*} 
{\em s-}OCN can form either on the grain via the reactions 
\begin{equation*}
s\mbox{-}\ce{CN} \xrightarrow{s\mbox{-}\ce{O}} s\mbox{-}\ce{OCN}
\end{equation*}
and
\begin{equation*}
s\mbox{-}\ce{NO} \xrightarrow{s\mbox{-}\ce{C}} s\mbox{-}\ce{OCN}, 
\end{equation*}
or in the gas phase via atom-radical and radical-radical reactions, 
such as, 
\begin{equation*}
\ce{N}  + \ce{HCO} \longrightarrow \ce{OCN} + \ce{H} 
\end{equation*}
and
\begin{equation*}
\ce{CN} + \ce{OH} \longrightarrow \ce{OCN} + \ce{H} 
\end{equation*}
whence it can freeze out onto dust grains.  
Ethane, {\em s-}\ce{C2H6}, forms on the grain 
in a similar manner to propyne, via the sequential hydrogenation 
of {\em s-}\ce{C2} and {\em s-}\ce{C2H}.   
Note that radicals, such as, \ce{CH3O} and \ce{CH2OH}, can also 
be formed via the photodissociation of larger species, in this case,  
\ce{CH3OH}, as well as on the grain and via ion-molecule chemistry.  

\begin{table}
\caption{Initial fractional abundances (with respect to $n_\mathrm{H}$)
and molecular binding (desorption) energies, $E_\mathrm{D}$ (K).}
\centering
\begin{tabular}{lcccc}
\hline\hline
Species       & Gas      & Ice      & $E_\mathrm{D}$ & References\\ 
\hline
\ce{CH4}   & 6.5(-08) & 2.5(-05) & 1300 & 1 \\
\ce{NH3}   & 8.1(-09) & 1.2(-05) & 5530 & 2 \\
\ce{OH}    & 9.0(-09) & 1.2(-15) & 2850 & 1 \\
\ce{H2O}   & 1.5(-08) & 1.3(-04) & 5700 & 3 \\
\ce{CN}    & 7.0(-10) & $\cdots$ & 1600 & 2 \\ 
\ce{HCN}   & 9.6(-10) & 1.5(-07) & 2050 & 1 \\ 
\ce{CO}    & 3.2(-06) & 3.1(-05) & 1150 & 2 \\ 
\ce{N2}    & 6.4(-07) & 3.5(-06) & 1000 & 2 \\ 
\ce{HCO}   & 3.3(-10) & 1.7(-13) & 1600 & 2 \\ 
\ce{HCO+}  & 6.6(-10) & $\cdots$ & $\cdots$ & $\cdots$ \\ 
\ce{N2H+}  & 2.6(-10) & $\cdots$ & $\cdots$ & $\cdots$ \\ 
\ce{C2H6}  & 1.5(-10) & 7.8(-08) & 4390 & 2 \\
\ce{NO}    & 1.3(-08) & 6.1(-14) & 1600 & 2 \\ 
\ce{HNO}   & 1.9(-09) & 2.0(-09) & 2050 & 2 \\
\ce{O2}    & 4.0(-07) & 5.6(-09) & 1000 & 3 \\
\ce{H2O2}  & 8.2(-10) & 1.8(-08) & 5700 & 2 \\
\ce{H2S}   & 3.0(-09) & 7.0(-09) & 2740 & 1 \\
\ce{C3}    & 4.3(-08) & $\cdots$ & 2400 & 2 \\
\ce{C3H2}  & 7.3(-11) & 3.1(-09) & 3390 & 2 \\
\ce{OCN}   & 5.2(-09) & 2.3(-16) & 2400 & 2 \\
\ce{CO2}   & 4.4(-08) & 5.1(-07) & 2580 & 1 \\
\ce{CS}    & 5.6(-11) & 1.5(-13) & 1900 & 2 \\
\ce{SiO}   & 5.0(-11) & 3.9(-10) & 3500 & 2 \\
\ce{NO2}   & 7.6(-11) & 1.8(-10) & 2400 & 2 \\
\ce{O3}    & 7.9(-08) & 4.2(-10) & 1800 & 2 \\
\ce{SO}    & 6.9(-10) & 2.1(-12) & 2600 & 1 \\
\hline
\multicolumn{5}{c}{Complex Species}    \\
\hline
\ce{H2CO}     & 2.2(-09) & 3.3(-06) & 2050 & 2 \\        
\ce{CH3NH2}   & 1.1(-11) & 8.3(-07) & 5130 & 2 \\        
\ce{CH3OH}    & 2.2(-10) & 8.3(-07) & 5530 & 1 \\  
\ce{CH3CCH}   & 4.8(-10) & 2.3(-06) & 4290 & 2 \\ 
\ce{CH3CN}    & 8.7(-13) & 3.7(-09) & 4680 & 1 \\
\ce{CH3CHO}   & 7.0(-12) & 3.0(-11) & 2780 & 2 \\      
\ce{NH2CHO}   & 1.6(-12) & 2.5(-08) & 5560 & 2 \\    
\ce{CH3OCH3}  & 5.1(-13) & 1.4(-16) & 3680 & 2 \\   
\ce{C2H5OH}   & 1.4(-14) & 8.0(-14) & 6260 & 2 \\   
\ce{HCOOH}    & 1.7(-12) & 1.5(-11) & 5570 & 1 \\   
\ce{HC3N}     & 2.7(-11) & 5.5(-12) & 4580 & 2 \\
\ce{CH3COCH3} & $\cdots$ & $\cdots$ & 3500 & 2 \\       
\ce{CH3COOH}  & $\cdots$ & $\cdots$ & 6300 & 2 \\    
\ce{HCOOCH3}  & 1.0(-15) & 1.0(-15) & 4100 & 2 \\                                                 
\ce{HOCH2CHO} & $\cdots$ & $\cdots$ & 6680 & 2 \\   
\hline
\end{tabular}
\label{table1}
\tablefoot{$a(b)$ represents $a\times10^b$ and
we restrict listed fractional abundances to those $\gtrsim$~10$^{-16}$.}
\tablebib{
(1) \citet{collings04}; 
(2) \citet{garrod06}; 
(3) \citet{cuppen07}. }
\end{table}

\subsection{Vertical abundance profiles at R~=~305~AU}
\label{verticalresults}

\begin{figure*}
\includegraphics[width=0.5\textwidth]{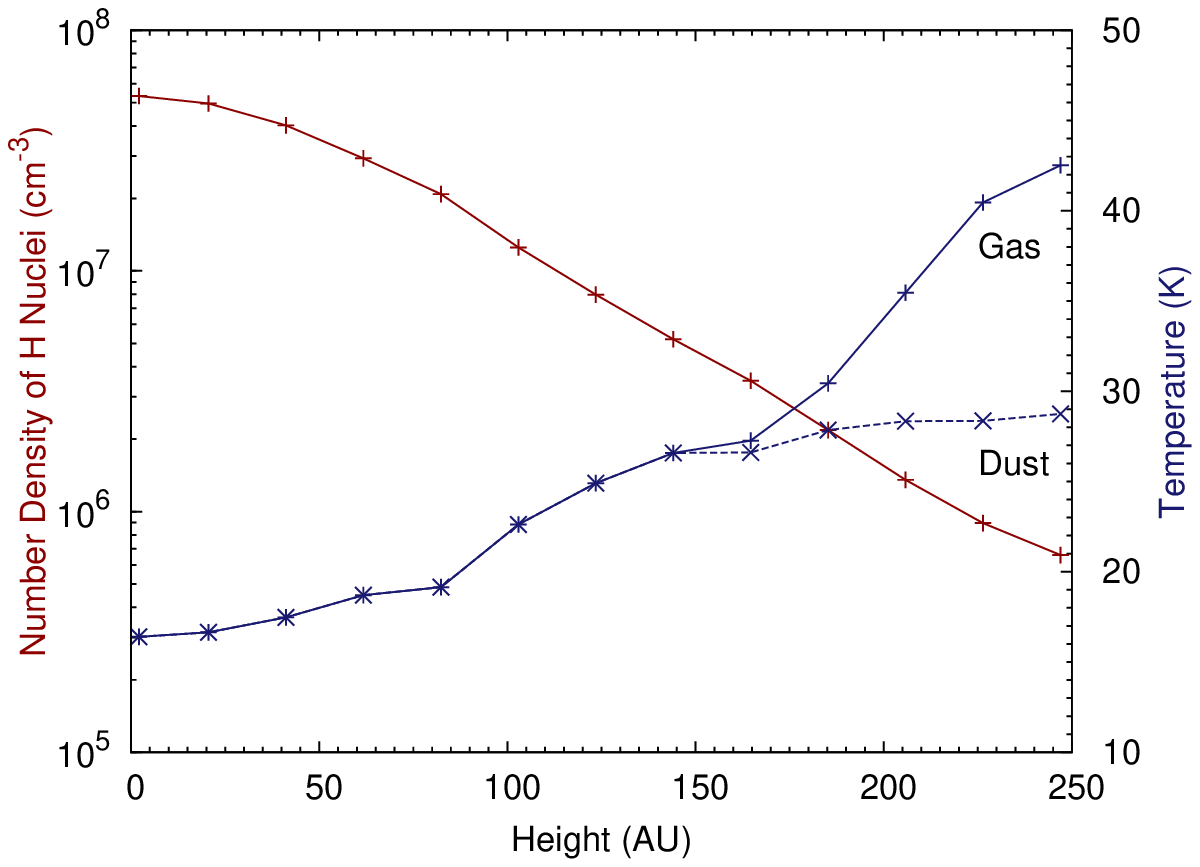}
\includegraphics[width=0.5\textwidth]{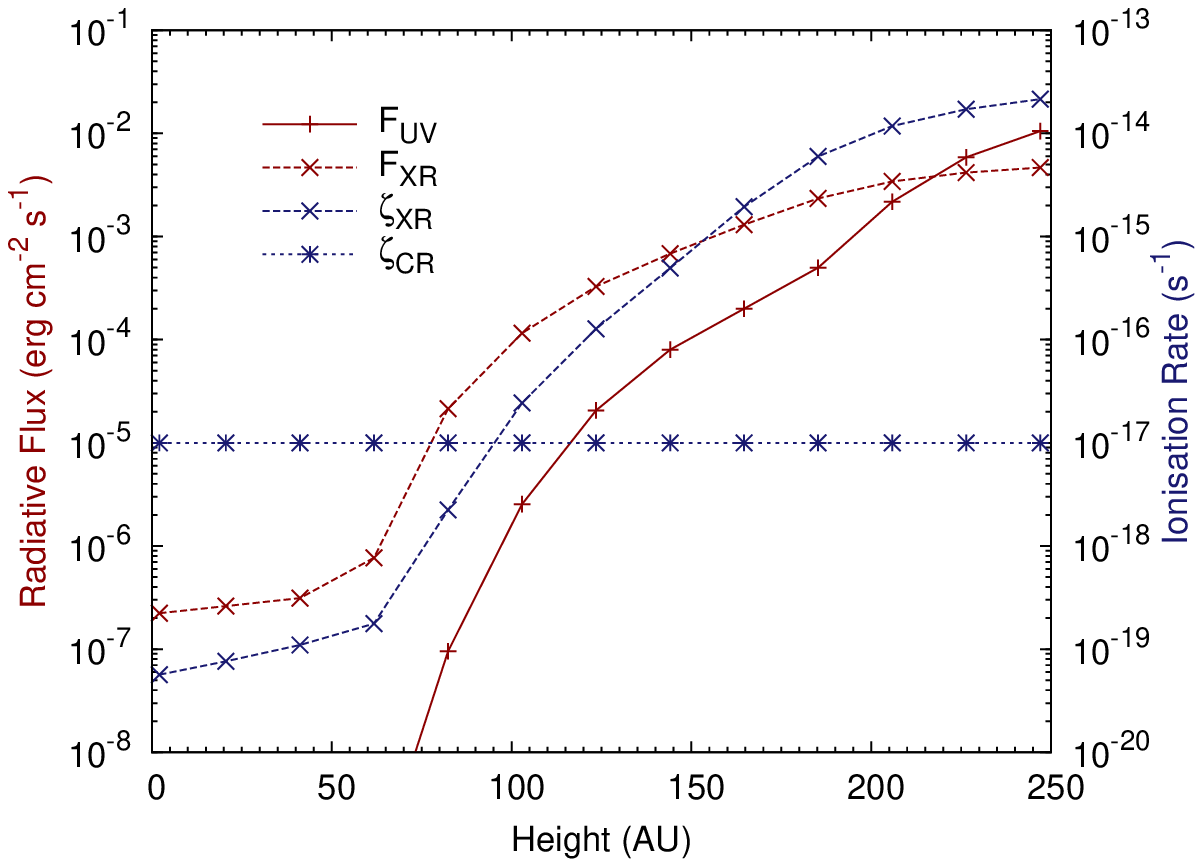}
\caption{Physical conditions as a function of disk height 
at a radius, $R$~=~305~AU.  
Number density of H nuclei (cm$^{-3}$) and gas and dust temperature (K) 
are shown in the left-hand panel (solid red lines, solid blue lines, and dashed blue lines, respectively).
UV and X-ray radiative fluxes (erg~cm$^{-2}$~s$^{-1}$) 
and X-ray and cosmic-ray ionisation rates (s$^{-1}$) 
are shown in the right-hand panel 
(solid red lines, dashed red lines, dashed blue lines, and dotted blue lines, respectively).}
\label{figure2}
\end{figure*}

Here, we present results from a series of 
reduced grids to investigate the 
particular chemical processes responsible for the production and destruction 
of COMs in the outer disk.  
We calculated the chemical evolution over a 13-point grid in a 
single vertical slice of the disk at a fixed radius of 305~AU. 
The physical conditions in this slice are presented in Fig.~\ref{figure2}. 
The H nuclei number density decreases from a maximum 
value of $\approx$~5~$\times$~10$^{7}$~cm$^{-3}$ 
in the disk midplane to $\approx$~6~$\times$~10$^{5}$~cm$^{-3}$ at the surface 
and the gas temperature increases from a minimum of $\approx$~16~K in the 
midplane to $\approx$~42~K at the disk surface.  
The gas and dust temperatures at the disk surface decouple 
above a height of $\approx$~150~AU such that 
the dust temperature on the disk surface reaches a 
maximum value of $\approx$~28~K.
The disk midplane is heavily 
shielded from both UV and X-ray photons; however, cosmic 
rays are able to penetrate the entire disk.  
As expected, the disk surface is heavily irradiated with the UV and X-ray 
fluxes reaching a value $\sim$~10$^{-2}$~erg~cm$^{-2}$~s$^{-1}$, 
an order of magnitude stronger than the integrated interstellar UV flux 
(1.6~$\times$~10$^{-3}$~erg~cm$^{-2}$~s$^{-1}$).  

We present results from five different models which increase incrementally
in complexity from Model 1 through to Model 5. 
Model 1 is the most simple and includes gas-phase chemistry, 
freezeout onto dust grains, and thermal desorption.  
In Model 2, we add cosmic-ray-induced thermal desorption and 
photodesorption by internal and external UV photons and X-ray photons.
In Model 3, we include grain-surface chemistry and in Model 4, we also 
add the cosmic-ray, X-ray, and UV photoprocessing of 
ice mantle material.  
The most complex model, Model 5, also includes reactive desorption 
(see Sect.~\ref{chemicalmodel}).  

\begin{figure*}[!ht]
\subfigure{\includegraphics[width=0.33\textwidth]{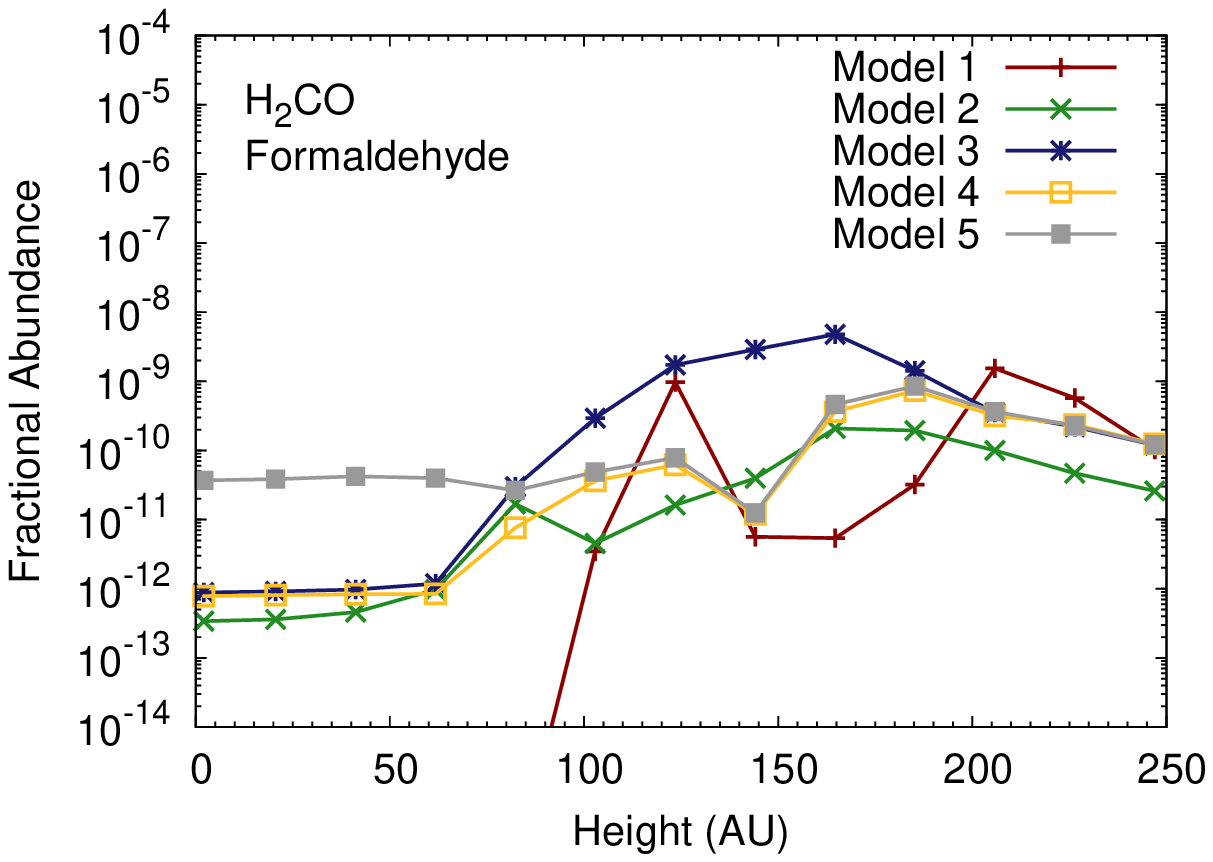}}    
\subfigure{\includegraphics[width=0.33\textwidth]{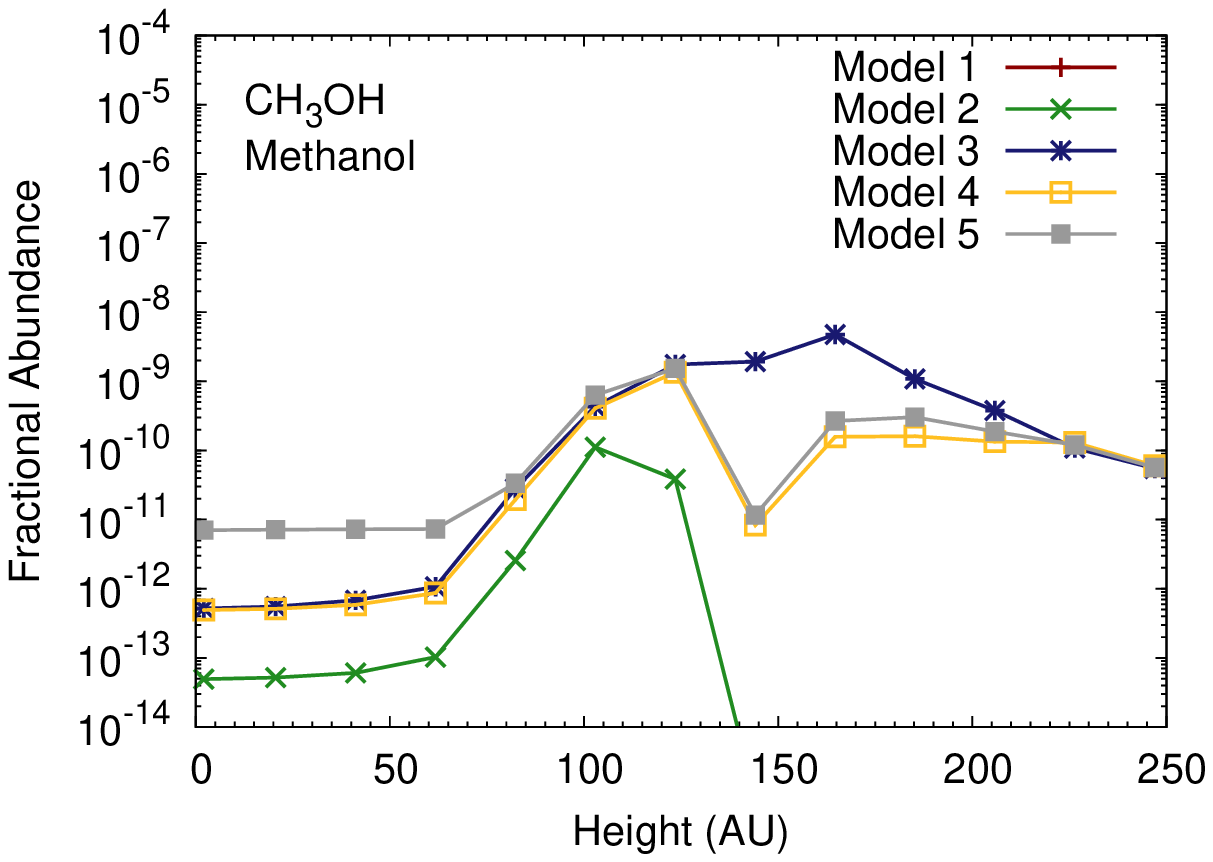}} 
\subfigure{\includegraphics[width=0.33\textwidth]{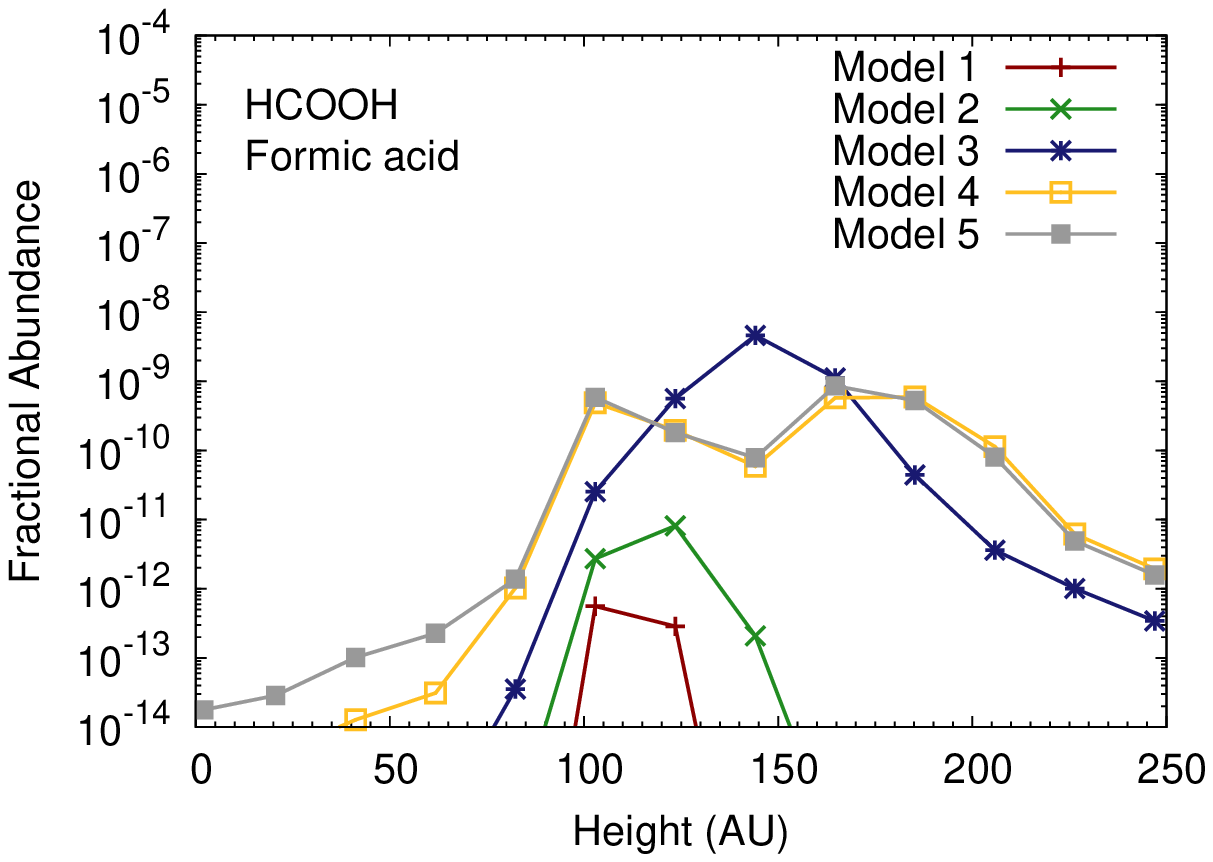}}   
\subfigure{\includegraphics[width=0.33\textwidth]{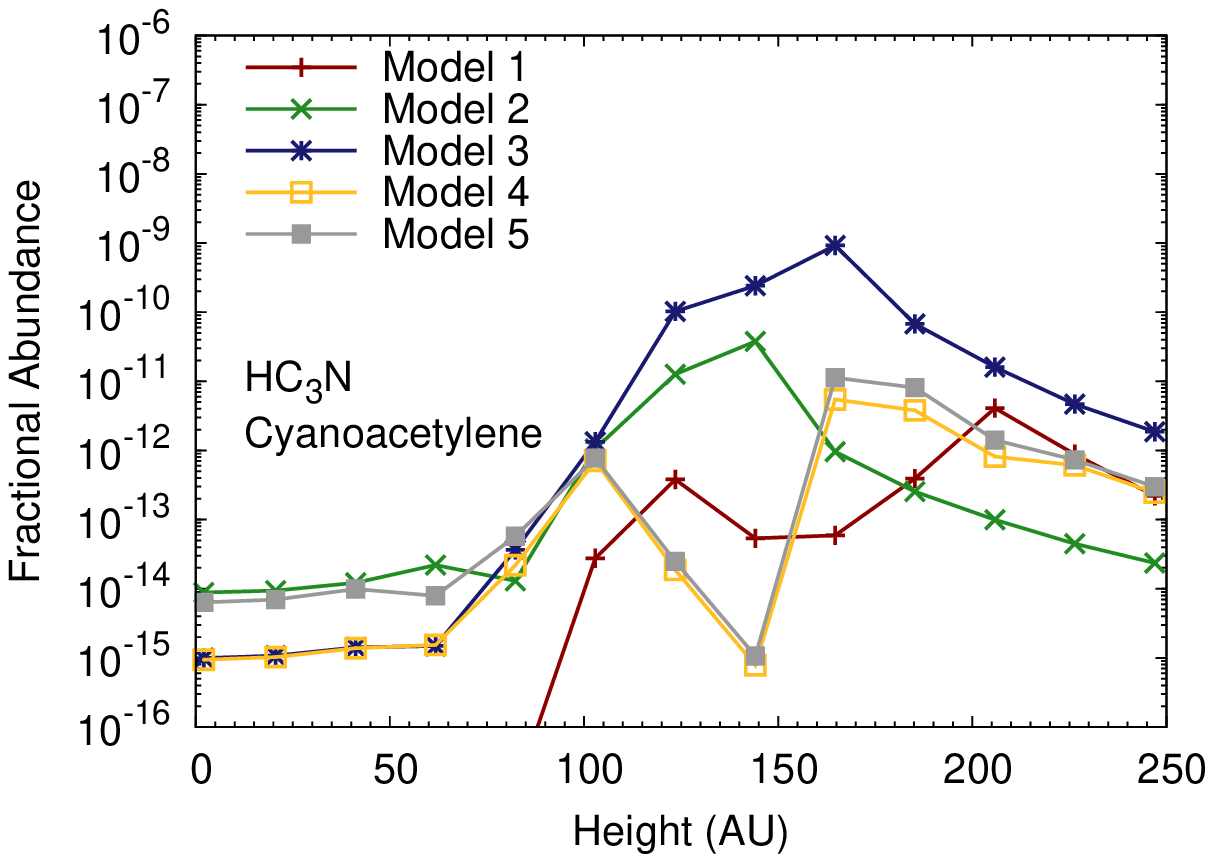}}    
\subfigure{\includegraphics[width=0.33\textwidth]{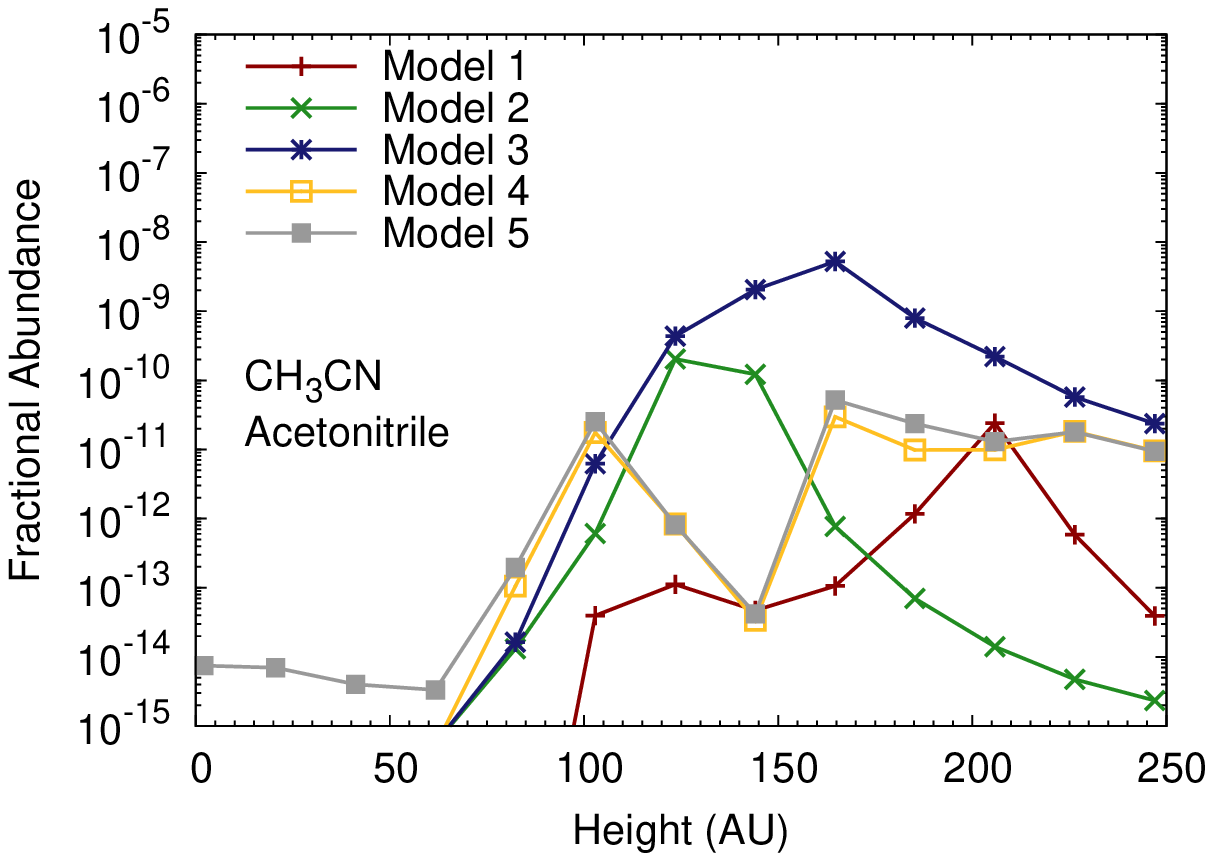}}   
\subfigure{\includegraphics[width=0.33\textwidth]{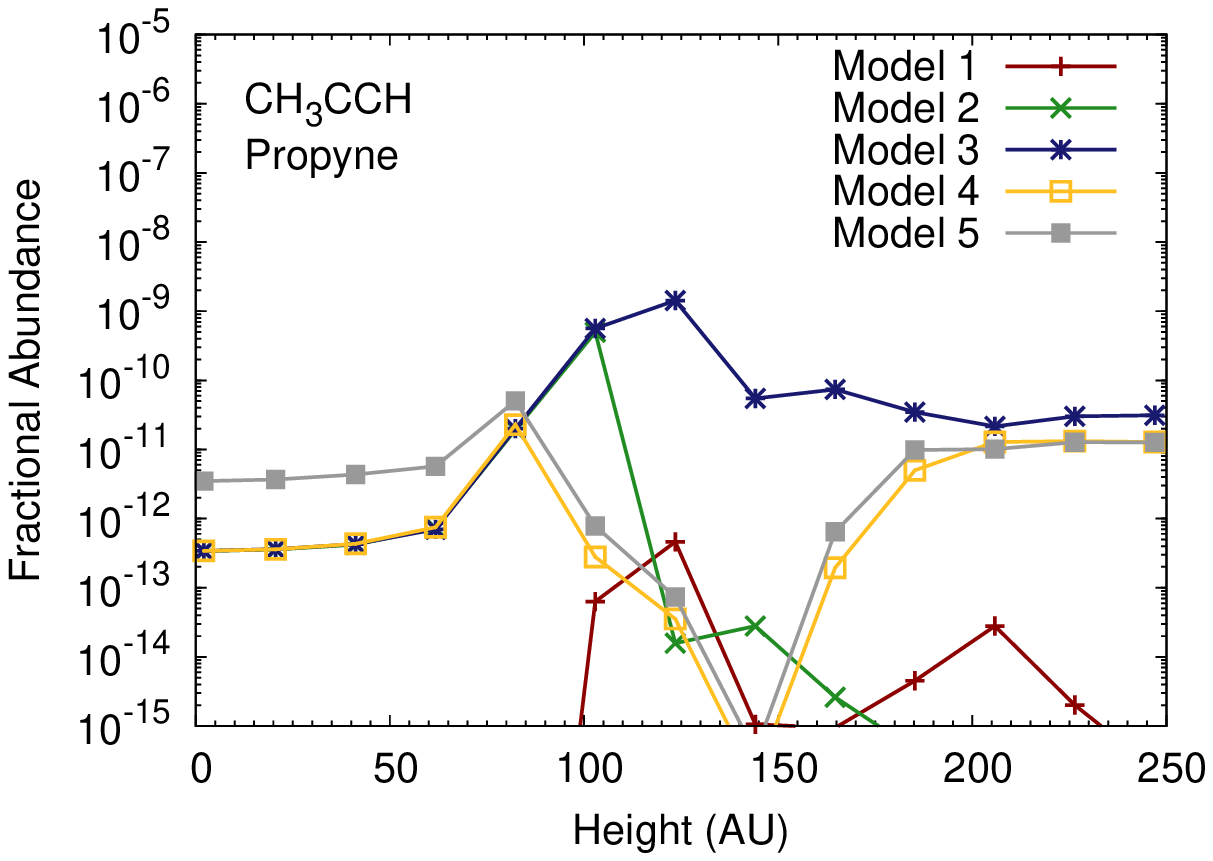}}    
\subfigure{\includegraphics[width=0.33\textwidth]{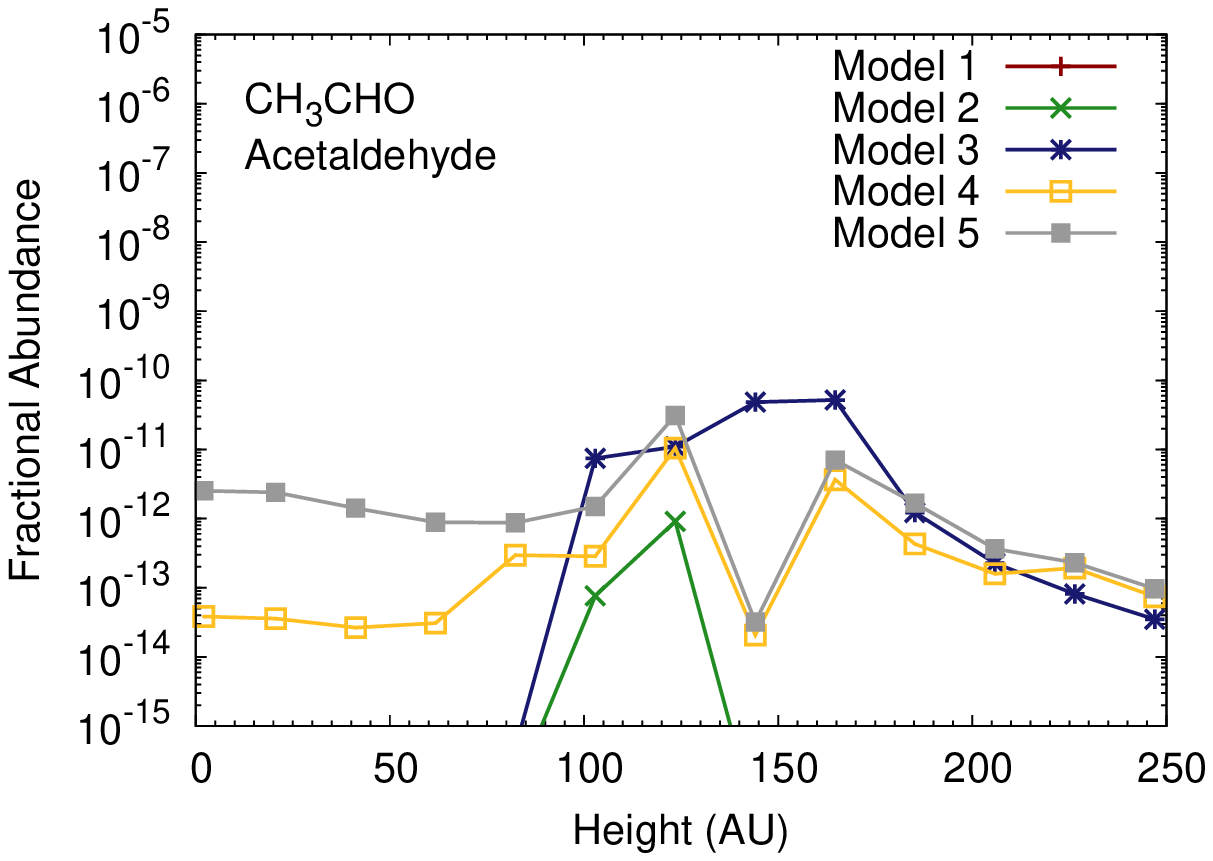}}  
\subfigure{\includegraphics[width=0.33\textwidth]{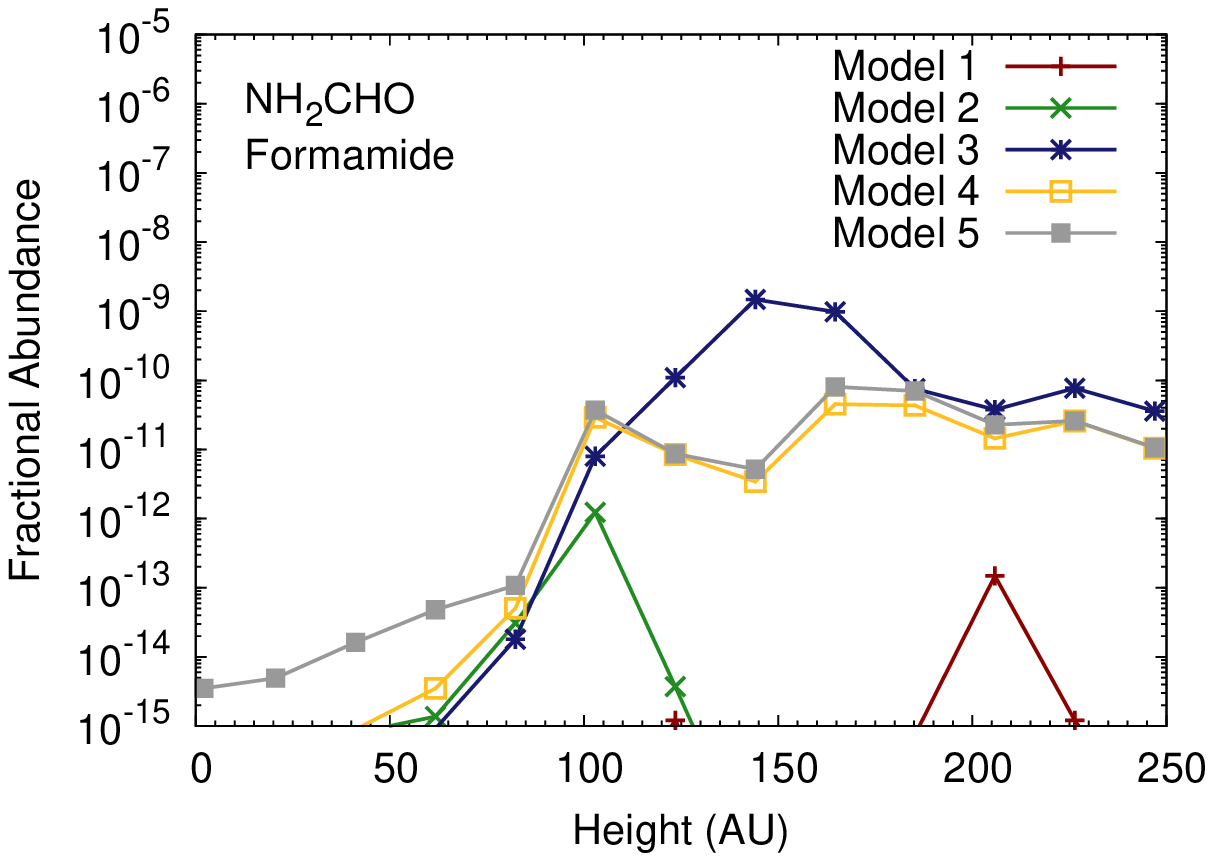}}  
\subfigure{\includegraphics[width=0.33\textwidth]{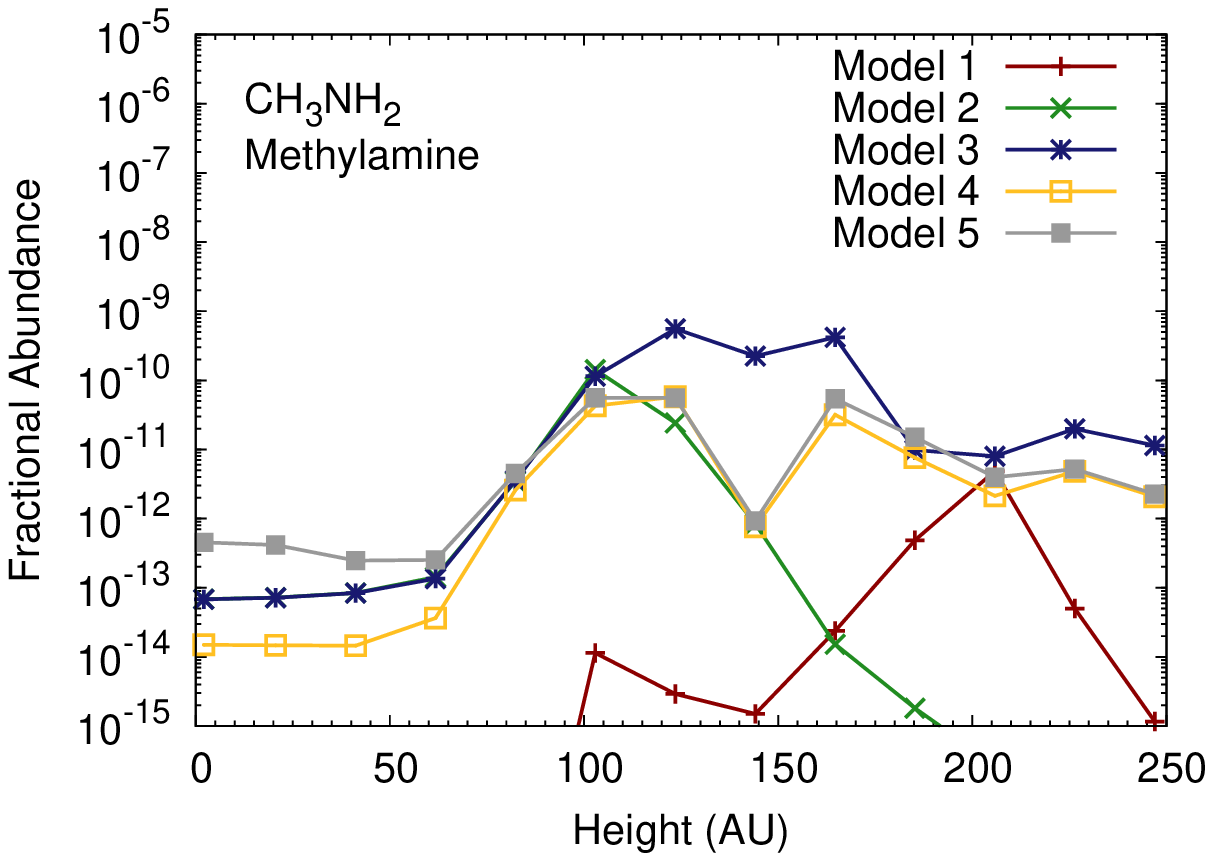}}  
\subfigure{\includegraphics[width=0.33\textwidth]{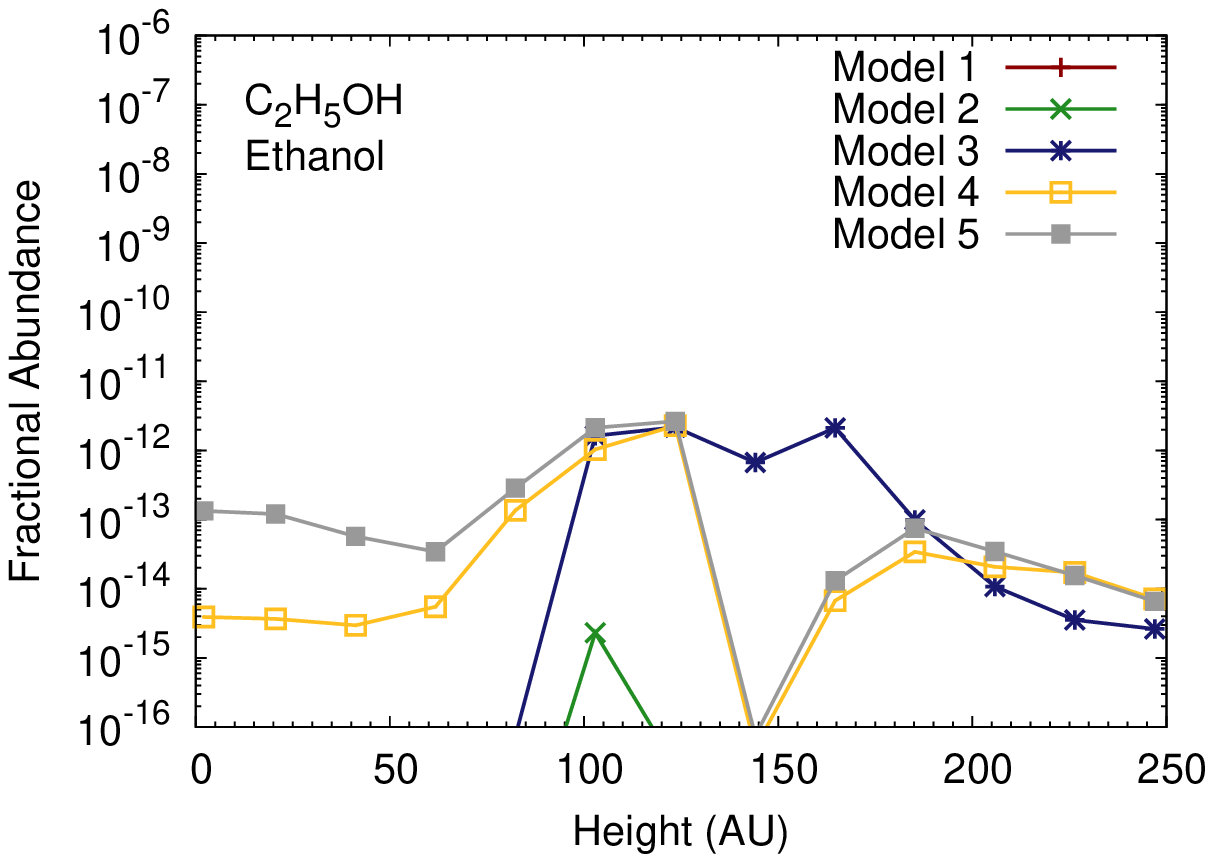}}  
\subfigure{\includegraphics[width=0.33\textwidth]{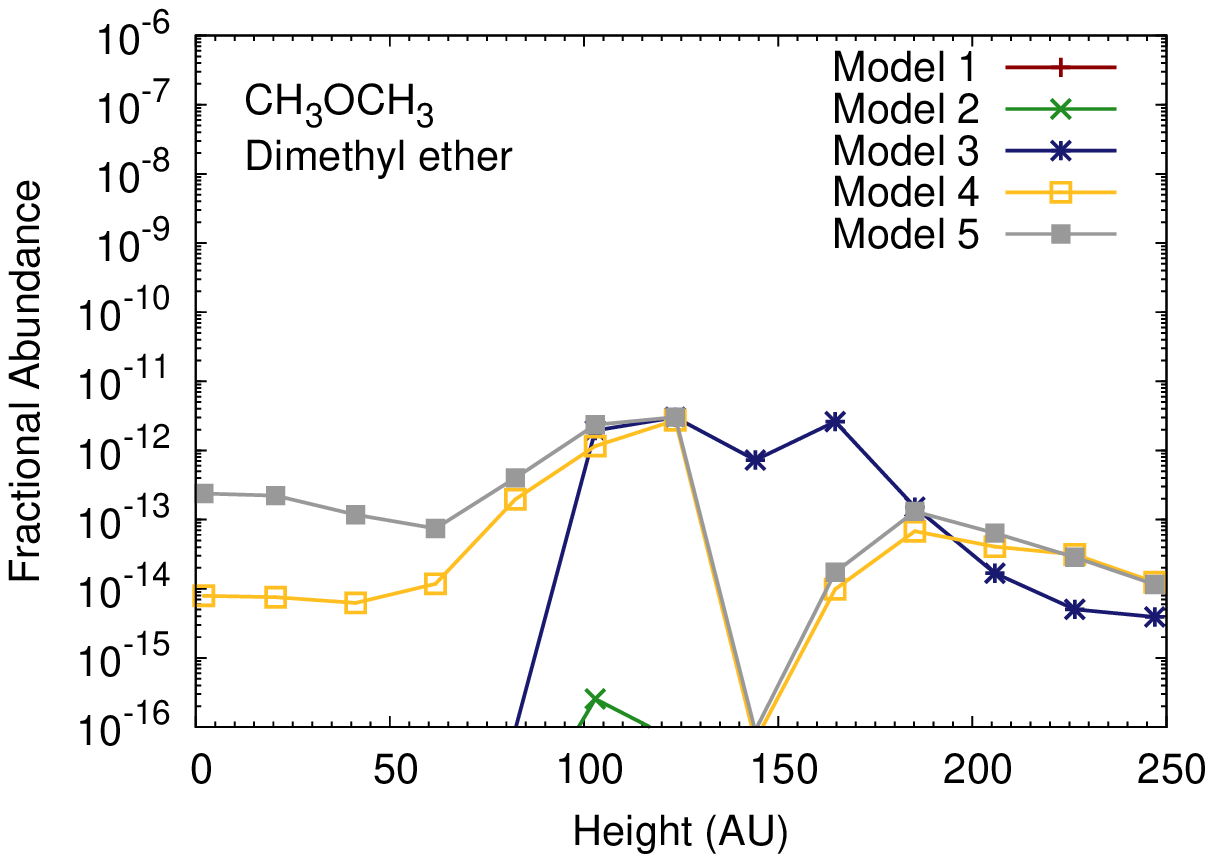}} 
\subfigure{\includegraphics[width=0.33\textwidth]{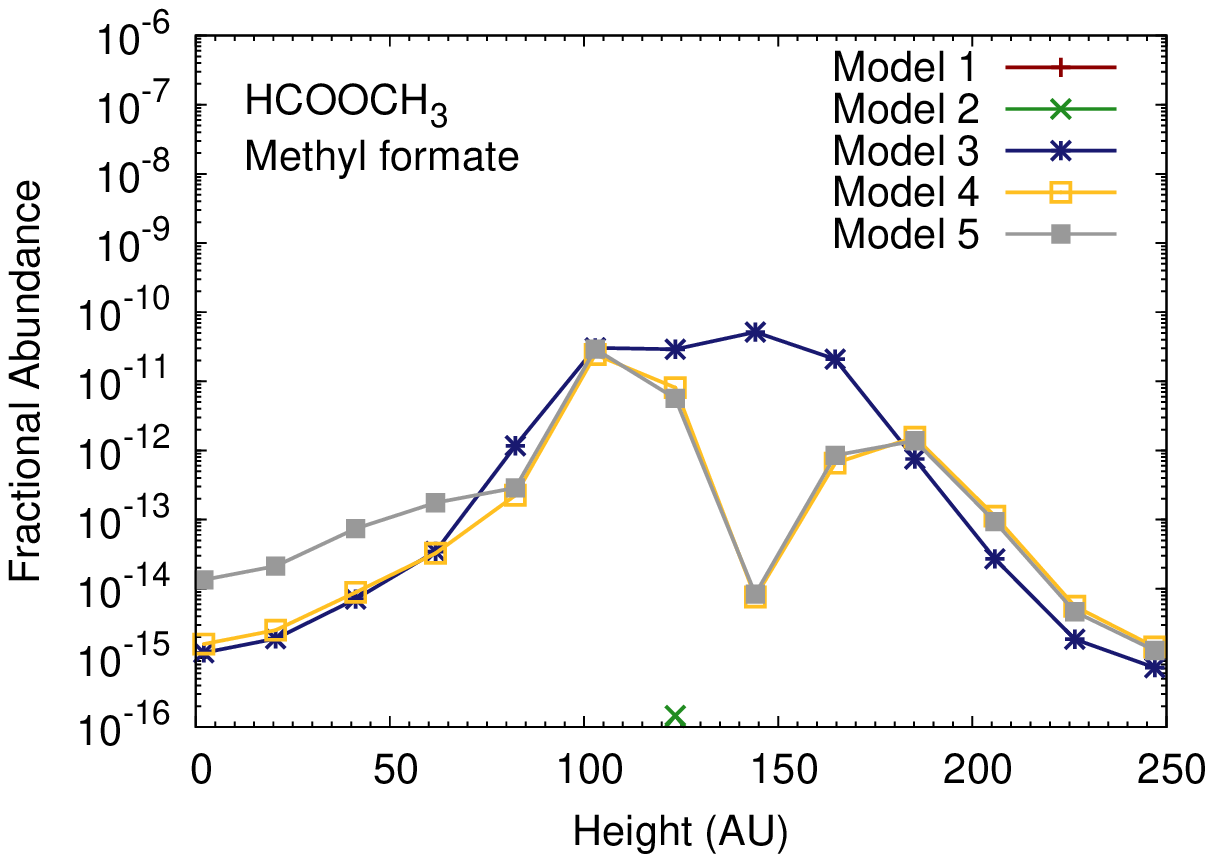}} 
\subfigure{\includegraphics[width=0.33\textwidth]{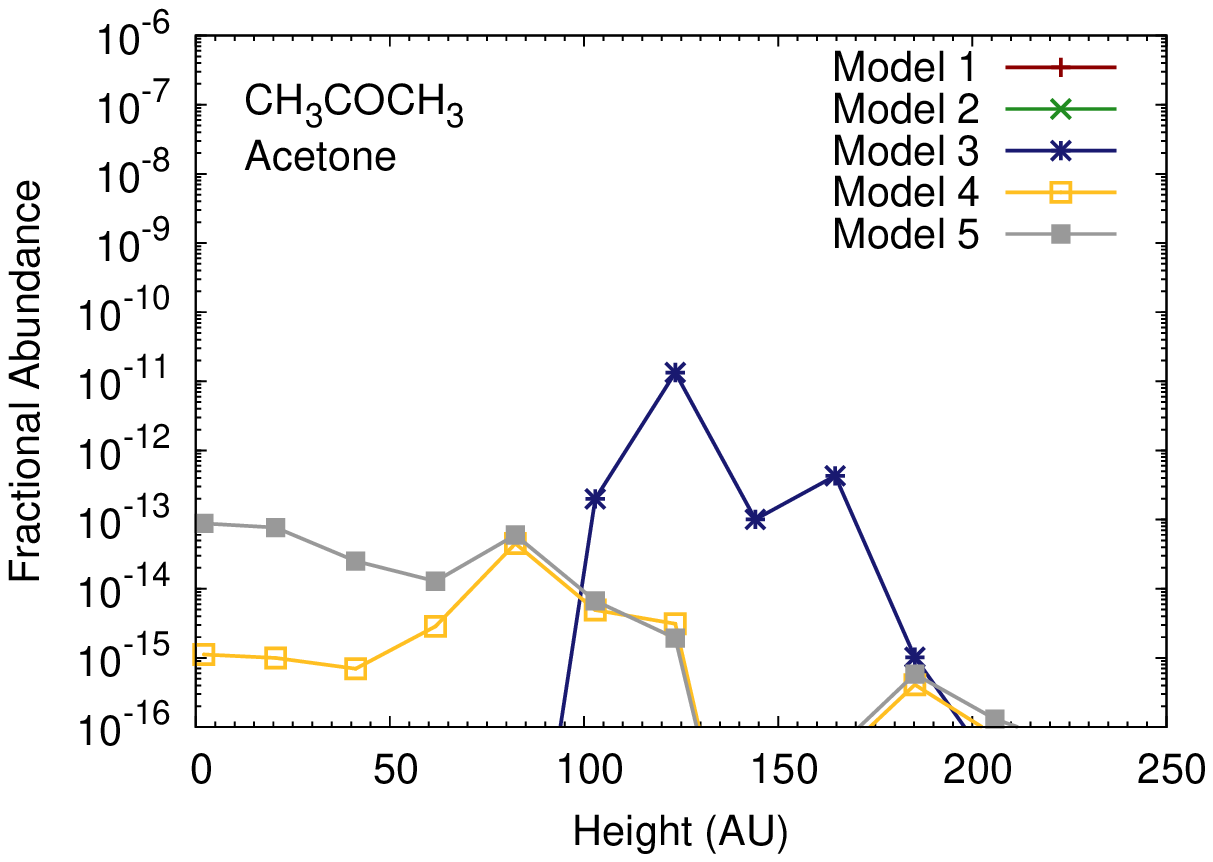}}  
\subfigure{\includegraphics[width=0.33\textwidth]{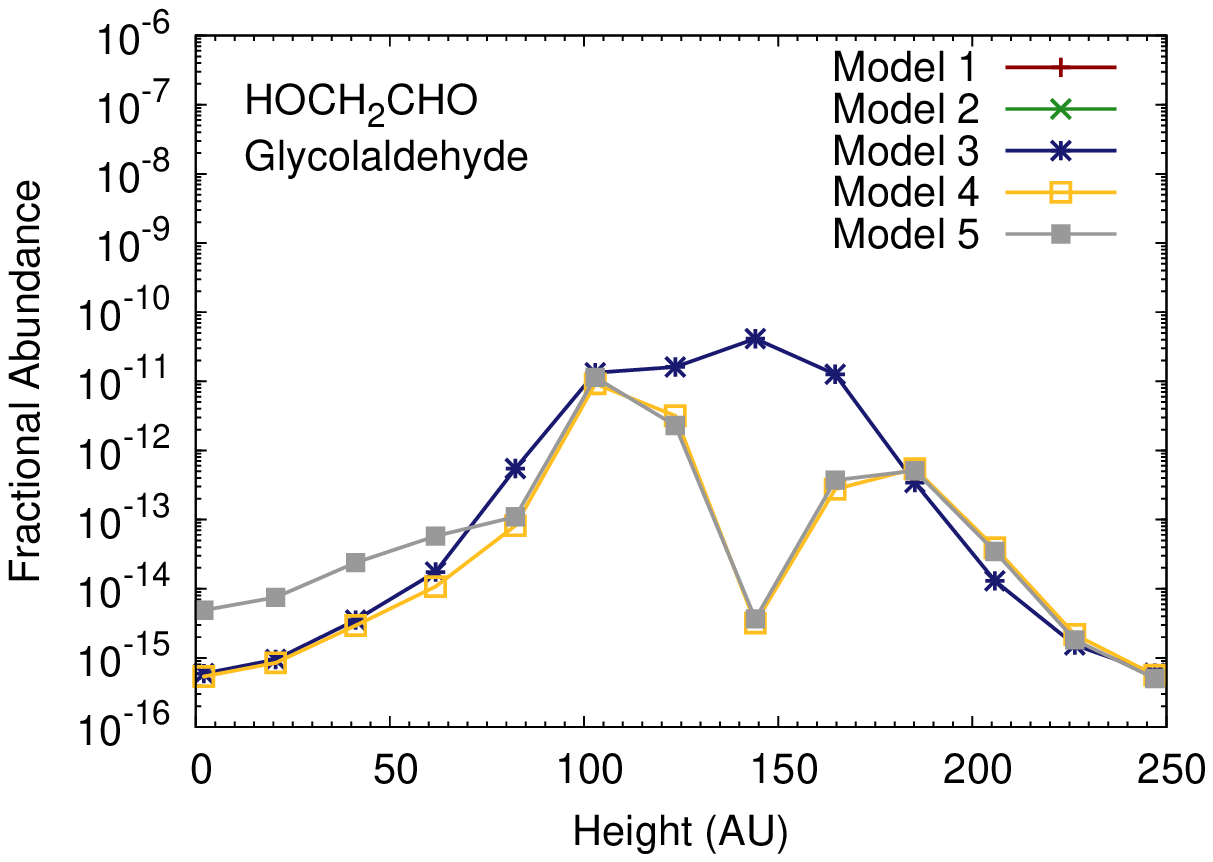}}    
\subfigure{\includegraphics[width=0.33\textwidth]{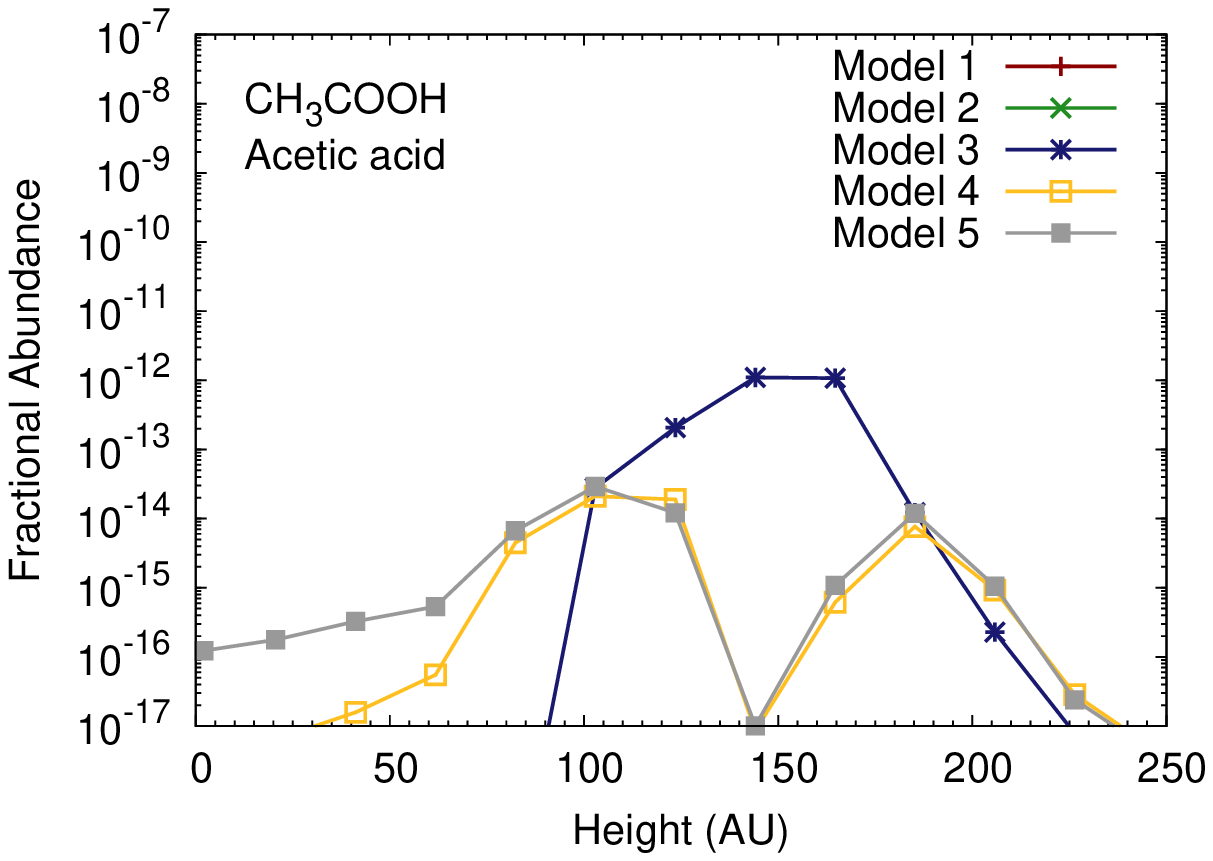}}  
\caption{Fractional abundance (with respect to H nuclei number density)
of gas-phase molecules as a function 
of disk height, $Z$ at a radius, $R$~=~305~AU. 
The chemical complexity in the model increases from Model 1 to Model 5 
(see Sect.~\ref{verticalresults} for details).}
\label{figure3}
\end{figure*}

\begin{figure*}[!ht]
\subfigure{\includegraphics[width=0.33\textwidth]{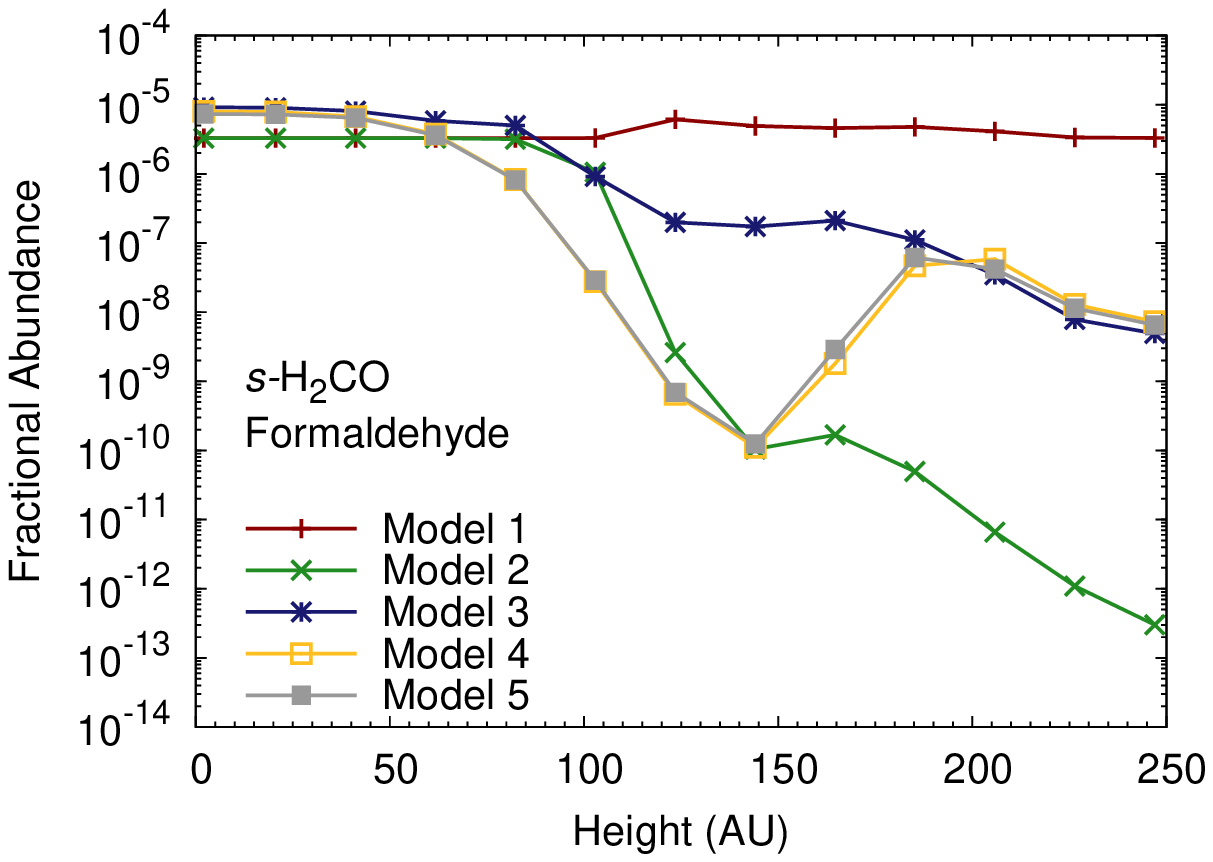}}    
\subfigure{\includegraphics[width=0.33\textwidth]{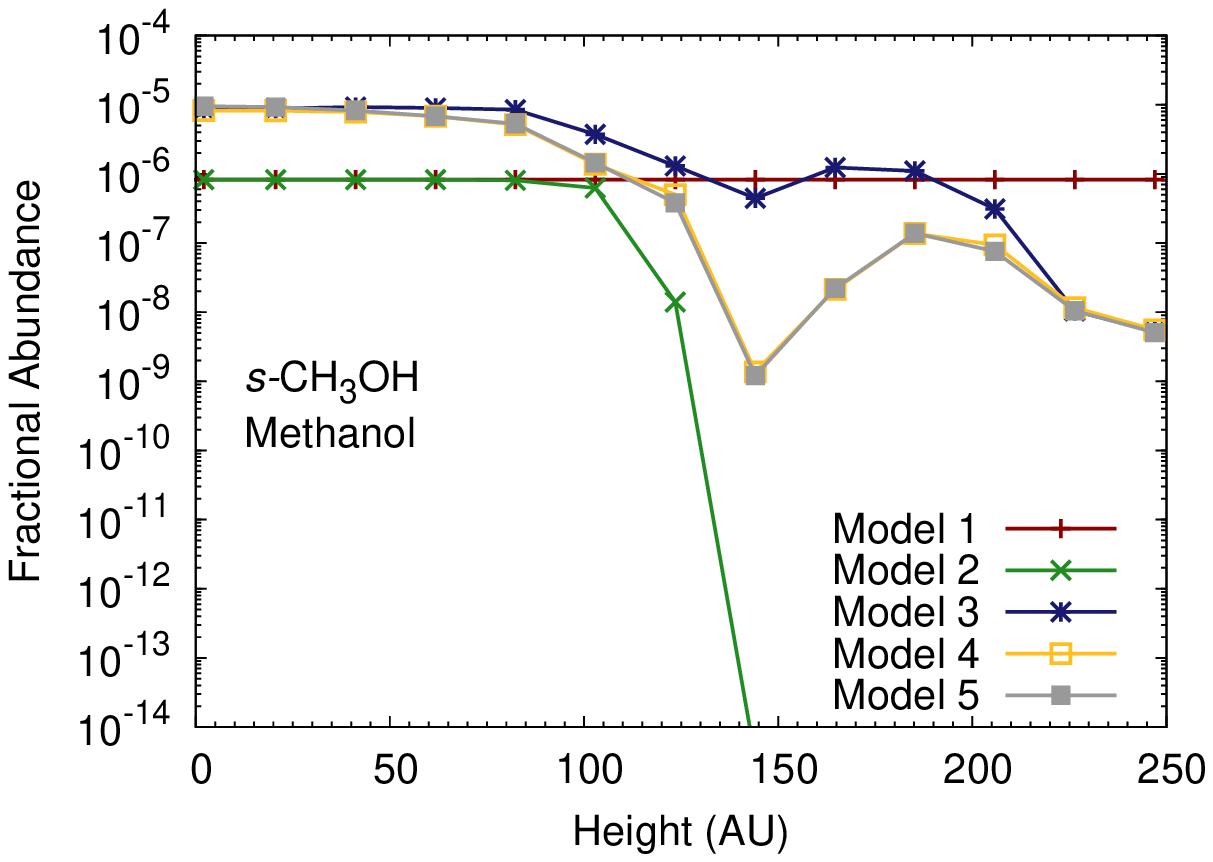}}   
\subfigure{\includegraphics[width=0.33\textwidth]{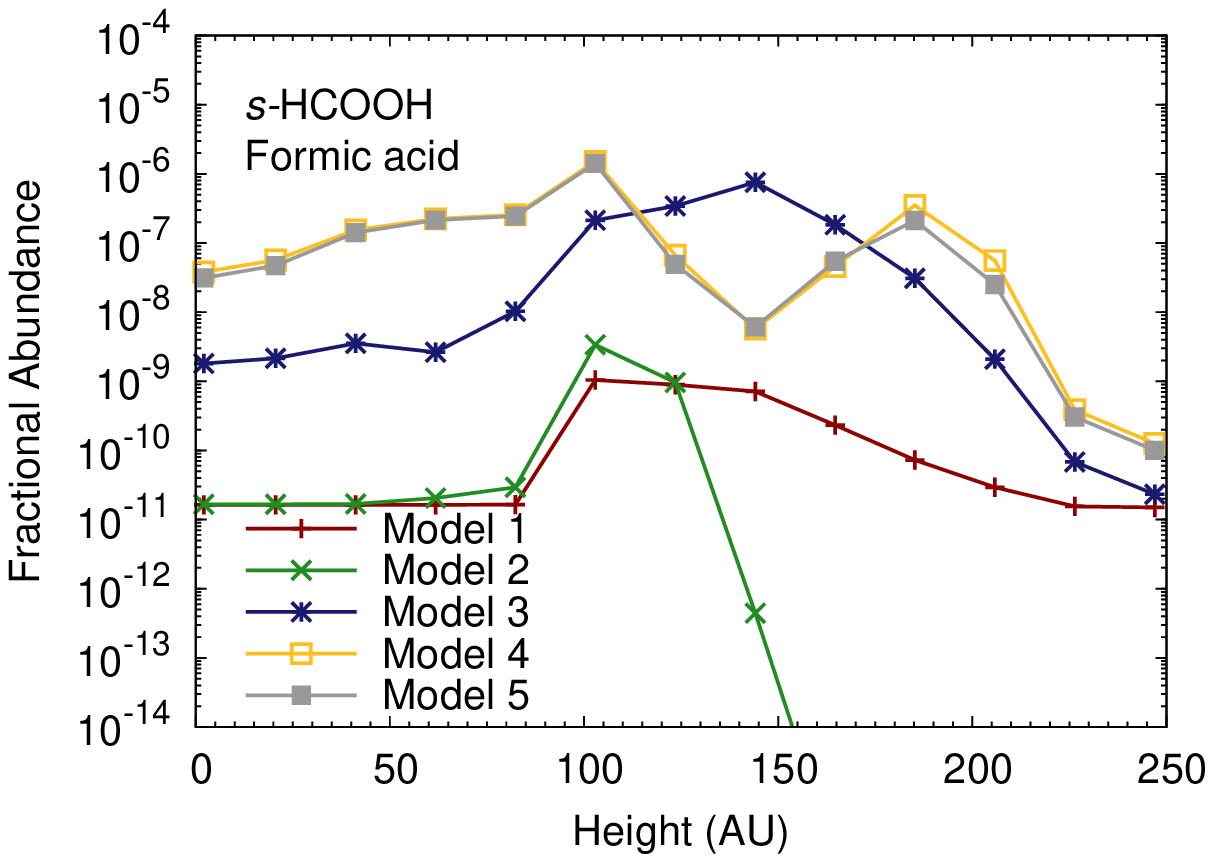}}   
\subfigure{\includegraphics[width=0.33\textwidth]{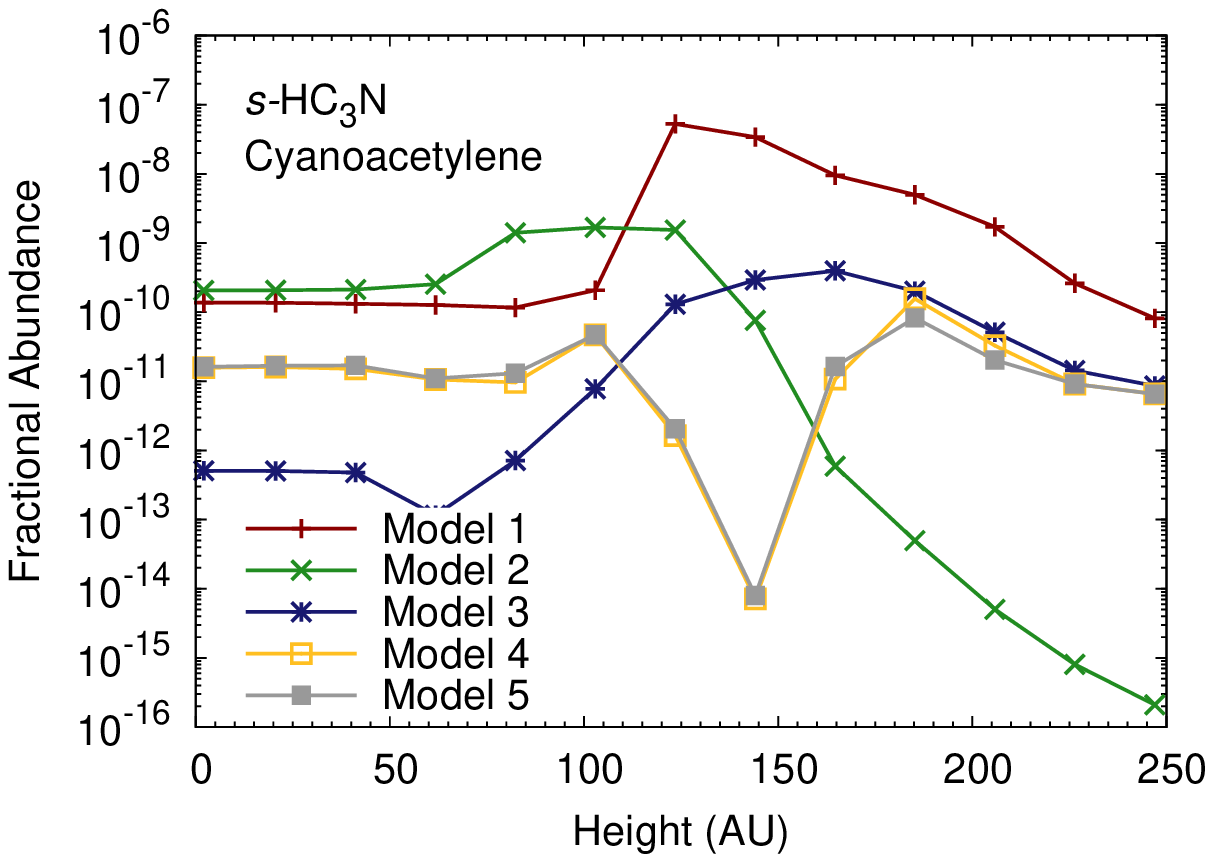}}    
\subfigure{\includegraphics[width=0.33\textwidth]{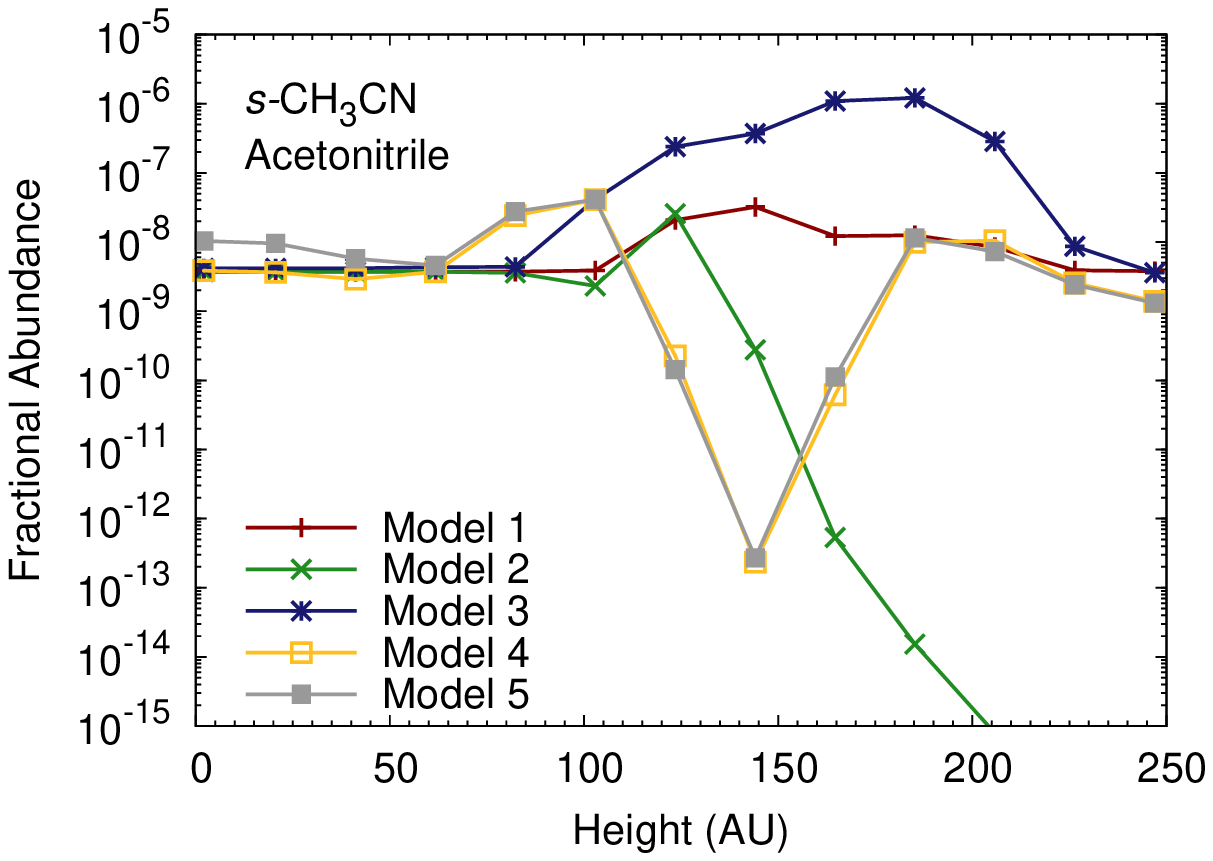}}   
\subfigure{\includegraphics[width=0.33\textwidth]{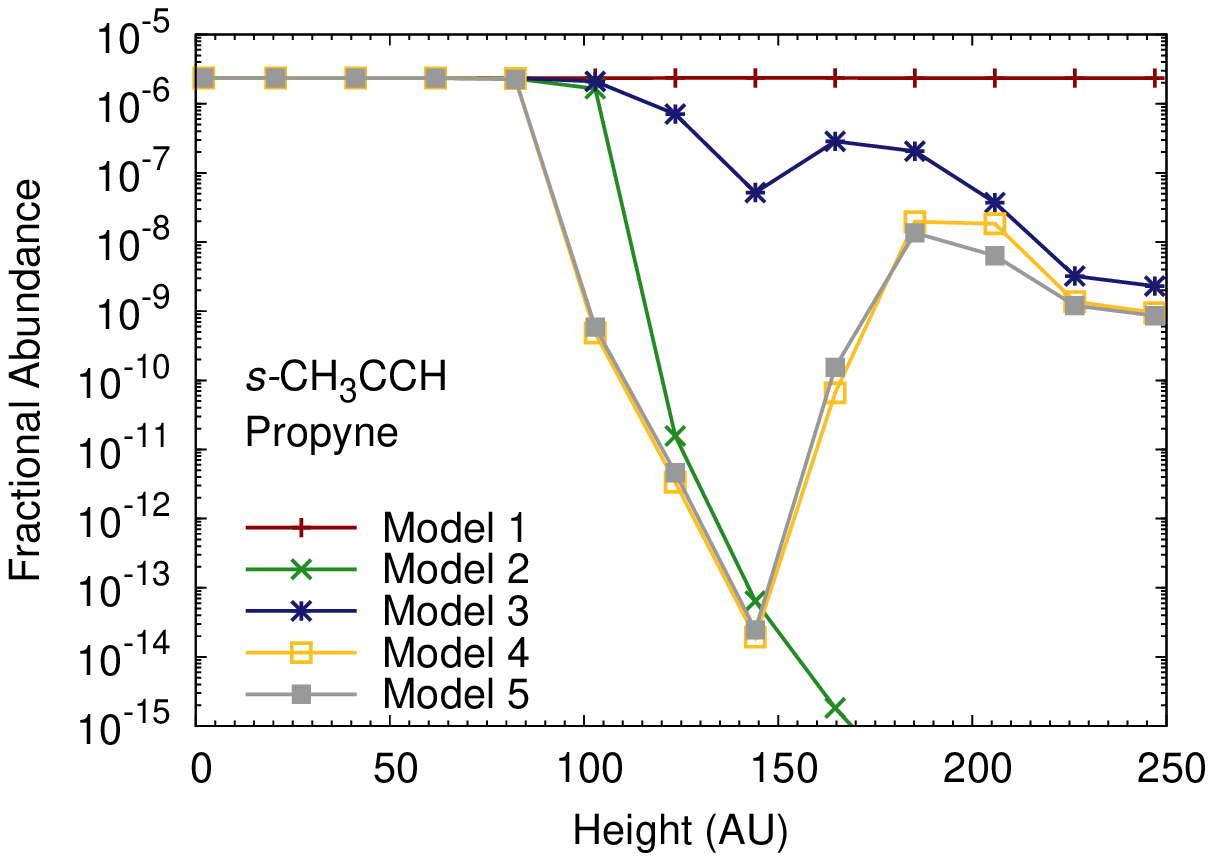}}    
\subfigure{\includegraphics[width=0.33\textwidth]{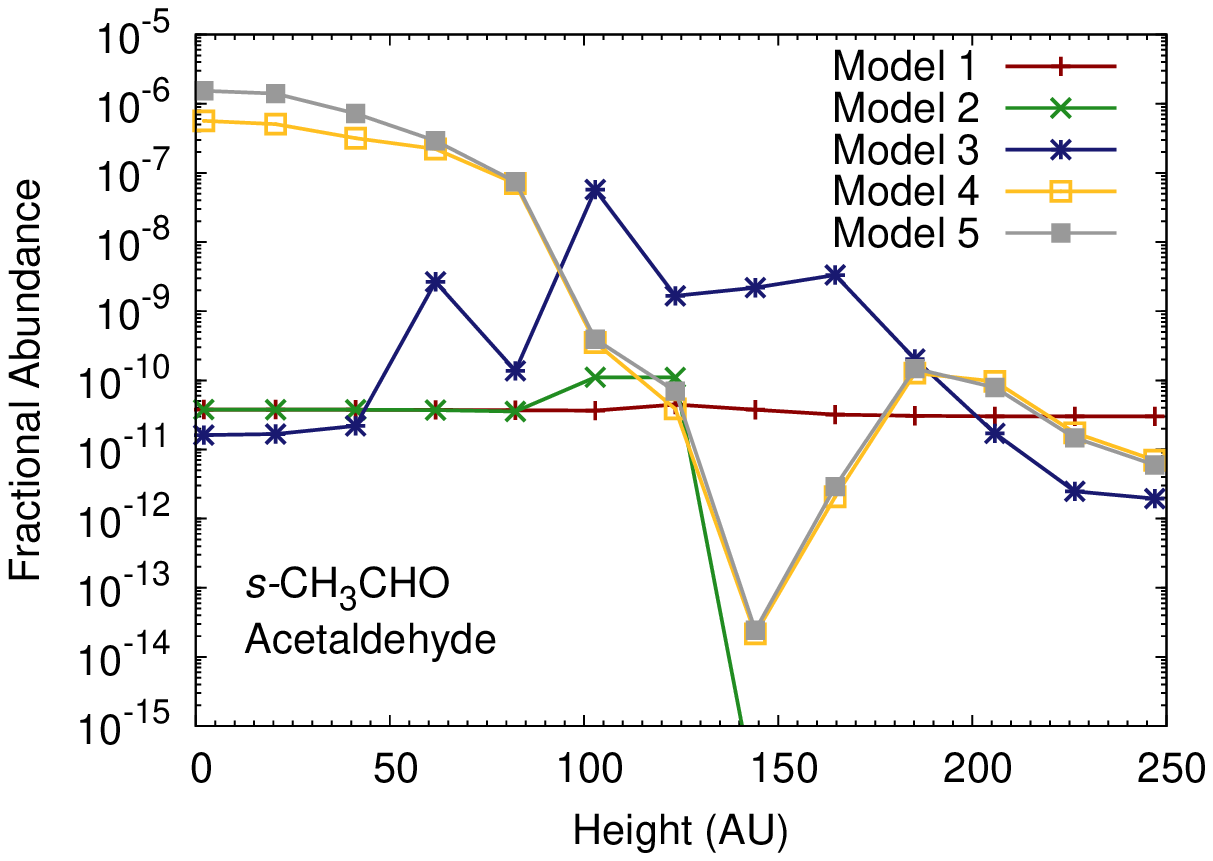}}  
\subfigure{\includegraphics[width=0.33\textwidth]{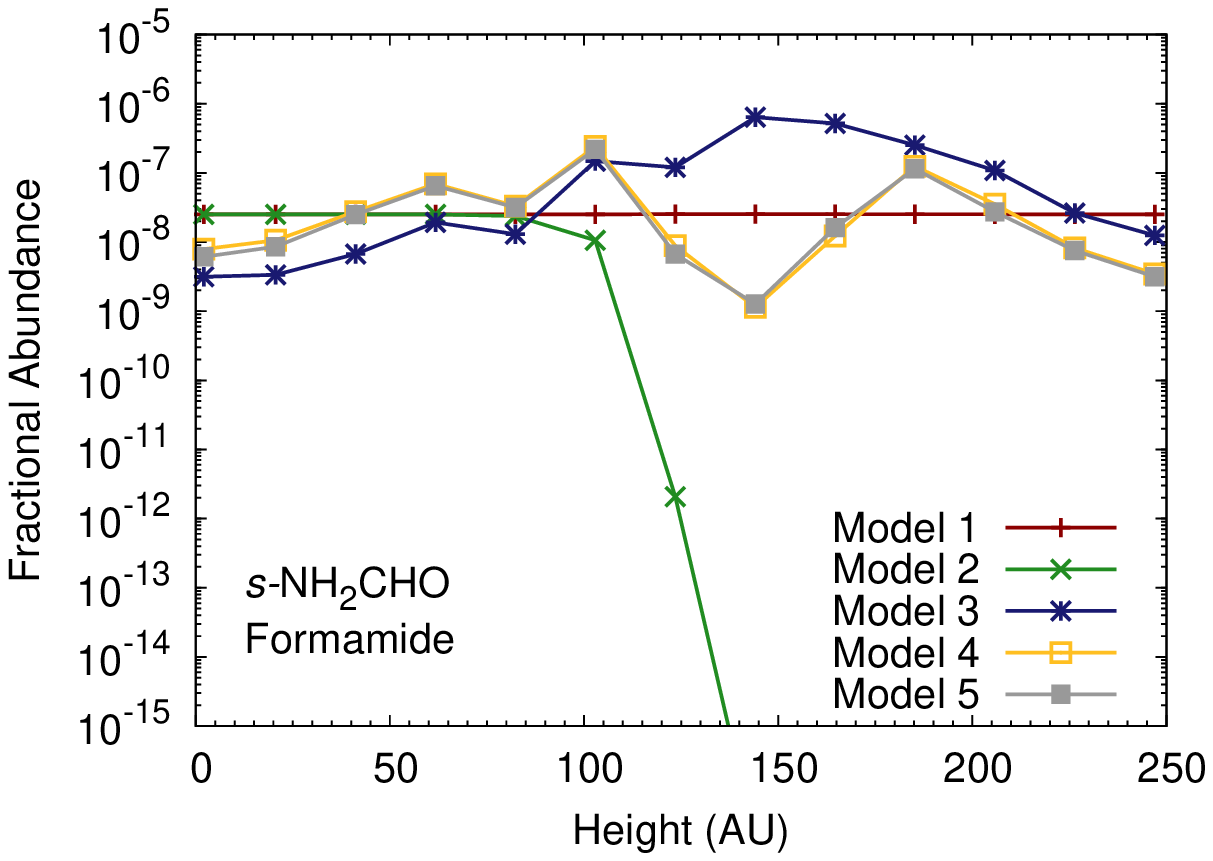}}  
\subfigure{\includegraphics[width=0.33\textwidth]{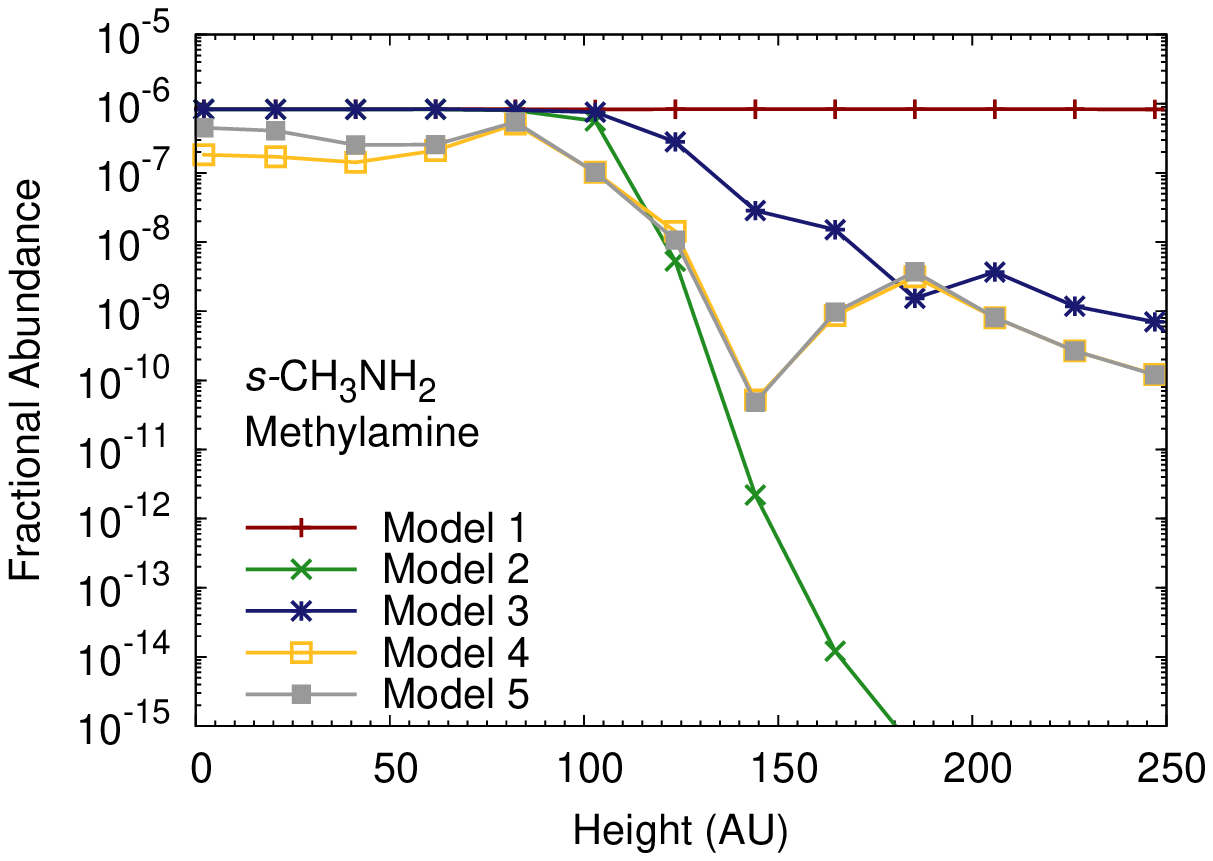}}  
\subfigure{\includegraphics[width=0.33\textwidth]{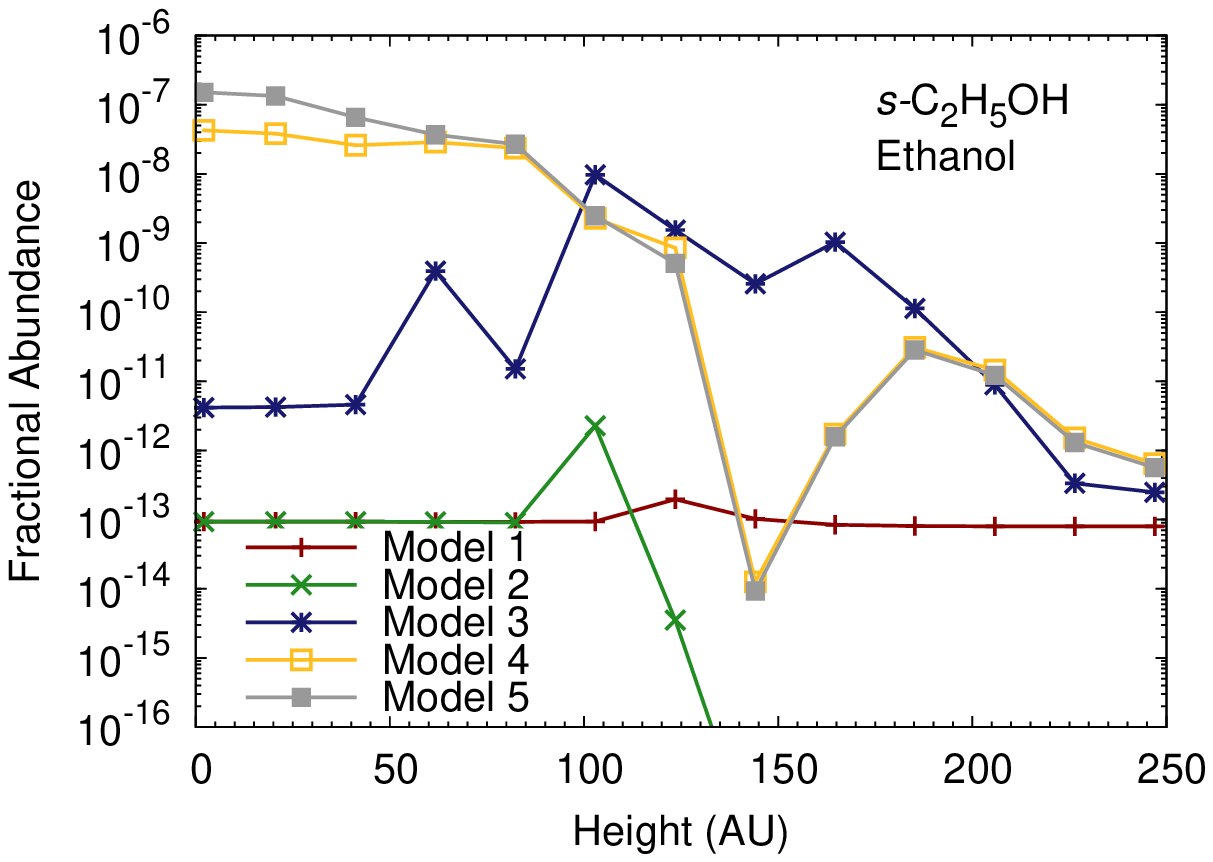}}  
\subfigure{\includegraphics[width=0.33\textwidth]{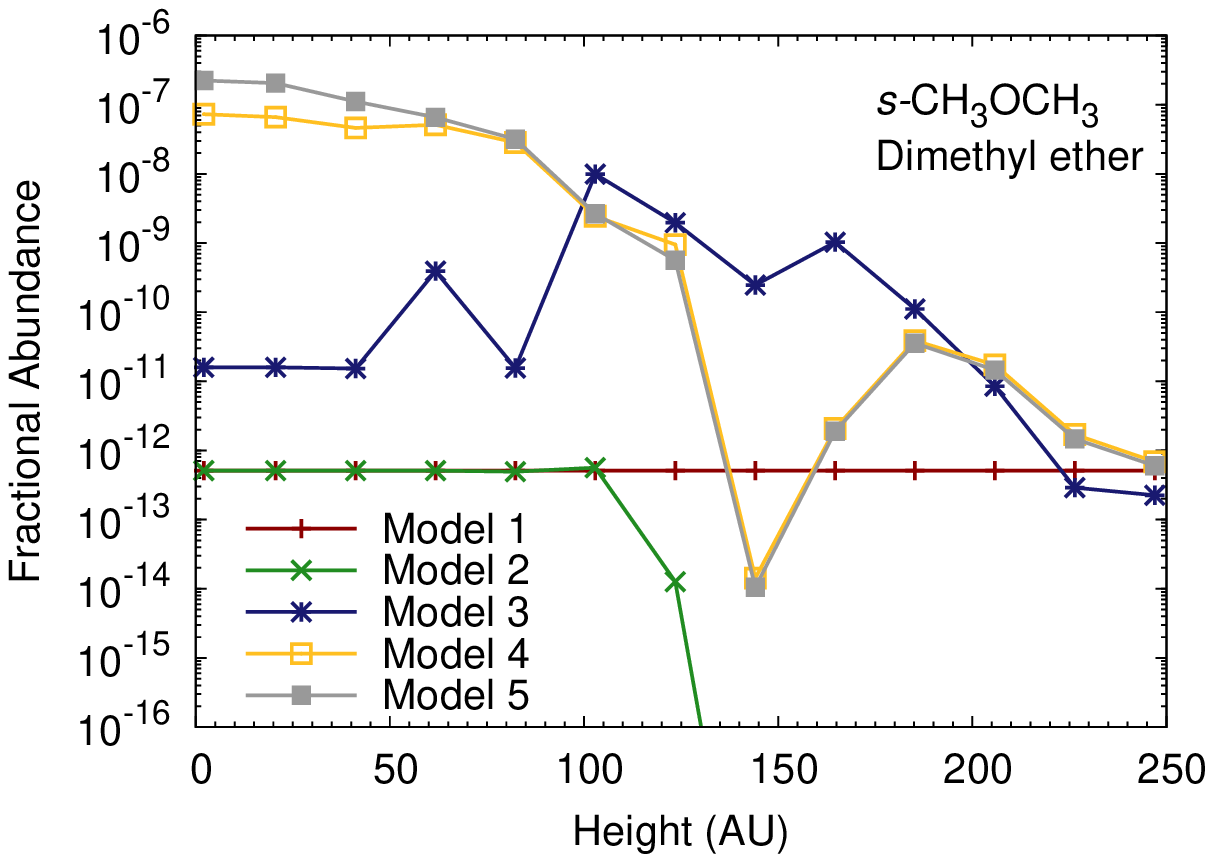}} 
\subfigure{\includegraphics[width=0.33\textwidth]{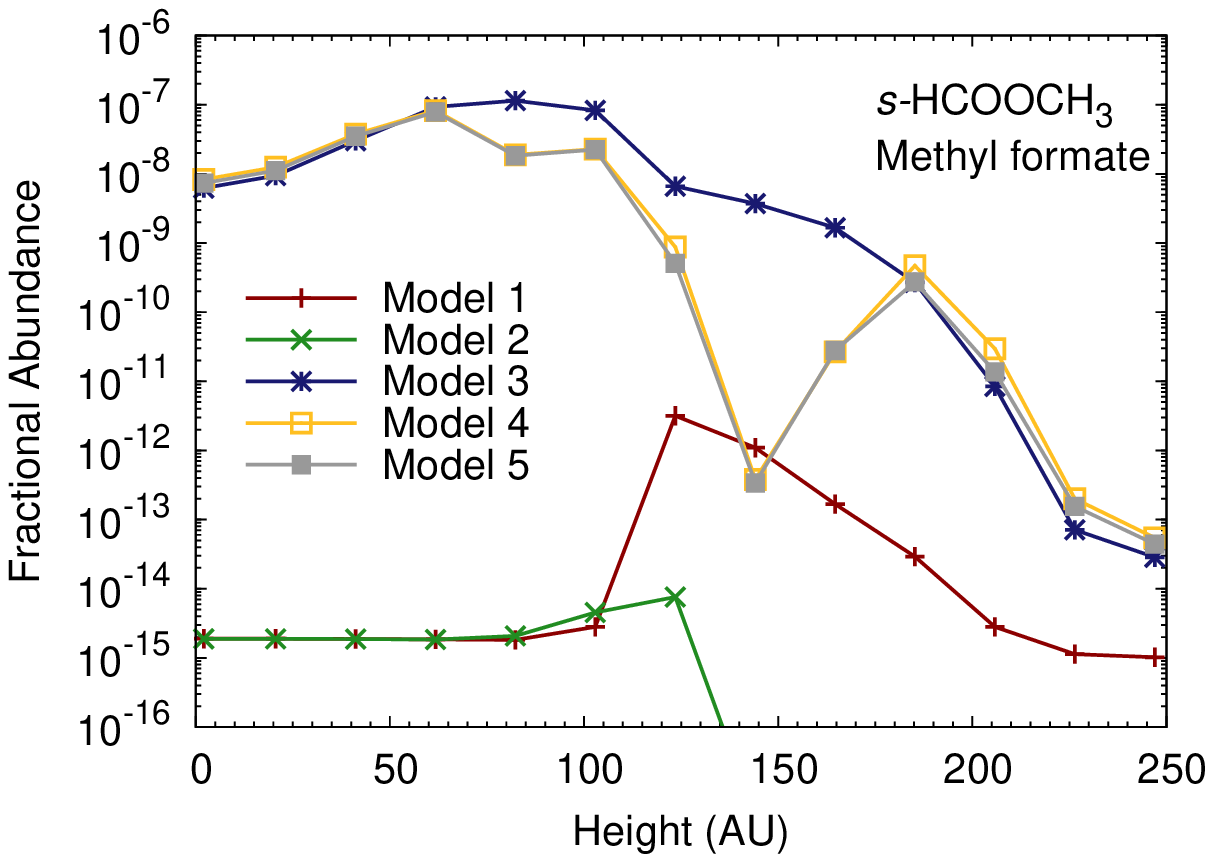}} 
\subfigure{\includegraphics[width=0.33\textwidth]{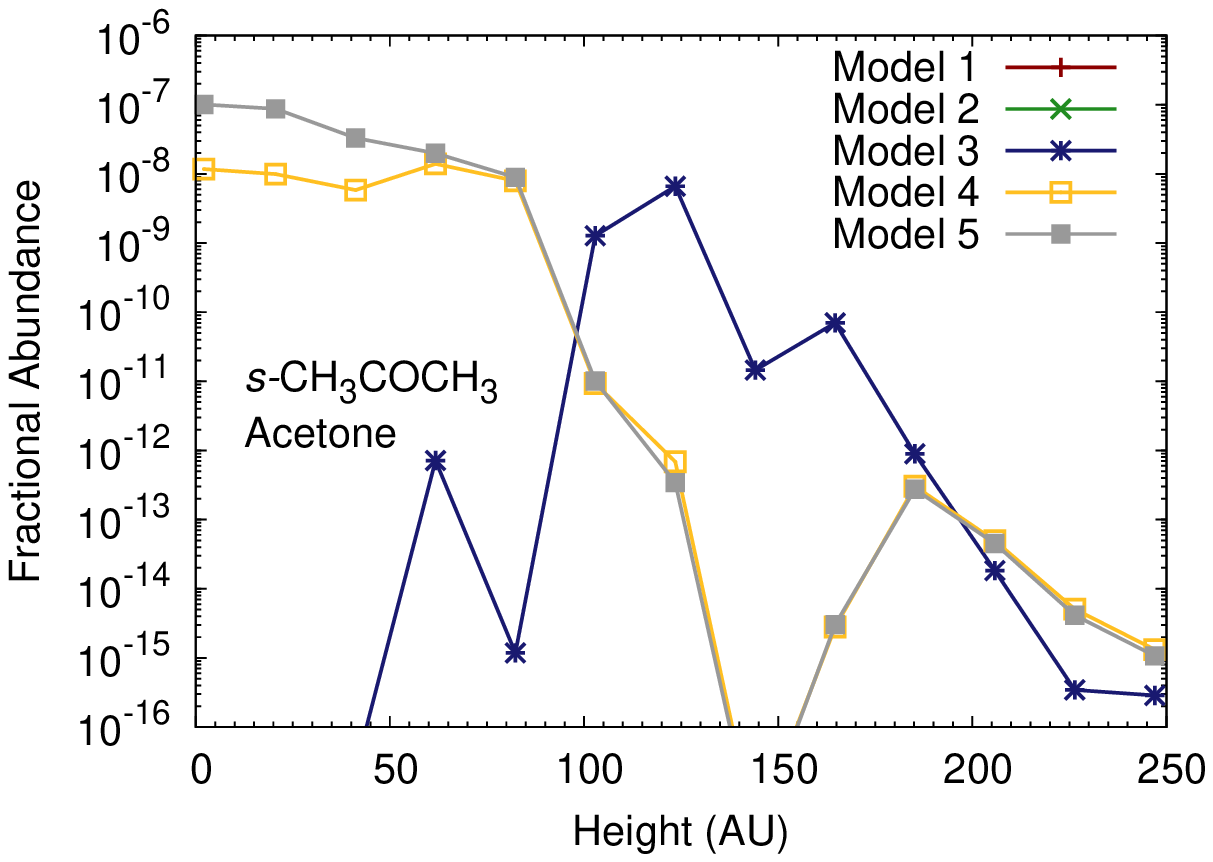}}  
\subfigure{\includegraphics[width=0.33\textwidth]{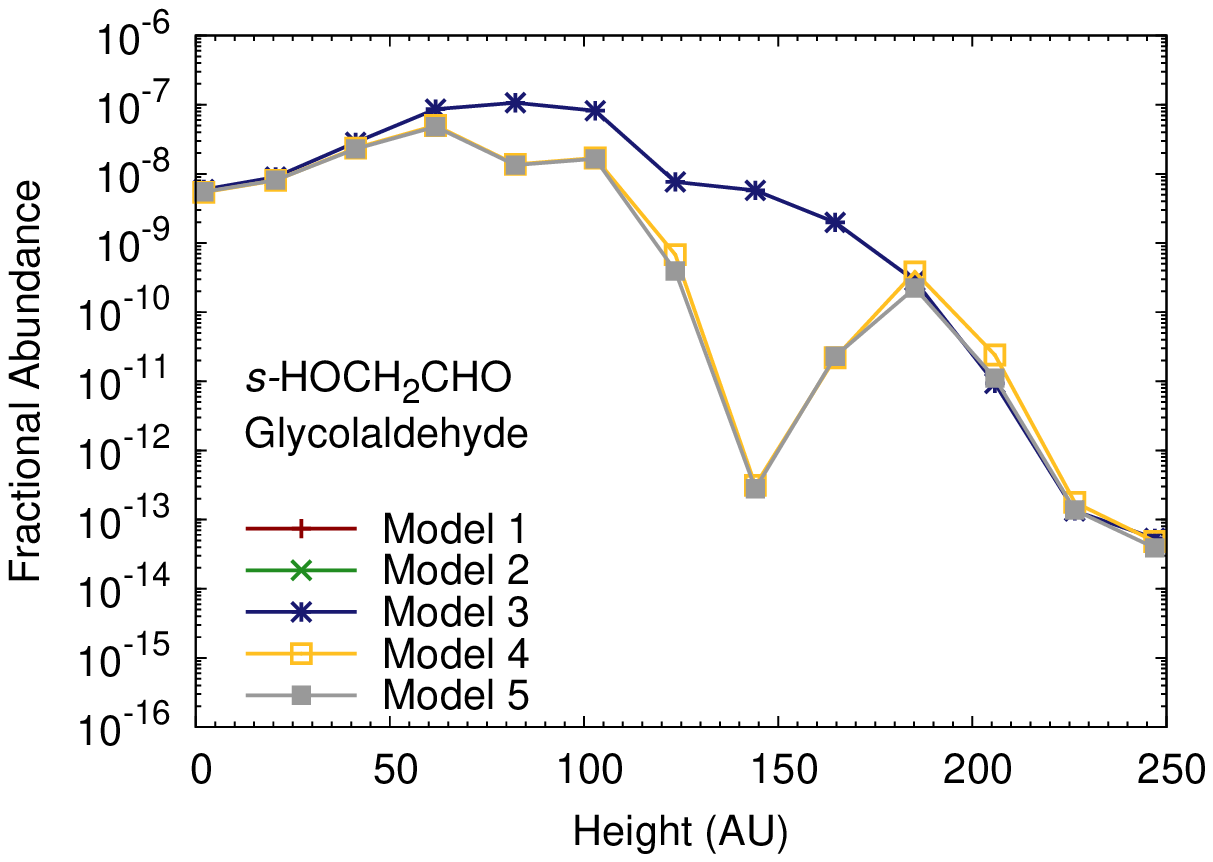}}    
\subfigure{\includegraphics[width=0.33\textwidth]{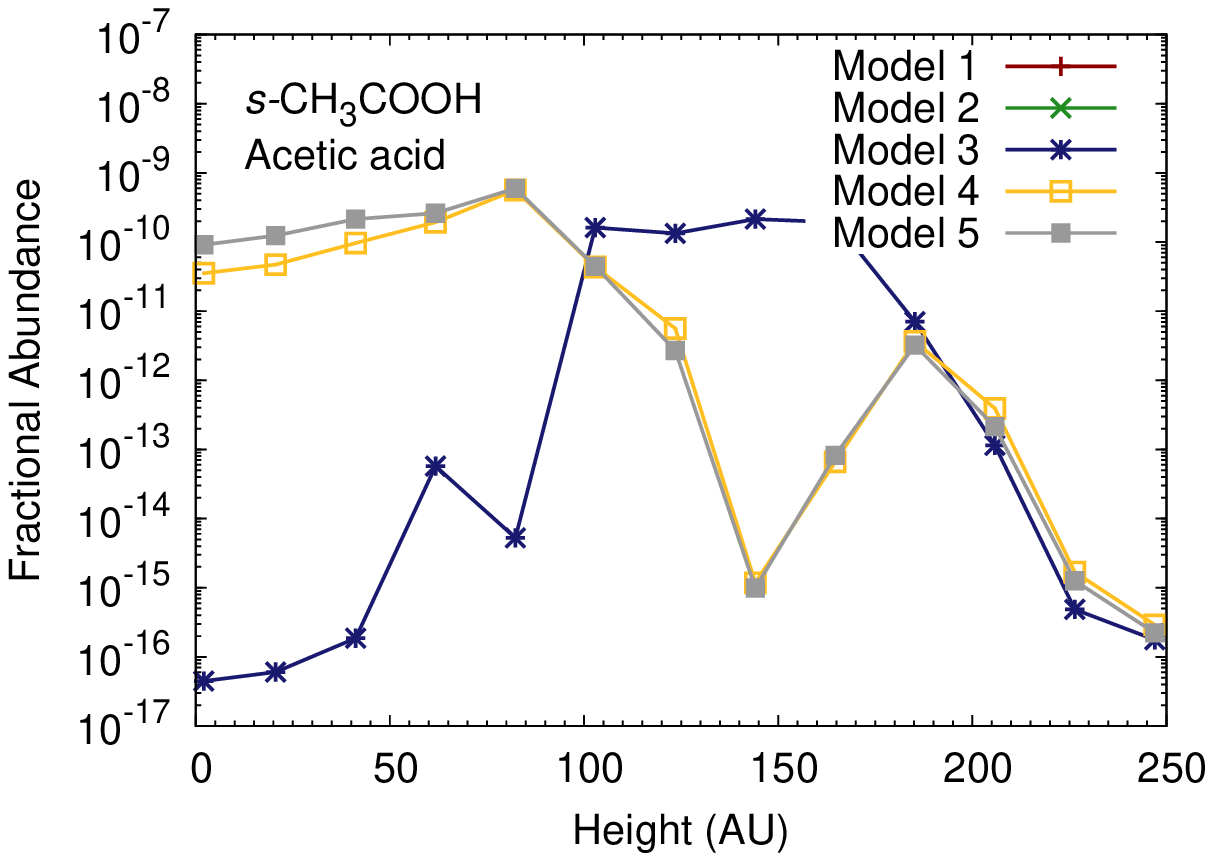}}  
\caption{Same as Fig.~\ref{figure3} for grain-surface (ice) species.}
\label{figure4}
\end{figure*}

In Figs.~\ref{figure3} and \ref{figure4} we present the 
fractional abundance (relative to H nuclei number density) 
as a function of disk height, $Z$, at a radius, $R$~=~305~AU,
of a selection of gas-phase and grain-surface (ice) COMs, respectively\footnotemark[10].  
Note that we have used the same scale for the abundance of 
each gas-phase molecule and analogous grain-surface species 
to ease the comparison between plots. 

\footnotetext[10]{The data used to plot Figs.~\ref{figure3} to \ref{figure7} are available upon request.}

\subsubsection{Model 1: Freezeout and thermal desorption}
\label{model1}

In Model 1 (red lines in Figs.~\ref{figure3} and \ref{figure4}), 
where we include freezeout and thermal desorption only, 
a handful of molecules achieve an appreciable fractional 
abundance ($\gtrsim$~10$^{-11}$) in the disk molecular layer: 
\ce{H2CO}, \ce{HC3N}, \ce{CH3CN}, and \ce{CH3NH2}.   
These species are depleted in the disk midplane below a 
height of $\approx$~100~AU due to efficient freezeout onto dust grains.  
Higher in the disk, gas-phase formation replenishes 
molecules lost via freezeout onto grain surfaces.    
These species generally retain 
a similar peak fractional abundance as that achieved under dark 
cloud conditions.  
The exception to this is acetonitrile (\ce{CH3CN}) which increases from 
an initial fractional abundance of $\sim$~10$^{-12}$ to reach a peak 
value of $\sim$~10$^{-10}$ at $Z$~$\approx$~200~AU.  
The fractional abundances decrease towards the disk surface due 
to increasing photodestruction.  
None of the other gas-phase species achieve significant peak 
fractional abundances in the molecular layer ($\lesssim$~10$^{-11}$), even those 
which begin with an appreciable initial abundance, i.e., \ce{CH3OH} and 
\ce{CH3CCH}.  
The fractional abundance of gas-phase \ce{CH3OH} remains $\lesssim$~10$^{-14}$ 
throughout the disk height. 
For these species, freezeout onto dust grains wins over gas-phase formation.  
The more complex molecules, which cannot form in the gas-phase under dark cloud 
conditions, are also unable to form under the conditions in the outer disk, in  
the absence of grain-surface chemistry.  

Regarding the grain-surface results for Model~1 (Fig.~\ref{figure4}), 
due to the higher binding energies of most species, thermal 
desorption alone is unable to remove significant fractions of the 
grain mantle.  
Hence, the ice remains abundant throughout 
the vertical extent of the disk with most species retaining their 
initial fractional abundance.  
We see enhancements in the fractional abundances of {\em s-}\ce{H2CO}, {\em s-}\ce{HCOOH}, 
{\em s-}\ce{HC3N} and {\em s-}\ce{CH3CN} around 100~AU.  
All four species have gas-phase routes to formation; however, under 
the conditions beyond $Z$~$\approx$~100~AU, 
additional molecules created in the gas phase can accrete onto dust grains 
thereby increasing their abundance on the grain mantle. 

\subsubsection{Model 2: Non-thermal desorption}
\label{model2}

In Model 2 (green lines in Figs.~\ref{figure3} and \ref{figure4}), 
we have added cosmic-ray-induced thermal desorption and photodesorption due 
to external and internal UV photons and X-ray photons.  
Non-thermal desorption has a powerful effect on both 
the gas-phase and grain-surface abundances as we also 
concluded in our previous work (WMN10).  
There are several noticeable effects: (i) the gas-phase abundances 
of many molecules are enhanced towards the disk midplane, 
relative to the results for Model 1, due to cosmic-ray-induced thermal desorption
and photodesorption, (ii) the abundances of grain-surface
molecules drop significantly towards the disk surface due to photodesorption 
by external UV photons, 
and (iii) there is a shift in the position of the gas-phase 
`molecular layer' towards the midplane. 
This latter effect is due to a combination of non-thermal desorption and 
enhanced gas-phase formation lower in the disk, and enhanced destruction higher in the 
disk due to the release of a significant fraction of the grain mantle back into 
the gas phase.  
Non-thermal desorption effectively `seeds' or replenishes the gas with 
molecules that otherwise would remain 
bound to the grain, e.g., \ce{H2O} and its protonated form, \ce{H3O+}, 
which then go on to take part in gas-phase 
reactions which can form (destroy) molecules which would otherwise be 
depleted (abundant). 

The fractional abundance of gas-phase \ce{H2CO} is enhanced to 
$\sim$~10$^{-12}$ in the disk midplane. 
However, its fractional 
abundance in the molecular layer and disk surface 
reaches values similar to that in Model 1 
($\sim$~10$^{-10}$~--~10$^{-9}$). 
\ce{CH3OH} reaches a peak abundance of $\sim$~10$^{-10}$ around a height of $\approx$~100~AU 
corresponding to the height where photodesorption by external photons begins to have an effect.  
Once released from the grain, methanol is efficiently destroyed under the conditions 
in the upper disk since methanol does not have efficient gas-phase routes to formation at 
low temperatures.   
Methanol is also significantly enhanced in the disk midplane by cosmic-ray-induced desorption, 
reaching a fractional abundance, $\sim$~10$^{-13}$.  
Other gas-phase species enhanced in the midplane and molecular layer due 
to non-thermal desorption and reaching appreciable 
fractional abundances ($\gtrsim$~10$^{-11}$) include 
\ce{HCOOH}, \ce{HC3N}, \ce{CH3CN}, \ce{CH3CCH}, and \ce{CH3NH2}. 

\subsubsection{Model 3: Grain-surface chemistry}
\label{model3}

In Model 3 (blue lines in Figs.~\ref{figure3} and \ref{figure4}), 
we have added thermal grain-surface chemistry \citep{hasegawa92}.  
Grain-surface chemistry is a very important process for 
building chemical complexity in the ice mantle under the conditions in the outer disk.  
There are several reasons for this.  
In addition to reactions involving atoms and other neutrals, which are well known in ice 
chemistry at 10~K, it appears that a reasonably high abundance of radicals, such 
as, {\em s-}OH and {\em s-}HCO, can be achieved without radiation processing of the ice mantle.  
They form via association reactions on the grain, e.g, 
{\em s-}H~+~{\em s-}CO, or accrete from the gas phase where they are 
formed via gas-phase chemistry or via the 
photodissociation of larger species. 
These radicals can diffuse and react at the dust temperatures 
in the outer disk (17~K~--~30~K).  
  
The major effect of the addition of thermal grain-surface chemistry 
is that the gas-phase fractional abundances 
of all molecules we consider in this section 
are enhanced (relative to the results for Model 1 and Model 2). 
We see around an order of magnitude increase in the peak abundance of formaldehyde in the  
molecular layer of the disk, from $\sim$~10$^{-9}$ to $\sim$~10$^{-8}$.  
Gas-phase methanol is enhanced throughout the molecular layer 
and disk surface also reaching 
a peak fractional abundance of $\sim$~10$^{-8}$ at a height of $\approx$~150~AU.   
We see a corresponding rise in the ice abundance indicative that grain-surface 
formation of methanol 
is replenishing the grain mantle molecules lost to the gas
by non-thermal desorption.  
Formic acid, HCOOH, reaches a peak abundance of $\sim$~10$^{-8}$ in the molecular layer.  
We see a dramatic rise in the fractional abundance of {\em s-}HCOOH throughout the disk height.  
{\em s-}HCOOH can be formed at low temperatures ($\approx$~20~K) on grain surfaces via the barrierless 
reactions,
\begin{equation*}
s\mbox{-}\ce{H} + s\mbox{-}\ce{COOH} \longrightarrow s\mbox{-}\ce{HCOOH}
\end{equation*}
and 
\begin{equation*}
s\mbox{-}\ce{OH} + s\mbox{-}\ce{HCO} \longrightarrow s\mbox{-}\ce{HCOOH}.
\end{equation*}
The former reaction requires {\em s-}COOH to be sufficiently abundant 
and this is formed on grain-surfaces via {\em s-}OH + {\em s-}CO which 
has a high reaction barrier ($\approx$~3000~K). 
Thus, the latter reaction is the main route to formation in the cold midplane.  
Although the mobility of both {\em s-}OH and {\em s-}HCO is relatively low 
at $\approx$~20~K, the long lifetime of the 
disk, $\sim$~10$^6$~years, and the high density in the midplane, 
$\sim$~10$^{7}$~cm$^{-3}$, allows the sufficient buildup of formic 
acid ice via this reaction.   

Gas-phase \ce{HC3N} experiences a order-of-magnitude enhancement 
in Model 3 relative to Model 2, 
reaching a peak fractional abundance of $\sim$~10$^{-9}$.  
However, we see a drop in the fractional abundance of grain-surface {\em s-}\ce{HC3N} 
below a height, $Z$~$\lesssim$~150~AU, relative to the results for Model~2.  
{\em s-}\ce{HC3N} can be sequentially hydrogenated up to 
{\em s-}\ce{C2H5CN} (propionitrile).  

We also see enhancements, related to grain-surface chemistry, 
in gas-phase \ce{CH3CN}, \ce{CH3CCH}, \ce{CH3CHO}, \ce{NH2CHO}, and \ce{CH3NH2} in the 
molecular layer and disk surface, 
reaching peak fractional abundances of $\sim$~10$^{-10}$~--~10$^{-8}$.
{\em s-}\ce{CH3CN} is formed via successive hydrogenation of {\em s-}\ce{C2N}, 
\begin{equation*}
s\mbox{-}\ce{C2N}   \xrightarrow{s\mbox{-}\ce{H}}
s\mbox{-}\ce{HCCN}  \xrightarrow{s\mbox{-}\ce{H}}
s\mbox{-}\ce{CH2CN} \xrightarrow{s\mbox{-}\ce{H}}
s\mbox{-}\ce{CH3CN}
\end{equation*}  
In turn, {\em s-}\ce{C2N} can form via {\em s-}C + {\em s-}CN and {\em s-}N + {\em s-}\ce{C2}, or freeze out 
from the gas phase, where it is formed via ion-molecule and radical-radical 
reactions. 
In warmer regions, {\em s-}\ce{CH3CN} can also form via the 
radical-radical reaction, {\em s-}CN~+~{\em s-}\ce{CH3}.   
Intermediate species in the above hydrogenation sequence can 
also form via grain-surface association reactions, e.g., 
\begin{equation*}
s\mbox{-}\ce{CH} + s\mbox{-}\ce{CN} \longrightarrow s\mbox{-}\ce{HCCN}
\end{equation*}
and
\begin{equation*}
s\mbox{-}\ce{CH2} + s\mbox{-}\ce{CN} \longrightarrow s\mbox{-}\ce{CH2CN}.
\end{equation*}
{\em s-}\ce{CH3CCH}, forms via the successive hydrogenation of {\em s-}\ce{C3} and {\em s-}\ce{C3H}, 
and in warmer gas can also form via the barrierless radical-radical reaction,
\begin{equation*}
s\mbox{-}\ce{CH} + s\mbox{-}\ce{C2H3} \longrightarrow s\mbox{-}\ce{CH3CCH}.
\end{equation*}
In turn, {\em s-}\ce{C2H3}, has several grain-surface formation routes, in addition 
to the hydrogenation of {\em s-}\ce{C2H2}, e.g., 
{\em s-}\ce{C}~+~{\em s-}\ce{CH3} and {\em s-}\ce{CH}~+~{\em s-}\ce{CH2}.  
Acetaldehyde ({\em s-}\ce{CH3CHO}) can form via the association of 
{\em s-}\ce{CH3} and {\em s-}\ce{HCO} or via {\em s-}\ce{CH3CO} 
which, in turn, forms on the grain via {\em s-}\ce{CH3}~+~{\em s-}\ce{CO}.  
This latter reaction has a barrier of $\approx$~3500~K and can only 
proceed on sufficiently warm grains.    
The formation of formamide ({\em s-}\ce{NH2CHO}) and 
methylamine ({\em s-}\ce{CH3NH2}) via atom-addition reactions was discussed 
previously in relation to dark cloud chemistry (see Sect.~\ref{initialabundances}).  
In warmer regions, again, there are formation routes via the 
association of the amine and formyl radicals ({\em s-}\ce{NH2} + {\em s-}\ce{HCO}), 
and the methyl and amine radicals ({\em s-}\ce{CH3} + {\em s-}\ce{NH2}), respectively.  
In all cases, these precursor radicals are either
formed on the grain via atom-addition reactions, or they form in the gas phase via ion-molecule chemistry 
or photodissociation  
and freeze out onto dust grains. 
    
Moving on to the more complex species, we see much the same effect as for 
those already discussed; gas-phase 
abundances are enhanced when grain-surface chemistry is included.  
Each gas-phase species reaches a peak abundance of 
$\sim$~10$^{-12}$~--~10$^{-10}$ in the molecular 
layer, which correlates with enhancements in the grain-surface abundances 
and hence are directly due to synthesis via grain-surface chemistry and 
subsequent release to the gas phase via non-thermal desorption.  
The peak fractional abundance attained by the more complex species 
on the grain surface ranges between $\sim$~10$^{-9}$~--~10$^{-7}$.
For those complex species for which we begin with negligible abundances on the grain 
($\lesssim$~10$^{-13}$), 
i.e., {\em s-}\ce{C2H5OH}, {\em s-}\ce{CH3OCH3}, {\em s-}\ce{CH3COCH3}, {\em s-}\ce{CH3COOH}, {\em s-}\ce{HCOOCH3}, and {\em s-}\ce{HOCH2CHO}, 
grain-surface chemistry is absolutely necessary for their formation.  
These species form predominantly via radical-radical association routes and, thus, 
require significant abundances of precursor radicals. 
For example, methyl formate 
({\em s-}\ce{HCOOCH3}) can form on the grain via the association reaction,
\begin{equation*}
s\mbox{-}\ce{HCO} + s\mbox{-}\ce{CH3O} \longrightarrow s\mbox{-}\ce{HCOOCH3}.
\end{equation*}
Both reactants are steps on the ladder of the sequential hydrogenation 
of CO to form \ce{CH3OH}.  
Similarly, 
\begin{equation*}
s\mbox{-}\ce{CH3} + s\mbox{-}\ce{CH2OH} \longrightarrow s\mbox{-}\ce{C2H5OH},
\end{equation*}
\begin{equation*}
s\mbox{-}\ce{CH3} + s\mbox{-}\ce{CH3O} \longrightarrow s\mbox{-}\ce{CH3OCH3},
\end{equation*}
\begin{equation*}
s\mbox{-}\ce{CH3} + s\mbox{-}\ce{CH3CO} \longrightarrow s\mbox{-}\ce{CH3COCH3}, 
\end{equation*}
\begin{equation*}
s\mbox{-}\ce{OH} + s\mbox{-}\ce{CH3CO} \longrightarrow s\mbox{-}\ce{CH3COOH},  
\end{equation*}
and, 
\begin{equation*}
s\mbox{-}\ce{CH2OH} + s\mbox{-}\ce{HCO} \longrightarrow s\mbox{-}\ce{HOCH2CHO}.  
\end{equation*}
Note that the isomers, {\em s-}\ce{HCOOCH3} and {\em s-}\ce{HOCH2CHO} (methyl formate 
and glycolaldehyde), are relatively 
abundant in the disk midplane.  
Both species are formed via reactants which are 
products of the hydrogenation of {\em s-}CO.  
In contrast, the third member of this family, {\em s-}\ce{CH3COOH}, 
is formed on the grain via {\em s-}\ce{CH3CO} which is formed via the association of 
{\em s-}\ce{CH3} and {\em s-}CO.  
This reaction has a barrier of around 3500~K and so cannot 
proceed at the low temperatures in the disk midplane.   
Radicals, such as, \ce{CH2OH} and \ce{CH3O}, are also formed in the gas via the 
photodissociation of larger molecules, e.g., \ce{CH3OH}.  
  
\subsubsection{Model 4: Radiation processing of ice}
\label{model4}

In Model 4 we added cosmic-ray, X-ray and UV photo-processing 
of ice mantle material.  
The behaviour of ice under irradiation is still very uncertain with only 
a handful of quantitative experiments conducted.  
Experiments on UV-irradiated pure methanol ice show that 
chemistry is induced by the process with many gas-phase products, other than 
methanol, observed \citep[see e.g.,][]{gerakines96,hudson00,oberg09c}.  
Experiments investigating soft X-ray irradiated ice also show a rich chemistry 
as discussed earlier in the context of X-ray desorption 
\citep[see e.g.,][]{andrade10,ciaravella12,jimenezescobar12}.  
Experiments have also been performed to simulate the chemical consequences 
of the direct impact of cosmic-ray particles on ice-covered dust grains 
\citep[see, e.g,][]{kaiser97,bennett07,bennett11}.  
High energy cosmic rays can fully penetrate dust grains creating suprathermal atoms 
which, in turn, transfer energy to the ice mantle and ionise molecules creating 
high-energy electrons ($\sim$~keV) which induce a further cascade of secondary electrons. 
\cite{bennett07} simulate this effect by irradiating 
astrophysical ices (consisting of {\em s-}CO and {\em s-}\ce{CH3OH}) 
with energetic electrons and find that complex molecules, such as 
glycolaldehyde and methyl~formate, can efficiently form, reaching 
relative abundances commensurate with those observed in hot cores.

The radiation processing of grain mantle material should provide an additional 
means for complex species to build up on the grain since the process 
allows the replenishment of grain-surface atoms and radicals 
to take part in further reactions.  
Indeed, to reproduce the observed 
abundances of gas-phase COMs in hot cores ($\sim$~10$^{-8}$~--~10$^{-6}$), 
models need to include radiative processing of ice to produce the 
necessary precursor molecules \citep[see e.g.,][]{garrod08}.  
For example, instead of {\em s-}\ce{CH3OH} effectively becoming the `end state'
of {\em s-}CO via hydrogenation, cosmic-rays, X-rays, and UV photons can break apart 
{\em s-}\ce{CH3OH} into {\em s-}\ce{CH3} and {\em s-}\ce{OH} which are then available to either reform methanol 
or form other species such as {\em s-}\ce{CH3CN} ({\em s-}\ce{CH3} + {\em s-}\ce{CN}) 
or {\em s-}HCOOH ({\em s-}\ce{OH} + {\em s-}\ce{HCO}). 
More complex molecules can also be synthesised by the association of surface 
radicals; a most important case in hot cores is the association of the {\em s-}\ce{HCO} 
and {\em s-}\ce{CH3O} radicals to form methyl formate ({\em s-}\ce{HCOOCH3}). 
\citet{laas11} found that the gas-phase formation of methyl 
formate contributes, at most, to 1.6 \% of the total abundance in hot cores.   
Another case is the surface formation of dimethyl ether ({\em s-}\ce{CH3OCH3}) 
via the association of the surface radicals {\em s-}\ce{CH3} and {\em s-}\ce{CH3O}.  
These and many other cases are discussed in Sect.~\ref{model3} and 
in further detail in \citet{garrod08}.

The results for Model 4 are represented by the yellow-orange lines in 
Figs.~\ref{figure3} and \ref{figure4}.  
In general, we see drops in abundances of gas-phase molecules in the 
molecular layer and disk surface ($\gtrsim$~100~AU) which correlates with 
drops in the abundances of grain-surface molecules at a similar height.  
Conversely, in the disk midplane ($\lesssim$~100~AU), we see an increase in 
the abundances of gas-phase and grain-surface species. 
This is most noticeable for those species which otherwise are unable to 
form efficiently on the grain at low temperatures: {\em s-}\ce{CH3CHO}, {\em s-}\ce{C2H5OH}, 
{\em s-}\ce{CH3OCH3}, {\em s-}\ce{CH3COCH3}, and, {\em s-}\ce{CH3COOH}.  
The grain-surface fractional abundances of these species are enhanced by between 
four (e.g., {\em s-}\ce{CH3CHO}) and nine orders of magnitude 
(e.g., {\em s-}\ce{CH3COCH3}) to values $\sim$~10$^{-10}$~--~10$^{-6}$ (relative to the 
results from Model 3).  
However, this dramatic increase in grain-surface abundance in the disk midplane 
does not necessarily translate to an `observable' gas-phase  
fractional abundance ($\sim$~10$^{-17}$~--~10$^{-13}$).

\begin{figure}
\centering
\subfigure{\includegraphics[width=0.5\textwidth]{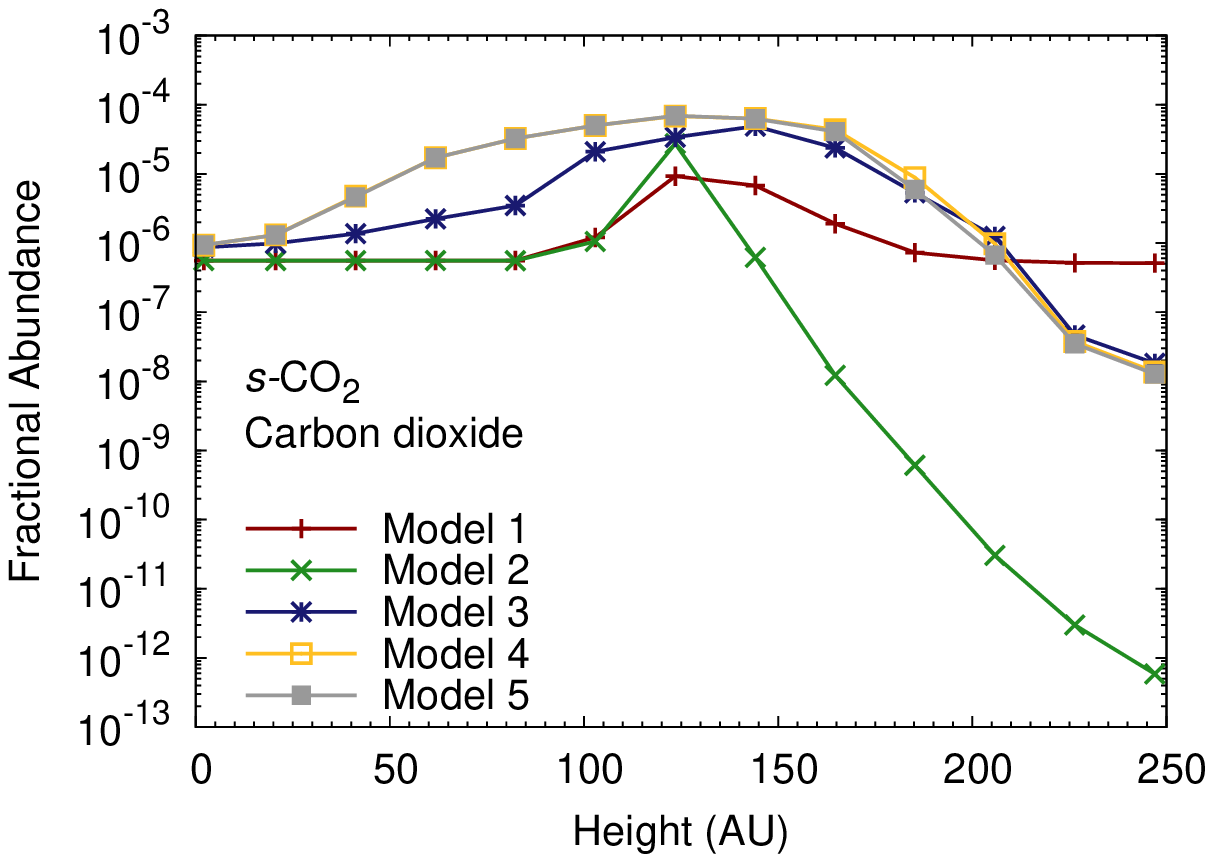}}  
\caption{Fractional abundance (with respect to H nuclei number density)
of {\em s-}\ce{CO2} as a function of disk height, $Z$ at a radius, $R$~=~305~AU. 
The chemical complexity in the model increases from Model 1 to Model 5 
(see Sect.~\ref{verticalresults} for details).}
\label{figure5}
\end{figure}

The internal cosmic-ray-induced photons help to build 
chemical complexity on the ice in the midplane by breaking 
apart the more simple species, e.g., methanol, generating 
radicals which can go on to create more complex species. 
However, the increasing strength of {\em external} 
UV photons and X-rays towards the disk surface acts to break down this complexity 
and the grain-surface chemistry favours the production of more simple 
ice species, e.g., {\em s-}\ce{CO2} and {\em s-}\ce{H2O}.  
In fact, we find that the main repository of carbon and oxygen 
in the molecular layer is {\em s-}\ce{CO2}.  
In Fig.~\ref{figure5} we present the fractional abundance of 
{\em s-}\ce{CO2} as a function of disk height for our reduced grid at 
R~=~305~AU for each of our chemical models. 
In Model~4, {\em s-}\ce{CO2} reaches a peak fractional abundance of 
$\sim$~10$^{-4}$ at a height of $\approx$~150~AU, 
which corresponds to the point where there is a sharp decrease in the 
abundance of other C- and O-containing complex molecules.
  
{\em s-}\ce{CO2} is formed on grain-surfaces via the reaction,
\begin{equation}
s\mbox{-}\ce{CO} + s\mbox{-}\ce{OH} \longrightarrow s\mbox{-}\ce{CO2} + s\mbox{-}\ce{H}.  
\end{equation} 
We find that the rate for this reaction, under the physical conditions in the molecular layer, 
is marginally faster than the rate for the rehydrogenation of {\em s-}\ce{OH}.  
Over time, {\em s-}\ce{CO2} grows at the expense of {\em s-}\ce{CO} and {\em s-}\ce{H2O}, and indeed, 
other O- and C-containing species. 
\ce{CO2} has a smaller cross-section for photodissociation than \ce{H2O} 
at longer wavelengths \citep[$>$~1200~$\AA$,][]{vandishoeck06} and so 
is more photostable in the molecular layer of the disk where the UV radiation 
field is softer than that in the upper disk.  
Historically, the above reaction has been included in chemical networks with a small 
reaction barrier of $\approx$~80~K based on the gas-phase reaction potential energy surface 
\citep[see, e.g.,][]{smith88,ruffle01,garrod08,garrod11}.  
Recent experiments suggest the effective barrier for this reaction is 
closer to $\approx$~400~K \citep{noble11}    
and we have adopted the higher barrier in the work presented here. 
Lowering the barrier to 80~K further increases the production of {\em s-}\ce{CO2}, 
further decreasing the abundances of other  O- and C-containing species.  
\ce{CO2} ice has been observed in many different environments with 
a typical abundance $\sim$~30\% that of water ice; however, its exact 
formation mechanism under cold interstellar conditions remains a puzzle.  
Recently, \citet{poteet13} observed \ce{CO2} ice in absorption towards 
the low-mass protostar, HOPS-68, and analysis of their data revealed the 
\ce{CO2} was contained within an ice matrix consisting of almost pure 
\ce{CO2} ice ($\approx$~90\%).  
The authors postulate that HOPS-68 has a flattened envelope morphology, with a  
high concentration of material within $\approx$~10~AU of the central star, thereby 
explaining the lack of primordial hydrogen-rich ices along the line of sight.  
They also propose a scenario where an energetic event has led to the evaporation 
of the primordial grain mantle and subsequent cooling and recondensation has 
led to the production of \ce{CO2} ice in a H-poor ice mantle.  
Certainly, our results suggest that 
the reprocessing of ice species by UV photons may play a role in driving 
the production of \ce{CO2} ice at the expense of other typical grain mantle species, 
such as, \ce{CO} and \ce{H2O}.  

Concerning commonly observed gas-phase species in disks, we find the 
inclusion of radiation processing of ice has an effect mainly on those 
species for which precursor species remain frozen out at this radius.
For CO, we do not see a strong effect because CO does not rely on grain-surface chemistry 
for its formation. 
In the molecular and surface layers of the disk, CO exists predominantly in the 
gas phase as it is able to thermally desorb from grain surfaces.   
We see a small decrease in the CO abundance (on the order of a factor of a few) 
in Model 4 relative to Model 3
between a height of 50 AU and 150 AU as CO is driven into {\em s-}\ce{CO2}(discussed above).    
We also do not see a strong effect on the abundance of carbon monosulphide, CS, since this 
species reaches its peak fractional abundance ($\sim$~10$^{-8}$) in the surface layers of 
the disk (above a height of 150~AU) only.

In contrast, for CN (and its precursor species, HCN),  
we do see a similar drop in abundance around 150~AU in Model 4 (compared with 
Model 3) since HCN mainly exists as ice on the grain at this disk radius.      
The ethynyl radical, \ce{C2H} exhibits a similar behaviour to CN which is related to 
acetylene, \ce{C2H2}, also existing primarily as ice, albeit at a lower 
abundance than for more abundant ice species, such as {\em s-}HCN and {\em s-}\ce{H2O}.  
{\em s-}\ce{C2H2} can be hydrogenated on the grain to form {\em s-}\ce{C2H3} (and beyond). 

\subsubsection{Model 5: Reactive desorption}
\label{model5}

In Model 5 we also include reactive desorption and 
the results are represented by the gray lines 
in Figs.~\ref{figure3} and \ref{figure4}.  
In general, we find reactive desorption helps release further 
gas-phase COMs in the disk midplane leading to an enhancement 
of around one to two orders of magnitude.    
However, this enhancement is not necessarily sufficient to 
increase the molecular column density to an observable value.  
For example, acetaldehyde (\ce{CH3CHO}) is 
enhanced from a fractional abundance of $\sim$~10$^{-13}$ in Model~4 
to $\approx$~2~$\times$~10$^{-12}$ in Model~5.  
We also see a slight increase (around a factor of a few) 
in the grain-surface abundances of several species in 
the disk midplane; {\em s-}\ce{CH3CHO}, {\em s-}\ce{CH3NH2}, {\em s-}\ce{C2H5OH}, {\em s-}\ce{CH3OCH3}, 
{\em s-}\ce{CH3COCH3}, and {\em s-}\ce{CH3COOH}.  
At heights $Z$~$\gtrsim$~100~AU, the results from Model 4 
and Model 5 are similar.  

The inclusion of reactive desorption also helps increase the abundance 
of gas-phase CS, CN, HCN, and \ce{C2H} in the disk midplane.  
However, this enhancement is not sufficient to significantly 
increase the observable column density.  
This is because these species can form in the gas phase and 
thus are not reliant on grain-surface association reactions.  
For CO, reactive desorption 
is unimportant as, again, grain-surface chemistry does not contribute 
significantly to the formation of CO on the grain.  
In the disk midplane, cosmic-ray-induced desorption is the 
primary mechanism for releasing {\em s-}CO into the gas phase.

\subsection{Full disk model}
\label{fulldiskmodel}

\begin{figure*}[!ht]
\subfigure{\includegraphics[width=0.33\textwidth]{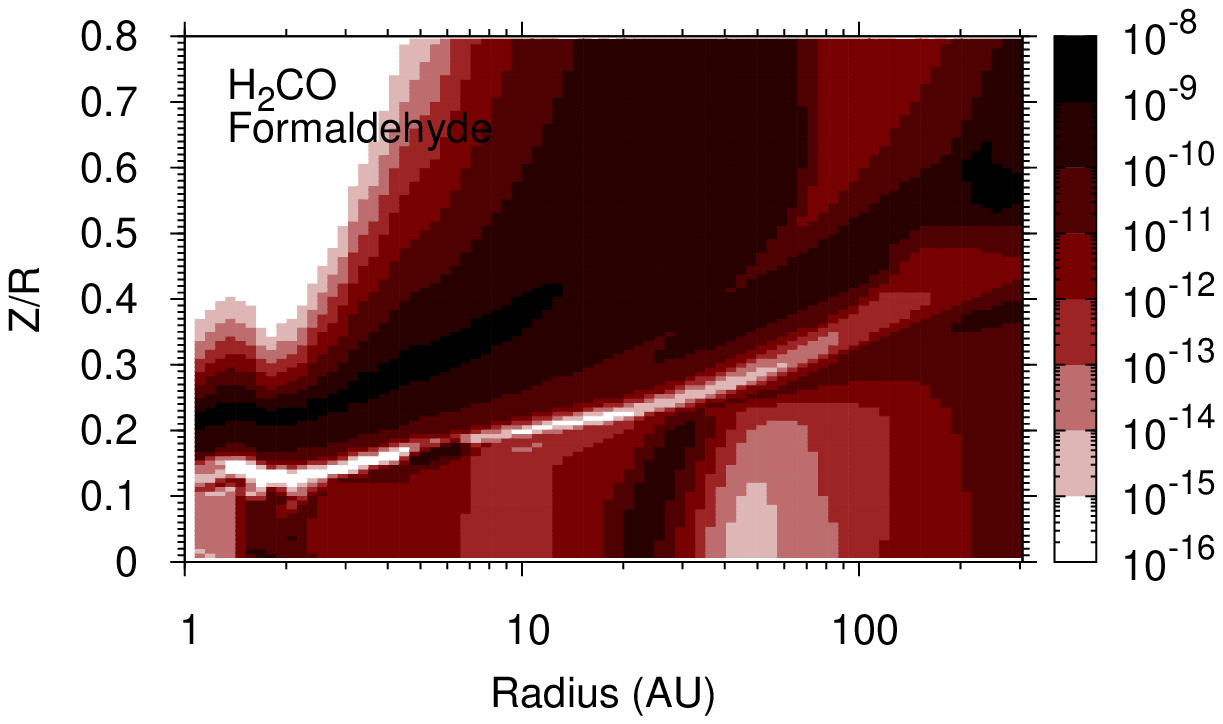}}
\subfigure{\includegraphics[width=0.33\textwidth]{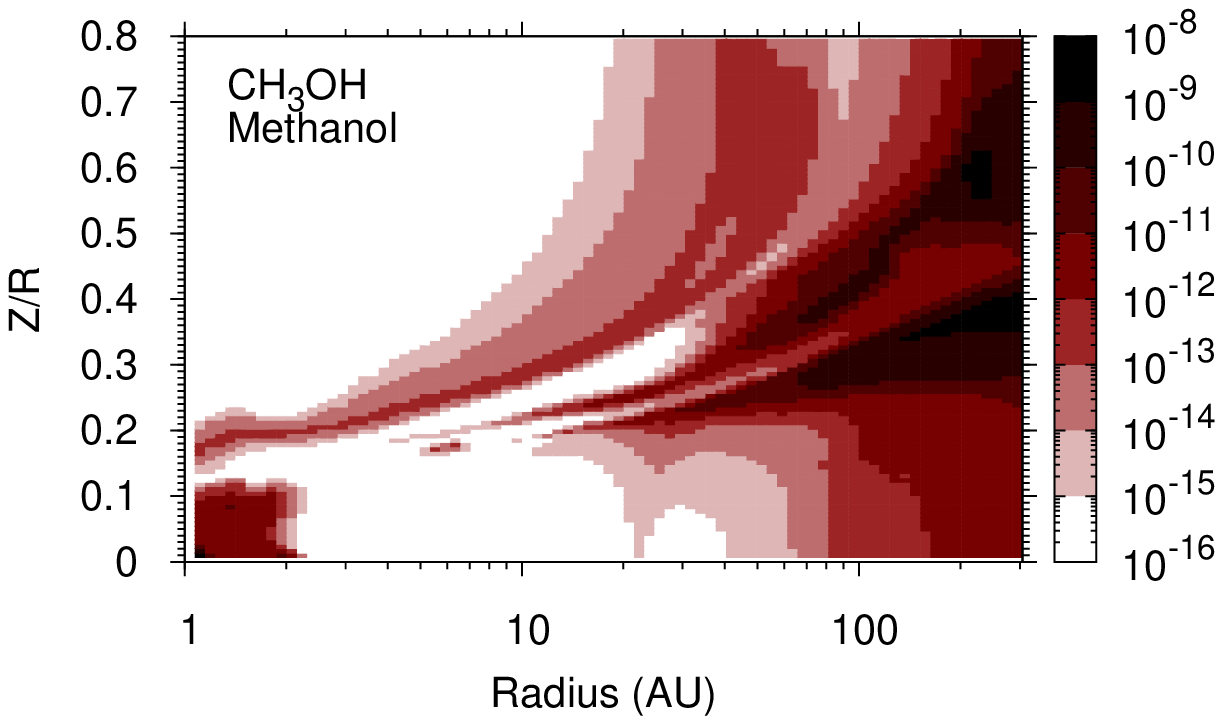}}
\subfigure{\includegraphics[width=0.33\textwidth]{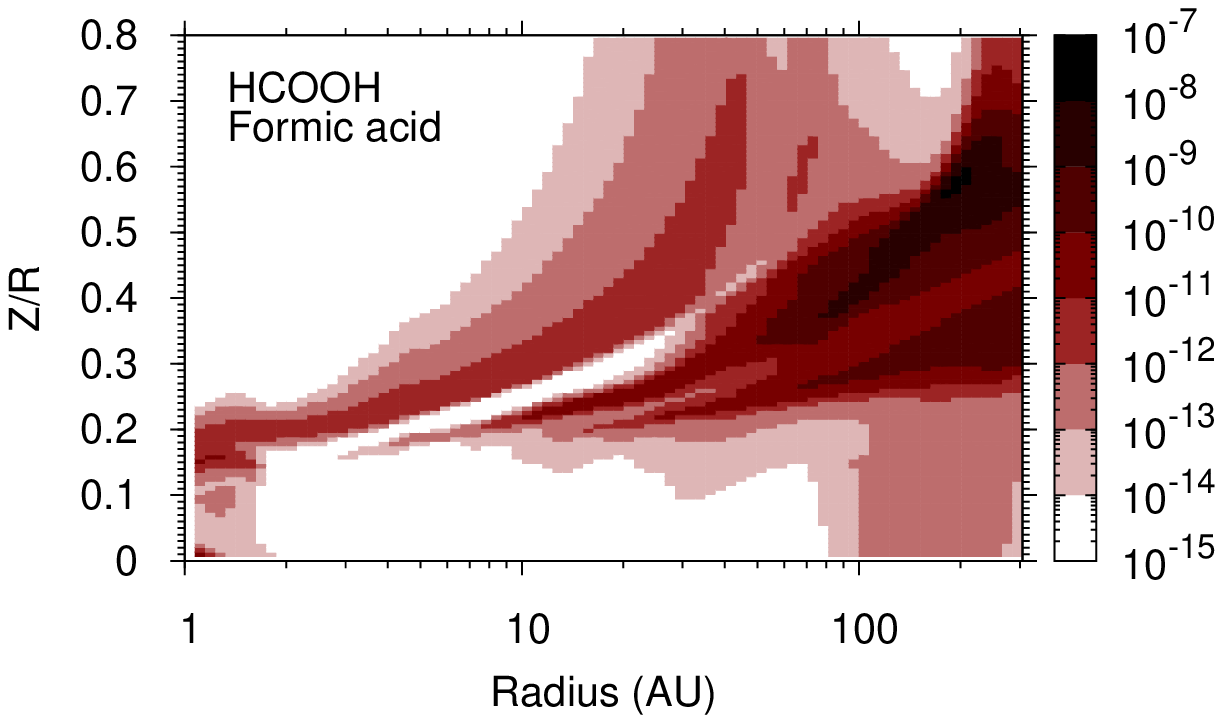}}
\subfigure{\includegraphics[width=0.33\textwidth]{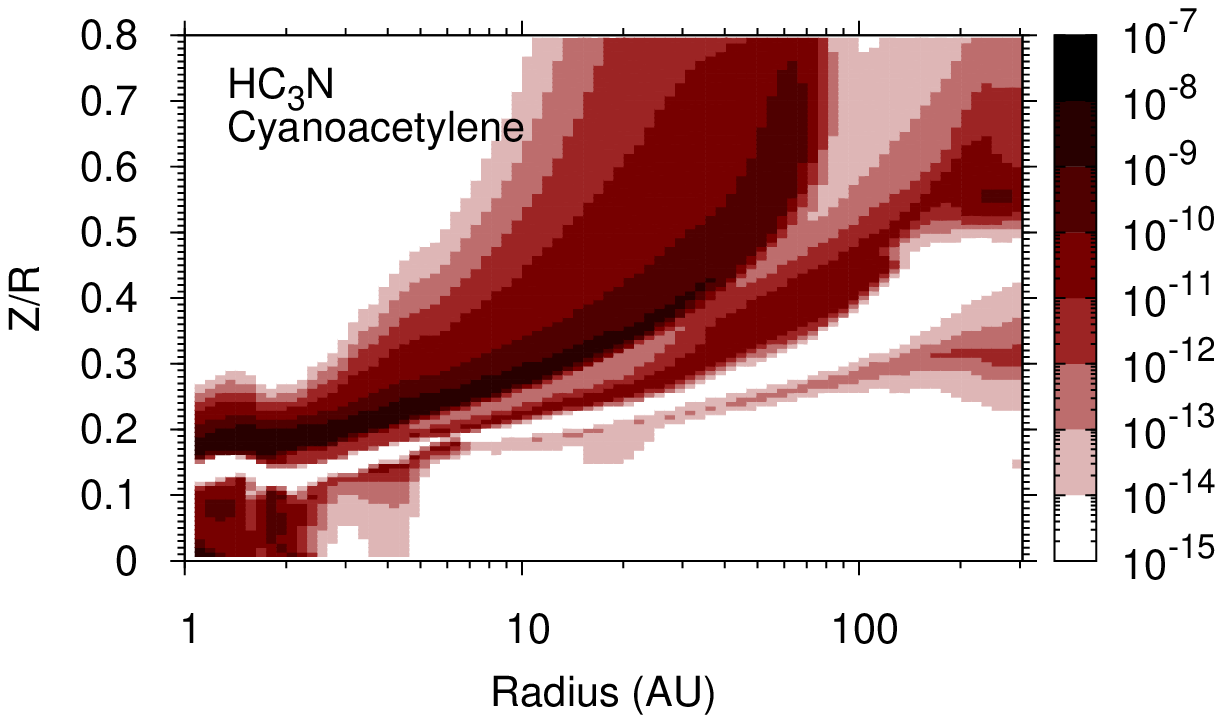}}
\subfigure{\includegraphics[width=0.33\textwidth]{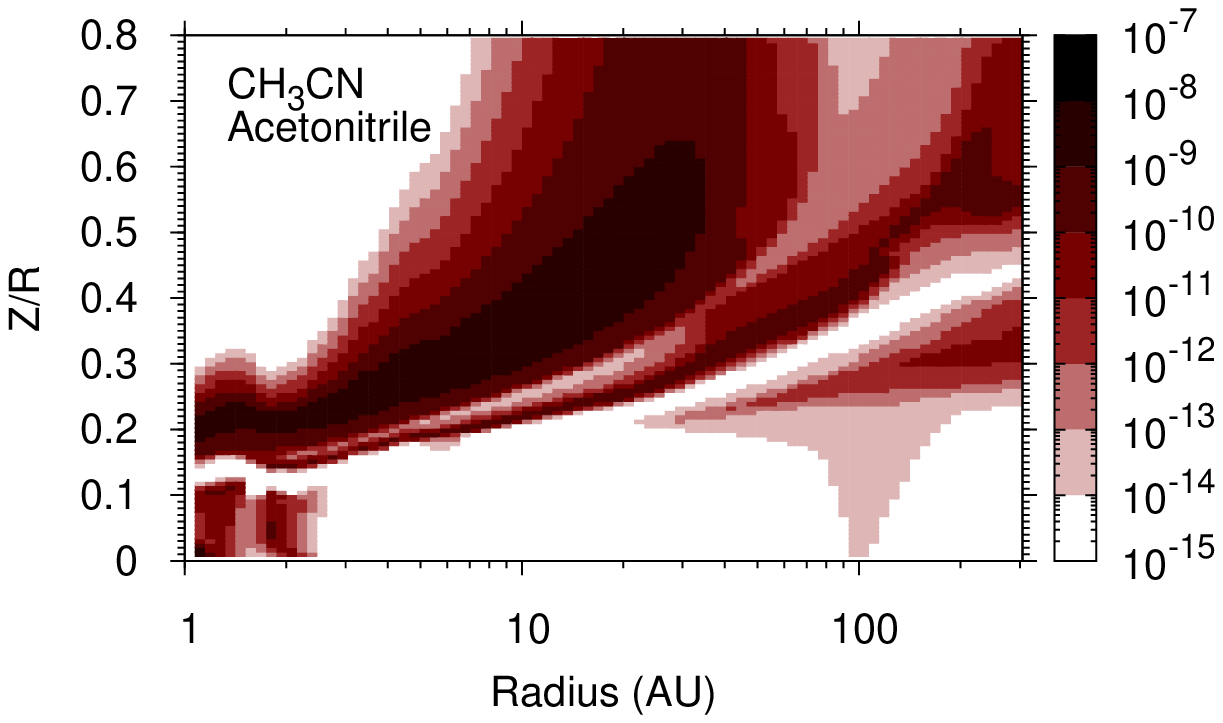}}
\subfigure{\includegraphics[width=0.33\textwidth]{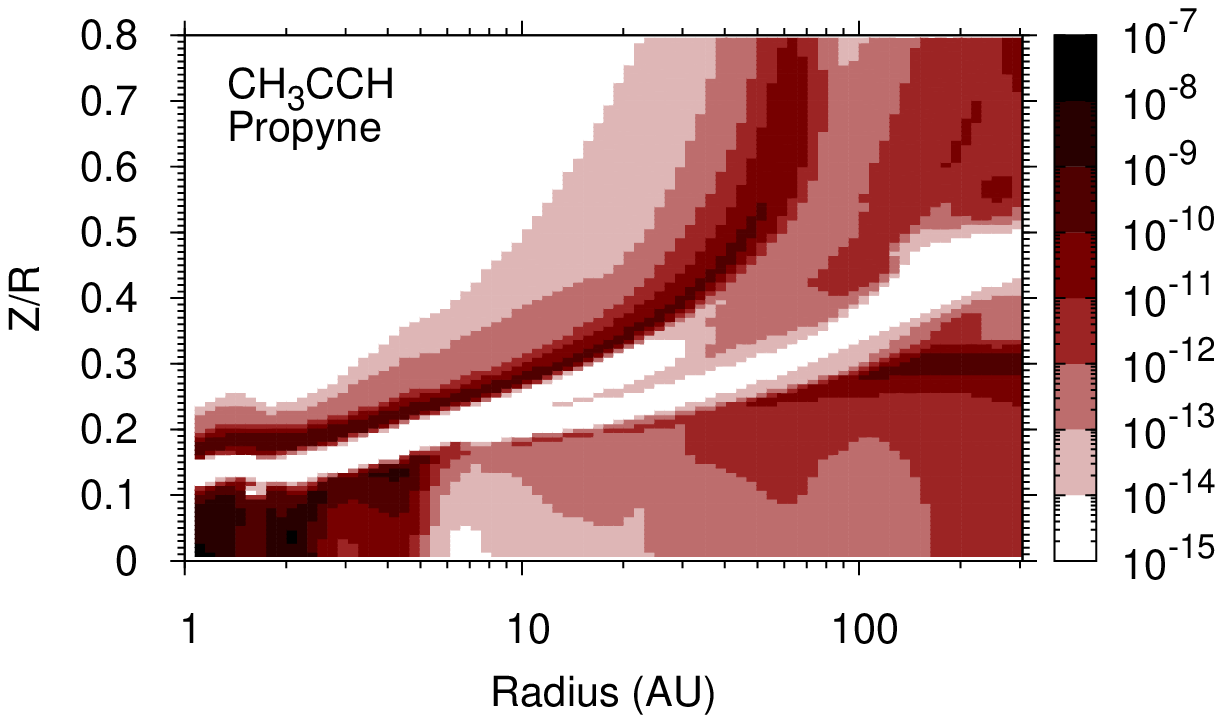}}
\subfigure{\includegraphics[width=0.33\textwidth]{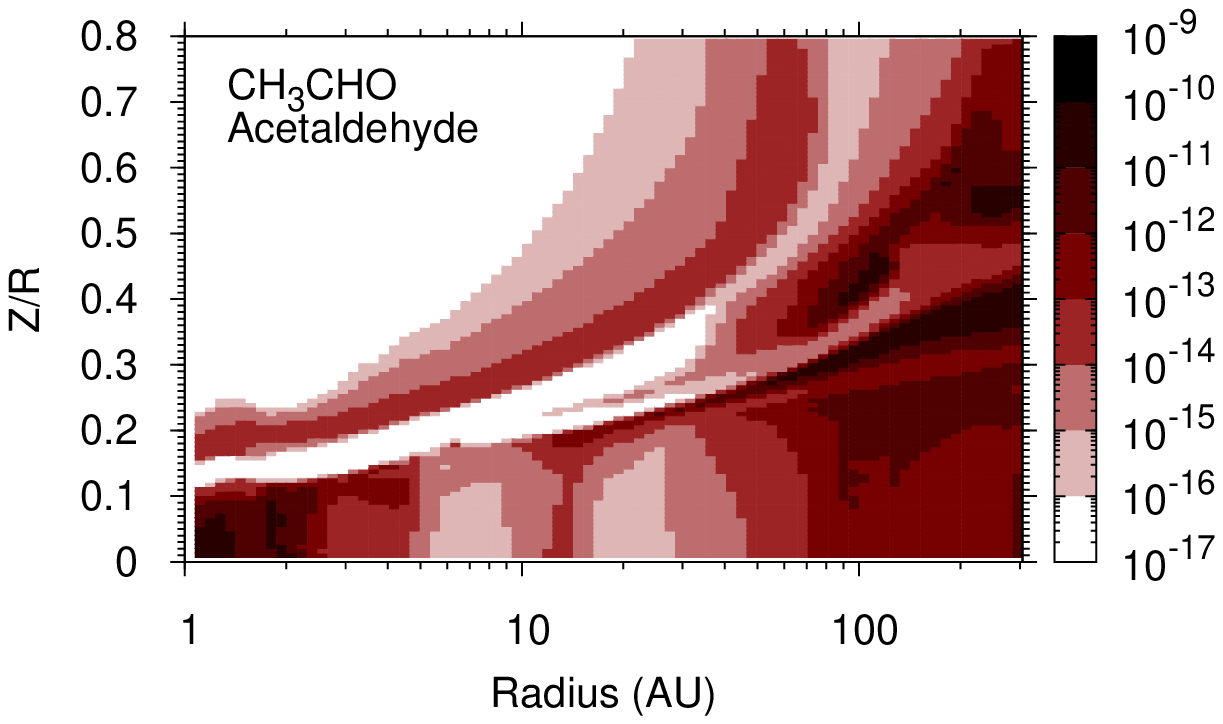}}
\subfigure{\includegraphics[width=0.33\textwidth]{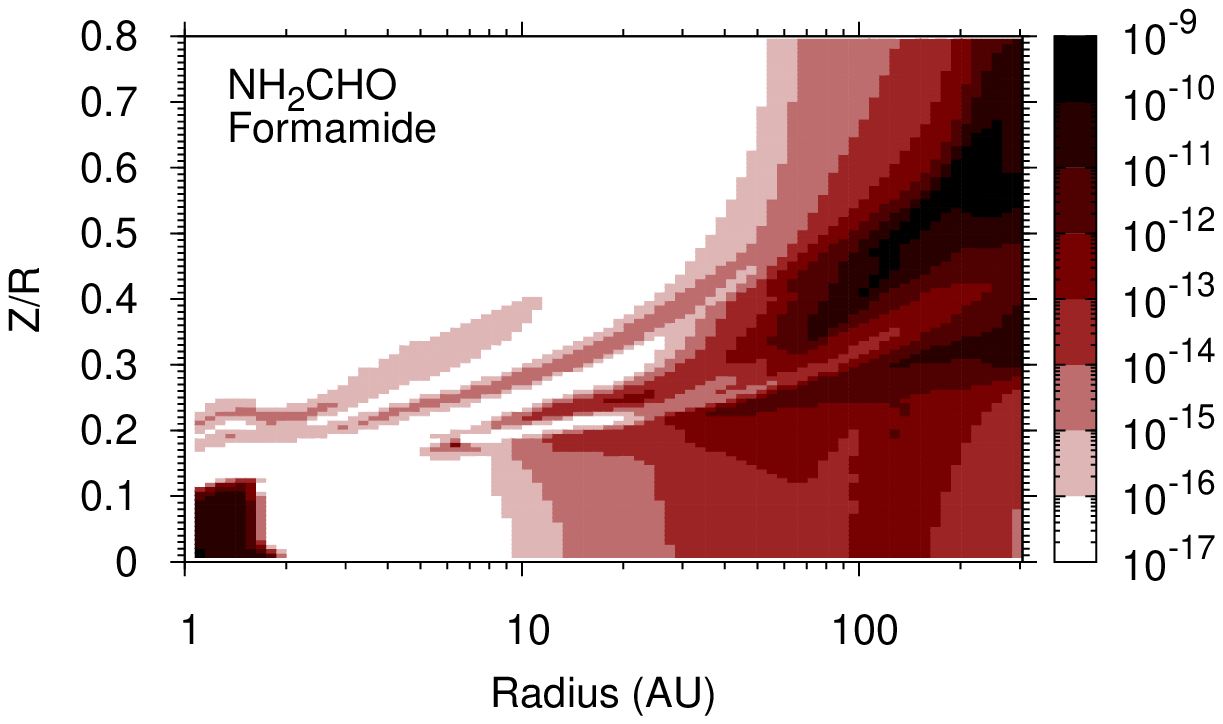}}
\subfigure{\includegraphics[width=0.33\textwidth]{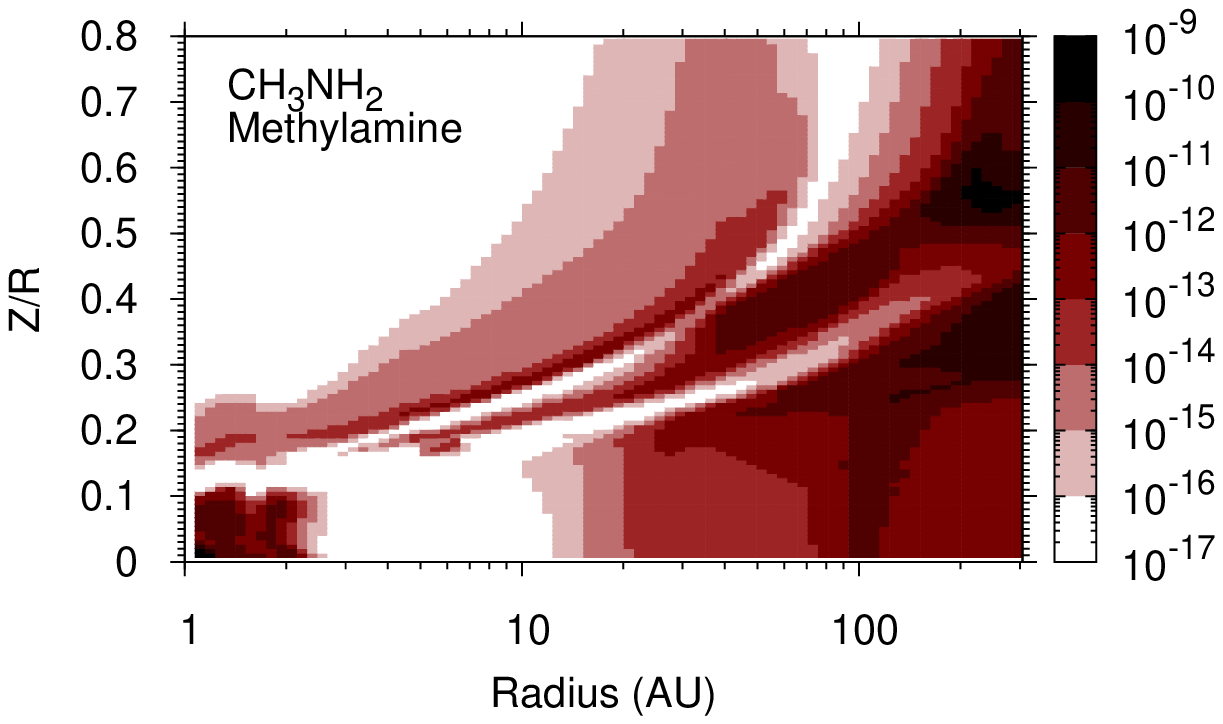}}
\subfigure{\includegraphics[width=0.33\textwidth]{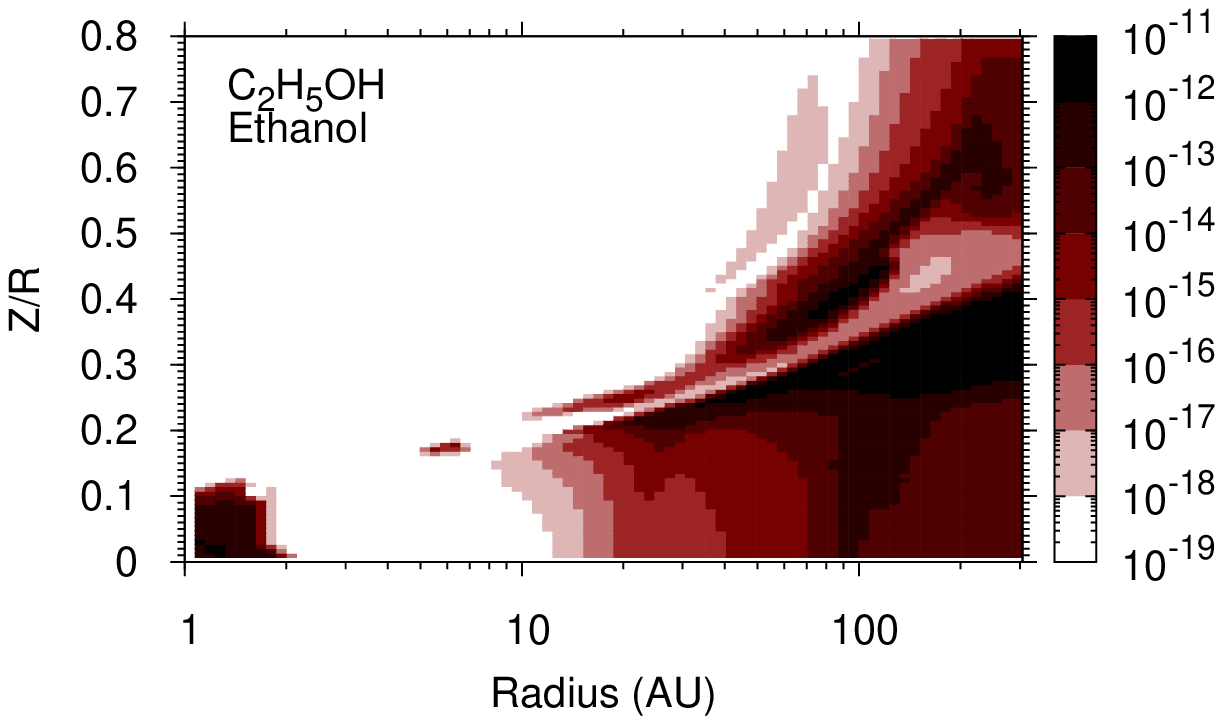}}
\subfigure{\includegraphics[width=0.33\textwidth]{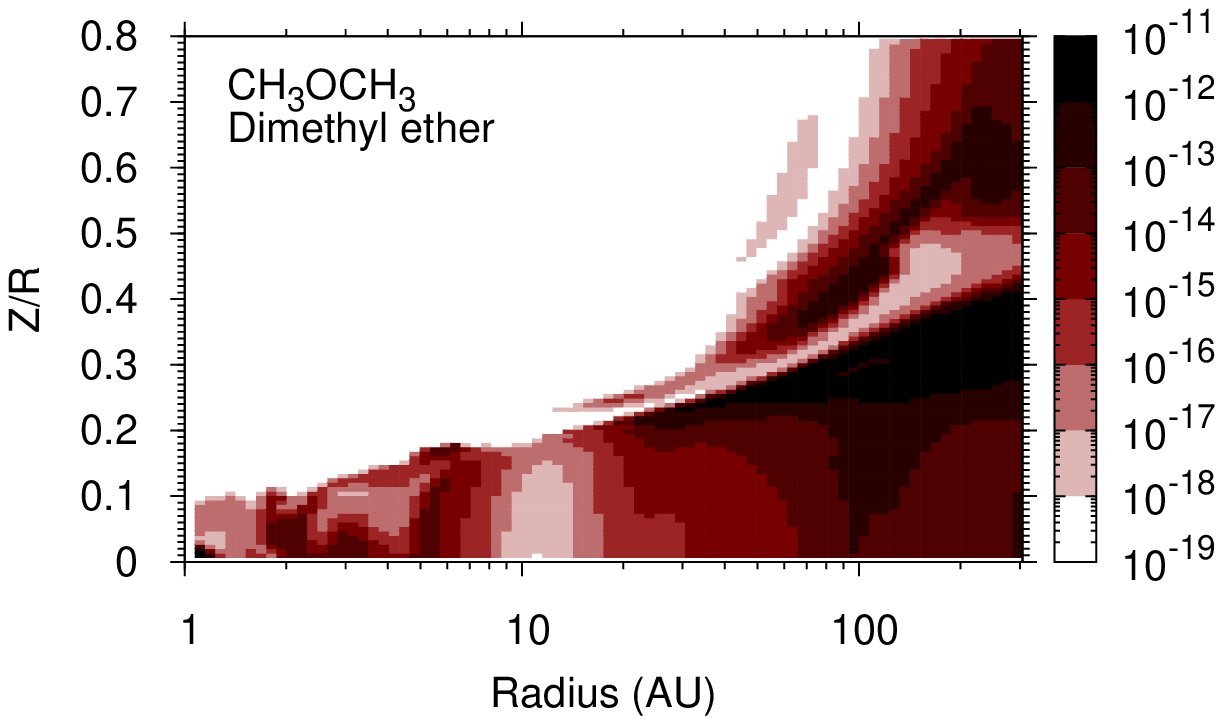}}
\subfigure{\includegraphics[width=0.33\textwidth]{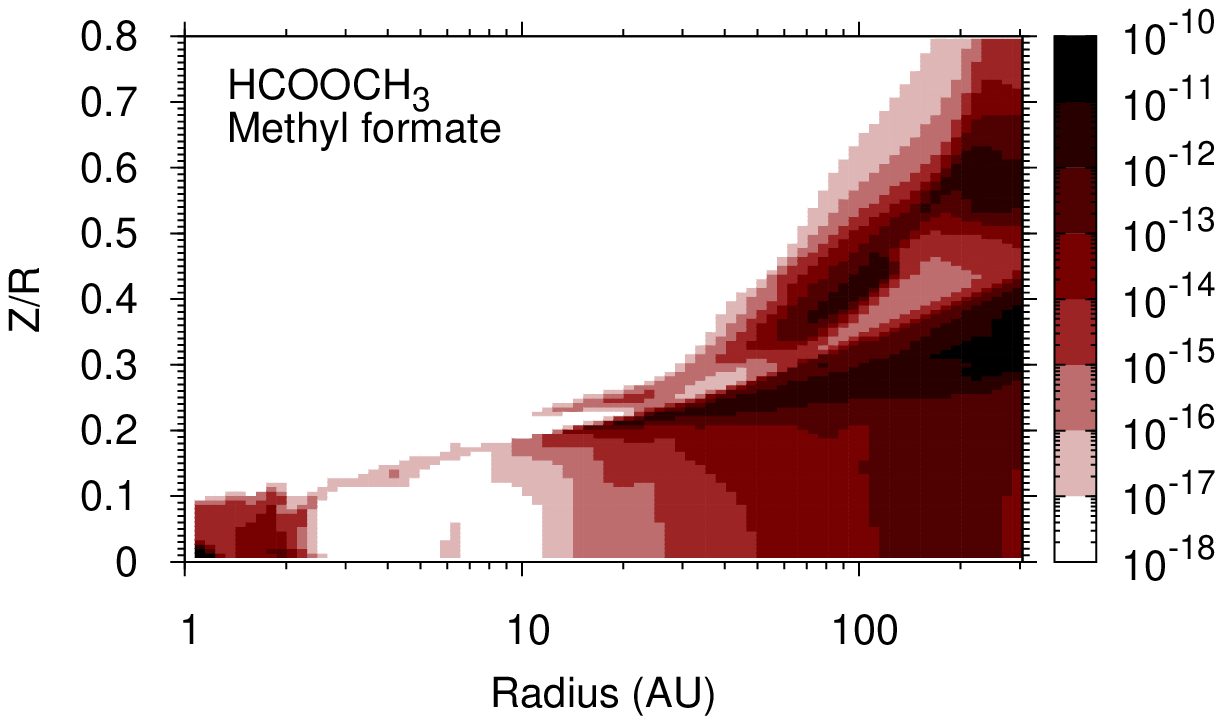}}
\subfigure{\includegraphics[width=0.33\textwidth]{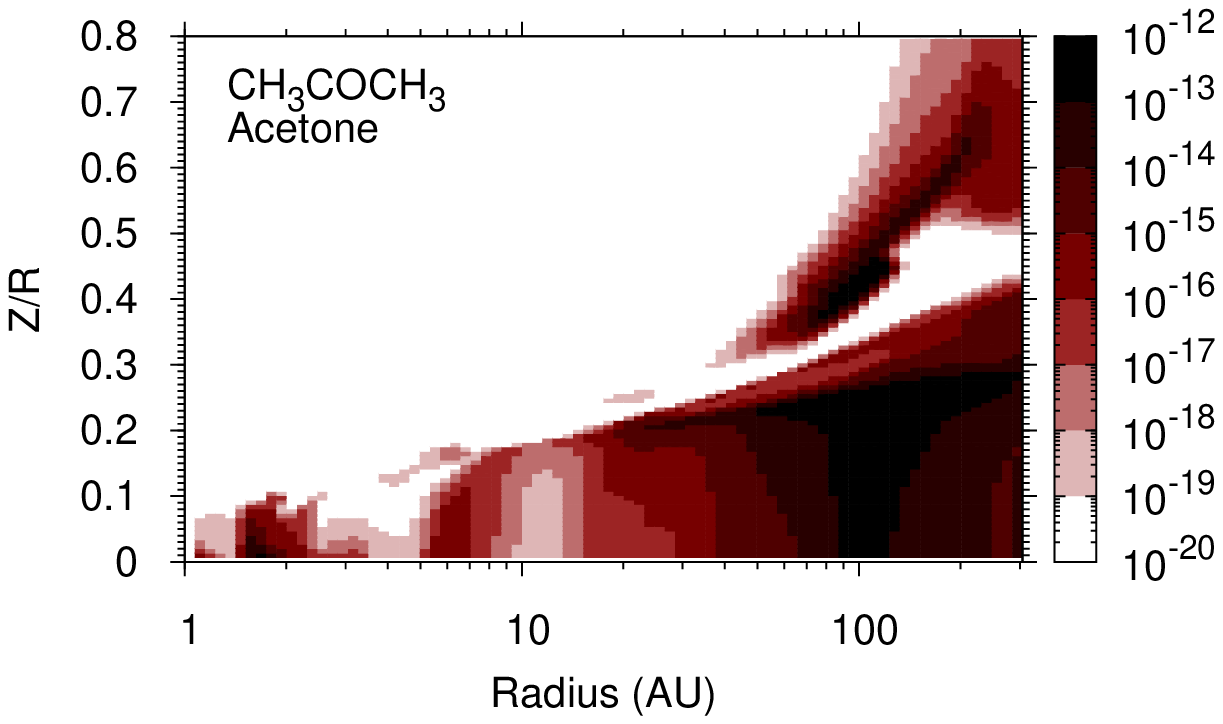}}
\subfigure{\includegraphics[width=0.33\textwidth]{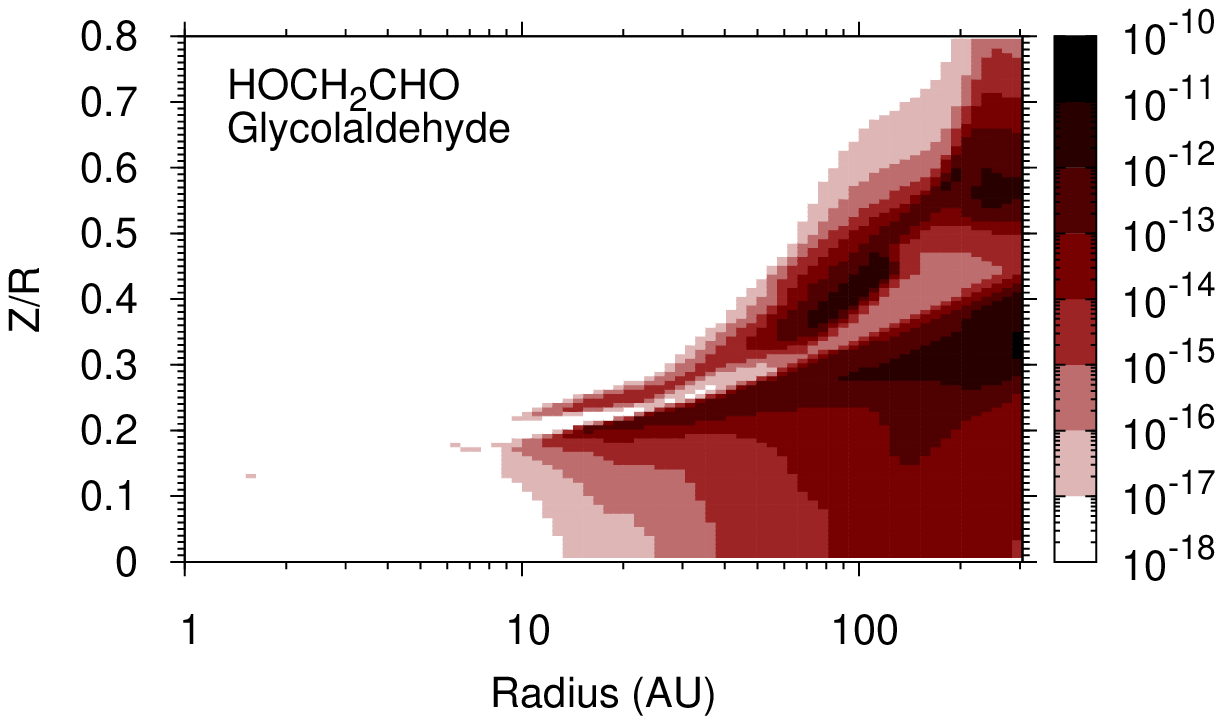}}
\subfigure{\includegraphics[width=0.33\textwidth]{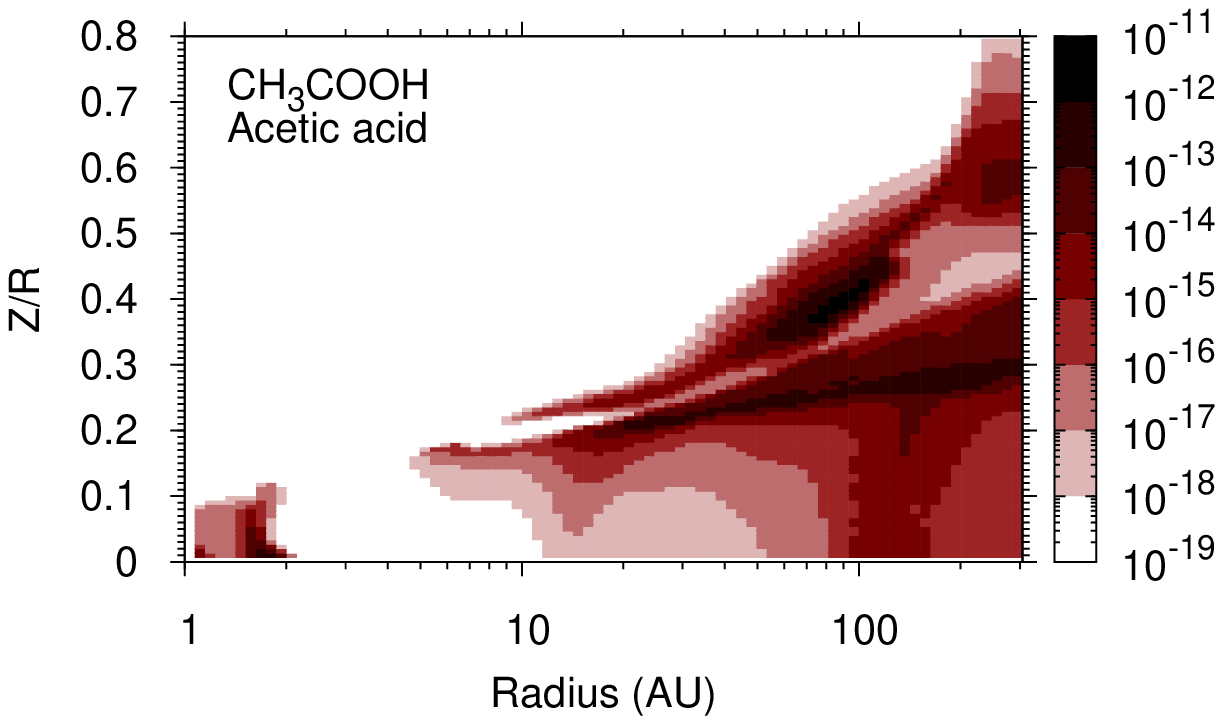}}
\caption{Fractional abundance of gas-phase molecules with respect to total H nuclei number density as a 
function of disk radius, $R$, and height, $Z$.}
\label{figure6}
\end{figure*}

\begin{figure*}[!ht]
\subfigure{\includegraphics[width=0.33\textwidth]{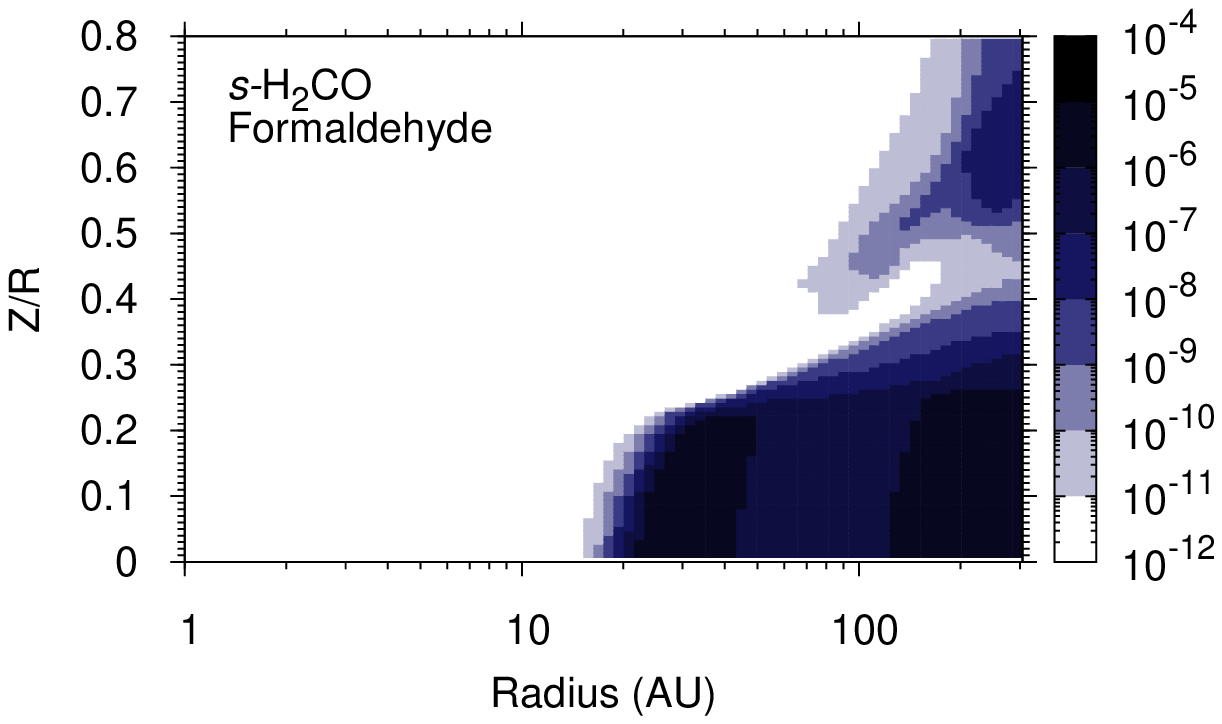}}
\subfigure{\includegraphics[width=0.33\textwidth]{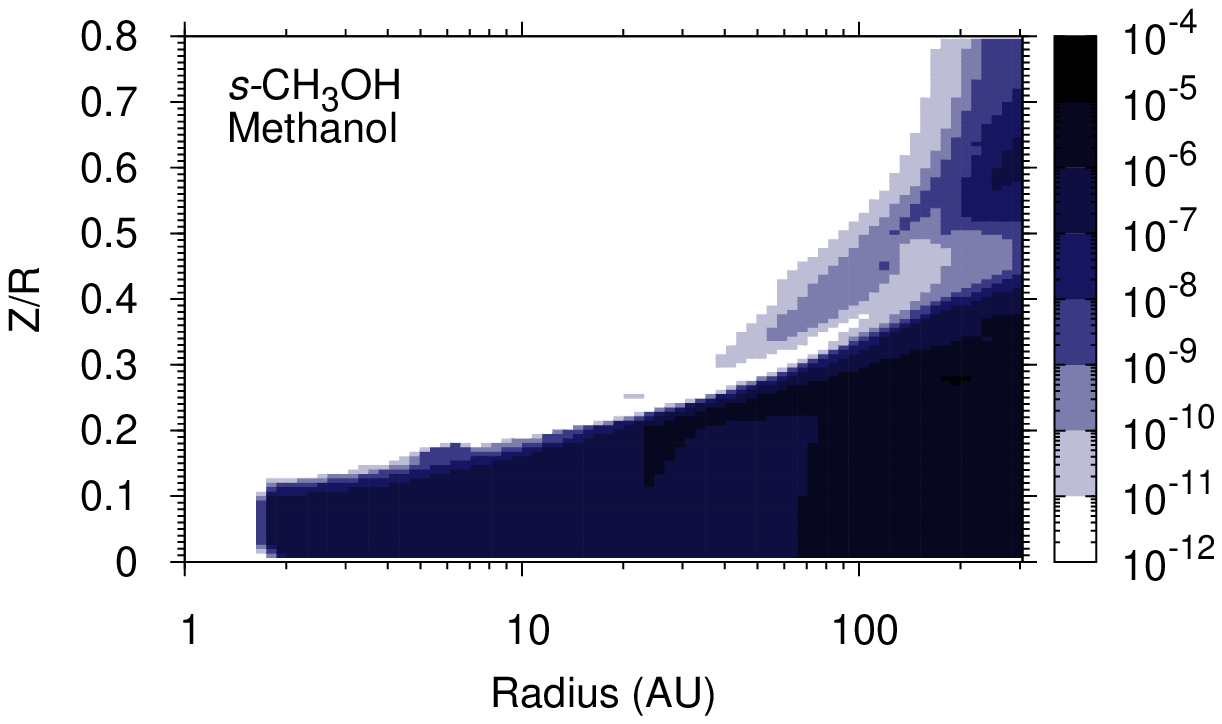}}
\subfigure{\includegraphics[width=0.33\textwidth]{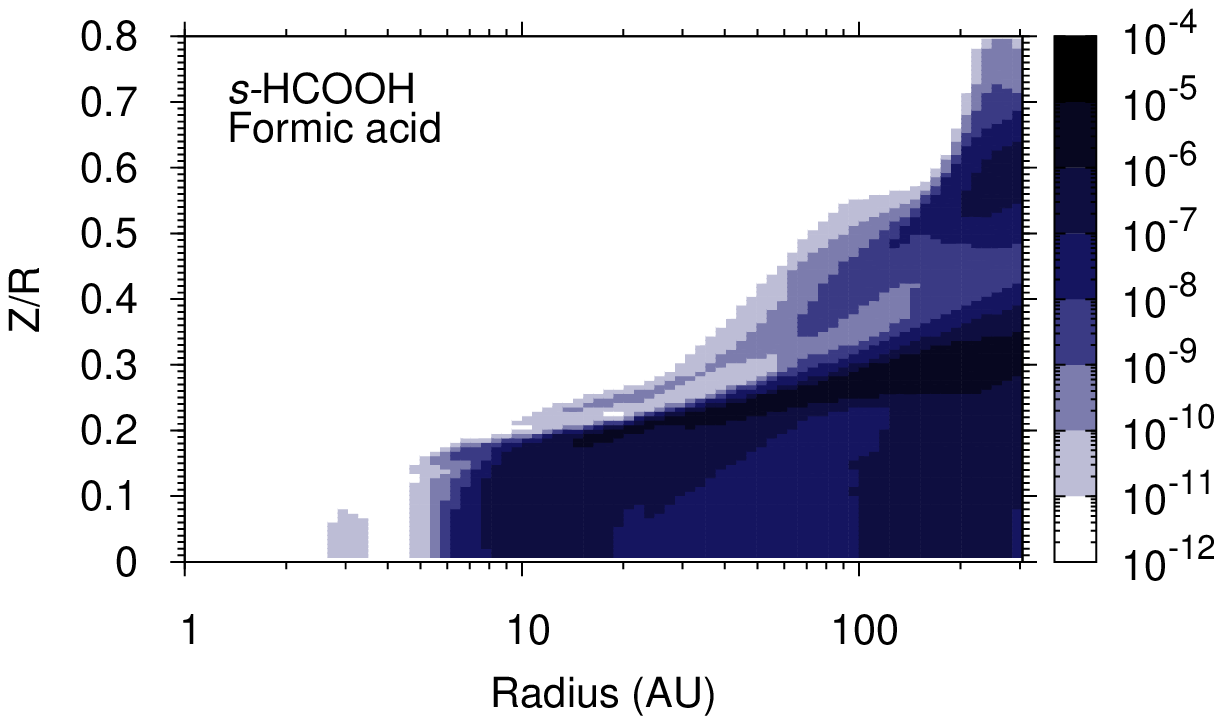}}
\subfigure{\includegraphics[width=0.33\textwidth]{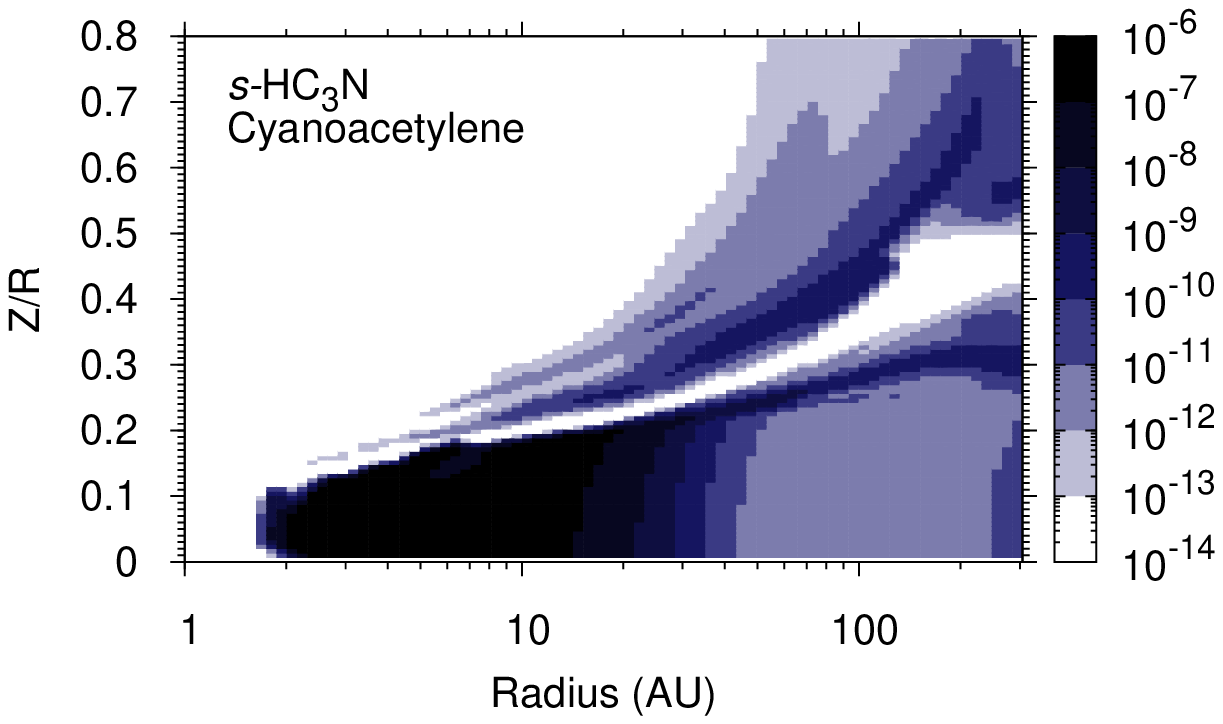}}
\subfigure{\includegraphics[width=0.33\textwidth]{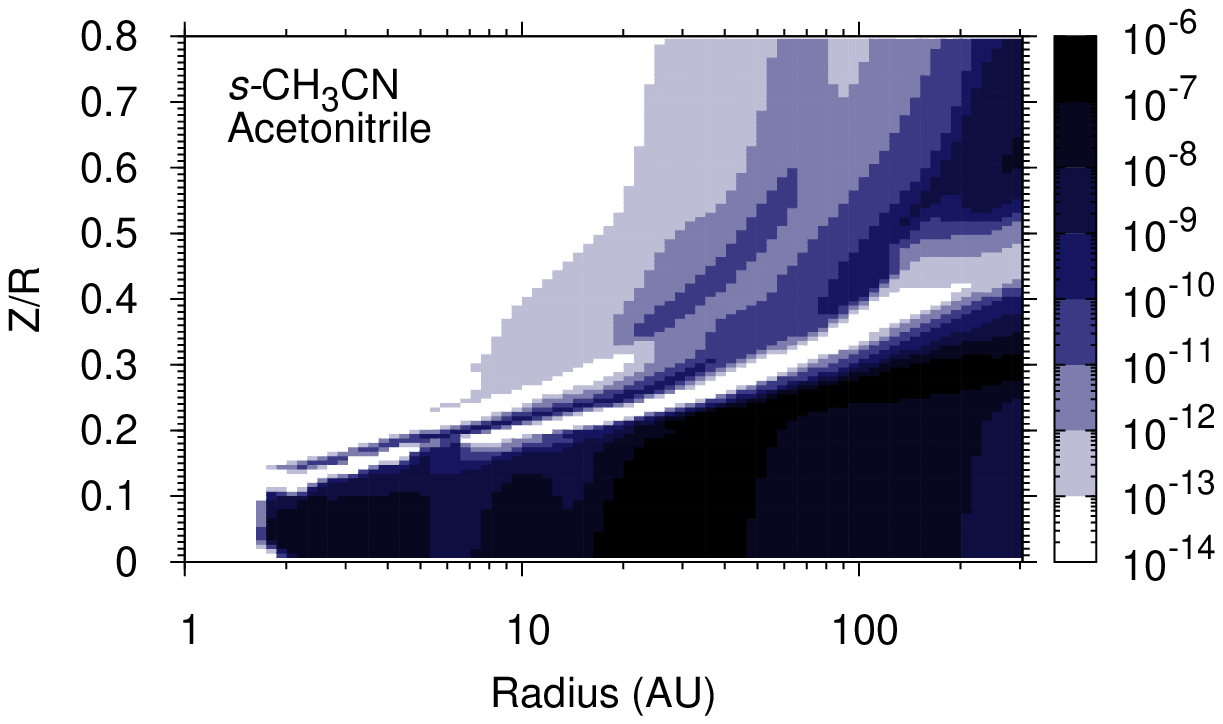}}
\subfigure{\includegraphics[width=0.33\textwidth]{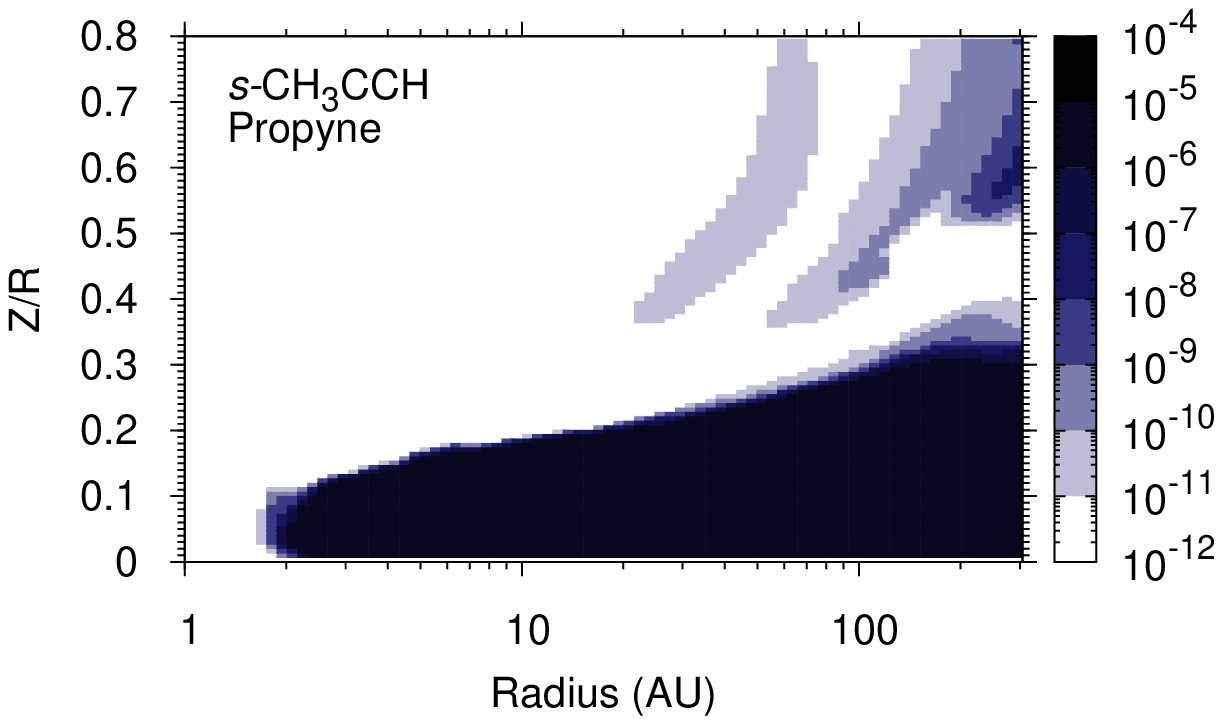}}
\subfigure{\includegraphics[width=0.33\textwidth]{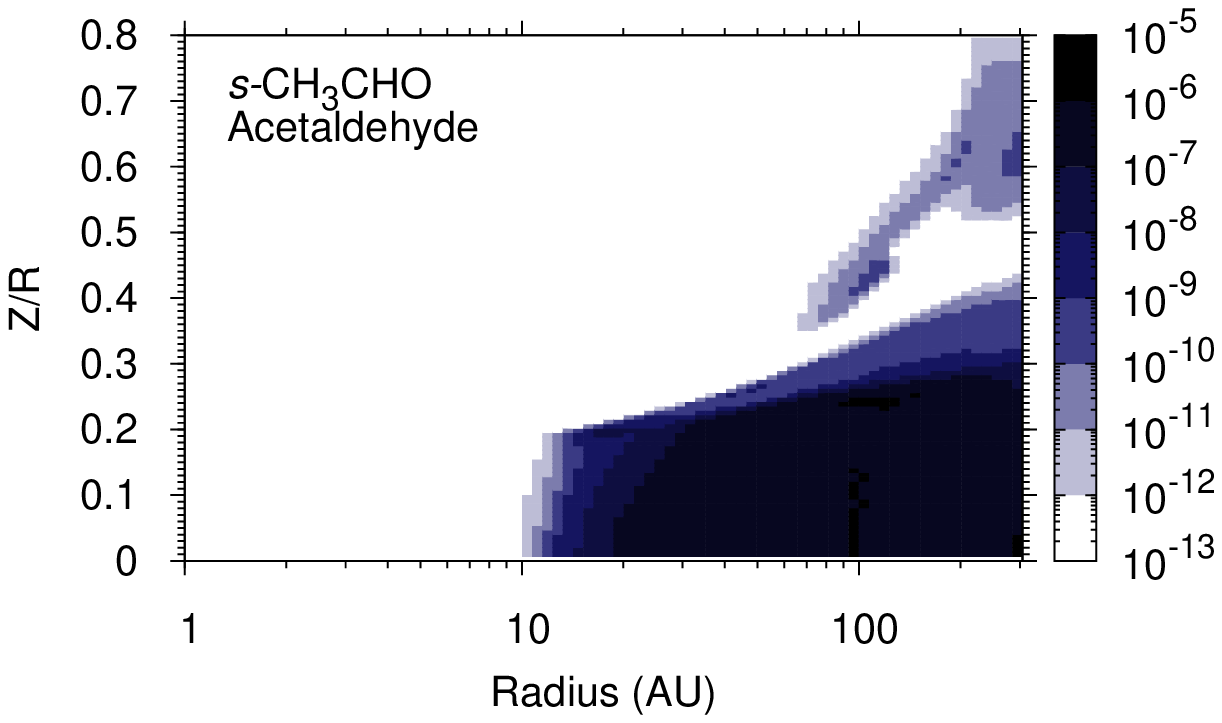}}
\subfigure{\includegraphics[width=0.33\textwidth]{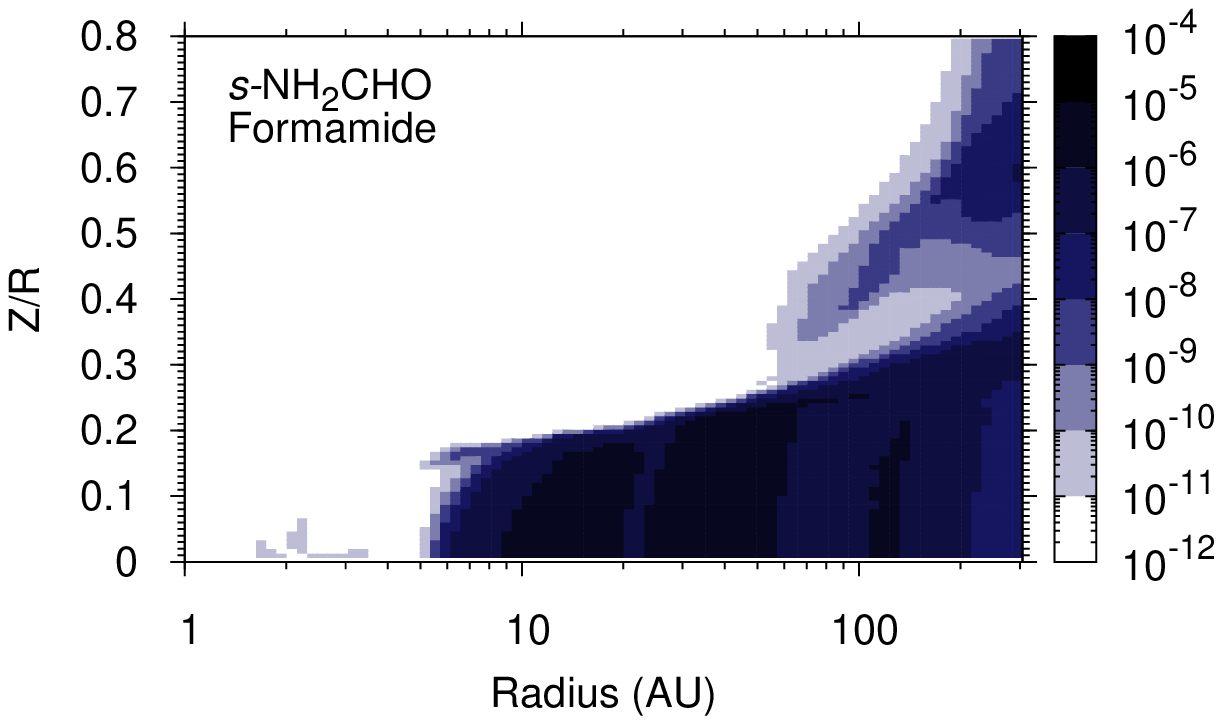}}
\subfigure{\includegraphics[width=0.33\textwidth]{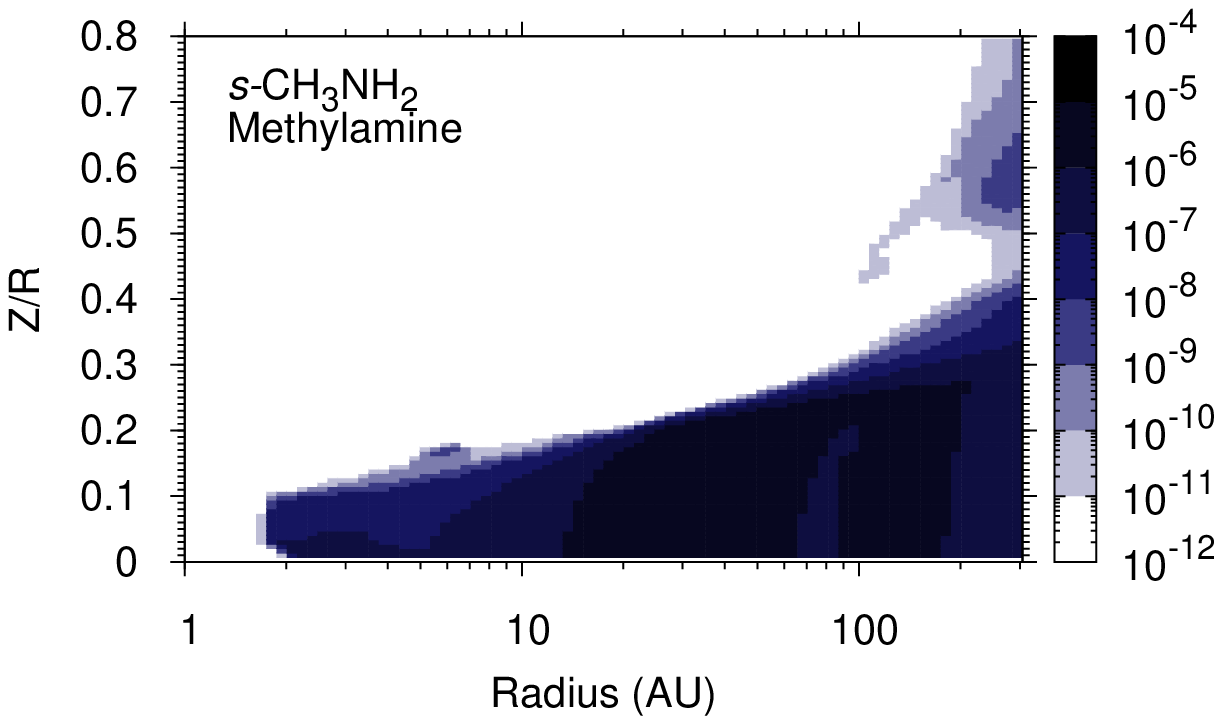}}
\subfigure{\includegraphics[width=0.33\textwidth]{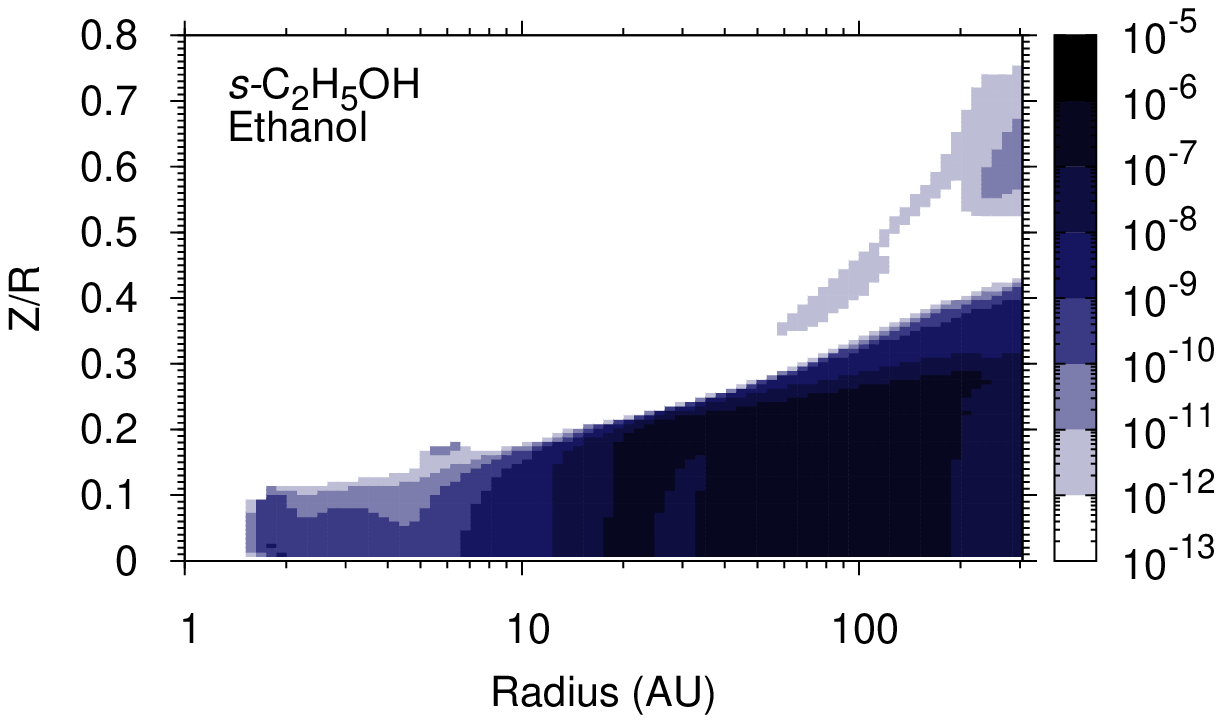}}
\subfigure{\includegraphics[width=0.33\textwidth]{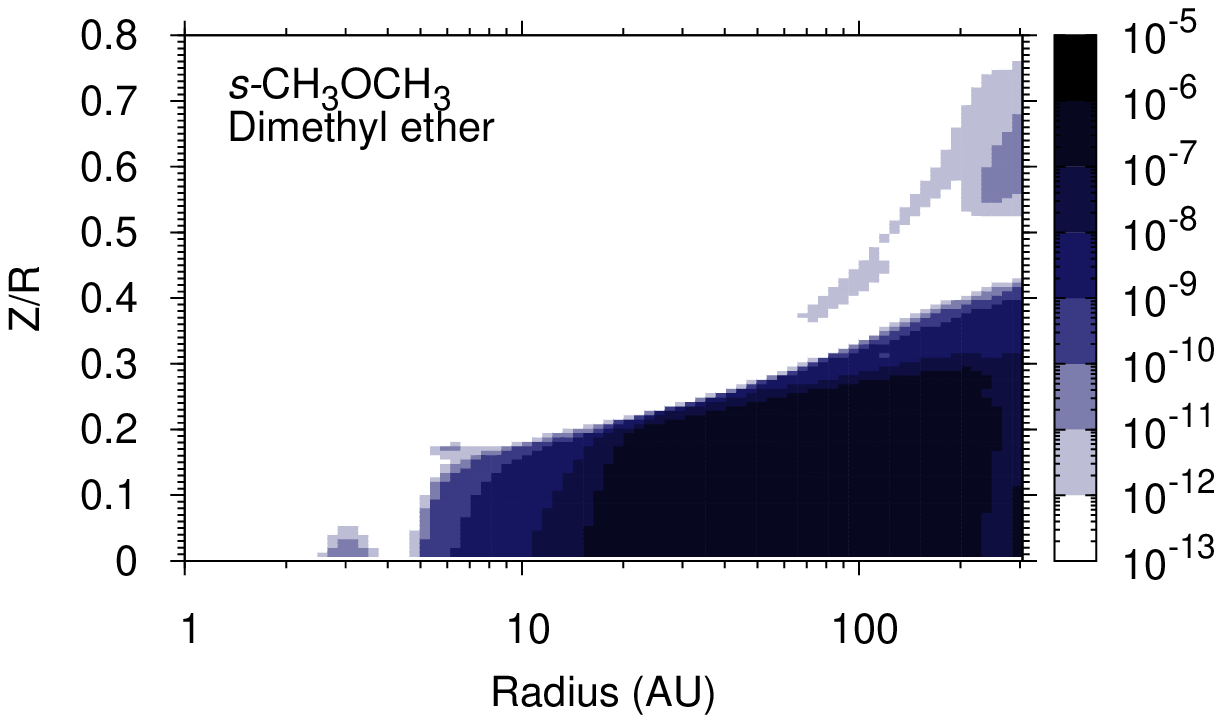}}
\subfigure{\includegraphics[width=0.33\textwidth]{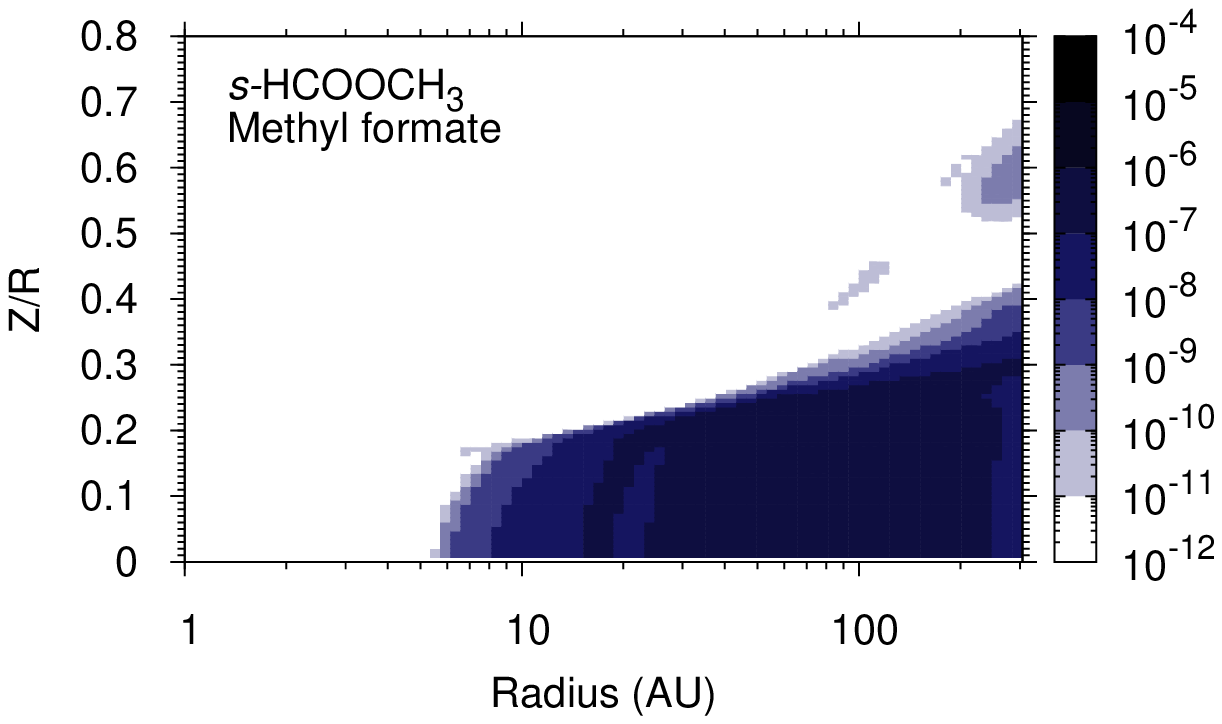}}
\subfigure{\includegraphics[width=0.33\textwidth]{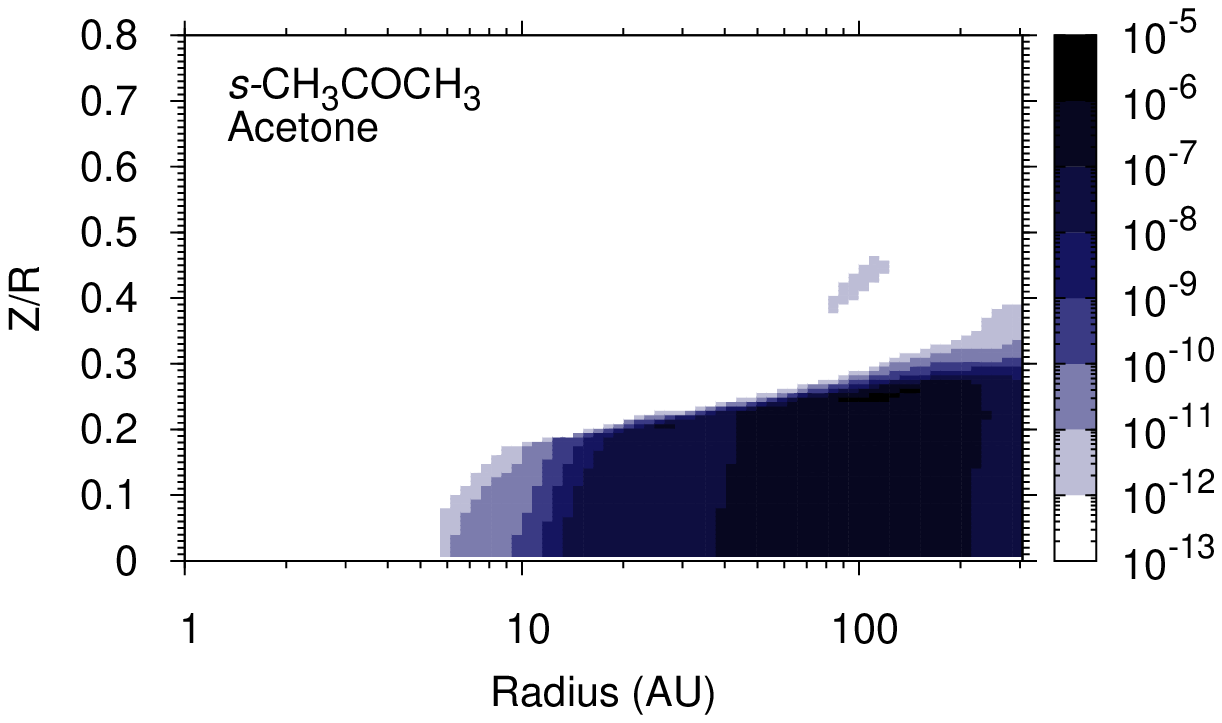}}
\subfigure{\includegraphics[width=0.33\textwidth]{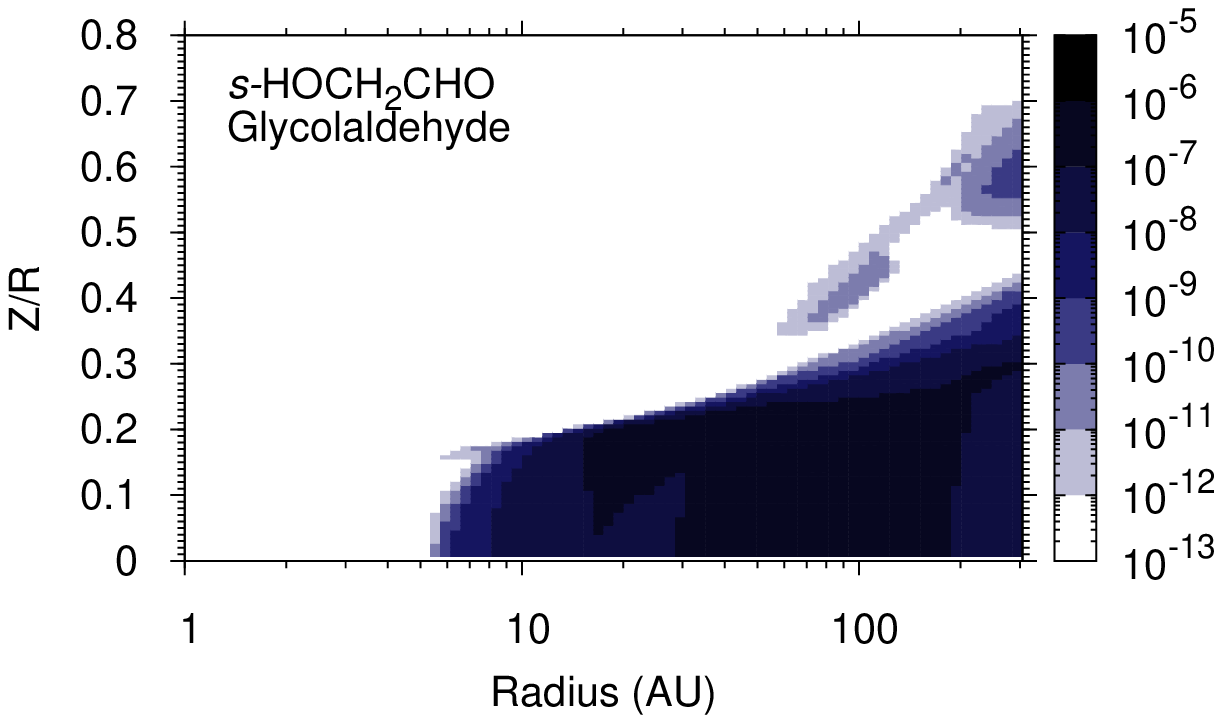}}
\subfigure{\includegraphics[width=0.33\textwidth]{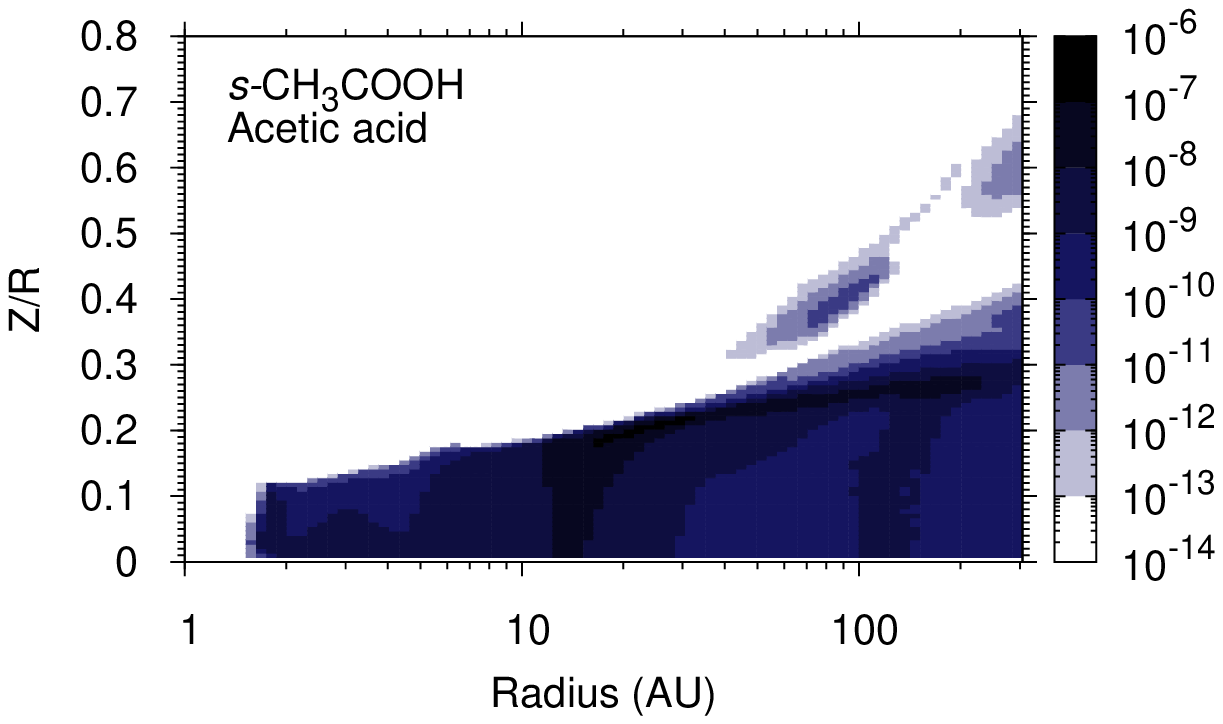}}
\caption{Same as Fig.~\ref{figure6} for grain-surface species.}
\label{figure7}
\end{figure*}

\begin{figure*}[!ht]
\subfigure{\includegraphics[width=0.33\textwidth]{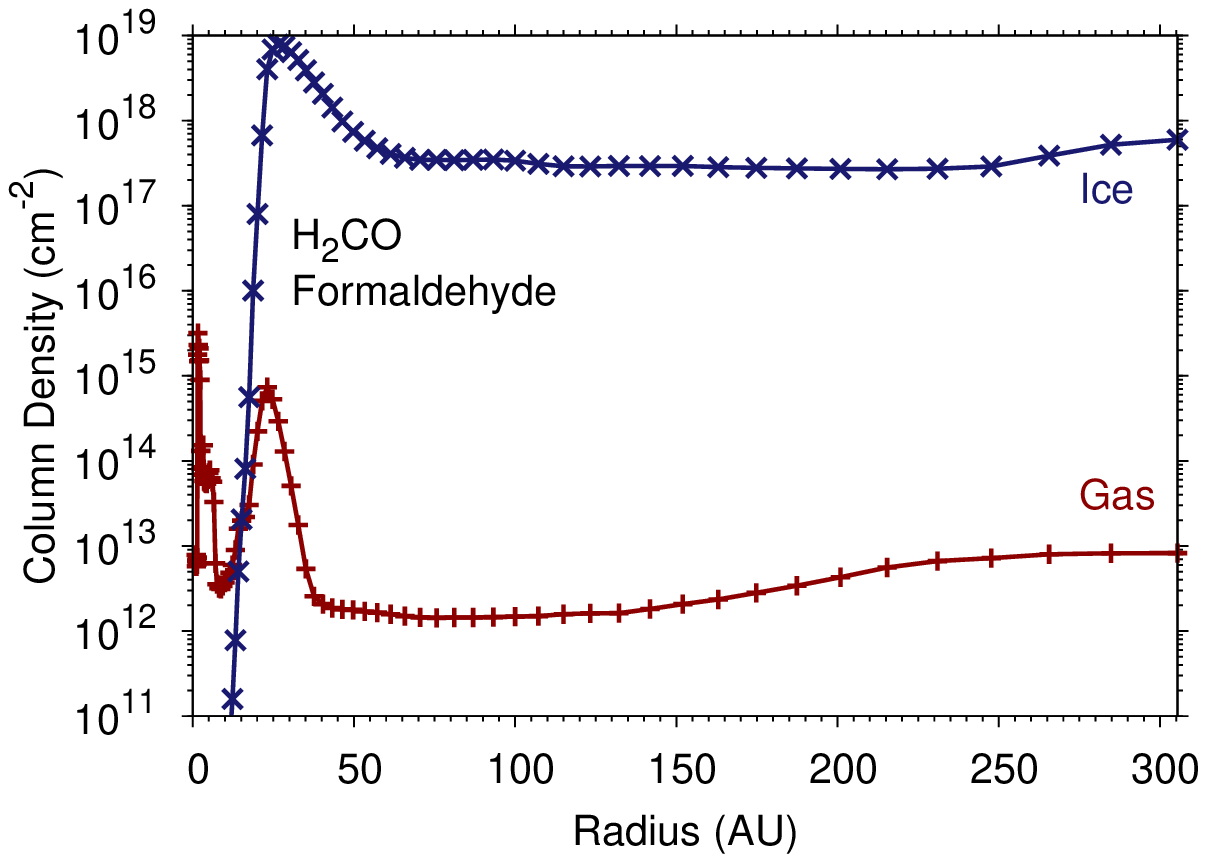}}
\subfigure{\includegraphics[width=0.33\textwidth]{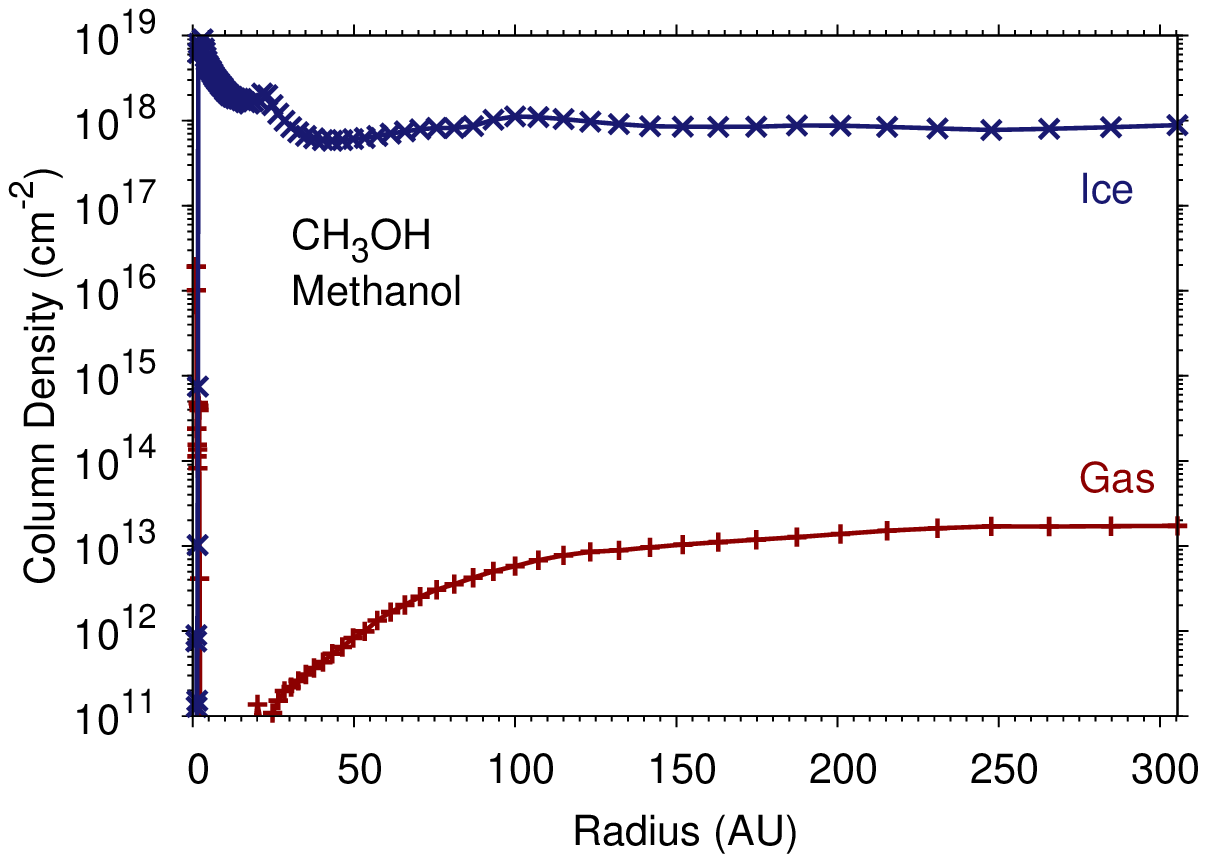}}
\subfigure{\includegraphics[width=0.33\textwidth]{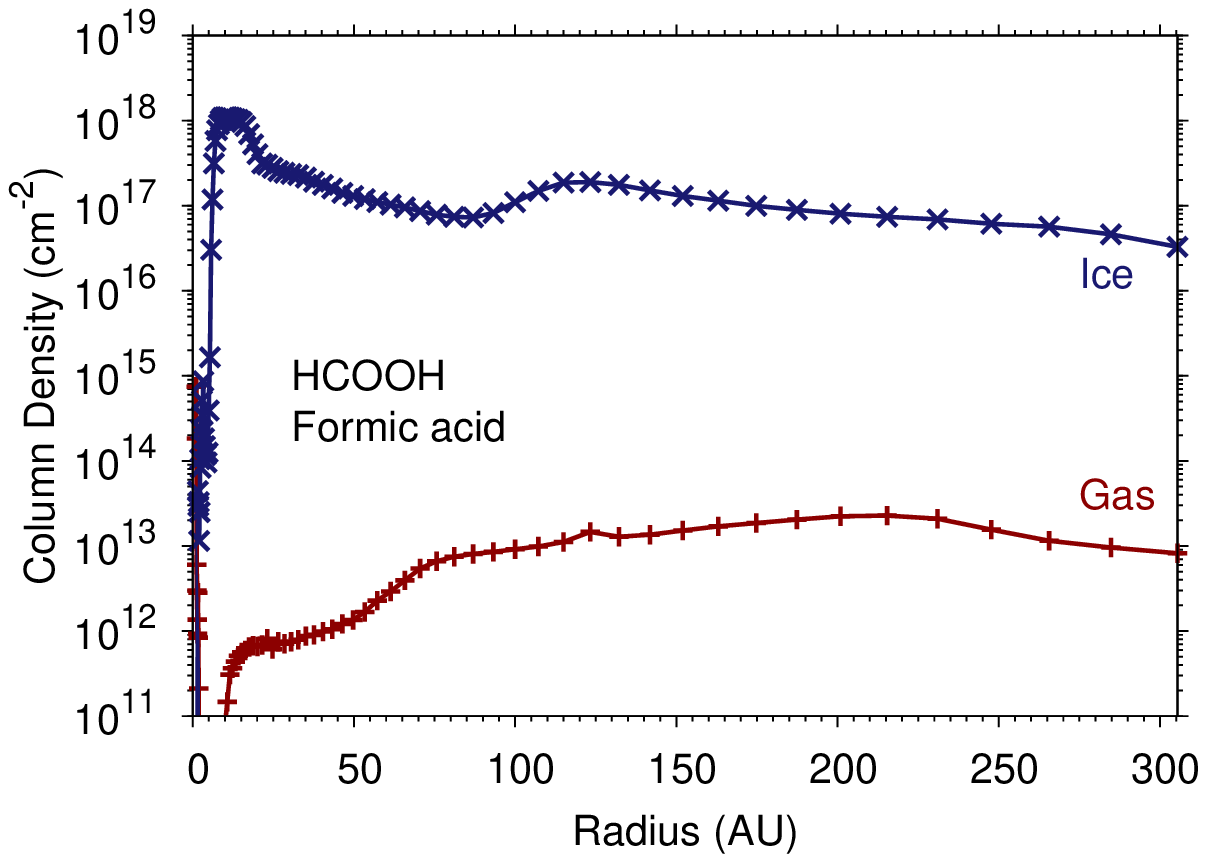}}
\subfigure{\includegraphics[width=0.33\textwidth]{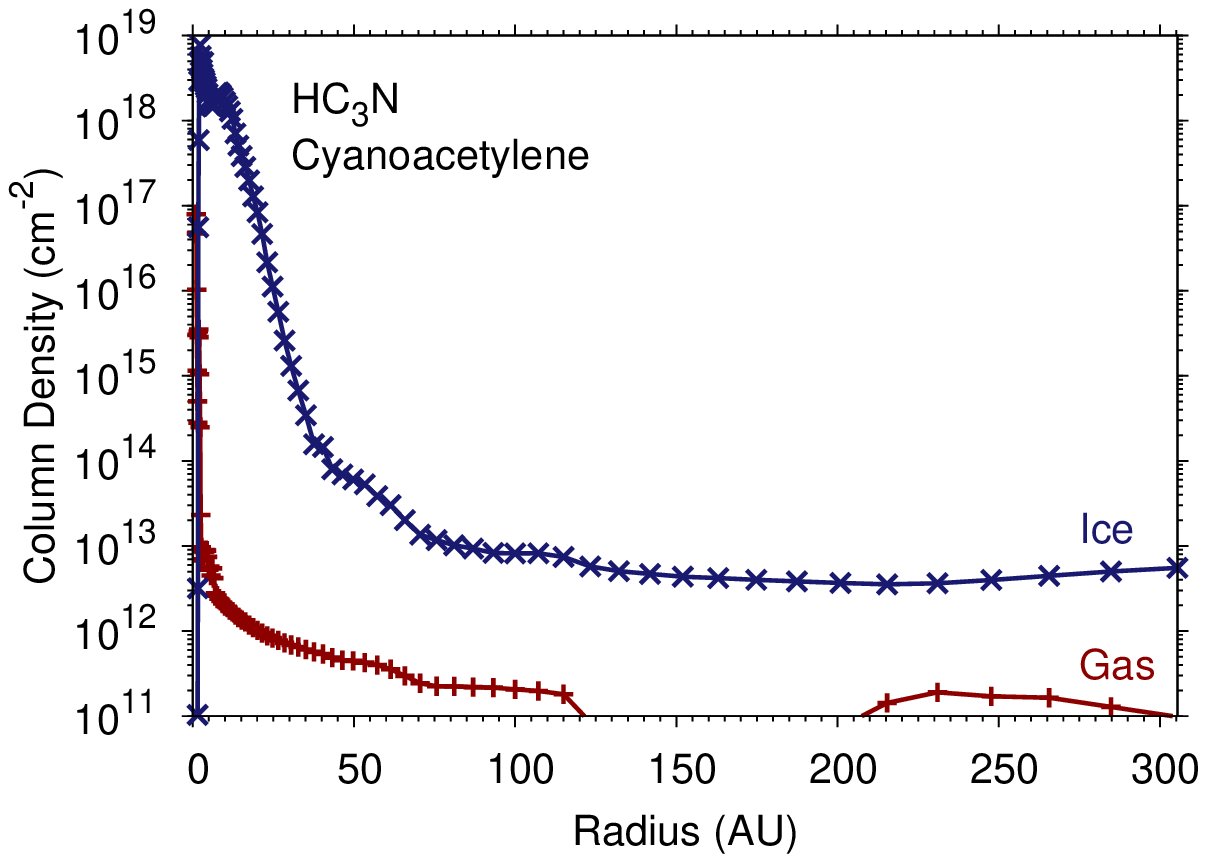}}
\subfigure{\includegraphics[width=0.33\textwidth]{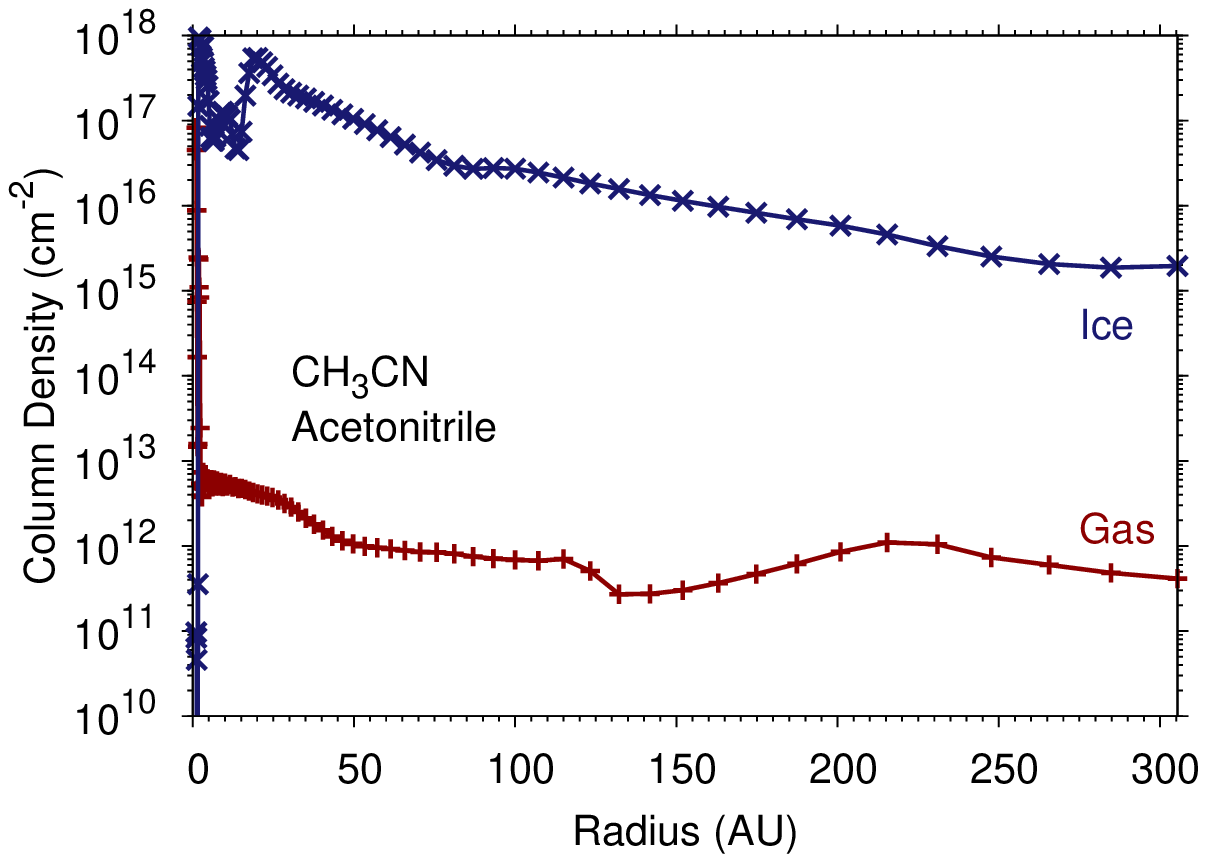}}
\subfigure{\includegraphics[width=0.33\textwidth]{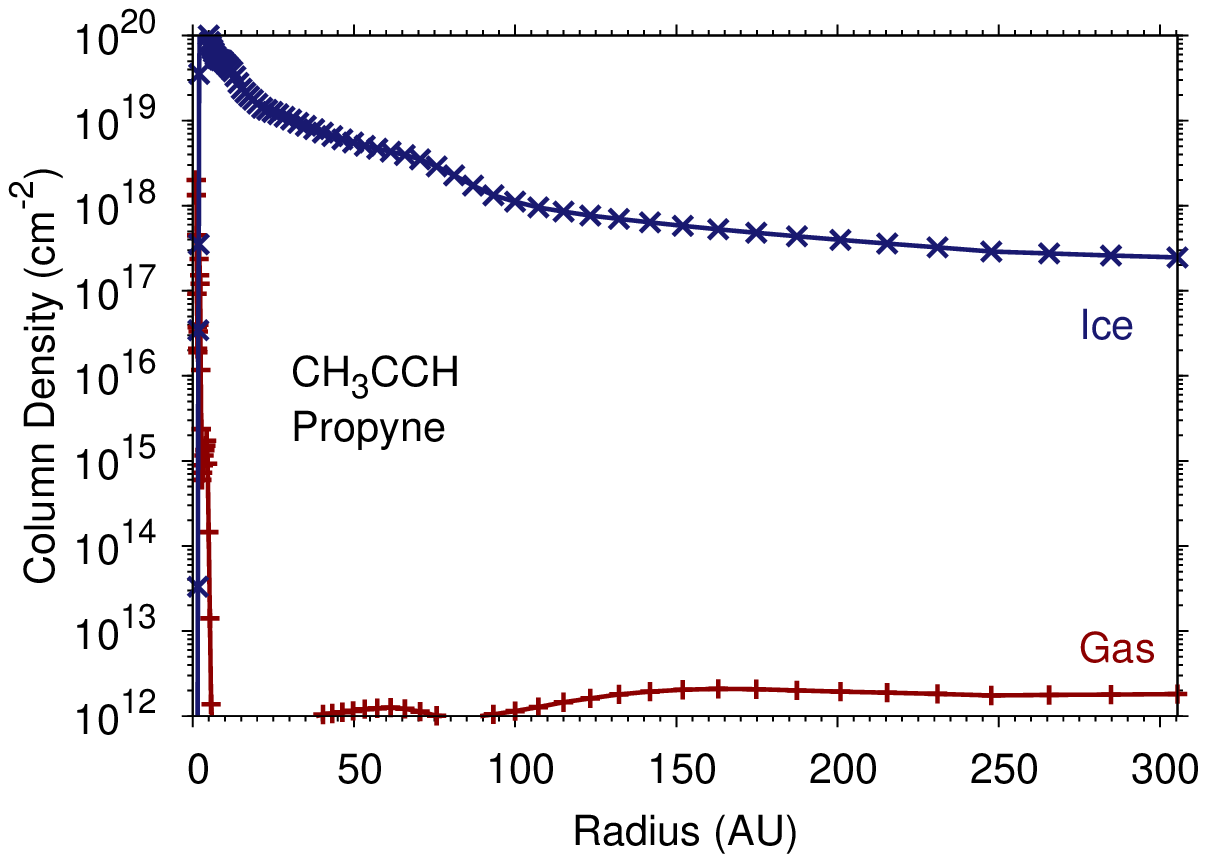}}
\subfigure{\includegraphics[width=0.33\textwidth]{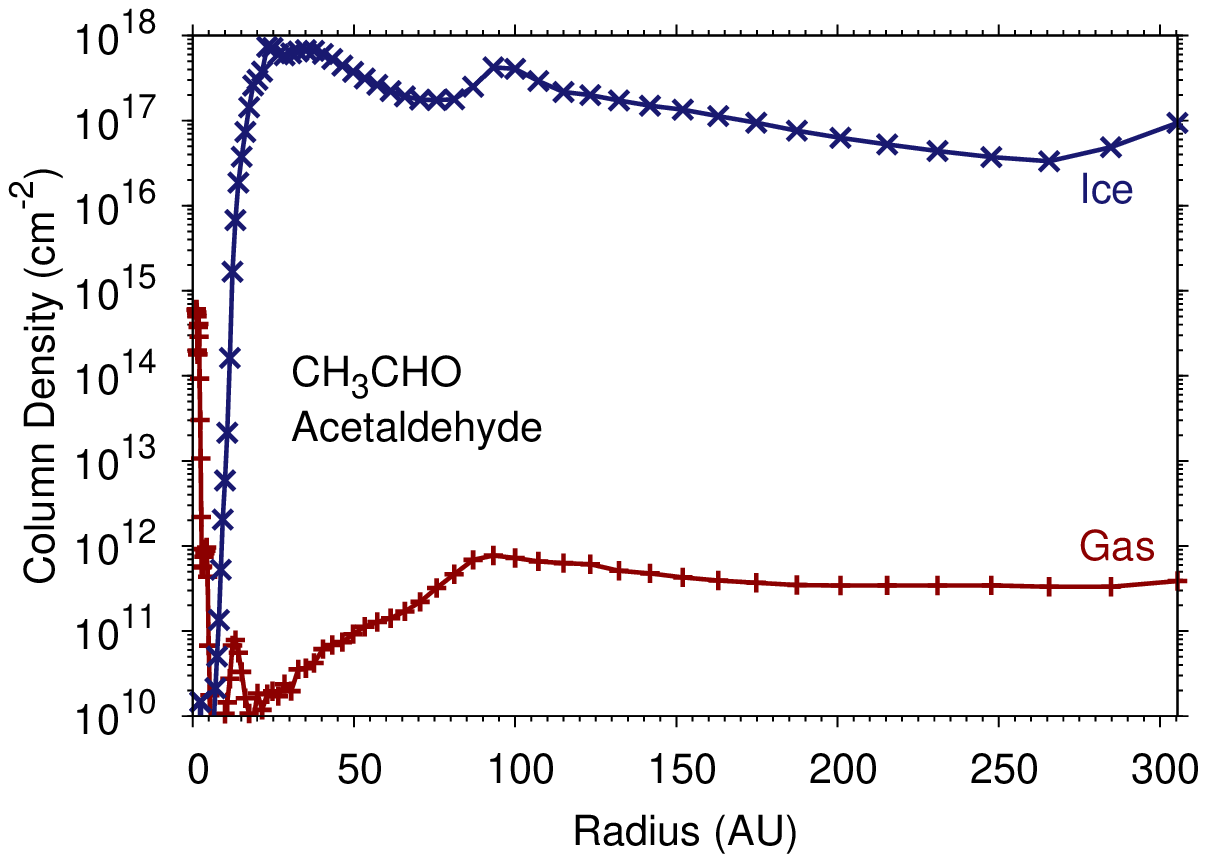}}
\subfigure{\includegraphics[width=0.33\textwidth]{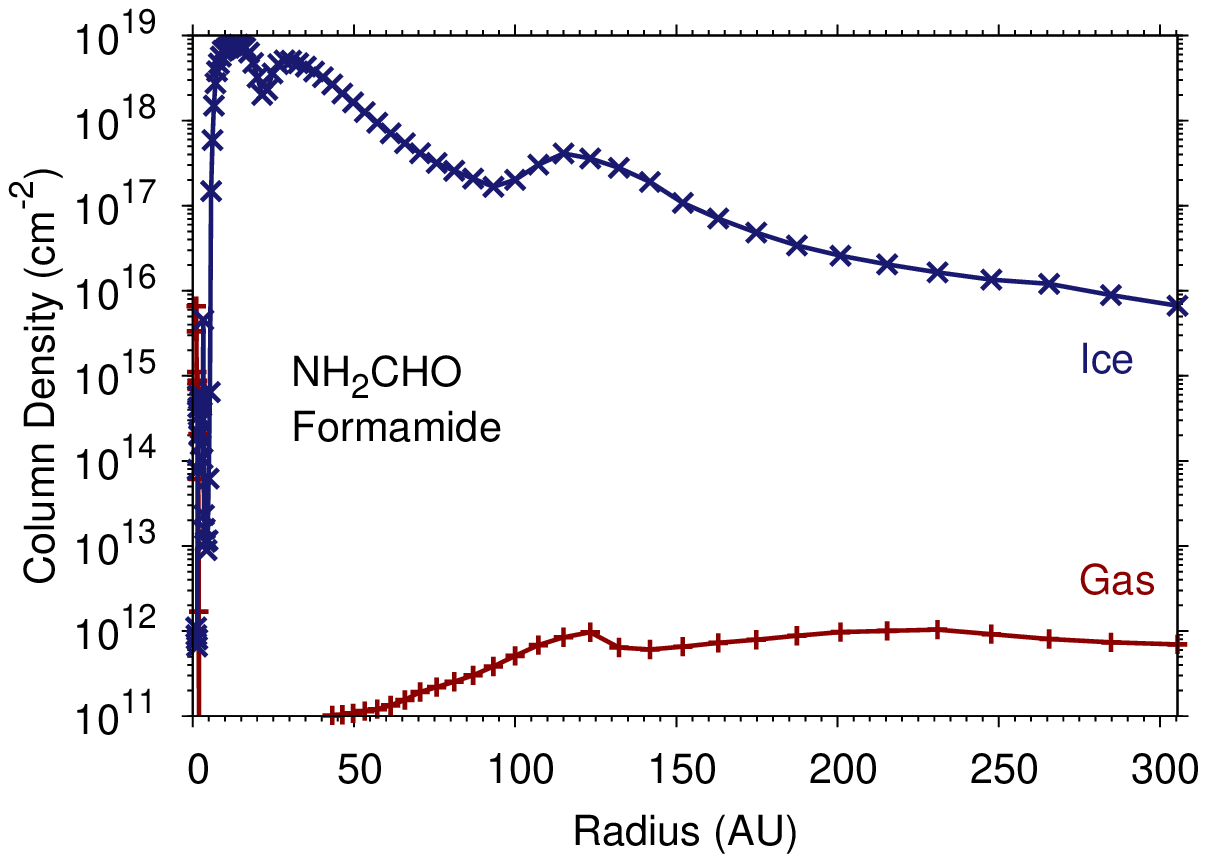}}
\subfigure{\includegraphics[width=0.33\textwidth]{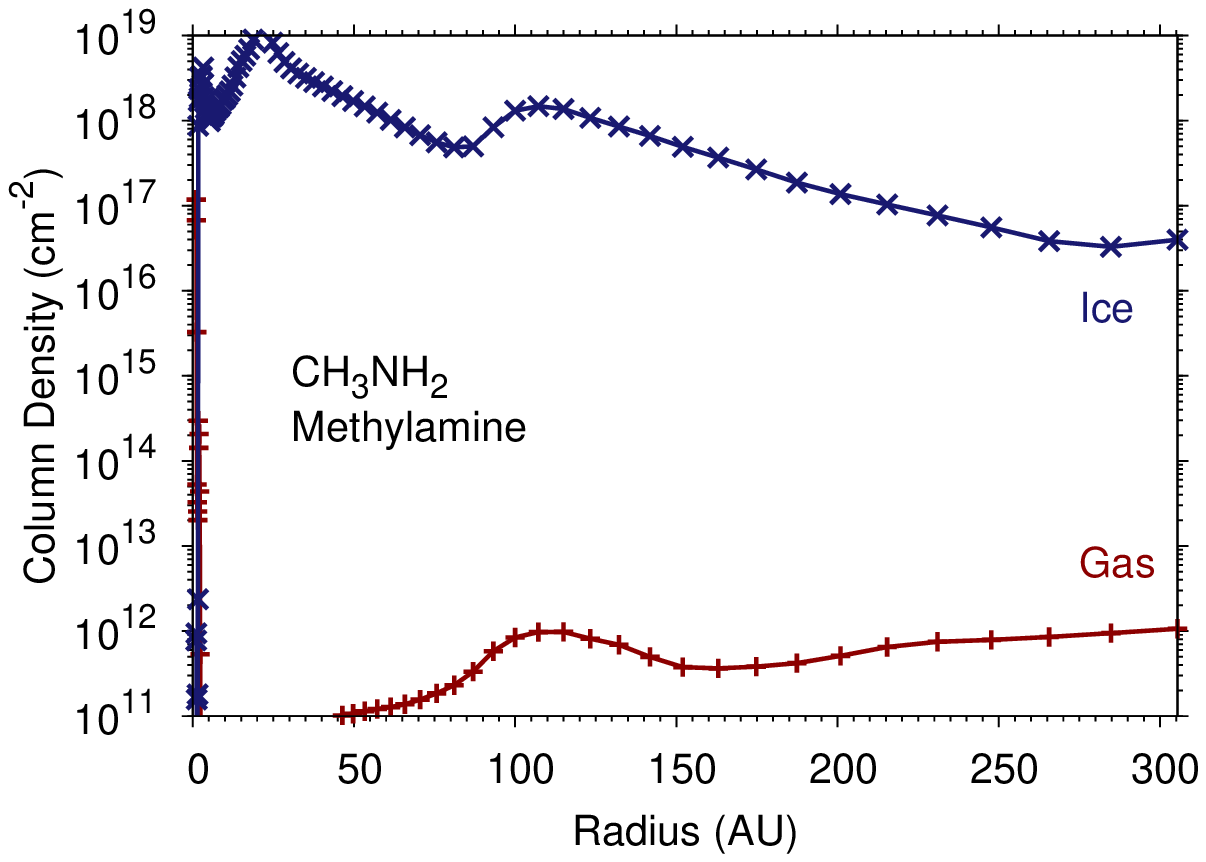}}
\subfigure{\includegraphics[width=0.33\textwidth]{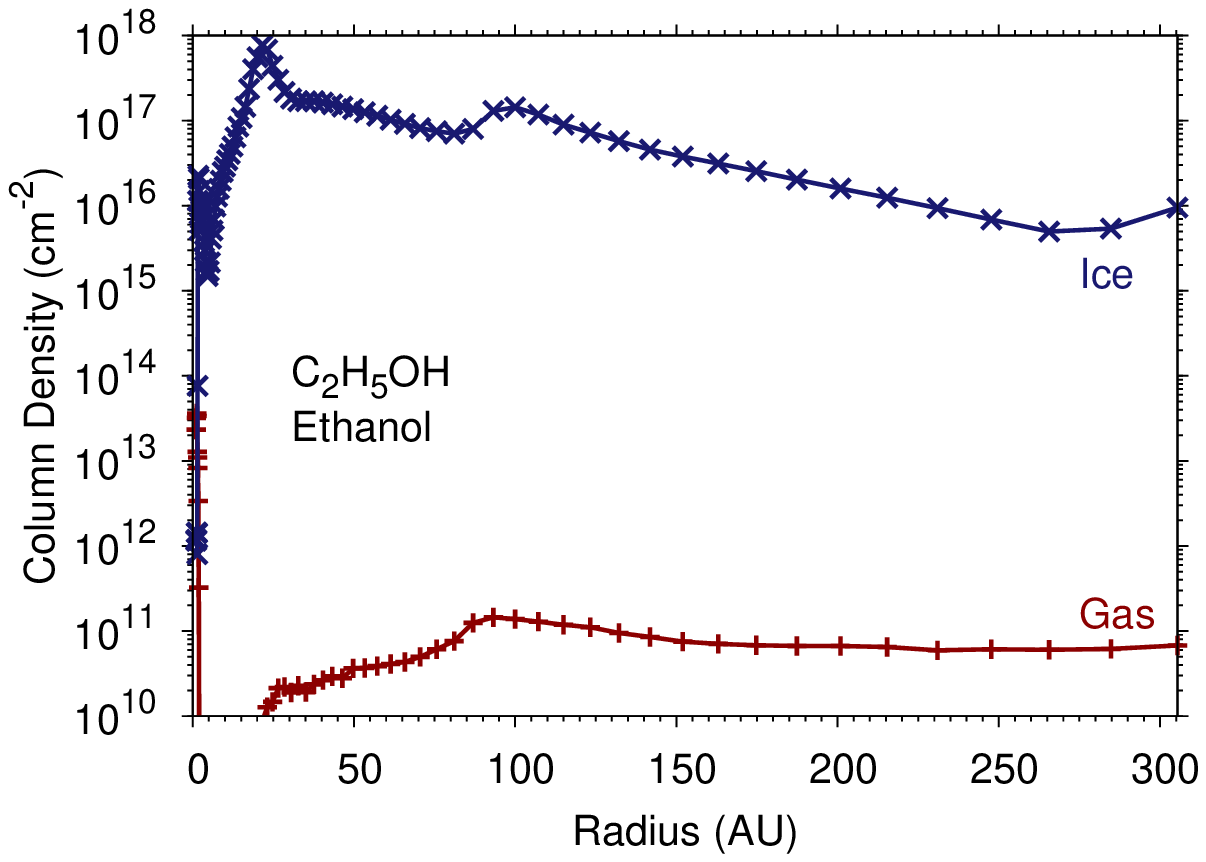}}
\subfigure{\includegraphics[width=0.33\textwidth]{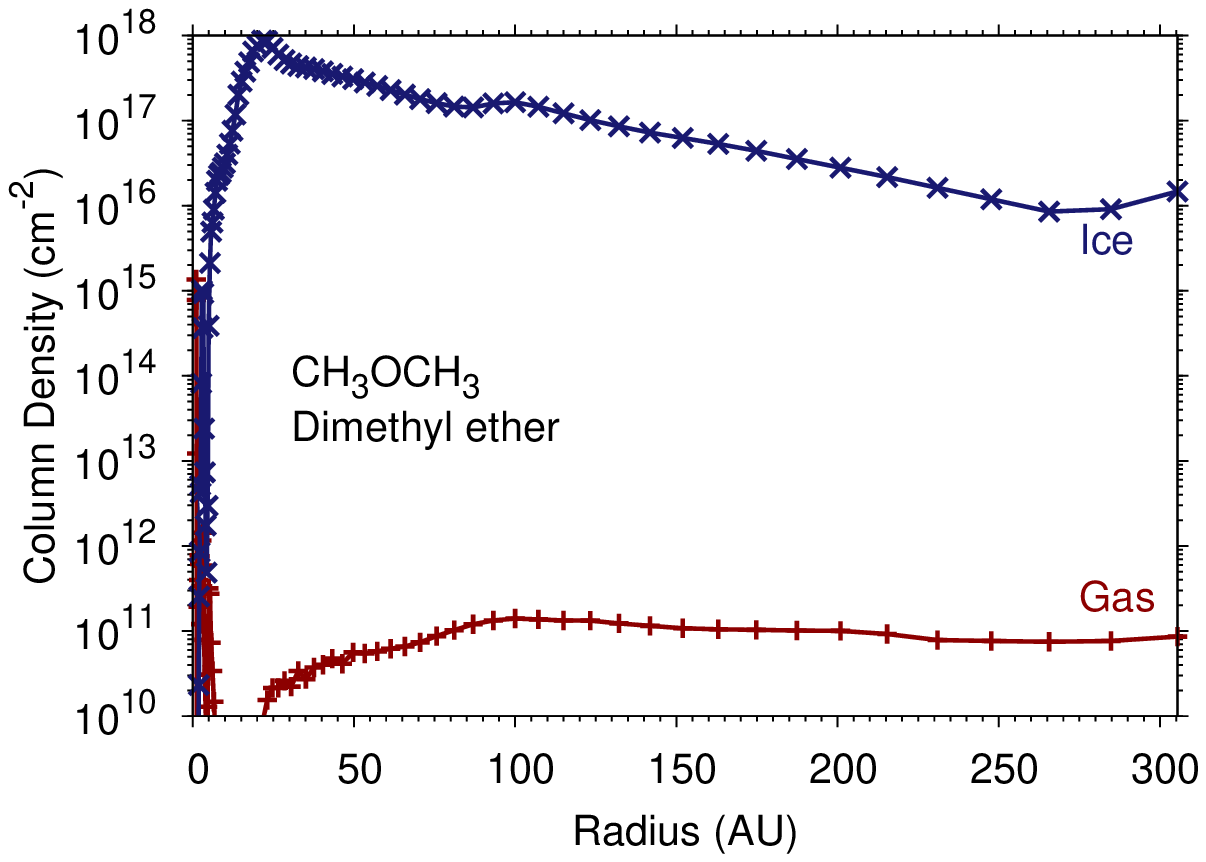}}
\subfigure{\includegraphics[width=0.33\textwidth]{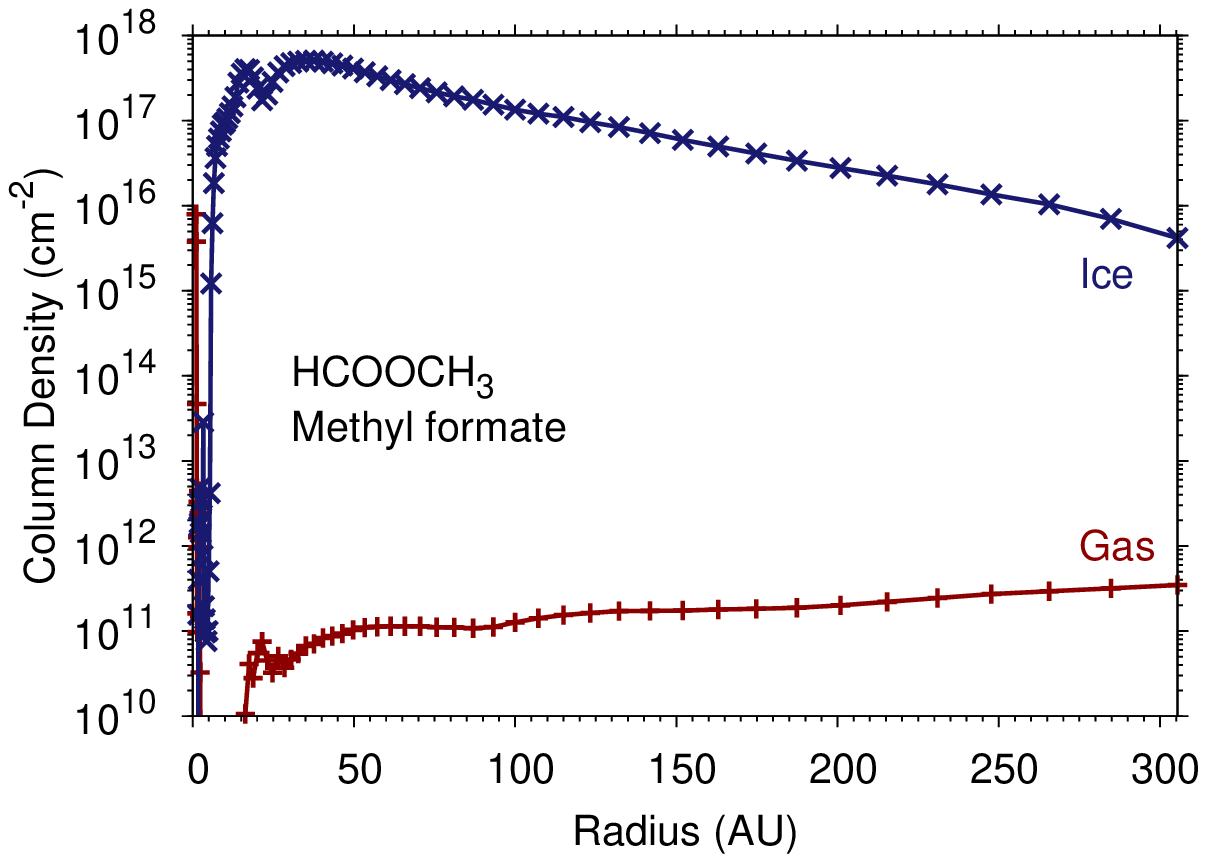}}
\subfigure{\includegraphics[width=0.33\textwidth]{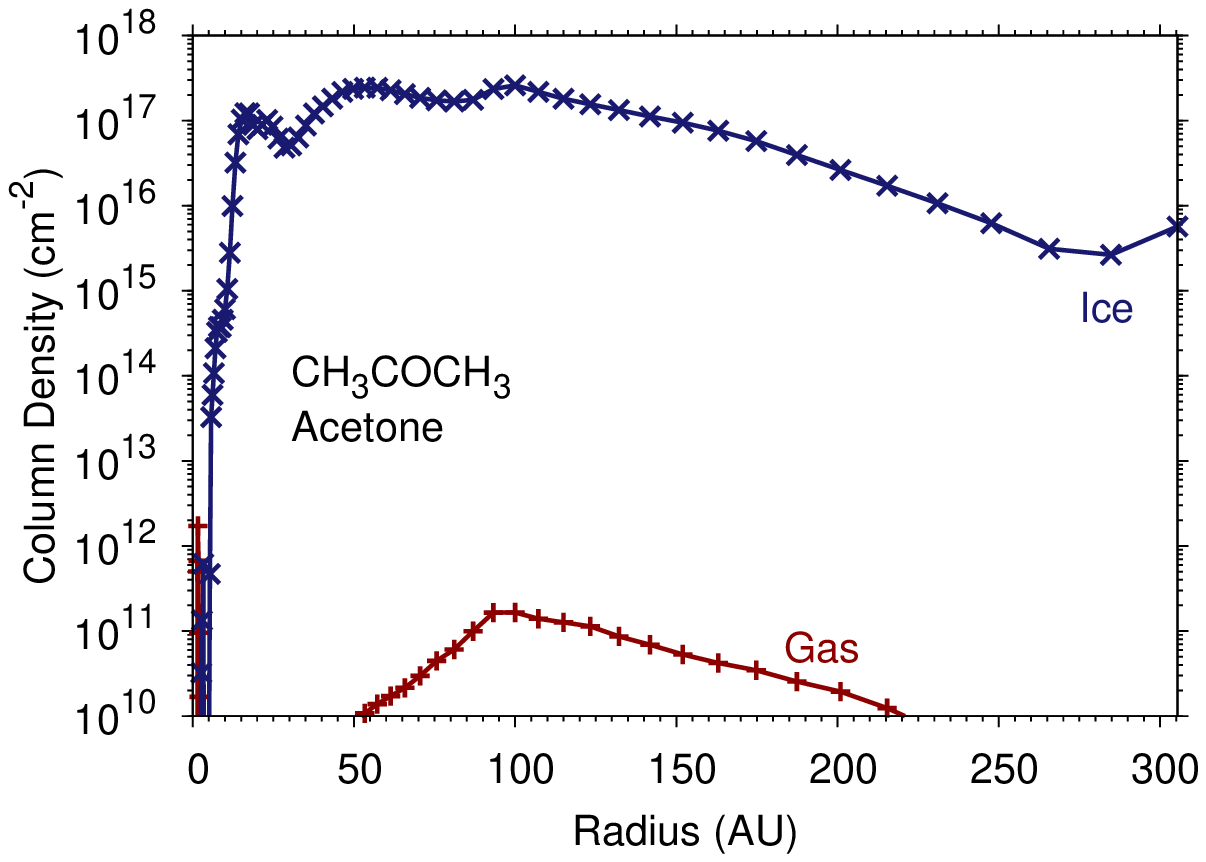}}
\subfigure{\includegraphics[width=0.33\textwidth]{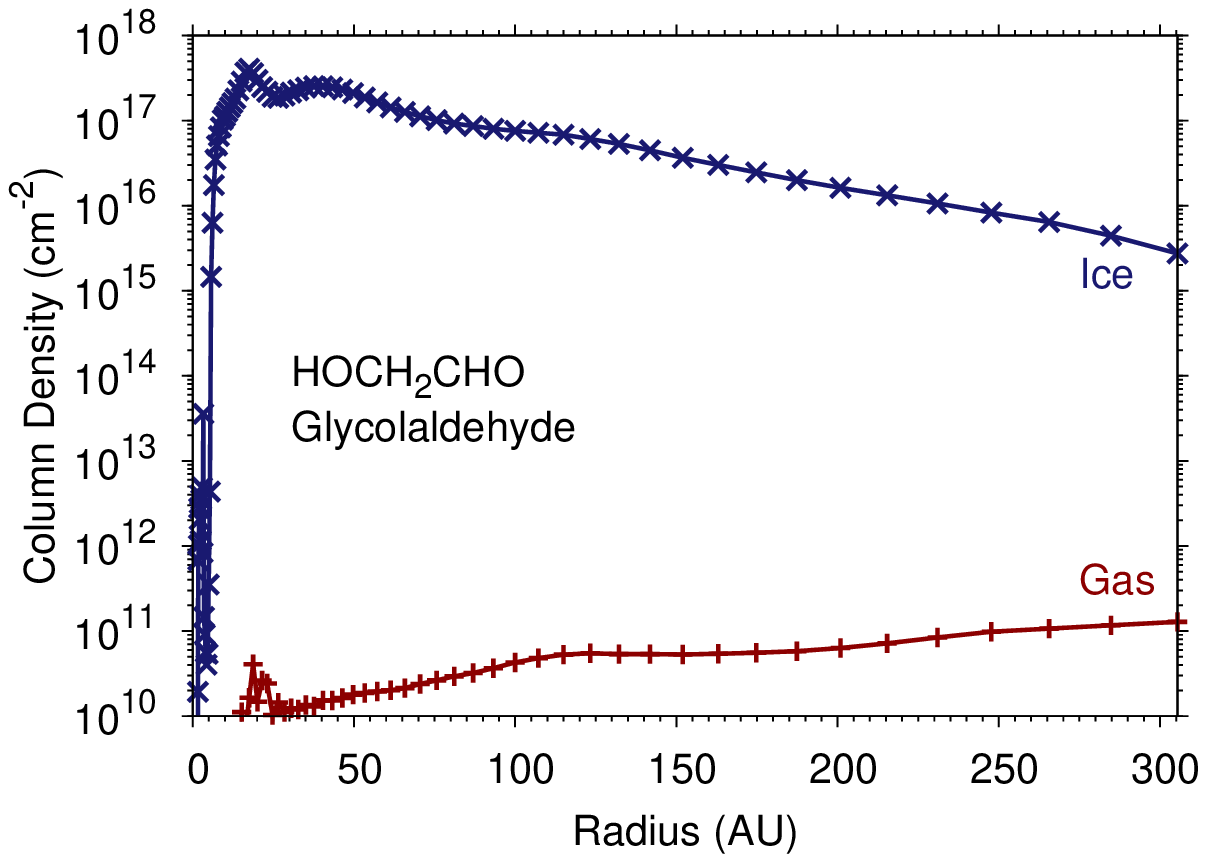}}
\subfigure{\includegraphics[width=0.33\textwidth]{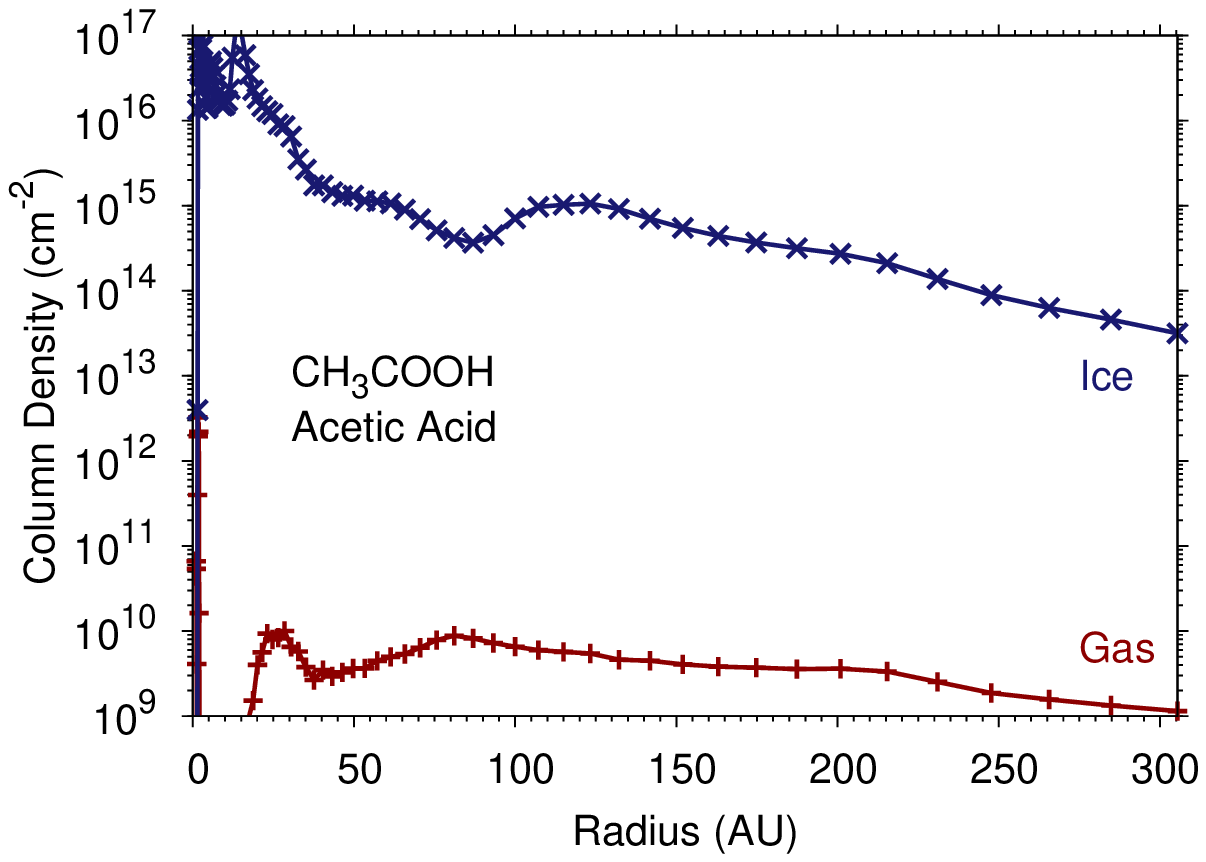}}
\caption{Column density (cm$^{-2}$) as a function of radius, $R$, for gas-phase (red lines) 
and grain-surface (blue lines) molecules.  The corresponding data can be found in Tables~\ref{table5} and \ref{table6}.}
\label{figure8}
\end{figure*}

In Figs.~\ref{figure6} and \ref{figure7}, we display the 
fractional abundances of COMs (relative to total number density) 
as a function of disk radius, $R$, and height, $Z/R$, for gas-phase 
and grain-surface (ice mantle) species, respectively.  
In Fig.~\ref{figure8}, we display the vertical 
column densities of gas-phase (red lines) and grain-surface (blue lines) 
species as a function of disk radius, $R$.  

Figures~\ref{figure6} to \ref{figure8} show that   
the grain-surface (ice mantle) abundances and resulting 
column densities are consistently higher than those of the corresponding gas-phase species.  
Most of the grain-surface COMs reach maximum fractional abundances of
$\sim$~10$^{-6}$ to 10$^{-4}$ relative to the local number density 
and are confined to a layer in the midplane,  
whereas, we see a large spread in 
the maximum fractional abundances of gas-phase species  
from $\sim$~10$^{-12}$ to $\sim$~10$^{-7}$.  
The grain-surface molecules also
reach their peak abundance over a much larger region in the 
disk than their gas-phase analogues.  
The exceptions to this are formaldehyde, \ce{H2CO}, cyanoacetylene, \ce{HC3N}, 
and acetonitrile, \ce{CH3CN}, 
all of which can be formed readily via gas-phase chemistry.
These species are abundant throughout most of the upper regions of the 
disk, whereas, the more complex species reach their peak abundance in the outer disk only
in a layer bounded below by freezeout and above by destruction via photodissociation 
and ion-molecule reactions.  
Generally, the more complex the molecule, the less 
abundant it is in the gas phase, and the smaller the extent over which 
the molecule reaches an appreciable abundance.  
This indicates that species which can form in the gas phase, 
e.g., \ce{H2CO} and \ce{HC3N}, are
less sensitive to the variation in disk physical conditions than the 
more complex species, such as, \ce{HCOOCH3} and \ce{HOCH2CHO}, 
which can only form via grain-surface chemistry.   
Given that the abundance of gas-phase complex species relies 
on efficient photodesorption and limited photodestruction over a narrow 
range of the disk, one can imagine that a small increase/decrease 
in UV flux or X-ray flux in the disk may either inhibit or enhance the 
abundance of gas-phase complex molecules 
to a much more significant degree than for the simpler species.  
In addition, the evolution of dust grains may also be important.  
Dust-grain settling (or sedimentation) towards the disk midplane can lead to the increased 
penetration of UV radiation, whereas dust-grain coagulation (or grain growth) 
leads to bigger grains and a smaller grain-surface area available 
for the absorption of UV photons \citep[see, e.g.,][]{dalessio01,dullemond04,dalessio06}.  
We discuss this issue further in Sect.~\ref{discussion}.  
We intend to explore the effects of dust evolution in future work. 

Several of the grain-surface species 
({\em s-}\ce{CH3OH}, {\em s-}\ce{HC3N}, {\em s-}\ce{CH3CN}, {\em s-}\ce{CH3CCH}, {\em s-}\ce{CH3NH2}, 
{\em s-}\ce{C2H5OH}, and {\em s-}\ce{CH3COOH}) 
remain frozen out down to $\approx$~2~AU. 
These species have desorption or binding energies higher than $\approx$~4500~K 
and so we would expect their snow lines to reside at a similar radius to that for 
water ice ($\approx$~2~AU).  
{\em s-}\ce{HCOOH}, {\em s-}\ce{NH2CHO}, {\em s-}\ce{HCOOCH3}, and {\em s-}\ce{HOCH2CHO}
also possess high binding energies ($\gtrsim$~4000~K); 
however, these species have snow lines at $\approx$~5~AU.
Within 5~AU, the dust temperature is $>$~70~K and 
radical-radical association reactions are more important than 
atom-addition reactions due to the fast desorption rates of 
atoms at these temperatures.  
Grain-surface species which depend on atom-addition routes to their formation  
are not formed as efficiently on warm grains.  
For example, {\em s-}\ce{HCOOCH3} is formed either via the hydrogenation of 
{\em s-}\ce{COOCH3} or via the reaction between {\em s-}\ce{HCO} and {\em s-}\ce{CH3O}. 
These latter two species, in turn, are formed via hydrogenation of {\em s-}CO on the grain.  
{\em s-}\ce{CH3O} is also formed via the photodissociation of {\em s-}\ce{CH3OH} by cosmic-ray-induced photons.   
The radical-radical formation routes of {\em s-}\ce{HCOOH}, {\em s-}\ce{NH2CHO}, and {\em s-}\ce{HOCH2CHO} 
all rely on the formation of {\em s-}\ce{HCO} which, in turn, is formed 
mainly via the hydrogenation of {\em s-}CO. 
In contrast, at warmer temperatures, {\em s-}\ce{CH3OH} can efficiently form via the 
association of {\em s-}\ce{CH3} and {\em s-}\ce{OH} rather than via the hydrogenation of {\em s-}CO.  
Both these radicals can form in the gas and accrete onto grains, or they 
are formed via the cosmic-ray induced photodissociation of grain-mantle molecules.  
A similar argument holds for 
{\em s-}\ce{CH3CN}  ({\em s-}\ce{CH3}  + {\em s-}\ce{CN}), 
{\em s-}\ce{CH3CCH} ({\em s-}\ce{C2H3} + {\em s-}\ce{CH}), 
{\em s-}\ce{CH3NH2} ({\em s-}\ce{CH3}  + {\em s-}\ce{NH2}), 
{\em s-}\ce{C2H5OH}  ({\em s-}\ce{CH3}  + {\em s-}\ce{CH2OH}), 
{\em s-}\ce{CH3COOH} ({\em s-}\ce{CH3}   + {\em s-}\ce{CH3CO}).  
{\em s-}\ce{CH2OH} and {\em s-}\ce{CH3CO} also have radical-radical association 
formation routes, i.e., 
{\em s-}\ce{CH2} + {\em s-}\ce{OH} and {\em s-}\ce{CH3} + {\em s-}\ce{CO}.
 
In Fig.~\ref{figure8}, we display the column density 
of gas-phase (red lines) and grain-surface (blue lines) species 
as a function of radius, $R$.  
In Table~\ref{table2}, we also display the calculated column densities 
at radii of 10, 30, 100, and 305~AU.  
An expanded version of Table~\ref{table2} which includes 
all data used to plot Fig.~\ref{figure8} is available in the 
electronic edition of the journal (Tables \ref{table5} and \ref{table6}).

Most of the gas-phase molecules  
reach values $\gtrsim$~10$^{12}$~cm$^{-2}$ throughout most of the outer disk 
($R$~$\gtrsim$~50~AU). 
The exceptions are \ce{HC3N} and the more complex species considered here, i.e., 
\ce{C2H5OH}, \ce{CH3OCH3}, \ce{HCOOCH3}, \ce{CH3COCH3}, \ce{HOCH2CHO}, 
and \ce{CH3COOH}. 
Most of these species achieve a column density $\sim$~10$^{11}$~cm$^{-2}$ 
throughout the outer disk.  
\ce{CH3COCH3} peaks around 100~AU before decreasing towards 
larger radii and \ce{CH3COOH} reaches a peak column density of $\sim$~10$^{10}$~cm$^{-2}$ 
between $\approx$~25 and 50~AU before steadily decreasing towards larger radii.  
Both these species form via {\em s-}\ce{CH3CO} which, in turn, forms on the grain via 
{\em s-}\ce{CH3} + {\em s-}\ce{CO}.  
This latter reaction has an appreciable reaction barrier ($\approx$~3500~K) 
and can only proceed on sufficiently warm grains.

We produce very similar gas-phase and grain-surface column densities
for the structural isomers, glycolaldehyde (\ce{HOCH2CHO}) and 
methyl formate (\ce{HCOOCH3}), which are formed primarily 
via the grain-surface association routes, {\em s-}\ce{CH2OH}~+~{\em s-}\ce{HCO} and 
{\em s-}\ce{CH3O}~+~{\em s-}\ce{HCO}, respectively.  
These species have been detected in the gas phase 
in the hot core, Sgr~B2(N), and in 
the low-mass protostar, IRAS~16293+2422 \citep[see, e.g., ][]{hollis00,jorgensen12}.  
In both sources, methyl formate is more than a factor of 10 more abundant 
than glycolaldehyde.  
In this work, we have assumed a branching ratio of 1:1 for the 
production of \ce{CH3O} and \ce{CH2OH} via the photodissociation of gas-phase and grain-surface methanol.    
Hence, the formation rates of both radicals via this mechanism 
are similar leading to similar abundances of methyl formate and glycolaldehyde.    
\citet{laas11} investigated various branching ratios for methanol photodissociation in hot cores 
and concluded that branching ratios for grain-surface cosmic-ray-induced photodissociation have an 
influence on the resulting gas-phase abundances.  
They found that models including ratios favouring the methoxy channel ({\em s-}\ce{CH3O}) agreed 
best with observed abundances of methyl formate; however, ratios favouring  the methyl channel 
({\em s-}\ce{CH3}) agreed best with the observed gas-phase abundance of glycolaldehyde.     
This is in contrast with laboratory experiments which show that formation of the hydroxymethyl radical
({\em s-}\ce{CH2OH}) is the dominant channel \citep{oberg09c}.  

The mobilities of the {\em s-}\ce{CH2OH} and {\em s-}\ce{CH3O} radicals can also influence the production 
rates of grain-surface methyl formate and glycolaldehyde.  
Here, we have followed \citet{garrod08} and assumed that {\em s-}\ce{CH2OH} is more strongly bound 
to the grain mantle than {\em s-}\ce{CH3O} ($E_{D}$~=~5080~K and 2250~K, respectively) due to the $-$OH 
group which allows hydrogen bonding with the water ice. 
Hence, we expect {\em s-}\ce{CH3O} to have higher mobility than {\em s-}\ce{CH2OH}.  
However, the reaction rates for the grain-surface formation of methyl formate 
and glycolaldehyde at the temperatures found in the disk midplane 
are dominated by the mobility of the {\em s-}\ce{HCO} radical.   
This radical has a significantly lower binding energy to the grain mantle ($E_D$~=~1600~K) 
leading, again, to similar grain-surface formation rates for methyl formate and glycolaldehyde.  

\begin{table*}
\footnotesize
\caption{Column density (cm$^{-2}$) of gas-phase and grain-surface organic molecules at radii of 10, 30, 100, and 305~AU from 
our full disk model.}
\centering
\begin{tabular}{llcccccccc}
\hline\hline
 & & \multicolumn{4}{c}{Gas phase} & \multicolumn{4}{c}{Grain surface} \\
\multicolumn{2}{c}{Species}                 & 10 AU   & 30 AU   & 100 AU  & 305 AU  & 10AU    & 30 AU   & 100 AU  & 305 AU  \\ 
\hline\\
Formaldehyde                & \ce{H2CO}     & 3.7(12) & 5.1(13) & 1.5(12) & 8.3(12) & 6.7(09) & 6.4(18) & 3.4(17) & 6.0(17) \\
Methanol                    & \ce{CH3OH}    & 1.0(09) & 2.2(11) & 5.8(12) & 1.7(13) & 2.3(18) & 8.4(17) & 1.1(18) & 8.8(17) \\
Formic acid                 & \ce{HCOOH}    & 8.1(10) & 7.5(11) & 9.1(12) & 8.2(12) & 1.1(18) & 2.4(17) & 1.1(17) & 3.3(16) \\
Cyanoacetylene              & \ce{HC3N}     & 2.0(12) & 6.9(11) & 2.1(11) & 9.8(10) & 1.7(18) & 1.3(15) & 8.2(12) & 5.5(12) \\
Acetonitrile                & \ce{CH3CN}    & 5.5(12) & 2.9(12) & 6.9(11) & 4.1(11) & 1.2(17) & 2.1(17) & 2.7(16) & 2.0(15) \\
Propyne                     & \ce{CH3CCH}   & 4.5(11) & 8.2(11) & 1.1(12) & 1.8(12) & 4.7(19) & 1.0(19) & 1.1(18) & 2.5(17) \\
Acetaldehyde                & \ce{CH3CHO}   & 1.1(10) & 2.0(10) & 7.2(11) & 3.9(11) & 5.9(12) & 6.2(17) & 4.1(17) & 9.4(16) \\
Formamide                   & \ce{NH2CHO}   & 1.7(09) & 5.6(10) & 5.1(11) & 7.0(11) & 6.9(18) & 5.1(18) & 2.0(17) & 6.7(15) \\
Methylamine                 & \ce{CH3NH2}   & 7.2(08) & 6.6(10) & 8.4(11) & 1.1(12) & 1.8(18) & 4.1(18) & 1.3(18) & 4.0(16) \\
Ethanol                     & \ce{C2H5OH}   & 3.8(06) & 1.9(10) & 1.4(11) & 6.8(10) & 2.9(16) & 1.8(17) & 1.4(17) & 9.5(15) \\
Dimethyl ether              & \ce{CH3OCH3}  & 2.2(07) & 2.2(10) & 1.4(11) & 8.6(10) & 3.1(16) & 4.7(17) & 1.6(17) & 1.5(16) \\
Methyl formate              & \ce{HCOOCH3}  & 5.8(07) & 4.6(10) & 1.3(11) & 3.5(11) & 9.5(16) & 4.8(17) & 1.3(17) & 4.2(15) \\
Acetone                     & \ce{CH3COCH3} & 5.8(06) & 1.5(09) & 1.7(11) & 5.7(09) & 6.1(14) & 5.1(16) & 2.6(17) & 5.7(15) \\
Glycolaldehyde              & \ce{HOCH2CHO} & 4.2(07) & 1.2(10) & 4.3(10) & 1.3(11) & 1.1(17) & 2.1(17) & 7.6(16) & 2.7(15) \\
Acetic acid                 & \ce{CH3COOH}  & 2.1(07) & 6.5(09) & 6.6(09) & 1.1(09) & 1.5(16) & 6.5(15) & 7.1(14) & 3.2(13) \\
\hline
\end{tabular}
\label{table2}
\tablefoot{$a(b)$ represents $a\times10^b$ \\An expanded version of this table is avialable in the electronic version of the journal}
\end{table*}

\subsection{Line spectra}
\label{linespectra}

\begin{figure*}
\subfigure{\includegraphics[width=0.5\textwidth]{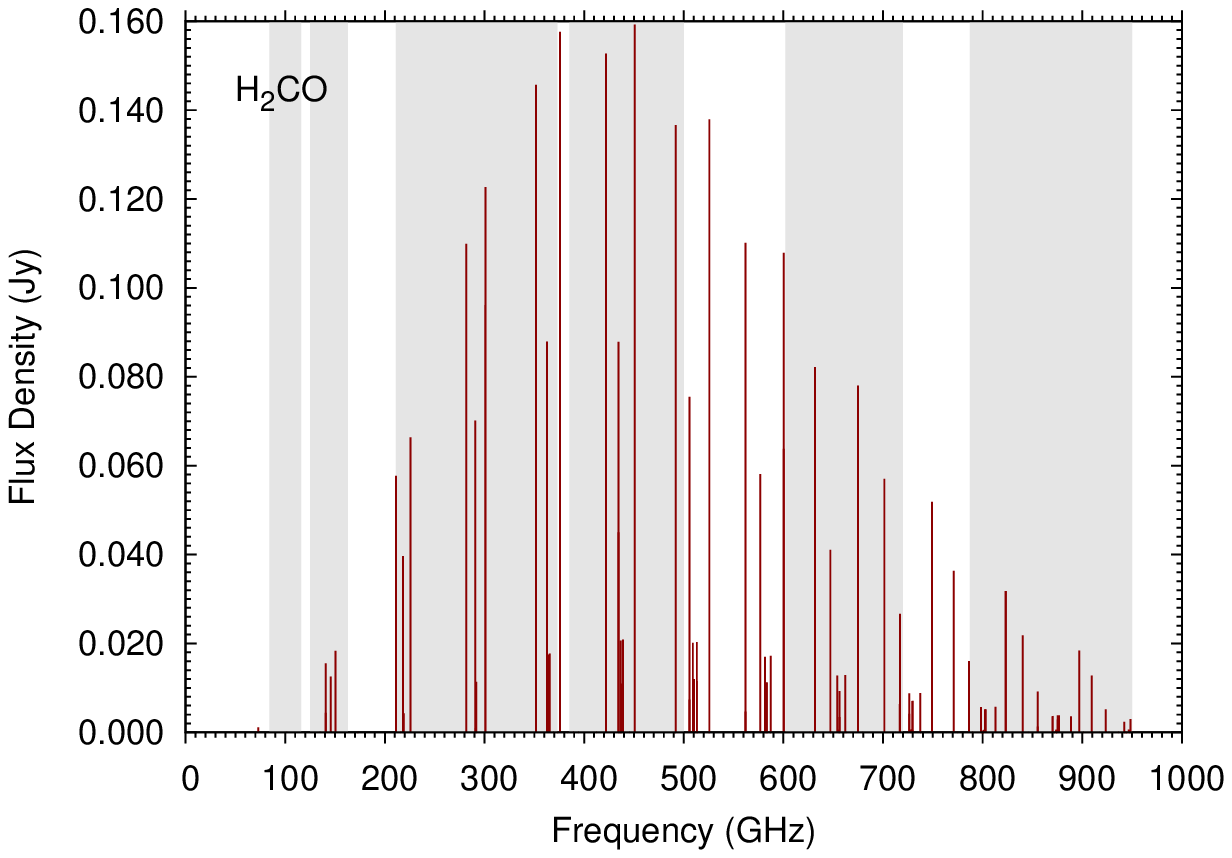}}
\subfigure{\includegraphics[width=0.5\textwidth]{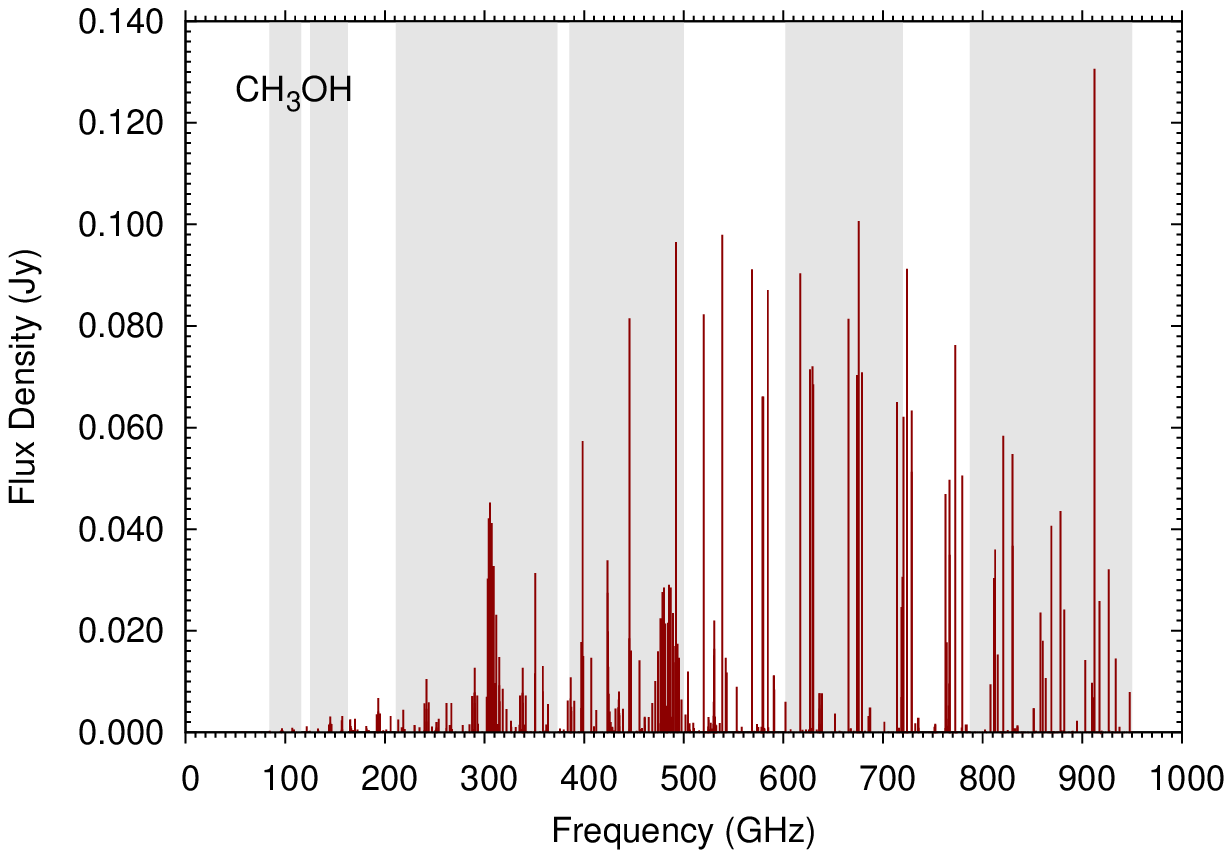}}
\caption{Disk-integrated line spectra (Jy) for \ce{H2CO} (left) and \ce{CH3OH} (right).  
The gray boxes indicate the frequency coverage of ALMA `Full Science' receivers: 
band 3 (84 to 116~GHz), band 4 (125 to 163~GHZ), band 6 (211 to 275~GHz), 
band 7 (275 to 373~GHz), band 8 (385 to 500~GHz), band 9 (602 to 720~GHz), and 
band 10 (787 to 950~GHz).}
\label{figure9}
\end{figure*}

In Fig.~\ref{figure9} we display our disk-integrated
line spectra for \ce{H2CO} and \ce{CH3OH} up to a frequency of 1000~GHz to cover 
the full frequency range expected for ALMA `Full Science' operations.  
We also highlight, in gray, the frequency bands expected to be available at 
the commencement of `Full Science':  
band 3 (84 to 116 GHz), band 4 (125 to 163~GHZ), 
band 6 (211 to 275 GHz), band 7 (275 to 373 GHz), 
band 8 (385 to 500~GHz), band 9 (602 to 720 GHz), and band 10 (787 to 950~GHz).  
In our calculations, we assume a distance to source of 140~pc.  
We find peak flux densities of $\approx$~160~mJy for \ce{H2CO} and 
$\approx$~120~mJy for \ce{CH3OH}.   

Our calculations suggest that reasonably strong lines of \ce{H2CO} and \ce{CH3OH} 
which fall into bands 7, 8, and 9 
are good targets for ALMA `Early Science' and `Full Science' capabilities.  
The strongest methanol line in band 7 is the 3$_{12}$-3$_{03}$ transition of 
A-\ce{CH3OH} at 305.474~GHz.  
The peak line flux density calculated is $\approx$~45~mJy.  
Using the full ALMA array (50 antennae) and a channel width of 0.2 km/s, a sensitivity of 
5~mJy can be reached in an integration time of $\approx$~30~min.  
In band 8, a good candidate line is the 2$_{12}$-1$_{01}$ transition of 
A-\ce{CH3OH} at 
398.447~GHz.  The peak line flux density calculated for this transition is 
$\approx$~60~mJy.  
Again, using the full ALMA array and a similar channel width, a sensitivity of 
10~mJy can be reached in $\approx$~60~min.  
Observations of weak lines in ALMA band 9 are more challenging due to the 
increasing influence of the atmosphere\footnotemark[11]
and the strong continuum emission at higher frequencies.
Nevertheless, several methanol line transitions in this band may 
also be accessible with full ALMA.  
Under the same conditions, a sensitivity of 25~mJy can be reached in 
$\approx$~120~mins at a frequency of 665.442~GHz, corresponding 
to the 5$_{24}$-4$_{14}$ transition of E-\ce{CH3OH}.  
We remind the reader that the line intensities calculated here can 
be considered lower limits to the potential intensities 
due to the truncation of our disk model at 305~AU. 
It is possible that methanol has not yet been observed in disks because lower 
frequency transitions, which we find to be relatively weak, have historically been targeted.  
A deep search for methanol in nearby well-studied objects should help ascertain whether 
this species is present: if so, this is a clear indication that our current 
grain-surface chemistry theory works across different physical regimes and sources. 

\footnotetext[11]{\url{http://almascience.eso.org/about-alma/weather/atmosphere-model}} 

We also calculated the line spectra for \ce{HCOOH}, \ce{HC3N}, \ce{CH3CN}, 
\ce{CH3CCH}, and \ce{NH2CHO} and found that the line intensities were 
negligible in all cases ($\ll$~10~mJy).  
For more complex molecules (e.g., \ce{HCOOCH3} and \ce{CH3OCH3}), 
line detection and identification in nearby protoplanetary disks 
may prove challenging, even with the superior sensitivity and spatial resolution 
of ALMA.  
Also, millimeter and (sub)millimeter observations of molecular line emission 
from warm, chemically rich sources (for example, hot cores and 
massive star-forming regions located towards the galactic centre), often suffer 
from line overlapping (or blending) which can hinder
the solid identification of specific COMs \citep[see the discussion in, e.g.,][]{herbst09}. 

It is also worth noting here that the Square Kilometre Array 
(SKA\footnotemark[12]), due for 
completion by 2020 and consisting of some 3000 dishes spread over numerous sites around the 
world, will have high sensitivity in the 70~MHz to 30~GHz frequency range.  
This may allow detection of line emission/absorption from complex molecules in 
nearby disks at lower frequencies than those achievable with ALMA 
\citep[see, e.g.,][]{lazio04}.

\footnotetext[12]{\url{http://www.skatelescope.org/}}

\section{Discussion}
\label{discussion}

In this section, we discuss and compare our results with observations of molecular 
line emission from protoplanetary disks and cometary comae, and with results from other models 
with similar chemical complexity.  
We also discuss the astrobiological significance of our work.   

\subsection{Comparison with observations}
\label{comparisonwithobservations}

Our exploratory calculations suggest that complex organic molecules may be efficiently 
formed on grain surfaces in protoplanetary disks. 
However, it is difficult to 
observe ice species in disks and indeed, this has only been achieved for water ice in 
a handful of (almost) edge-on systems \citep[see, e.g.,][]{terada07}.  
Instead, in the cold outer regions of disks, we are limited to observing 
gas-phase molecules which possess a permanent dipole moment, only.  
This also presents difficulties if the gas-phase form of the molecule is not present in sufficient quantities 
and/or also possesses a complex rotational spectrum.  

The only relatively complex molecules detected in disks, to date, are 
formaldehyde, \ce{H2CO}, cyanoacetylene, \ce{HC3N}, and cyclopropenylidene, $c$-\ce{C3H2}.
\citet{dutrey97} detected several rotational lines of \ce{H2CO} in the disks of DM~Tau 
and GG~Tau deriving a column density of $\sim$~10$^{12}$~cm$^{-2}$.  
\citet{aikawa03} and \citet{thi04} present detections 
of formaldehyde in the disk of LkCa~15 determining 
column densities of 7.2~--~19~$\times$~10$^{12}$~cm$^{-2}$ 
and 7.1~--~51~$\times$~10$^{11}$~cm$^{-2}$, respectively.  
The large spread in column density is due to the difficulty in using a simple model 
to derive column densities from observations, 
even when several lines of the species are detected.  
From Fig.~\ref{figure8} and Table~\ref{table2}, we can see the 
column density in the outer disk i.e., $>$~10~AU (10$^{12}$~--~10$^{13}$ cm$^{-2}$) 
compares well with those values constrained from observation.  

More recently, \citet{oberg10,oberg11} and \citet{qi13a} present detections 
of \ce{H2CO} using the SMA in a selection of protoplanetary disks in the 
well-studied Taurus region and in the Southern sky.  
They confirmed the previous detections of \ce{H2CO} in the disks of DM~Tau and LkCa~15, 
and they also present new detections: one line in the disk of AA~Tau, two lines 
in GM~Tau, and two lines in TW~Hya.  
They also detected formaldehyde in the disks of IM~Lup, V4046~Sgr, 
and HD~142527.  
Their detected lines and line intensities towards T~Tauri disks 
are listed in Table~\ref{table3}. 
Their values range from 
$\sim$~100~mJy~km~s$^{-1}$ to $\sim$~1~Jy~km~s$^{-1}$ depending on the 
source and transition.   
The authors do not infer any column densities using their data 
and explain the difficulties in doing so.  
Instead, they present integrated intensities with which we can compare 
our calculated line intensities.  

We also note here the first detection of 
\ce{HC3N} in a selection of T~Tauri disks by \citet{chapillon12}.  
The authors state that their observations are most sensitive to a radius of 
around 300~AU and derive column densities of $\lesssim$~3.5~$\times$~10$^{11}$~cm$^{-2}$, 
$\approx$~8~$\times$~10$^{11}$~cm$^{-2}$ and $\approx$~13~$\times$~10$^{11}$~cm$^{-2}$ 
for DM~Tau, LkCa~15 and GO~Tau, respectively.  
Comparing this with our calculations at 305~AU in Table~\ref{table2}, 
we determine a column density of 1~$\times$~10$^{11}$~cm$^{-2}$, which is 
within the upper limit derived for DM~Tau, but around one order of 
magnitude lower than the column densities for the remaining two sources.   
We also calculated the rotational line spectra for \ce{HC3N} and found the  
lines were much weaker than the observed line intensities 
($\ll$~10~mJy~km~s$^{-1}$ versus 60~--~100~mJy~km~s$^{-1}$).  
We note here that LkCa~15 is a particularly peculiar object: 
the discovery of a large cavity in continuum emission within a radius of 
$\approx$~50~AU has reclassified this 
object as a transition disk \citep{pietu06} in which planet formation is likely at an 
advanced stage \citep{kraus12}.   
Analysis of CO line observations also identified the lack 
of a vertical temperature gradient in this disk \citep[see, e.g.,][]{pietu07}.  
In addition, GO~Tau hosts a particularly large, massive molecular disk 
\citep[$R_\ce{CO}$~$\sim$~900~AU,][]{schaefer09}. 
Hence, our disk model is likely not a good analogue for both these sources, 
providing further explanation for the disagreement between our model 
results and observations.

Recently, \citet{qi13b} reported the detection of cyclopropenylidene, $c$-\ce{C3H2}, 
in a protoplanetary disk for the first time.  
The authors identified several lines of this species in ALMA Science Verification observations 
of the disk of HD~163296, a Herbig~Ae star.  
This allowed the authors to derive a column density $\sim$~10$^{12}$~--~10$^{13}$~cm$^{-2}$.    
Herbig Ae/Be stars are more massive and luminous than T~Tauri stars, hence, our disk model 
is not a suitable analogue for this source.  
However, it is interesting to consider whether this species may also be detectable in 
disks around T~Tauri stars.  
Our model predicts a column density of $\approx$~1~$\times$~10$^{11}$~cm$^{-2}$ at a radius of 
100~AU and $\approx$~3~$\times$~10$^{11}$~cm$^{-2}$ at 305~AU, around two orders 
of magnitude lower than that derived for HD~163296.    

Gas-phase methanol, \ce{CH3OH}, has not yet been detected in a 
protoplanetary disk.  
However, there have been multiple searches in several well-studied objects 
giving well-constrained upper limits to the line intensities and 
column densities.  
\citet{thi04} searched for four lines of methanol 
(2$_{02}$-1$_{01}$ A, 4$_{22}$-3$_{12}$ E, 5$_{05}$-4$_{04}$ A, 7$_{07}$-6$_{06}$ A) 
in the disks of LkCa15 and TW Hya using the IRAM 30~m and JCMT single-dish telescopes.   
In all cases, upper limits only were determined, leading to derived 
upper column densities between $\approx$~1~$\times$~10$^{13}$ and 
$\approx$~4~$\times$~10$^{14}$~cm$^{-2}$.  
Again, our calculated column densities agree with these values in that 
we predict column densities generally lower than the upper limits derived from 
the observations.  
\citet{oberg10,oberg11} also included a line transition of methanol 
in their SMA line survey of protoplanetary disks.  
They targeted the 4$_{22}$-3$_{12}$ transition of E-type \ce{CH3OH} at 
218.440~GHz in a range of T~Tauri and Herbig~Ae/Be disks and 
were unable to detect the line in all cases.    

In Table~\ref{table3}, we compare our modelled line intensities 
with observations towards sources in which \ce{H2CO} has been detected and in which 
\ce{H2CO} and \ce{CH3OH} upper limits have been determined.  
We restrict this list to T~Tauri stars 
which possess a substantial gaseous disk.  
We have rescaled our modelled intensities by the disk size and 
distance to source using the values listed in Table~\ref{table3}.  
We have listed the sources roughly in order of decreasing spectral type, 
from K3 (GM Aur) to M1 (DM Tau). 
We have converted the IRAM 30~m and JCMT line intensities from \citet{thi04} 
using the standard relation 
\begin{equation}
\left ( \frac{T}{1\,\mathrm{K}} \right ) 
= \left ( \frac{S_\nu}{1\,\mathrm{Jy}\,\mathrm{beam}^{-1}} \right ) 
\left [ 13.6 \left ( \frac{300\,\mathrm{GHz}}{\nu} \right )^{2} 
\left ( \frac{1"}{\theta^2} \right ) \right ]
\end{equation}
where $T$ is the line intensity in K, $S_\nu$ is the line intensity
in Jy~beam$^{-1}$, $\nu$ is the line frequency in GHz and 
$\theta$ is the beam size in arcseconds.  

The modelled line intensities for \ce{H2CO} agree reasonably well 
(within a factor of three) with most transitions towards most sources.  
For the hotter stars (GM~Tau, LkCa~15, V4046~Sgr, and TW Hya) 
there is better agreement for the higher frequency transitions 
than for the lower frequency transitions.  
For the cooler stars (DM Tau and GG~Tau), 
there is also reasonable agreement with the lower frequency transitions. 
The change in line intensity ratios moving from hotter stars to 
cooler stars reflects the change in disk temperature structure and thus 
excitation conditions.  
For the lines in which we see poor agreement, the calculations 
tend to underestimate the observed line intensities.  
We would not expect absolute agreement with any particular source because we have 
adopted `typical' T~Tauri star-disk parameters in our model.  
However, the level of agreement between our calculations 
and observations is sufficient for us to conclude that our model is providing 
a reasonable description of the formation and distribution of \ce{H2CO} in 
protoplanetary disks around T~Tauri stars and the 
resulting line emission expected from these objects.

Comparing the methanol upper limits and calculated line 
intensities, we see that our calculations fall well within the 
upper limits for all sources.  
Our calculations suggest that the lines of methanol targeted in previous 
surveys of disks are likely too weak to have been observable.  
However, our calculations also suggest several potential candidate lines we expect to 
be strong enough for detection with ALMA 
(see~Sect.~\ref{linespectra} and Fig.~\ref{figure9}).  

\begin{table*}
\footnotesize
\caption{\ce{H2CO} and \ce{CH3OH} rotational transitions in protoplanetary disks.}
\centering
\begin{tabular}{lcccccccc}
\hline\hline
Object     & Distance & $R(\ce{CO})$ & $i$(CO) & Transition & Frequency & Observed intensities & Modelled intensities & References \\
           & (pc)     &  (AU)        & (deg)   &          & (GHz)     & (mJy~km~s$^{-1}$)    &  (mJy~km~s$^{-1}$)   & \\
\hline
\multicolumn{9}{c}{\ce{H2CO}}\\
\hline
GM Aur    & 140 & 630 & 49 & 3$_{03}$-2$_{02}$ & 218.222 &      560 & 130  & 1,2 \\
          &     &     &    & 4$_{14}$-3$_{13}$ & 281.527 &      570 & 410  & 2   \\
LkCa 15   & 140 & 905 & 52 & 2$_{12}$-1$_{11}$ & 140.839 &      844 & 110  & 3,4 \\  
          &     &     &    &                   &         &      820 &      & 5   \\
          &     &     &    & 3$_{03}$-2$_{02}$ & 218.222 &      696 & 280  & 4   \\   
          &     &     &    &                   &         &      660 &      & 2   \\
          &     &     &    & 3$_{22}$-2$_{21}$ & 218.475 &  $<$ 498 & 30   & 4   \\   
          &     &     &    & 3$_{12}$-2$_{11}$ & 225.697 &      498 & 480  & 4   \\   
          &     &     &    & 4$_{14}$-3$_{13}$ & 281.527 &     1120 & 810  & 2   \\ 
          &     &     &    & 5$_{15}$-4$_{14}$ & 351.768 &     5264 & 1100 & 4   \\  
AS 205    & 125 & 250 & 25 & 3$_{03}$-2$_{02}$ & 218.222 &  $<$ 160 & 27   & 2,6 \\
          &     &     &    & 4$_{14}$-3$_{13}$ & 281.527 &  $<$ 300 & 78   & 2   \\
AS 209    & 125 & 340 & 38 & 3$_{03}$-2$_{02}$ & 218.222 &  $<$ 210 & 49   & 2,6 \\
          &     &     &    & 4$_{14}$-3$_{13}$ & 281.527 &  $<$ 480 & 140  & 2   \\
V4046 Sgr & 73  & 370 & 33 & 3$_{03}$-2$_{02}$ & 218.222 &     1001 & 170  & 2,7 \\
          &     &     &    & 4$_{14}$-3$_{13}$ & 281.527 &      950 & 500  & 2   \\
TW Hya    & 56  & 215 & 6  & 3$_{12}$-2$_{11}$ & 225.697 & $<$ 1026 & 170  & 4,8 \\  
          &     &     &    & 4$_{14}$-3$_{13}$ & 281.527 &     1220 & 290  & 9   \\
          &     &     &    & 5$_{15}$-4$_{14}$ & 351.768 &  $<$ 726 & 390  & 4   \\  
          &     &     &    &                   &         &      540 &      & 9   \\
AA Tau    & 140 & 995 & 75 & 3$_{03}$-2$_{02}$ & 218.222 &  $<$ 520 & 340  & 2,10 \\
          &     &     &    & 4$_{14}$-3$_{13}$ & 281.527 &      160 & 980  & 2    \\           
DM Tau    & 140 & 890 & 35 & 2$_{12}$-1$_{11}$ & 140.839 &      300 & 110  & 3,11 \\  
          &     &     &    & 2$_{02}$-1$_{01}$ & 145.602 &      110 & 83   & 11   \\  
          &     &     &    & 3$_{13}$-2$_{12}$ & 211.211 &      480 & 400  & 11   \\  
          &     &     &    & 3$_{03}$-2$_{02}$ & 218.222 &      350 & 270  & 2    \\  
          &     &     &    & 4$_{14}$-3$_{13}$ & 281.527 &      290 & 790  & 2    \\   
GG Tau    & 140 & 800 & 37 & 2$_{12}$-1$_{11}$ & 140.839 &      340 & 85   & 11,12 \\ 
          &     &     &    & 2$_{02}$-1$_{01}$ & 145.602 &      420 & 67   & 11    \\ 
          &     &     &    & 3$_{13}$-2$_{12}$ & 211.211 &      790 & 320  & 11    \\ 
IM Lup    & 190 & 900 & 54 & 3$_{03}$-2$_{02}$ & 218.222 &      530 & 150  & 2,13 \\
          &     &     &    & 4$_{14}$-3$_{13}$ & 281.527 &     1370 & 440  & 2    \\
\hline
\multicolumn{9}{c}{\ce{CH3OH}}\\
\hline   
GM Aur    & 140 & 630 & 49 & 4$_{22}$-3$_{12}$ E & 218.440 & $<$~90   &  15      & 1,14   \\
LkCa 15   & 140 & 905 & 52 & 2$_{02}$-1$_{01}$ A &  96.741 & $<$~247  & $<$ 1.00 & 3,4    \\   
          &     &     &    & 4$_{22}$-3$_{12}$ E & 218.440 & $<$~498  & 30       &  4     \\   
          &     &     &    &                     &         & $<$~150  &          & 14     \\
          &     &     &    & 5$_{05}$-4$_{04}$ A & 241.791 & $<$~497  & 71       &  4     \\   
AS 205    & 125 & 250 & 25 & 4$_{22}$-3$_{12}$ E & 218.440 & $<$~180  & 2.9      & 2,6,14 \\
AS 209    & 125 & 340 & 38 & 4$_{22}$-3$_{12}$ E & 218.440 & $<$~250  & 5.3      & 2,6,14 \\
V4046 Sgr & 73  & 370 & 33 & 4$_{22}$-3$_{12}$ E & 218.440 & $<$~380  & 18       & 7,14   \\
TW Hya    & 56  & 215 & 6  & 7$_{07}$-6$_{06}$ A & 338.409 & $<$~362  & 31       & 8,14   \\     
AA Tau    & 140 & 995 & 75 & 4$_{22}$-3$_{12}$ E & 218.440 & $<$~180  & 36       & 10,14  \\
DM Tau    & 140 & 890 & 35 & 3$_{03}$-2$_{02}$ A & 145.103 & $<$~240  & 20       & 3,15   \\   
          &     &     &    & 4$_{22}$-3$_{12}$ E & 218.440 & $<$~100  & 29       & 14     \\   
IM Lup    & 190 & 900 & 54 & 4$_{22}$-3$_{12}$ E & 218.440 & $<$~310  & 16       & 13,14  \\
\hline
\end{tabular}
\label{table3}
\tablebib{
 (1) \citet{dutrey08}; 
 (2) \citet{oberg10}; 
 (3) \citet{pietu07};  
 (4) \citet{thi04};  
 (5) \citet{aikawa03};  
 (6) \citet{andrews09}; 
 (7) \citet{rodriguez10}; 
 (8) \citet{andrews12}; 
 (9) \citet{qi13a}; 
(10) \citet{kesslersilacci04}; 
(11) \citet{dutrey97};  
(12) \citet{guilloteau99};  
(13) \citet{panic09};  
(14) K. \"{O}berg (2012, private communication);  
(15) \citet{dutrey00}.} 
\end{table*}

\subsection{Comparison with other models}
\label{comparisonwithothermodels}

Here, we compare our results with other protoplanetary disk 
models, concentrating on work which has published lists of column densities 
or fractional abundances for relatively complex species.  
Historically, chemical models of disks have concentrated on simple, abundant 
species (and isotopologues), since these are readily observed in many systems 
(e.g., CO, \ce{HCO+}, CN, CS, and HCN).  
As we enter the era of ALMA, it is likely that the molecular inventory 
of protoplanetary disks will significantly increase, requiring much more
sophisticated complex chemical models, such as that presented here.  

In Table~\ref{table4}, we compare column densities of various complex 
molecules at a radius of $\approx$~250~AU with other protoplanetary disk 
models of comparable chemical complexity and with similar chemical ingredients.  
\citet[][W07]{willacy07} presented a chemically complex model of a protoplanetary disk, 
including a comprehensive deuterium chemistry.  
We compare our column densities with Model~C in that work, 
which includes both grain-surface chemistry and non-thermal 
desorption.  
\citet[][SW11]{semenov11} present results from a disk model which 
uses a network with a number of chemical reactions ($\approx$~7300) 
approaching the number in the network presented here ($\approx$~9300). 
We compare our results with their `laminar' model in which they 
neglect turbulent mixing, since we do not consider mixing in this 
work.  
We also list the column densities from our most recent paper, WNMA12, 
which is most similar to the work presented here in that the 
disk physical model is identical as are the methods used to 
calculate the chemistry.  
In WNMA12 we used a chemical network based on `{\sc Rate}06', the most recent release of the 
UMIST Database for Astrochemistry (UDfA) available at that time, whereas, here, 
our network is derived from the Ohio State University (OSU) network and 
includes a vastly more comprehensive grain-surface chemical network to 
simulate the build up of complex molecules. 
The network used in W07 is also derived from UDfA, 
albeit an earlier version \citep[{\sc Rate}95,][]{millar97}, whereas,   
the network used by SW11 is also based on the OSU network.  
Care must be taken when comparing results from different 
protoplanetary disk models, as they often differ in physical ingredients 
as well as the chemistry.  

The work presented here generally predicts higher column densities 
for COMs than those presented in W07 and 
SW11 despite relatively similar (within an order of magnitude) 
column densities for CO, \ce{H2CO}, and \ce{HC3N}.  
In this work, we calculate significantly higher column densities 
for \ce{CH3OH}, \ce{HCOOH}, \ce{CH3CN}, \ce{CH3CHO}, \ce{NH2CHO}, 
\ce{HCOOCH3}, \ce{C2H5OH}, \ce{CH3OCH3}, and \ce{CH3COCH3}.  
The network used by SW11 is based on that presented in \citet{garrod06} 
which does not contain many pathways to the larger species introduced 
in \citet{garrod08}.    
Also, they adopt $E_{d}$~=~0.77~$E_{D}$ for their 
grain-surface diffusion rates \citep[which originates from][]{ruffle00}, 
where $E_{D}$ is the binding energy of the molecules to the grain surface.  
This is a rather conservative value and partly 
explains their much lower abundances of complex species.  
In addition, they do not consider quantum tunnelling of H atoms 
on grain surfaces, nor through reaction energy barriers 
\citep[a full description of the chemical model is provided in][]{semenov10}. 
The neglect of H atom tunnelling through reaction energy barriers explains 
the particularly low column density of methanol in SW11 ($\sim$~10$^{8}$~cm$^{-2}$). 
W07 include atom-addition grain-surface reactions only and thus 
neglect radical-radical pathways to form larger COMs.

Comparing our results with those from our previous work (WNMA12), 
we see a significant increase in the column density of 
\ce{CH3OH}, \ce{HCOOCH3}, \ce{CH3OCH3}, and \ce{CH3COCH3}
when using the gas-grain network presented here.   
The higher column density of grain-surface methanol, 
{\em s-}\ce{CH3OH}, can be attributed to the higher 
binding energy of CO adopted here.  
In previous work, we used the value measured for pure CO ice (855~K) as 
opposed to the value measured in water ice (1150~K).  
The binding energy regulates the abundance of {\em s-}CO on the 
grain and thus the amount available for conversion to {\em s-}\ce{CH3OH}, 
as well as the grain-surface radicals, {\em s-}\ce{HCO}, {\em s-}\ce{CH3O}, and {\em s-}\ce{CH2OH}.  
Regarding the formation of {\em s-}\ce{HCOOCH3}, 
the grain-surface association reaction,
{\em s-}\ce{HCO}~+~{\em s-}\ce{CH3O}, is included in both models.  
The difference in column density is due, again, to the different sets of binding energies 
adopted.  
The results from our exploratory calculations presented 
in Sect.~\ref{verticalresults} demonstrate the 
importance of radiation processing for the production of 
{\em s-}\ce{CH3OCH3} and {\em s-}\ce{CH3COCH3} in the disk midplane.  
The midplane is the densest region of the disk and thus 
contributes significantly to the vertical column density. 
In previous work we did not include the processing of ice mantle material 
by UV photons and X-rays.  
We also see a decrease in the column density of gas phase \ce{HC3N}, 
and a corresponding increase in the grain-surface column density, 
compared with our previous values.  
This is due to the increased binding energy for \ce{HC3N} adopted here 
(4580~K compared with 2970~K).  
Our previous value shows better agreement with the column densities constrained 
from observations (~$\sim$~10$^{12}$~cm$^{-2}$).

Protoplanetary disks are turbulent environments and the effects of vertical 
turbulent mixing on disk chemical structure has been investigated by 
multiple groups \citep[see, e.g.,][]{ilgner04,willacy06,semenov06,aikawa07,heinzeller11,semenov11}.  
\citet{semenov11} conducted a comprehensive investigation of disk chemical structure 
with and without turbulent mixing and identified a plethora of species 
which are sensitive to mixing.  
\citet{semenov11} also used a chemical network including several complex molecules 
(see~Table~\ref{table4}).  
Of the gas-phase molecules of interest here, 
they found that the column densities of \ce{HCOOH}, \ce{HC3N}, and \ce{CH3CN}, were 
significantly affected by the inclusion of turbulent mixing.  
However, they concentrated their discussions on species which reached column densities 
$\gtrsim$~10$^{11}$~cm$^{-2}$.  

In this work, we assume the dust grains are well mixed with the gas and, 
for the calculation of the chemical structure, we assume the grains are 
compact spherical grains with a fixed radius.  
In reality, the dust grains will have both a size distribution and variable 
dust-to-gas mass ratio 
caused by gravitational settling towards the midplane and dust-grain coagulation (grain growth).  
Several groups have also looked at the effects of dust-grain evolution on protoplanetary disk 
chemistry \citep[see, e.g.,][]{aikawa06,fogel11,vasyunin11,akimkin13}.  
A parameterised treatment of grain growth affects the geometrical height of the 
molecular layer but appears to have little effect on the column densities of 
gas-phase molecules \citep{aikawa06}.  
Larger grains may lead to a reduced volume grain-surface area (for a fixed dust-to-gas 
mass ratio) which will lower the accretion rate of molecules onto dust grains, thereby 
potentially lowering the formation rate of COMs.  
However, this effect depends on the assumed morphology and porosity of the grains.  
Grain coagulation models suggest that the growing dust grains retain a 
`fluffy' morphology (with a low filling factor, $\ll$~1) such that the volume 
grain-surface area may not significantly 
decrease before compression occurs \citep[see, e.g.,][]{ossenkopf93,ormel07,kataoka13}.
Grain settling towards the midplane allows the deeper penetration of UV radiation 
leading to warmer grains in the disk midplane. 
This subsequently results in a smaller depletion (freezeout) zone and a 
larger molecular layer situated closer to the midplane \citep{fogel11}.
\citet{akimkin13} performed a coupled calculation of the structure 
of a protoplanetary disk including dust evolution and radiative transfer, 
and subsequently calculated the chemical evolution.  
They find that the abundances of several species are enhanced in models 
including dust evolution, including the relatively complex species, 
\ce{NH2CN} and \ce{HCOOH}.   
We intend to explore the effects of grain evolution on the formation and distribution of 
COMs in future models.

A final issue to consider is the validity of our set of initial abundances.  
Disk formation is thought to be a vigorous, energetic, and potentially destructive process. 
Accretion shocks are thought to occur as material falls from the envelope onto the disk.  
Heating by the shock may raise the temperature of the dust grains above the sublimation temperature of 
ices and energised ions may sputter ices from dust grain surfaces 
\citep[see, e.g.,][]{neufeld94,tielens94}. 
Hence, using initial abundances representative of dark cloud (or prestellar) conditions may not 
be realistic because dust grains may be stripped of ices as they pass through an accretion shock 
during the disk formation stage.  
\citet{visser09} studied the 2D chemical evolution during the protostellar collapse phase 
to determine the chemical history of simple ices contained within the disk at the end of collapse.  
They concluded that accretion shocks that occur as material falls from 
the envelope onto the disk are much weaker than commonly assumed.  
For the outer disk, the main contribution to heating is via stellar heating 
\citep[see Fig.~3 in][]{visser09}.   
Sputtering of dust grains by energetic ions can also occur.  
\citet{visser09} also considered this and concluded that the shock velocities 
experienced by the gas, $\approx$~8~km~s$^{-1}$, are not sufficient to energise 
ions, such as \ce{He+}, to energies required for the removal of water molecules from grain surfaces. 
As a result, much of the material contained within the outer disk ($\gtrsim$~10~AU) 
at the end of collapse consists primarily of ``pristine'' interstellar ice 
\citep[see Fig. 4 in][]{visser11}.   
Hence, beginning our simulations with initial molecular abundances 
representative  of prestellar conditions is an appropriate assumption. 

\begin{table*}
\footnotesize
\caption{Calculated column densities (cm$^{-2}$) at $R$~=~250~AU.}
\centering
\begin{tabular}{llcccccc}
\hline\hline
                 &                 & \multicolumn{4}{c}{Gas phase} & \multicolumn{2}{c}{Grain surface} \\
\multicolumn{2}{c}{Species}        & W07$^1$  & SW11$^2$ & WNMA12$^3$ & This work & WNMA12$^3$ & This work \\ 
\hline\\
Carbon monoxide  & \ce{CO}         & 2.9(17)  & 1.2(18)  & 1.3(18)    & 3.1(17)   & 2.1(14)    & 2.4(17)  \\
Formaldehyde     & \ce{H2CO}       & 3.9(12)  & 2.4(13)  & 5.9(12)    & 7.2(12)   & 1.7(14)    & 2.9(17)  \\
Methanol         & \ce{CH3OH}      & 1.5(11)  & 6.1(08)  & 1.9(12)    & 1.7(13)   & 8.4(16)    & 7.7(17)  \\
Formic acid      & \ce{HCOOH}      & $\cdots$ & 1.4(11)  & 9.8(13)    & 1.6(13)   & 2.1(16)    & 6.1(16)  \\
Cyanoacetylene   & \ce{HC3N}       & 7.0(11)  & 2.9(11)  & 2.5(12)    & 1.7(11)   & 2.9(08)    & 4.0(12)  \\
Acetonitrile     & \ce{CH3CN}      & 4.9(10)  & 6.0(10)  & 5.6(12)    & 7.3(11)   & 5.8(15)    & 2.5(15)  \\
Acetaldehyde     & \ce{CH3CHO}     & $\cdots$ & 8.0(07)  & 6.2(10)    & 3.4(11)   & 2.2(15)    & 3.7(16)  \\
Formamide        & \ce{NH2CHO}     & $\cdots$ & 2.0(10)  & $\cdots$   & 9.2(11)   & $\cdots$   & 1.4(16)  \\
Methyl formate   & \ce{HCOOCH3}    & $\cdots$ & 8.8(04)  & 3.5(08)    & 2.7(11)   & 1.8(11)    & 1.1(16)  \\
Ethanol          & \ce{C2H5OH}     & $\cdots$ & 4.4(06)  & 1.4(12)    & 6.1(10)   & 1.8(16)    & 6.9(15)  \\
Dimethyl ether   & \ce{CH3OCH3}    & $\cdots$ & 4.2(02)  & 7.0(09)    & 7.6(10)   & 1.4(14)    & 1.2(16)  \\
Acetone          & \ce{CH3COCH3}   & $\cdots$ & 1.3(03)  & 2.8(06)    & 4.2(09)   & 2.6(09)    & 6.2(15)  \\
\hline
\end{tabular}
\label{table4}
\tablefoot{$a(b)$ represents $a\times10^b$}
\tablebib{
 (1) \citet[][W07]{willacy07}; 
 (2) \citet[][SW11]{semenov11}; 
 (3) \citet[][WNMA12]{walsh12}}  
\end{table*}

\subsection{Complex molecules in comets}
\label{comets}

It is generally accepted that minor bodies in the Solar 
System, such as asteroids and comets, likely formed in conjunction with the planets 
in the Sun's primordial disk and can be considered remnant material left 
over from the process of planet formation.  
When a comet's orbit is perturbed in such a way that it is injected into the inner Solar System, 
the gradual warming of the nearing Sun evaporates solid surface material and creates 
an expansive cometary coma of gaseous volatile material enveloping the comet nucleus.  
Photolysis of the sublimated material (termed `parent' species) and subsequent chemistry 
creates ions and radicals and new molecules (termed `daughter' species).  
It is now understood that comets are complex objects composed of ice  
(mainly \ce{H2O}, \ce{CO2}, and CO), refractory material (such as silicates), and 
organic matter.  
To date, more than 20 parent molecules have been observed in cometary comae including 
the relatively complex species, \ce{H2CO}, \ce{CH3OH}, 
HCOOH, \ce{CH3CHO}, \ce{HC3N}, \ce{CH3CN},
\ce{NH2CHO}, \ce{HCOOCH3}, and \ce{(HOCH2)2} (ethylene glycol), which are relevant to this work.  
Of these species, \ce{CH3CHO}, \ce{NH2CHO}, \ce{HCOOCH3}, and \ce{(HOCH2)2} have been 
observed in only a single object, comet Hale-Bopp, with percentage abundances 
of 0.02~\%, 0.015~\%, 0.08~\%, and 0.25~\% (relative to \ce{H2O}), respectively 
\citep[see, e.g.,][]{bockelee04,crovisier04a,crovisier04b,crovisier06,mumma11}. 

\begin{figure*}
\subfigure{\includegraphics[width=1.0\textwidth]{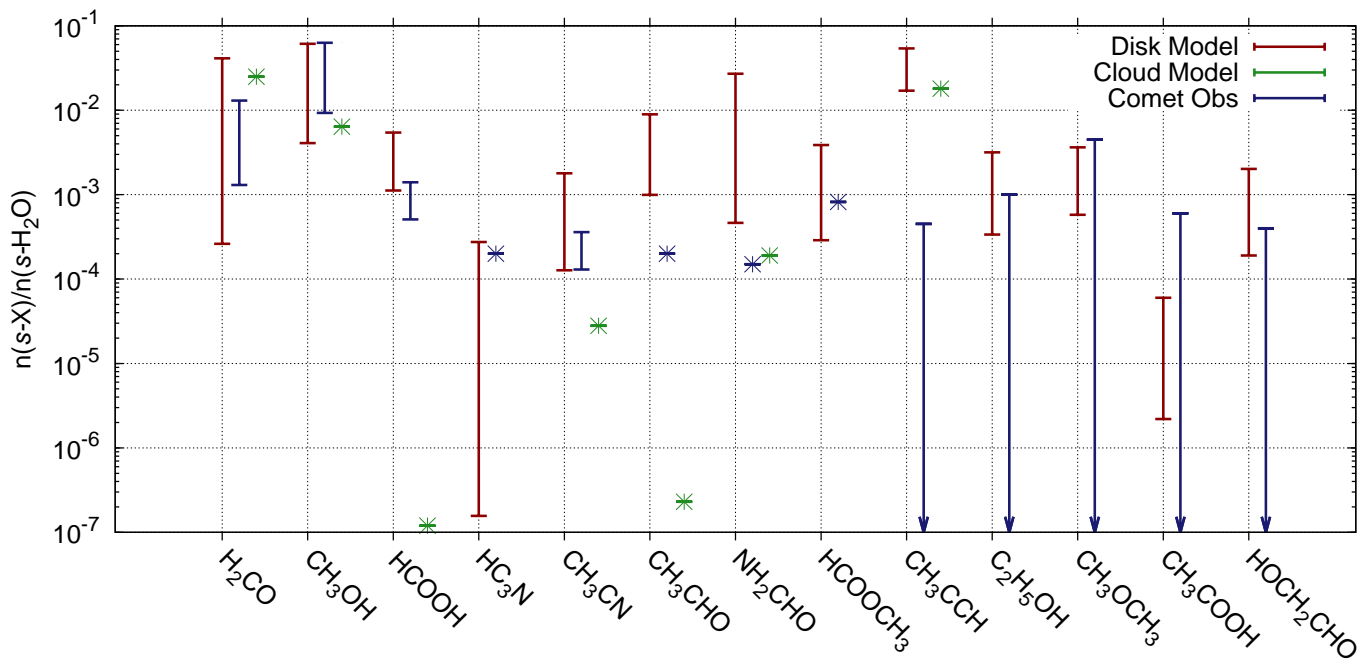}}
\caption{Range of abundances of grain-surface complex molecules relative to water ice from our 
model calculations (red lines) compared with those derived from observations of cometary comae 
(blue asterisks and lines) and our initial dark cloud ice ratios (green asterisks).  
The comet data is from \citet{bockelee04} and \citet{crovisier06}. 
The single points and upper limits for the comet ratios represent data derived from observations of 
comet Hale-Bopp \citep{crovisier04b}.}
\label{figure10}
\end{figure*}

In Fig.~\ref{figure10} we present the range of calculated abundances for 
each of these grain-surface species relative to water ice (red lines) and compare these with 
our initial adopted dark cloud ice ratios (green asterisks) and 
data derived from cometary comae observations (blue asterisks and lines).  
The fractional abundances from the disk model are determined by the relative vertical 
column densities at each radius. 
We restrict our data to $R$~$\gtrsim$~20~AU which corresponds to the radius 
beyond which grain-surface COMs achieve significant column densities 
(see Fig.~\ref{figure8}). 
This also correlates with the region where 
comets are postulated to have originally formed and resided in modern dynamical models of the 
Solar System \citep[see, e.g.,][]{gomes05,walsh11}.    
The single points and upper limits for the comet observations 
refer to data derived from observations of comet Hale-Bopp.  
We find that our range of calculated abundances (relative to water ice) are consistent 
with those derived from observations,  with some overlap between the two datasets for most species.  
Exceptions include {\em s-}\ce{CH3CHO} and {\em s-}\ce{NH2CHO} for which a single observation only is 
available.  
In both cases, our data range is larger than the observed ratio, with the lower limit of our data 
within a factor of a few of the measured ratio.  
Another exception is {\em s-}\ce{CH3CCH}, for which an upper limit towards Hale-Bopp only has been
derived \citep[][]{crovisier04b}.  
Again, we find our calculated ratio range is larger than the upper limit.  
In this case, the lower limit of our data is much further away from that derived from observation, 
by a factor of $\approx$~30.  
It is also interesting to compare our range of calculated abundances in the disk model 
with our initial abundances adopted from a dark cloud model (see Table~\ref{table1}).
The {\em s-}\ce{H2CO}/{\em s-}\ce{CH3OH} ratio indicates there is significant chemical processing 
of the dark cloud grain-surface material within the disk with this ratio decreasing from cloud to disk.  
For all other species (except {\em s-}\ce{CH3CCH}) the dark cloud abundance is lower than the lower 
limit reached in the disk model indicating that disk physical conditions are necessary 
for thermal grain-surface chemistry to efficiently form the 
complex molecules observed in cometary comae. 

It certainly appears that our grain-surface chemistry is appropriate for describing the grain-surface 
formation of most COMs observed in cometary comae, supporting the postulation that comets 
are formed via the coagulation and  growth of icy dust grains within the Sun's protoplanetary disk.   
One outstanding issue is the high abundance of ethylene glycol (\ce{(HOCH2)2}) observed towards 
comet Hale-Bopp, with a percentage abundance of 0.25\% relative to water ice.  
This ratio is similar to that observed for \ce{H2CO} and around an order of magnitude higher 
than the ratio derived for \ce{CH3CHO} and \ce{NH2CHO}.  
Also, \ce{(HOCH2)2} is observed to be at least 5 times more abundant than the chemically-related species, 
\ce{HOCH2CHO} \citep{crovisier04b}.
In this network, we include a single barrierless route to the formation of 
{\em s-}\ce{(HOCH2)2} via the grain-surface association of two {\em s-}\ce{CH2OH} radicals.  
Under the conditions throughout much of the disk, the mobility of 
this radical is significantly 
slower than smaller radicals, such as, {\em s-}\ce{CH3} and {\em s-}\ce{HCO}, due to 
its significantly larger binding energy to the grain mantle ($\approx$~5000~K).  
The large binding energy is due to the presence of the -OH functional group 
allowing hydrogen bonding of this species with the bulk water ice \citep[see, e.g.,][]{garrod08}.  
Hence, the reaction forming {\em s-}\ce{(HOCH2)2} cannot compete with other barrierless 
radical-radical association reactions which form, for example, 
{\em s-}\ce{C2H5OH} and {\em s-}\ce{HOCH2CHO}.  
We find a negligible abundance of {\em s-}\ce{(HOCH2)2} is produced throughout our disk model.  

In the network used here, radical-radical association pathways only have been included for 
the formation of many COMs, in addition to pathways involving sequential hydrogenation 
of precursor molecules.  
However, an alternative grain-surface route to the production of \ce{(HOCH2)2} (and other COMs) 
has been proposed by \citet{charnley97} involving sequential atom-addition reactions.  
For example, \ce{(HOCH2)2} is postulated to form via the hydrogenation of {\em s-}\ce{OCCHO}, which in turn is 
formed from {\em s-}\ce{CO} via atom-addition reactions, i.e., 
\begin{equation*}
s\mbox{-}\ce{CO}   \xrightarrow{s\mbox{-}\ce{H}}
s\mbox{-}\ce{HCO}  \xrightarrow{s\mbox{-}\ce{C}}
s\mbox{-}\ce{HC2O} \xrightarrow{s\mbox{-}\ce{O}}
s\mbox{-}\ce{OCCHO} 
\end{equation*}
followed by
\begin{equation*}
s\mbox{-}\ce{OCCHO}    \xrightarrow{s\mbox{-}\ce{H}} 
s\mbox{-}\ce{CHOCHO}   \xrightarrow{2s\mbox{-}\ce{H}}
s\mbox{-}\ce{HOCH2CHO} \xrightarrow{2s\mbox{-}\ce{H}}
s\mbox{-}\ce{(HOCH2)2}.
\end{equation*}
In this sequence, 2{\em s-}\ce{H} implies a barrier penetration reaction by a hydrogen atom
followed by the exothermic addition of an additional H atom.  
This sequence of atom-addition reactions is postulated to lead to different ratios of resultant 
grain-surface COMs relative to the radical-radical network used here and may provide a route to the 
formation of {\em s-}\ce{(HOCH2)2}.  
However, as discussed in \cite{herbst09}, many of these reaction rates 
and reaction barriers remain unmeasured.  
The efficacy of this type of formation route to COMs under protoplanetary disk 
conditions is yet to be studied and we intend to explore this in future work.  
Of course, it is also possible that significant processing of the cometary surface 
by UV photons (and potentially, cosmic rays) over the comet's lifetime may 
lead to a surface composition which differs from the initial grain mantle composition 
in the protoplanetary disk.  
In addition, thermally driven chemical processing of the comet's interior 
may occur.  This may be caused by heating due to radioactive decay of 
radionuclides, such as $^{26}$Al \citep[see, e.g.,][]{wallis80,prialnik87}.

\subsection{Implications for astrobiology}
\label{astrobiology}

One of the most complex molecules detected to date is aminoacetonitrile, 
\ce{NH2CH2CN}, which was observed towards the hot core in the 
massive star-forming region, Sgr B2(N), with 
a fractional abundance $\sim$~10$^{-9}$ \citep{belloche08}.  
\ce{NH2CH2CN} has been postulated as a potential precursor to the simplest amino acid, glycine, 
\ce{NH2CH2COOH}.  
In turn, amino acids are the building blocks of proteins, considered 
a key component for the commencement of life.    
Multiple routes to the formation of glycine (and other simple amino acids) 
under interstellar conditions 
have been proposed including via Strecker synthesis \citep[see, e.g.,][]{peltzer84}, 
UV-irradiated ice mantles \citep[see, e.g.,][]{bernstein02,munoz-caro02}, and gas-phase chemistry 
\citep[see, e.g,][]{blagojevic03}.  

Recently, \citet{garrod13} investigated the formation of glycine in hot cores 
via grain-surface radical-radical reactions, i.e., an extension to the reaction scheme used here, 
incorporating the ice chemistry proposed in \citet{woon02} to describe the formation of 
glycine in UV-irradiated ices.  
\citet{garrod13} calculated a peak fractional abundance for gas-phase glycine 
$\sim$~10$^{-10}$~--~10$^{-8}$ with the molecule returned to the gas phase 
at temperatures $\gtrsim$~200~K. 
He also included gas-phase formation of glycine \citep{blagojevic03} 
and determined it to have a negligible effect on the resulting abundances.  
The detection of glycine is considered one of the holy grails of astrochemistry and astrobiology; 
however, searches for gas-phase glycine, thus far, have been unsuccessful 
\citep[see, e.g.,][]{snyder05}.  
The predictions from \citet{garrod13} are consistent with upper limits derived 
from these observations.  
He proposes that due to the high binding energy of glycine, the 
emission from hot cores is expected to be very compact, and thus, an ideal 
target for detection with ALMA.   

Certainly, a similar grain-surface formation route to {\em s-}\ce{NH2CH2CN} and 
thus, {\em s-}\ce{NH2CH2COOH}, may be possible under protoplanetary disk conditions 
and should be explored in future models,   
particularly considering the recent identification of 
glycine in a sample returned from comet 81P/Wild 2 \citep{elsila09} and the detection of   
numerous amino acids in meteorites, some of which are either very rare on 
Earth or, indeed, unknown in terrestrial biochemistry \citep[for an overview, see, e.g.,][]{ehrenfreund00}.  
Models would help ascertain whether it is possible for simple amino acids to form 
on dust grains in the Sun's protoplanetary disk and become incorporated into comets 
and asteroids.  
Such models could also provide further evidence for 
the delivery of prebiotic molecules to Earth via asteroid and/or cometary impact, rather 
than forming `in situ' early in the Earth's evolution.  
However, based on our molecular line emission calculations 
(see Sect.~\ref{linespectra}), even if such prebiotic molecules 
were present in quantities similar to that expected in hot cores, 
the detection of the gas-phase form of these species in protoplanetary disks 
would be incredibly challenging, if not impossible, 
even with ALMA full capabilities.  

\section{Summary}
\label{summary}

In this work, we have presented the results of exploratory models 
investigating the synthesis of COMs
in a protoplanetary disk around a typical T~Tauri star.  
We used an extensive chemical network, typically adopted in chemical models 
of hot cores, which includes gas-phase chemistry, gas-grain interactions 
(freeze out and desorption), grain-surface chemistry, and 
the irradiation of ice mantle material.  

We summarise the main results of this work below.  
\begin{enumerate}
\item COMs can form efficiently on the grain mantle 
under the physical conditions in the disk midplane
via grain-surface chemical reactions, reaching peak fractional abundances,
$\sim$~10$^{-6}$ to $\sim$~10$^{-4}$ that of the H nuclei number density.  
\item Gas-phase COMs are released to the gas phase 
via non-thermal desorption, with photodesorption via external 
photons being the most important 
process for increasing the abundances in the `molecular layer', and 
cosmic-ray-induced desorption being most important in the disk midplane.
\item This mechanism is different to that in hot core models which require a
`warm up' phase in which the temperature increases from 10~K to $\gtrsim$~100~K 
over a time scale of $\approx$~10$^{5}$~years \citep[see, e.g.,][]{garrod06}.  
\item Most gas-phase COMs
reside in a narrow region within the `molecular layer' ranging in peak fractional 
abundance from $\sim$~10$^{-12}$ (e.g., \ce{CH3COCH3}) to $\sim$~10$^{-7}$ (e.g., \ce{HCOOH}).  
Generally, the more complex the species, the lower the 
peak gas-phase fractional abundance and column density.  
\item \ce{H2CO}, \ce{HC3N}, and \ce{CH3CN}, are exceptions to the above statement.  
These species have gas-phase and grain-surface routes to formation 
and so are relatively abundant throughout the 
molecular layer and upper disk.  
\item Including the irradiation of ice mantle material 
allows further complexity to build in the ice mantle
through the generation of grain-surface radicals which are available 
for further molecular synthesis.  
This process increases the abundances of more complex molecules 
in the disk midplane, further enhancing the abundance of several COMs, 
e.g., \ce{CH3CHO}, \ce{C2H5OH}, \ce{CH3OCH3}, \ce{CH3COCH3}, and \ce{CH3COOH}.  
However, this increase in grain-surface abundances does not necessarily translate 
to an `observable' abundance in the gas phase.    
\item Reactive desorption provides an additional means for molecules 
to return to the gas phase in the disk midplane,  
e.g., \ce{CH3CHO}, \ce{C2H5OH}, and \ce{CH3OCH3}.  
\item The calculated column densities for \ce{H2CO} and \ce{CH3OH} 
are consistent with values and upper limits derived from observations.  
\item There is reasonably good agreement 
between the majority of our calculated line intensities for \ce{H2CO} and those observed 
towards nearby T~Tauri stars. 
For the hotter stars, we get better agreement with the higher frequency transitions than the 
lower frequency transitions. 
For the cooler stars, we also get reasonable agreement with the lower frequency transitions.  
\item There is poor agreement with observed \ce{HC3N} line intensities towards 
LkCa~15 and GO~Tau, which is attributed to 
our lower calculated column density for this species. 
This disagreement may also be due to the particularites of these two sources: 
LkCa~15 is now considered a transition disk with a large gap in continuum emission 
within $\approx$~50~AU and 
GO~Tau hosts a particularly large, massive molecular disk ($R_\mathrm{CO}$~$\approx$~900~AU). 
\item The predicted line intensities for methanol line emission lie well below the upper limits 
determined towards all sources.  
We suggest strong lines of methanol around 
$\approx$~300~GHz (and higher frequencies) are excellent candidates for observation 
in nearby protoplanetary disks using ALMA 
( for details see Sect.~\ref{linespectra} and Fig.~\ref{figure9}).
\item The molecular line emission calculations put interesting constraints on the 
observability of COMs in protoplanetary disks. 
The calculations suggest that the detection of more complex species, 
especially those typically observed in hot cores, 
e.g., \ce{CH3CN} and \ce{HCOOCH3}, 
may prove challenging, even with ALMA `Full Science' operations.  
Detections of COMs of prebiotic significance, e.g., \ce{NH2CH2CN} and \ce{NH2CH2COOH}, 
in protoplanetary disks, may provide additional challenges, 
remaining beyond the reach of ALMA.
\item Our grain-surface fractional abundances (relative to water ice) for the outer 
disk ($R$~$\gtrsim$~20~AU) are consistent with abundances derived for comets, suggesting 
a grain-surface route to the formation of COMs observed in cometary comae.  
This lends support to the idea that comets formed via the coagulation and growth of 
icy grains in the Sun's natal protoplanetary disk.  
\end{enumerate}

Two of the most complex molecules observed in disks, \ce{H2CO} and \ce{HC3N}, can both 
be efficiently synthesised by gas-phase chemistry alone and, thus, are not 
currently validations of the efficacy of grain-surface chemistry in protoplanetary disks.  
Observations of molecules which can only be efficiently formed on the grain, e.g., 
\ce{CH3OH}, are required in order to determine the degree to which grain-surface chemistry 
contributes to the chemical content in protoplanetary disks.  
Methanol is an important molecule in that it is essentially the next `rung' on the `ladder' of 
molecular complexity following \ce{H2CO}.  
It is also a parent molecule of many more complex species.  
The calculations suggest that the expected line intensities of transitions of methanol 
lie well below current observational upper limits.  
Utilising ALMA, with its unprecedented sensitivity and spectral and spatial resolution, and performing 
a deep search for the strongest transitions 
of methanol that fall with the observing bands, would 
confirm whether grain-surface chemistry is an important mechanism 
in protoplanetary disks.   

\begin{acknowledgements}
The authors thank an anonymous referee for his or her useful comments 
which greatly improved the discussion in the paper.  
C.W. acknowledges support from the European Union A-ERC grant 291141
CHEMPLAN and financial support (via a Veni award)
from the Netherlands Organisation for Scientific Research (NWO). 
Astrophysics at QUB is supported by a grant from the STFC.  
E.~H.~thanks the National Science Foundation for support of his program in astrochemistry.  
He also acknowledges support from the NASA Exobiology and Evolutionary program through a 
subcontract from Rensselaer Polytechnic Institute. 
H.~N.~acknowledges the Grant-in-Aid for Scientific Research 23103005 and 25400229.  
She also acknowledges support from the Astrobiology Project of the CNSI, NINS (Grant Number 
AB251002, AB251012). 
S.L.W.W. and J.C.L. acknowledge support from S.L.W.W.'s  startup funds provided by Emory University.
The authors thank K.~\"{O}berg and D.~Semenov for providing observational data and model calculations  
for inclusion in this paper.  
\end{acknowledgements}

\clearpage

\Online

\appendix

\section{On the assumed parameters}

In this work, we have used a fixed set of parameters for the 
calculation of the thermal grain-surface reaction rates and desorption rates.  
Two parameters which may have a strong influence on the grain-surface abundances and 
subsequent gas-grain balance are the diffusion barriers between surface sites, $E_{b}$, and 
the probability for reactive desorption, $P_{rd}$.  
We assume the grain-surface diffusion barrier for each species is proportional to its 
binding (desorption) energy to the grain surface, $E_{D}$.  
Here, we assume an optimistic value, $E_{b}$/$E_{D}$~=~0.3.  
This value allows the reasonably efficient diffusion of radicals within the grain mantle 
when $T$~$\gtrsim$~15~K, which helps to build chemical complexity in the outer disk. 
\citet{vasyunin13a} recently explored the effects of the value assumed for $E_{b}$/$E_{D}$ 
in a macroscopic Monte Carlo model of hot core chemistry, using a `layer-by-layer' approach 
to calculate the grain mantle composition.  
They explored a range of values for 
$E_{b}$/$E_{D}$: 0.3 \citep{hasegawa92}, 0.5 \citep{garrod06}, and 0.77 \citep{ruffle00}.
They concluded models using the intermediate value, $E_{b}$/$E_{D}$~=~0.5, produced 
ice compositions in better agreement with observations; however, models 
with $E_{b}$/$E_{D}$~=~0.3 also gave reasonable agreement for the warm-up phase.  

In addition, we assume a conservative value for the probability for 
reactive desorption, $P_{rd}$~=~0.01.  
This value is that constrained in investigations into the efficacy of reactive 
desorption in dark cloud chemical models \citep{garrod07}.  
Recently, reactive desorption has been postulated as a potential mechanism for 
the release of precursor COMs (e.g., \ce{H2CO} and \ce{CH3OH}) 
in cold, dark clouds where they eventually form larger complex organic 
molecules (e.g., \ce{HCOOCH3} and \ce{CH3OCH3}) in the gas phase  
via radiative association \citep{vasyunin13b}.  
In addition, as discussed in the main body of the paper, 
recent experiments suggest that reactive desorption is particularly efficient 
for the reformation of doubly-deuterated water (\ce{D2O}) and \ce{O2} via the 
surface reactions, {\em s-}D + {\em s-}OD and {\em s-}O + {\em s-}O, 
with efficiencies, $>$~90~\% and $\approx$~60~\%, respectively \citep{dulieu13}.  

In this appendix, we present results from additional exploratory models to investigate the 
effects of a higher diffusion barrier and a higher probability for 
reactive desorption.  
In Table~\ref{tablea1}, we list the parameters adopted in four models.  
In Model A, we assume $E_{b}$/$E_{D}$~=~0.3 and $P_{rd}$~=~0.01.  
This model corresponds to our fiducial model, the full results for which are 
discussed in the main section of this paper.  
In Model B, we adopt a higher diffusion barrier, $E_{b}$/$E_{D}$~=~0.5, and in Model 
C we adopt a higher probability for reactive desorption, $P_{rd}$~=~0.1.  
We present results from an additional model, Model D, in which we have adopted 
the higher values for both parameters.  

\begin{figure*}
\subfigure{\includegraphics[width=0.33\textwidth]{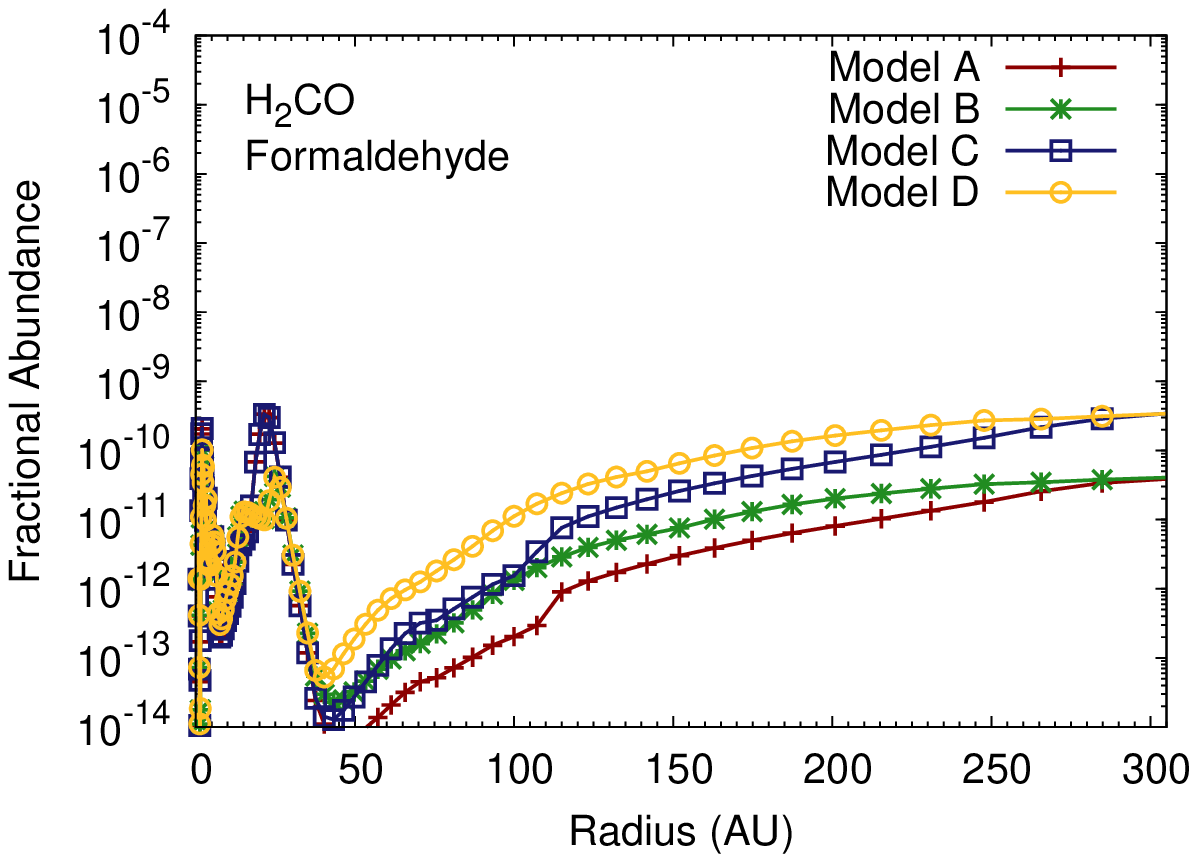}}
\subfigure{\includegraphics[width=0.33\textwidth]{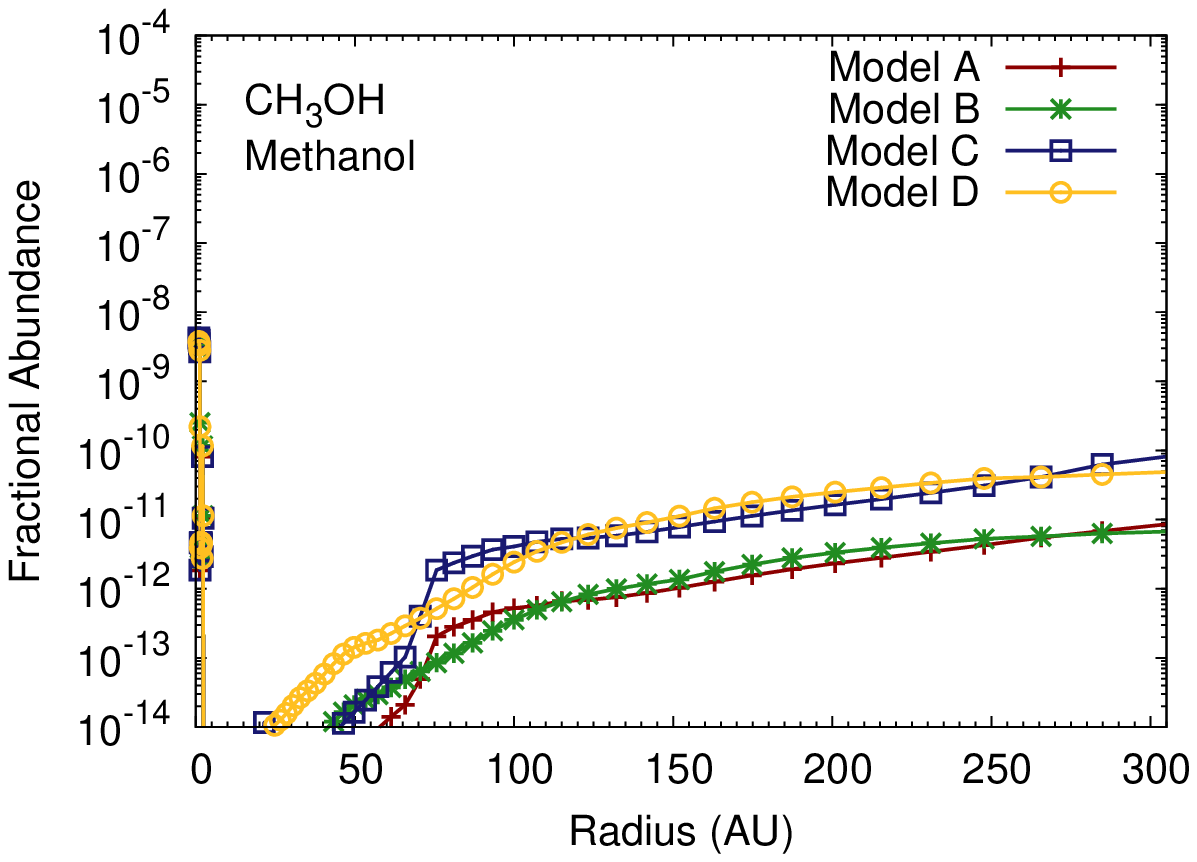}}
\subfigure{\includegraphics[width=0.33\textwidth]{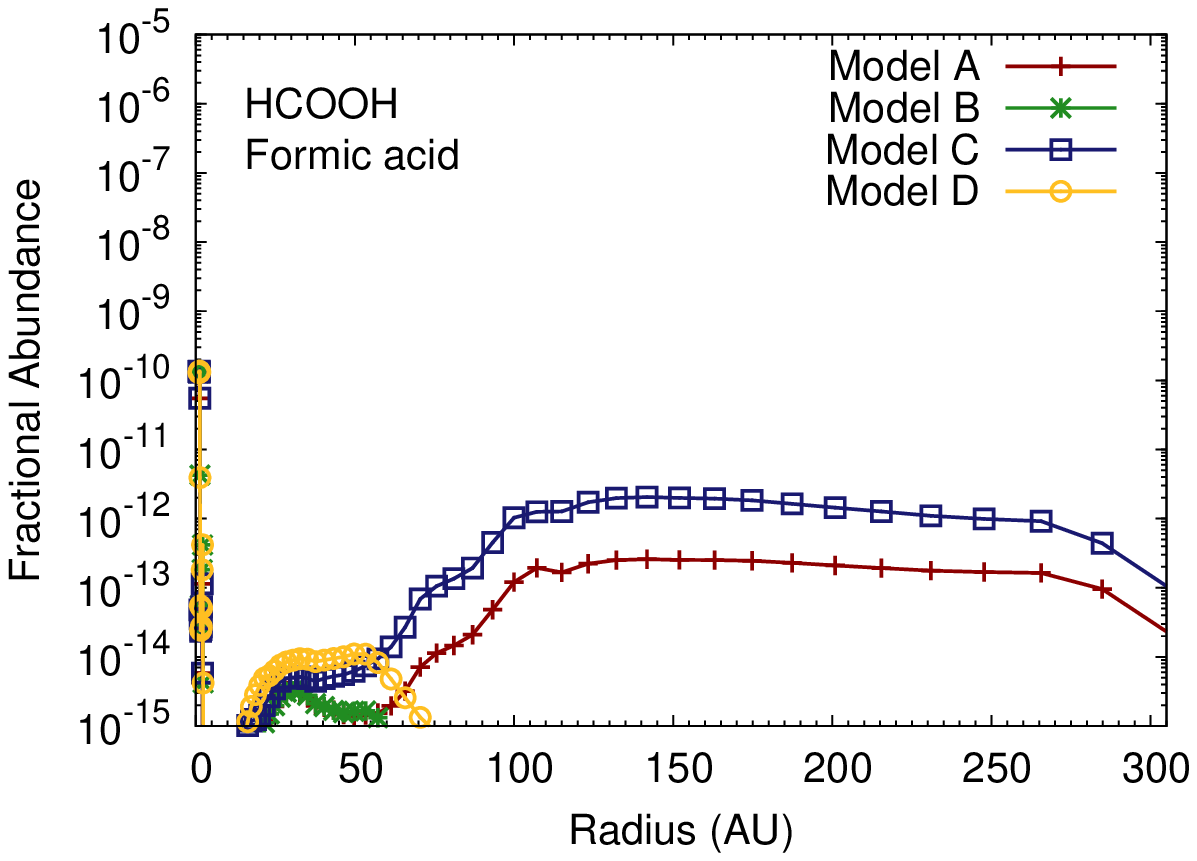}}
\subfigure{\includegraphics[width=0.33\textwidth]{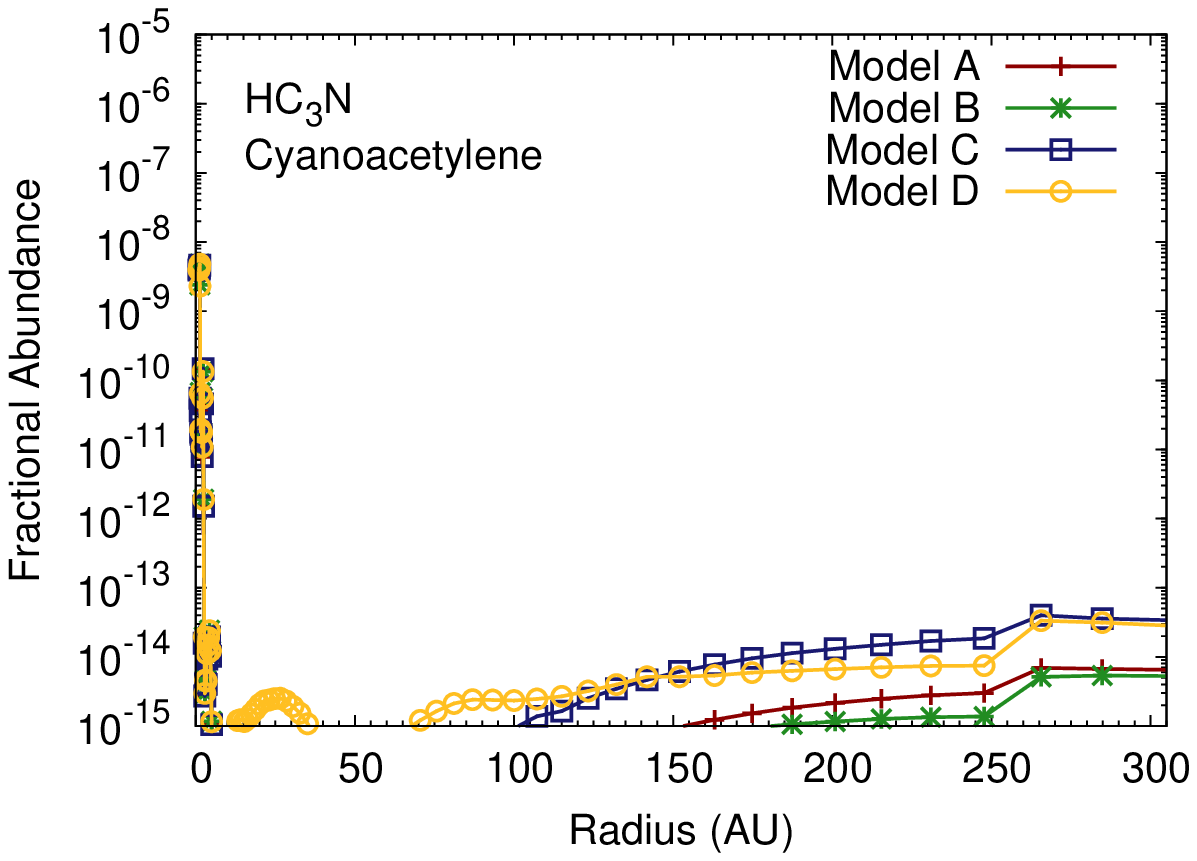}}
\subfigure{\includegraphics[width=0.33\textwidth]{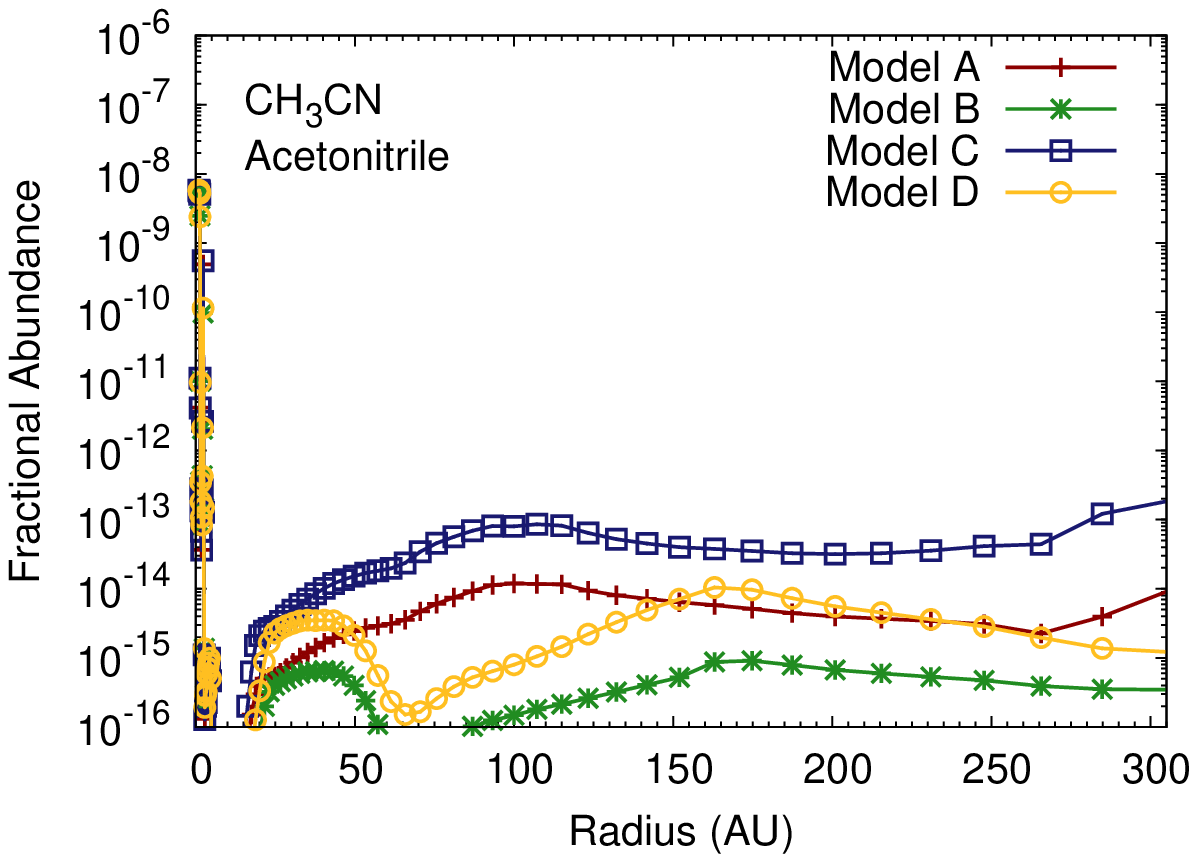}}
\subfigure{\includegraphics[width=0.33\textwidth]{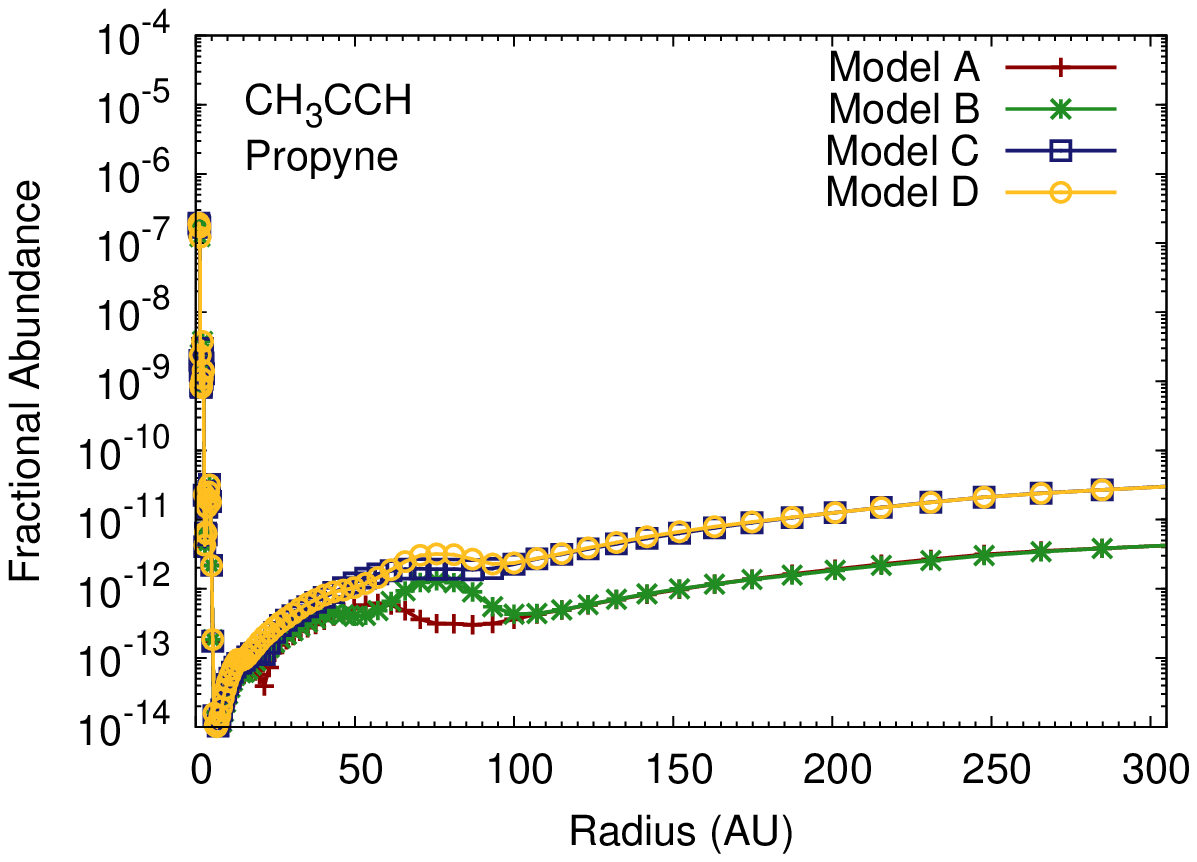}}
\subfigure{\includegraphics[width=0.33\textwidth]{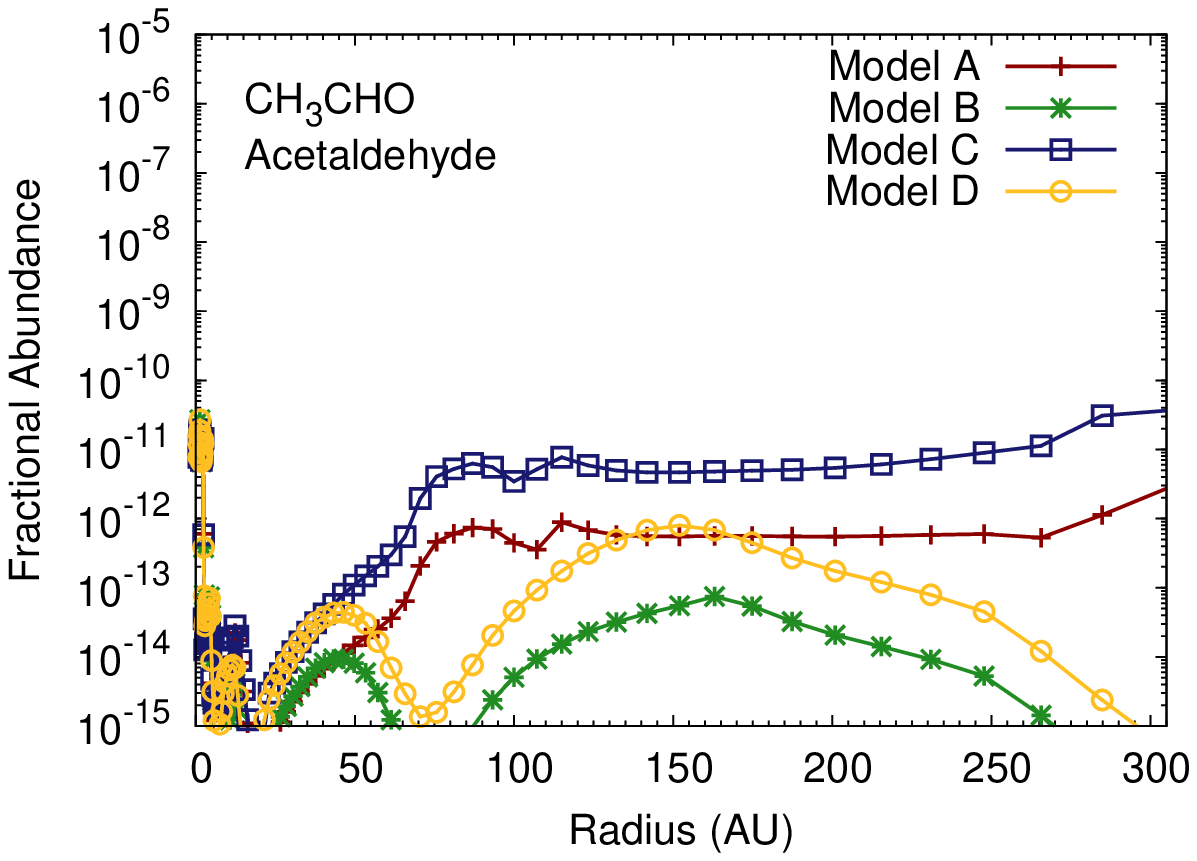}}
\subfigure{\includegraphics[width=0.33\textwidth]{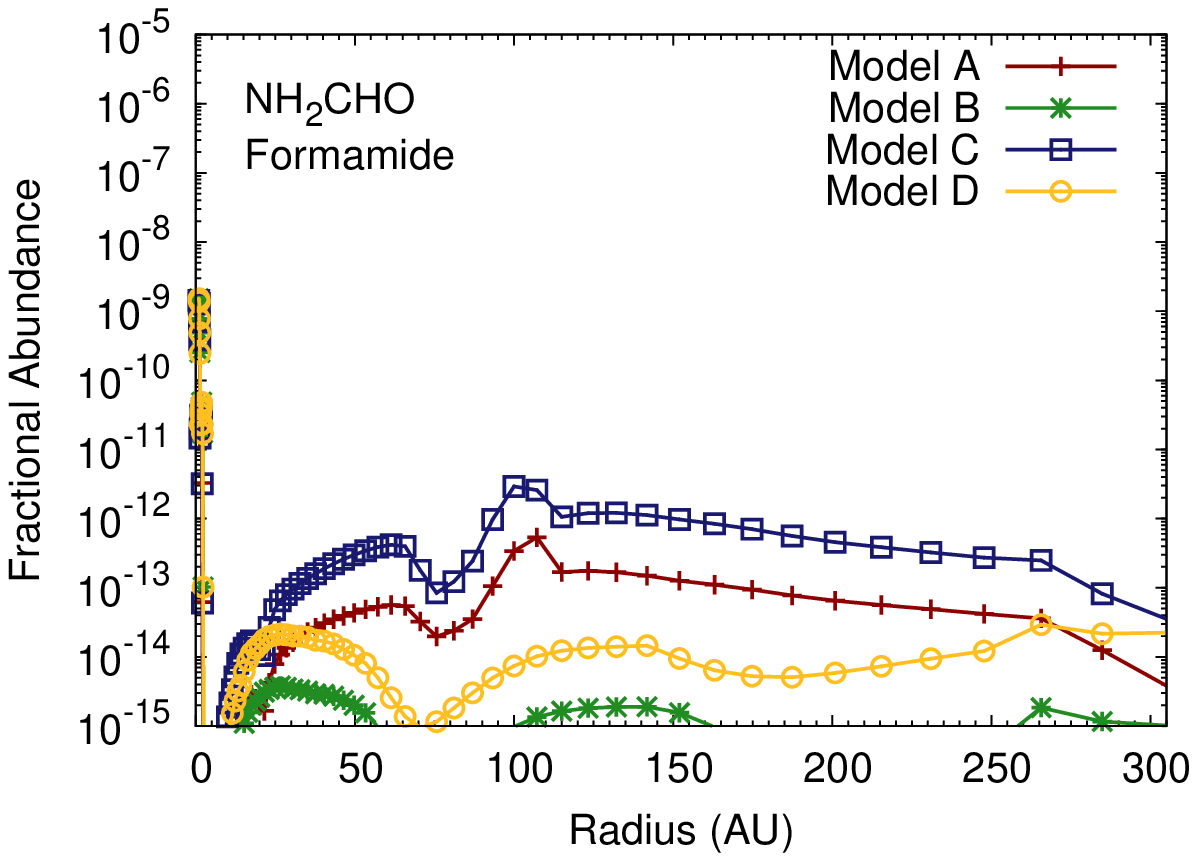}}
\subfigure{\includegraphics[width=0.33\textwidth]{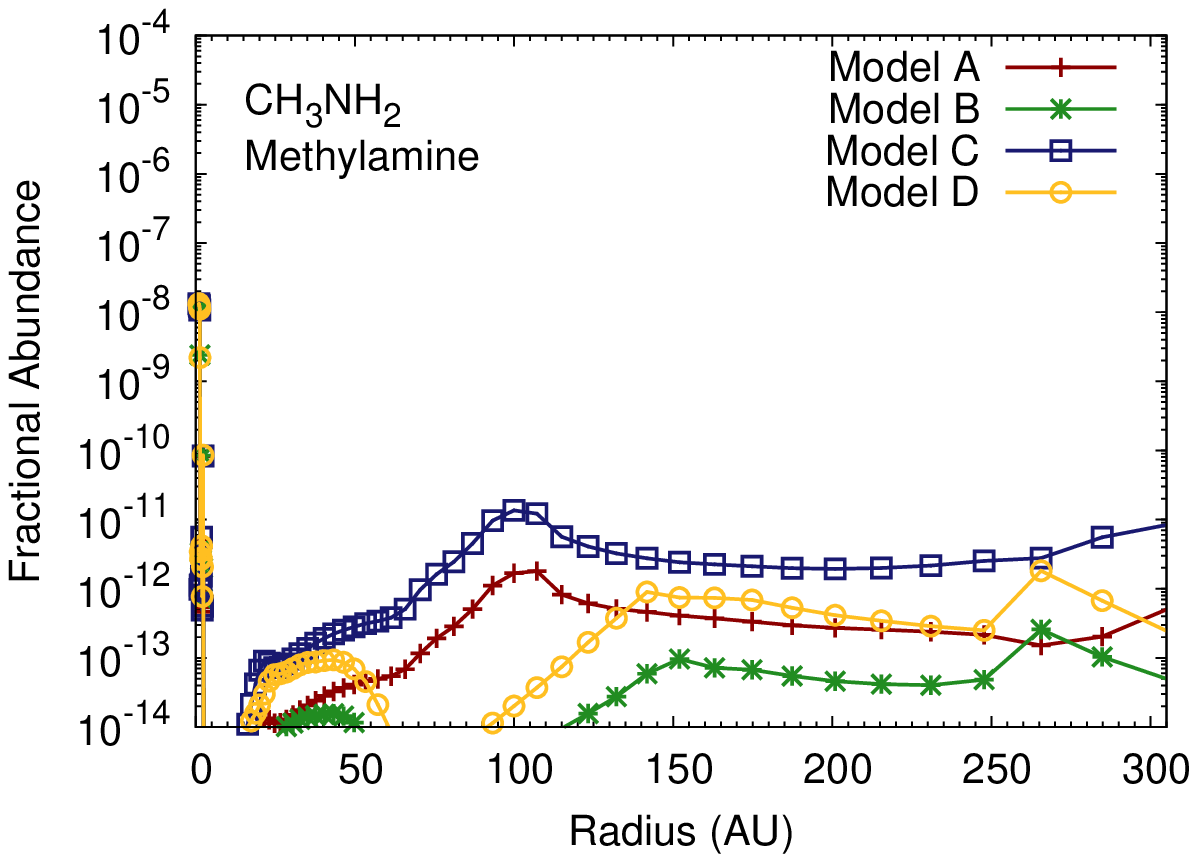}}
\subfigure{\includegraphics[width=0.33\textwidth]{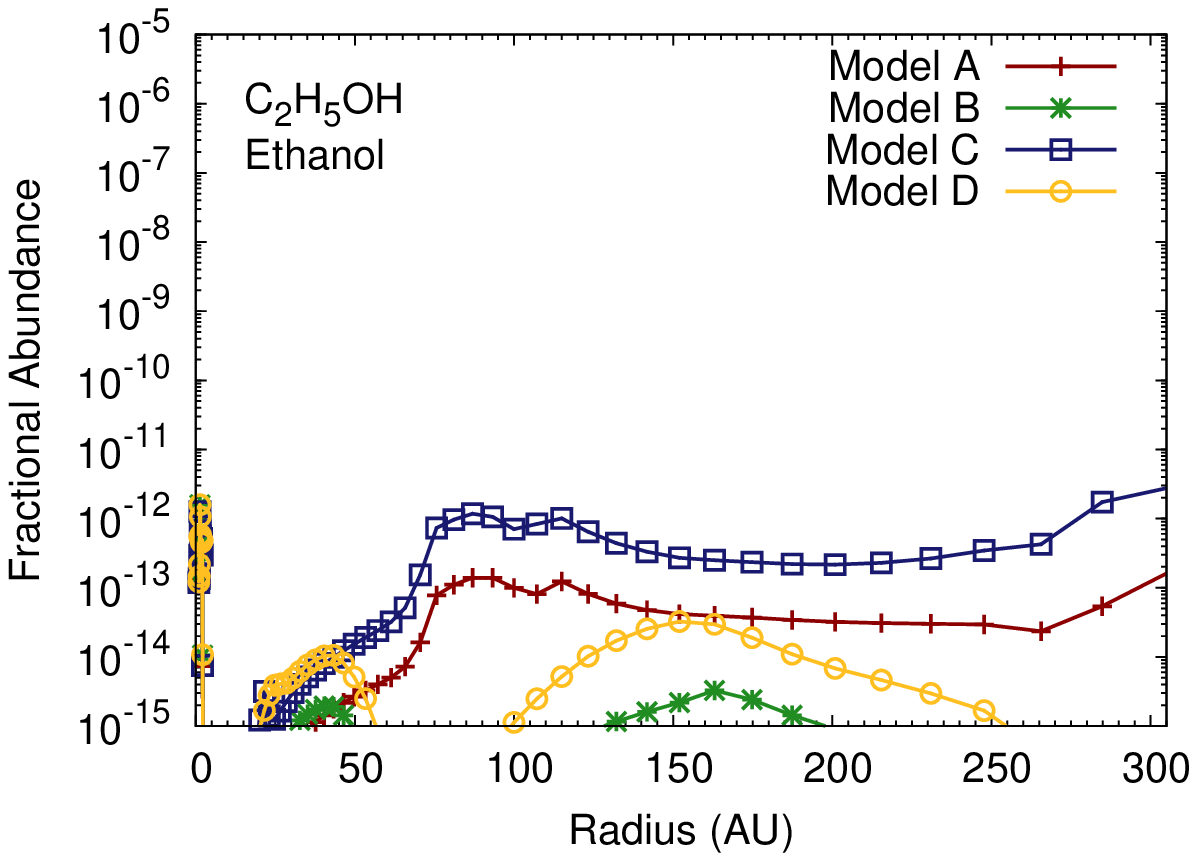}}
\subfigure{\includegraphics[width=0.33\textwidth]{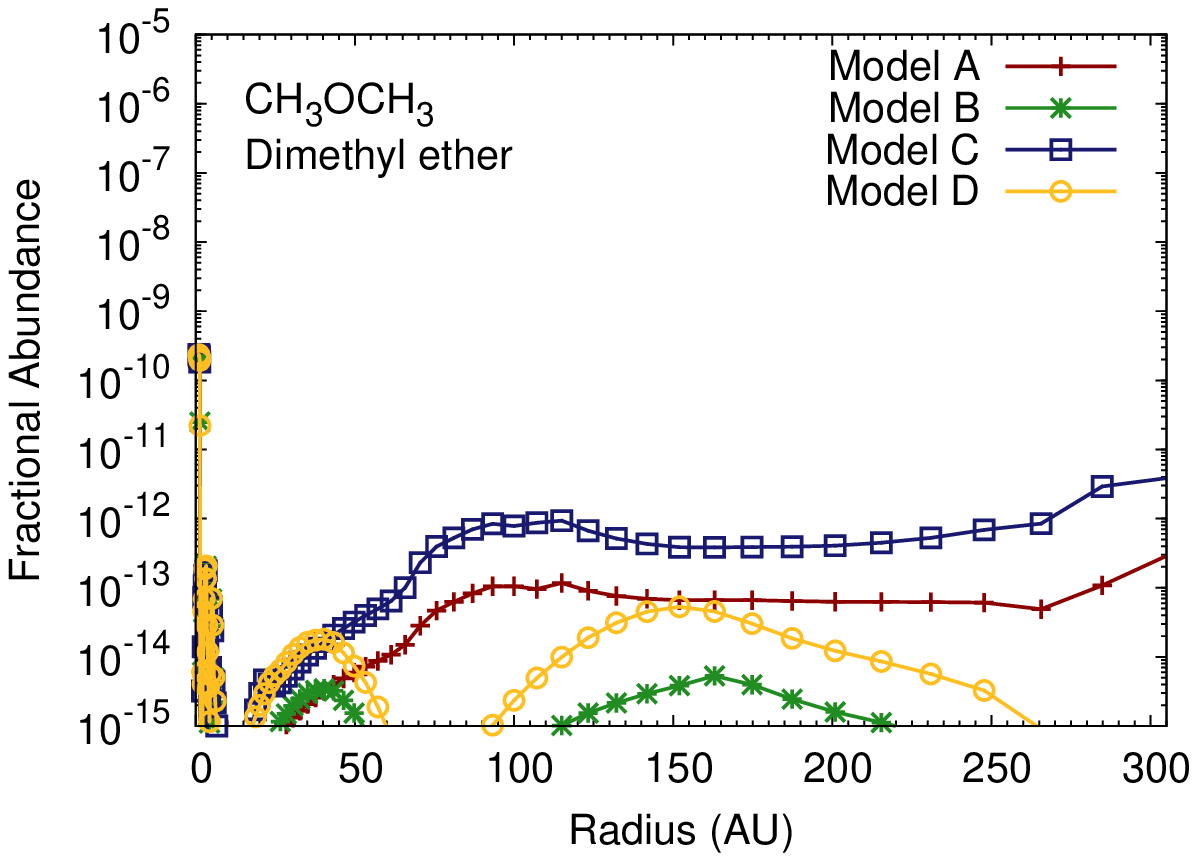}}
\subfigure{\includegraphics[width=0.33\textwidth]{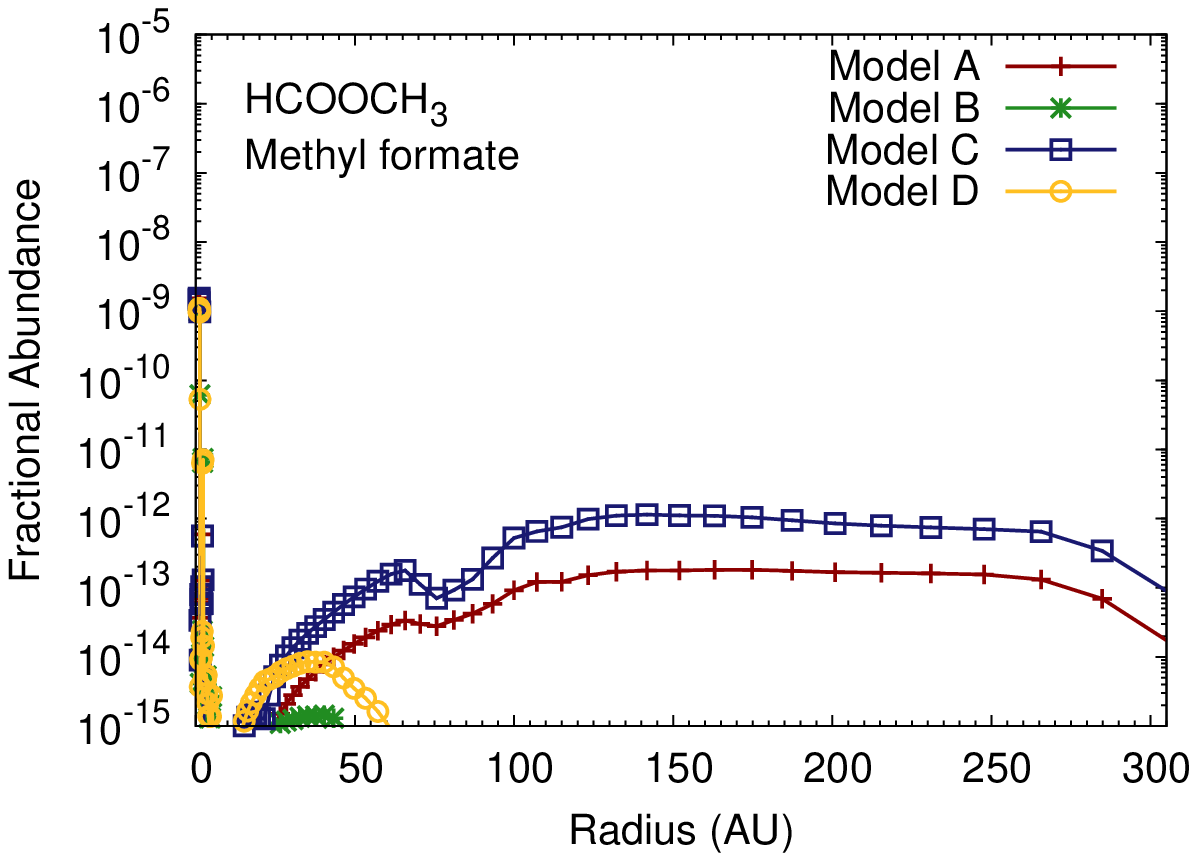}}
\subfigure{\includegraphics[width=0.33\textwidth]{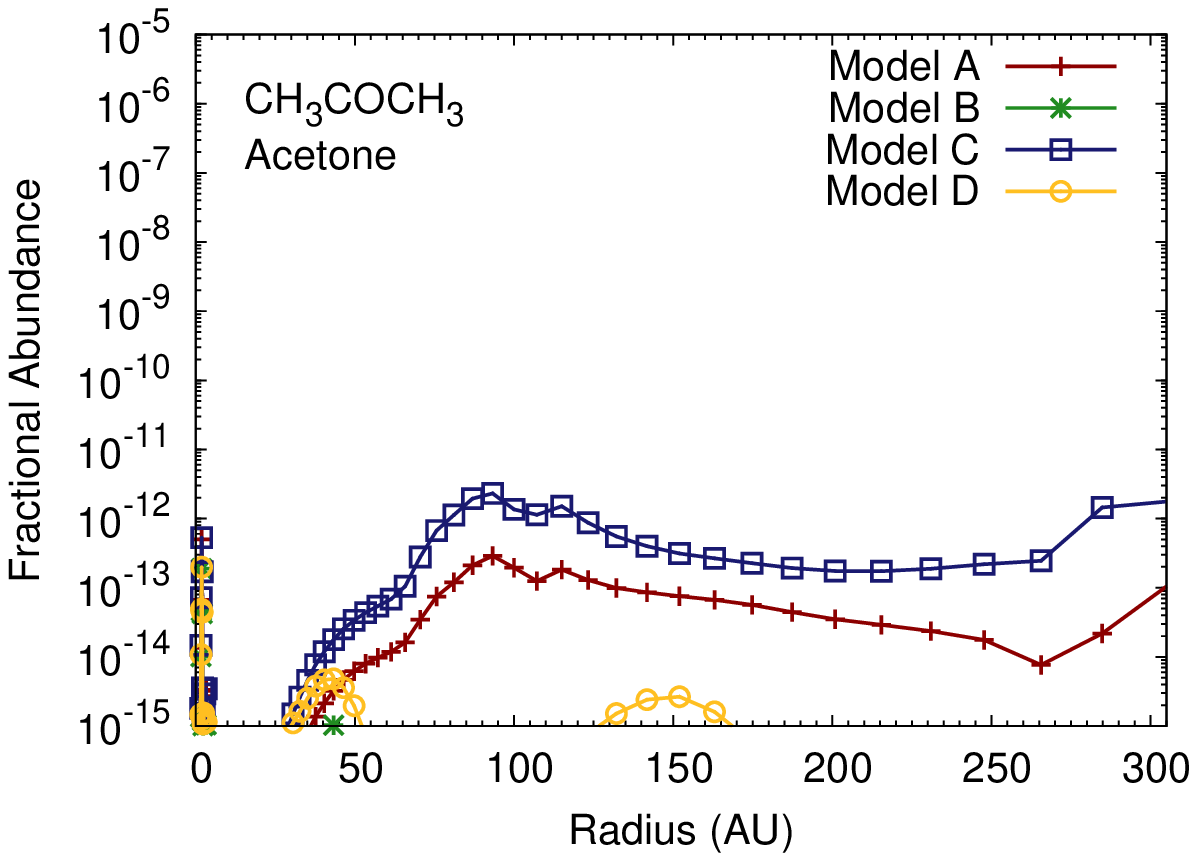}}
\subfigure{\includegraphics[width=0.33\textwidth]{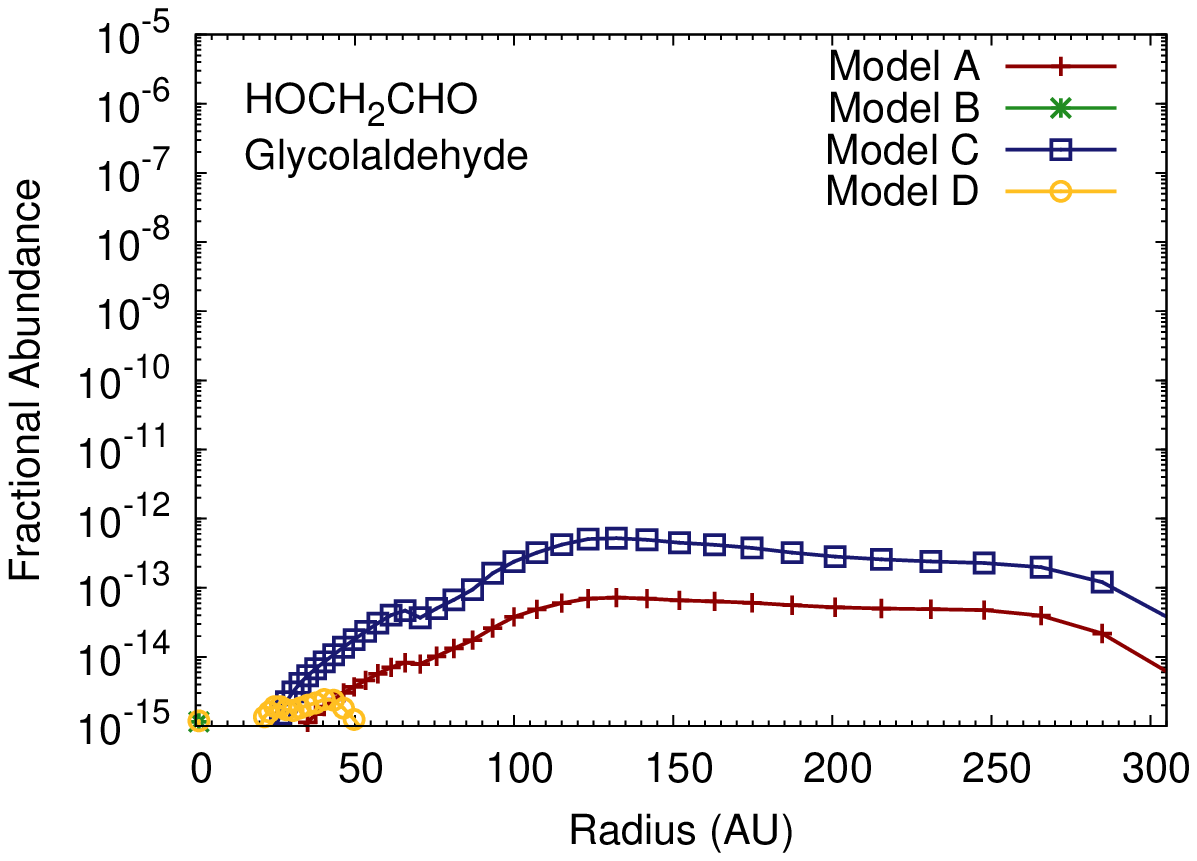}}
\subfigure{\includegraphics[width=0.33\textwidth]{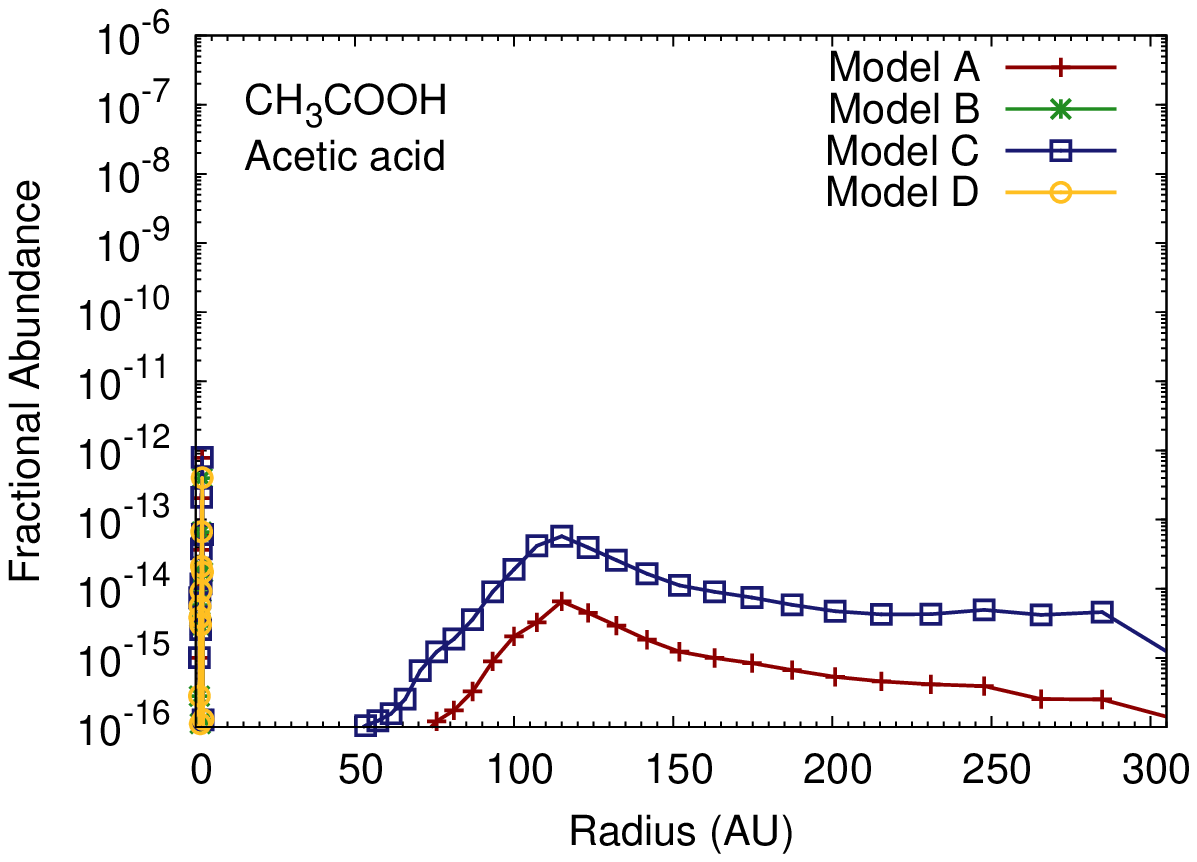}}
\caption{Fractional abundance (with respect to H nuclei number density) of gas-phase molecules
as a function of radius, $R$, along the disk midplane.  
The differences between Models A to D are described in the text and listed in 
Table~\ref{tablea1}.}
\label{figurea1}
\end{figure*}

\begin{figure*}
\subfigure{\includegraphics[width=0.33\textwidth]{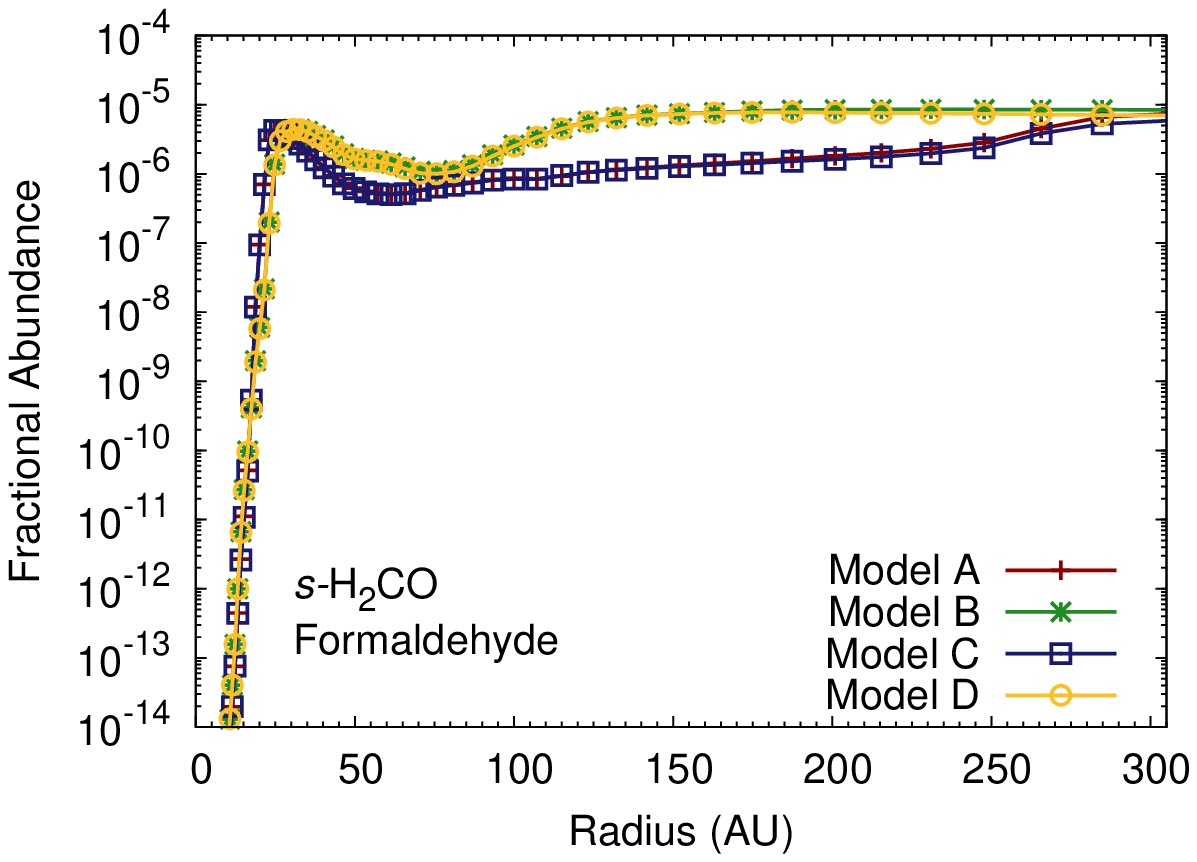}}
\subfigure{\includegraphics[width=0.33\textwidth]{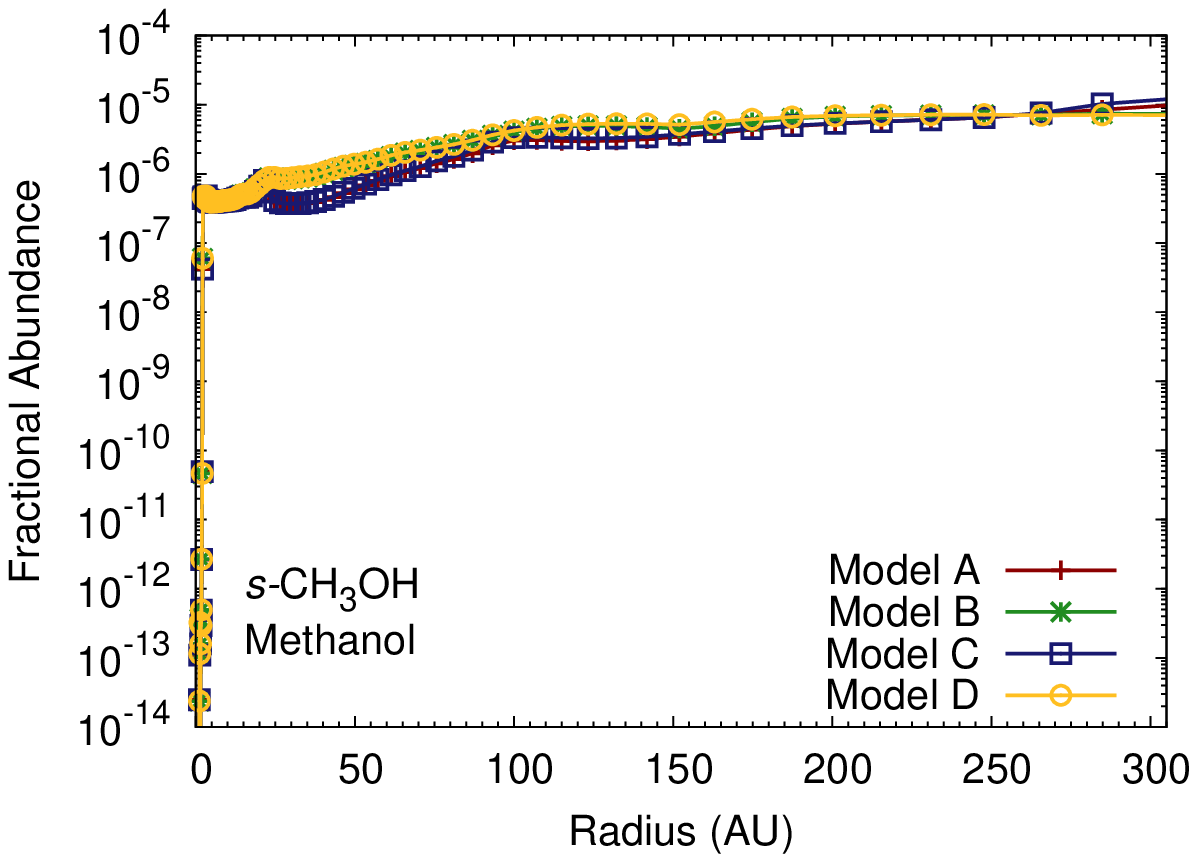}}
\subfigure{\includegraphics[width=0.33\textwidth]{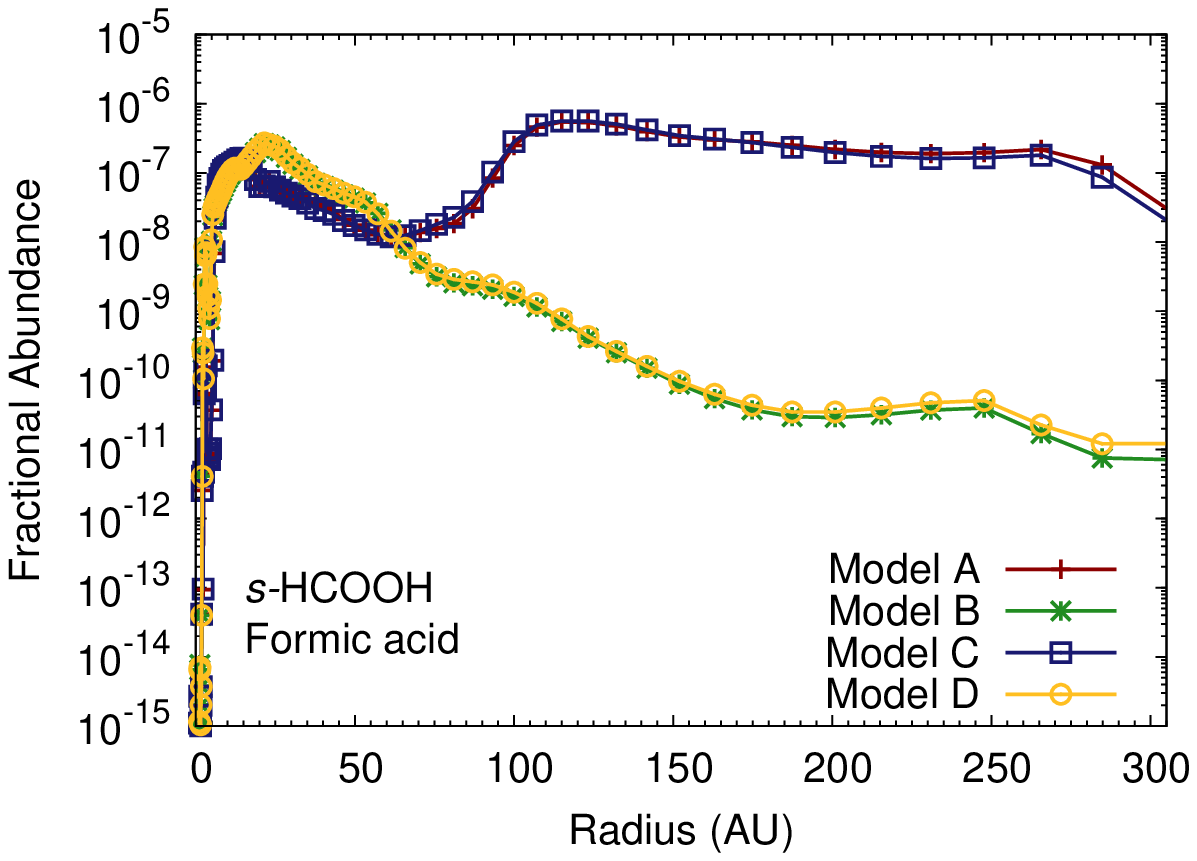}}
\subfigure{\includegraphics[width=0.33\textwidth]{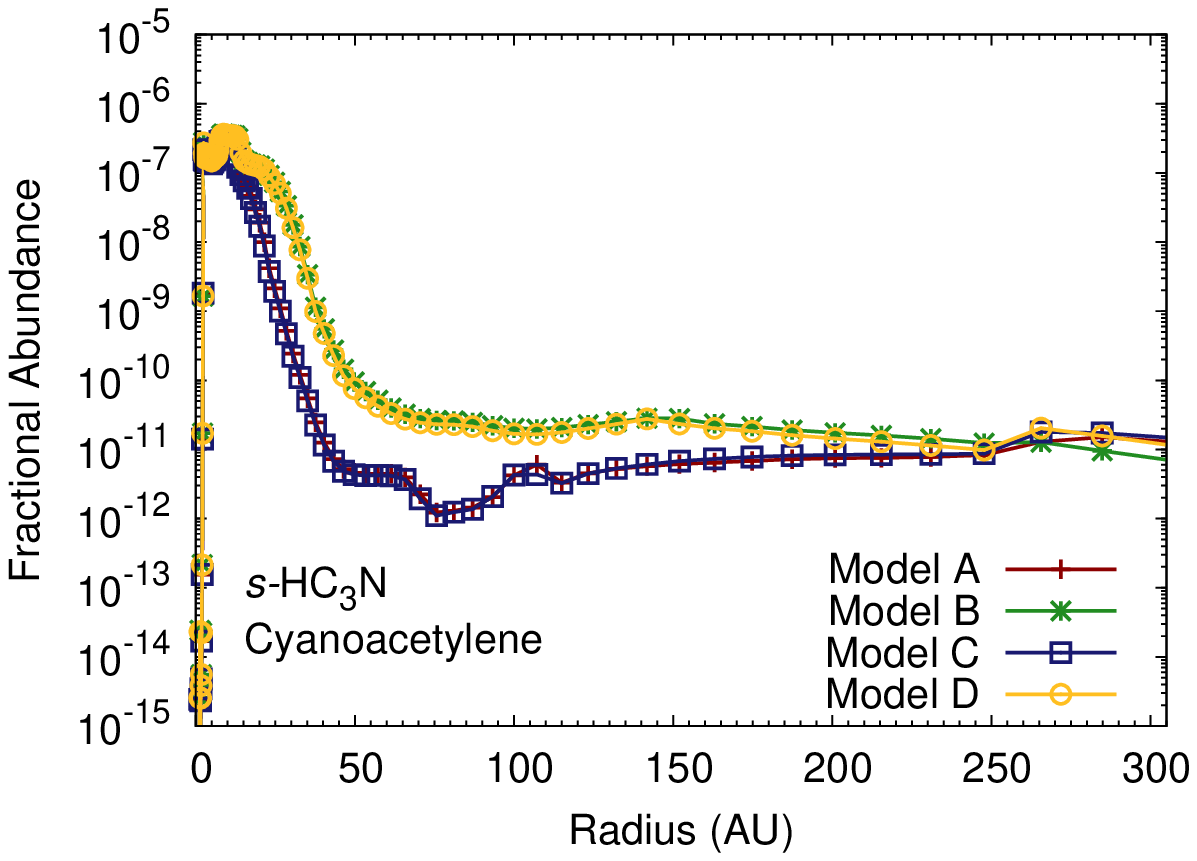}}
\subfigure{\includegraphics[width=0.33\textwidth]{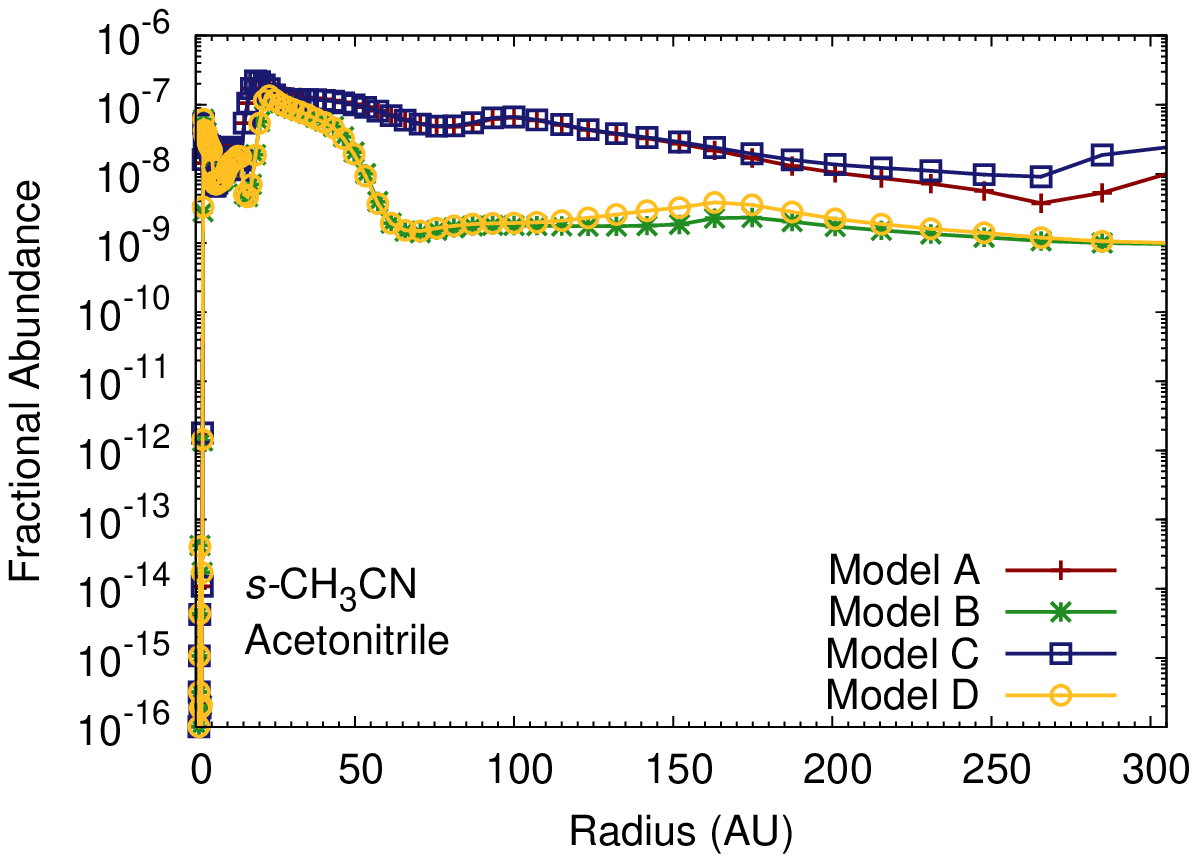}}
\subfigure{\includegraphics[width=0.33\textwidth]{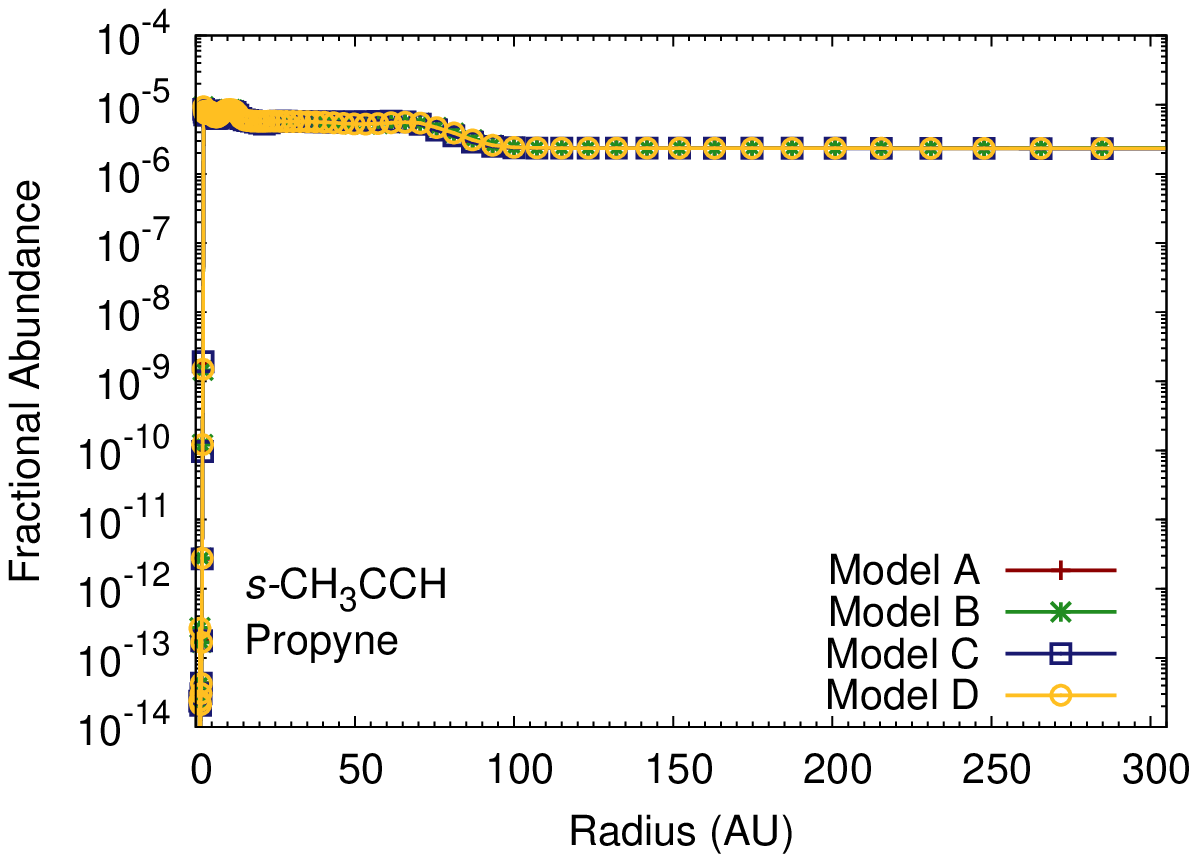}}
\subfigure{\includegraphics[width=0.33\textwidth]{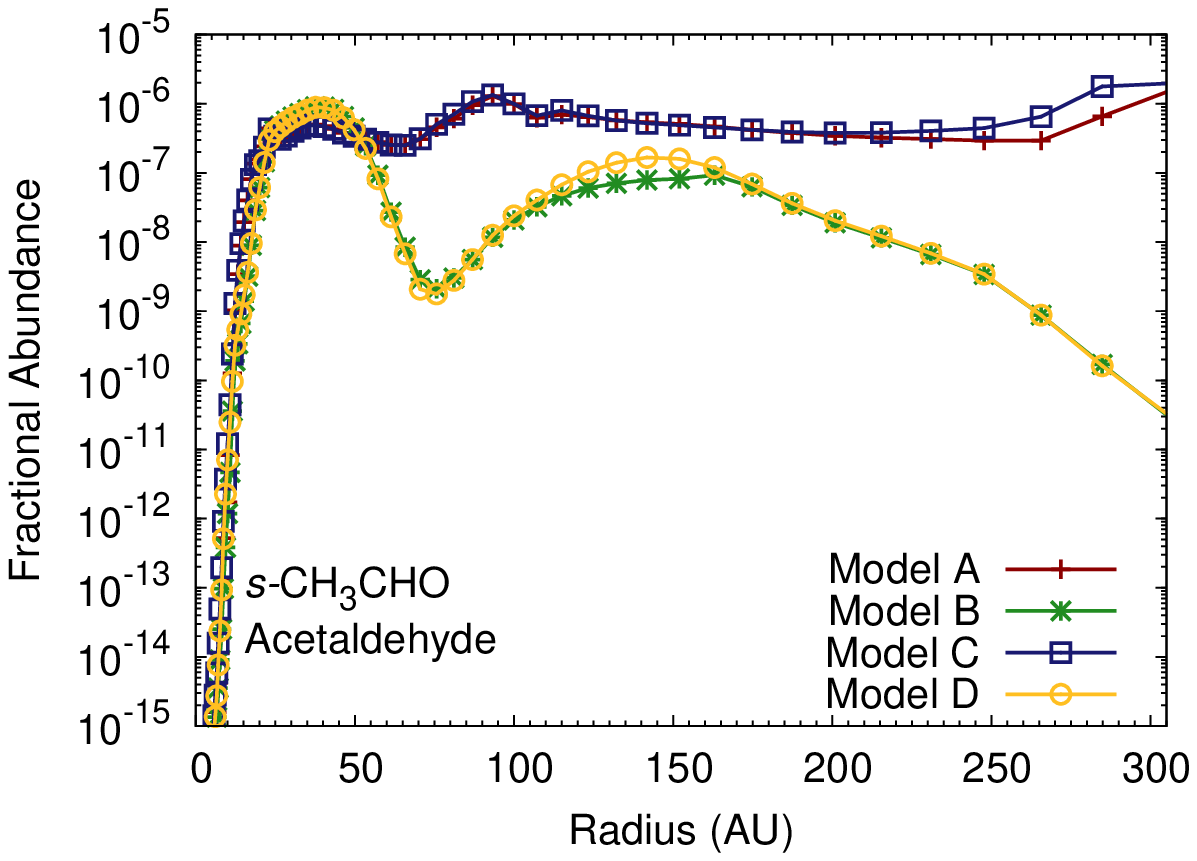}}
\subfigure{\includegraphics[width=0.33\textwidth]{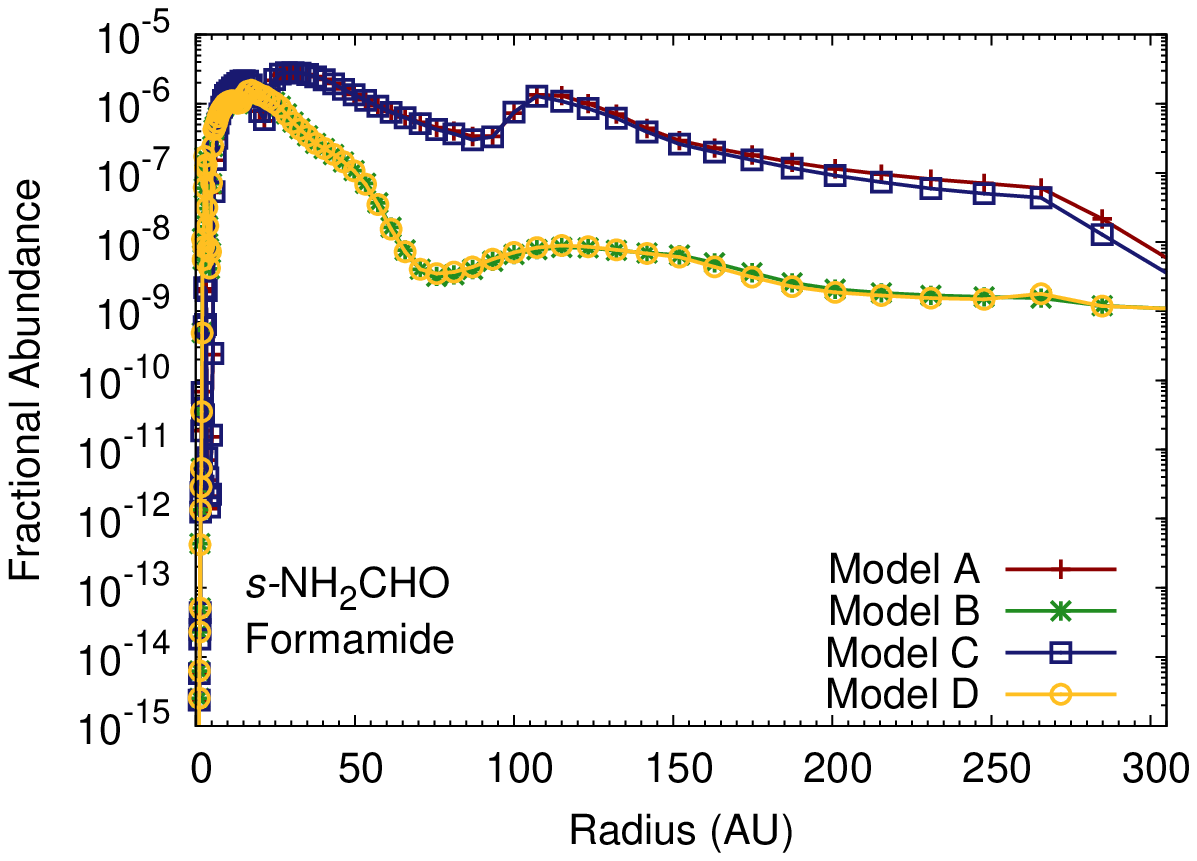}}
\subfigure{\includegraphics[width=0.33\textwidth]{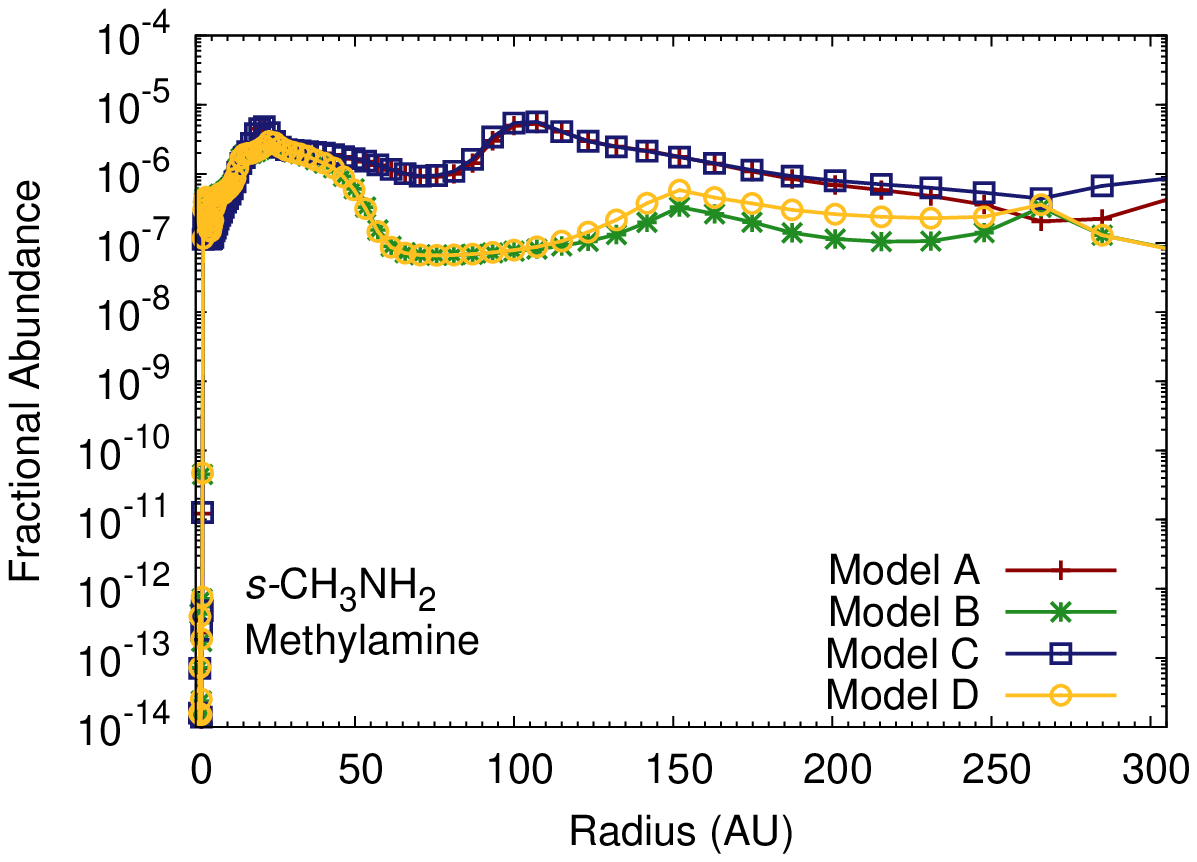}}
\subfigure{\includegraphics[width=0.33\textwidth]{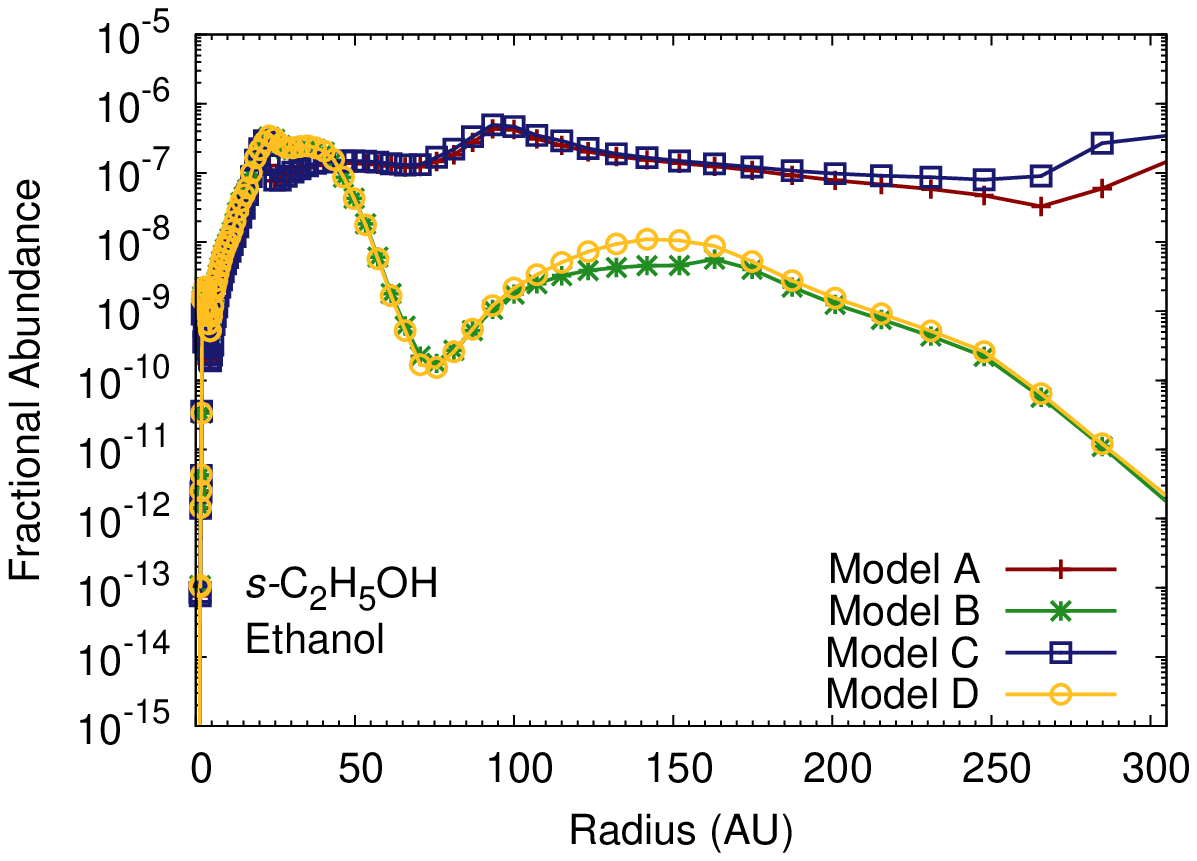}}
\subfigure{\includegraphics[width=0.33\textwidth]{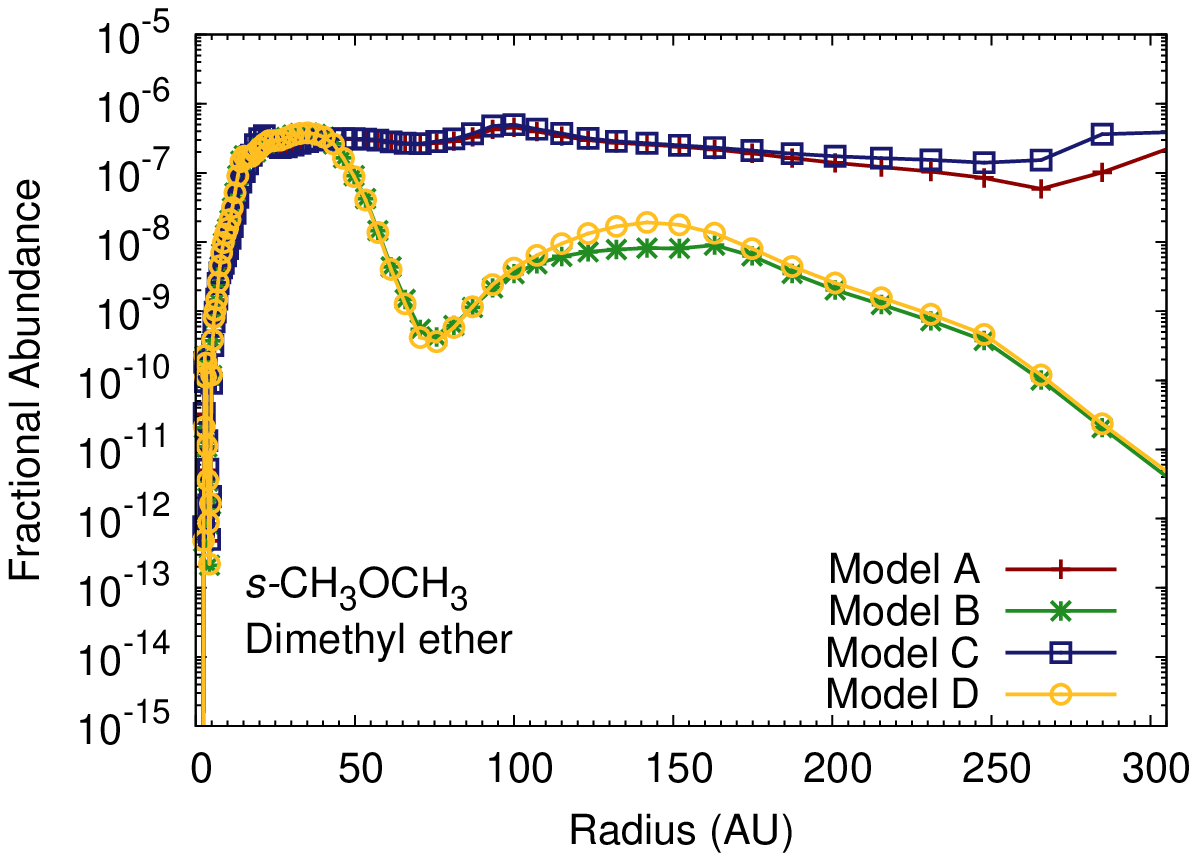}}
\subfigure{\includegraphics[width=0.33\textwidth]{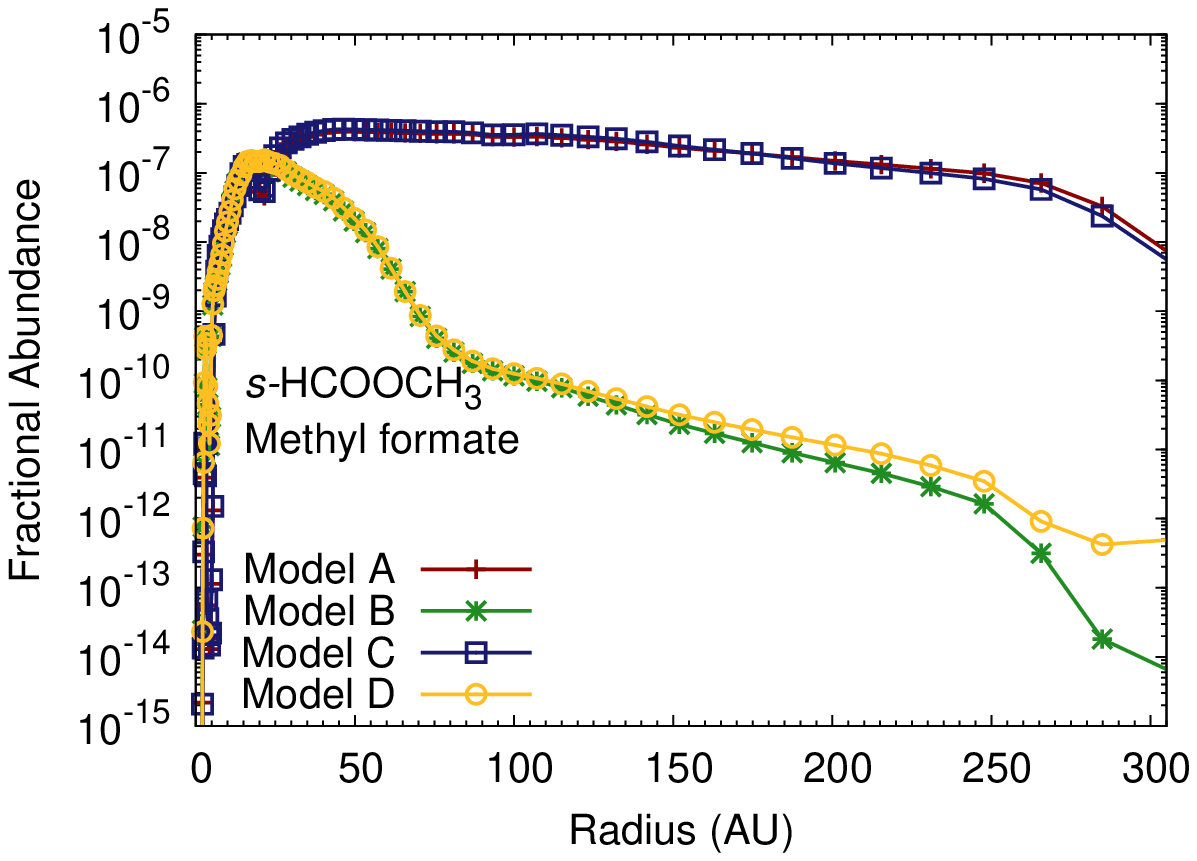}}
\subfigure{\includegraphics[width=0.33\textwidth]{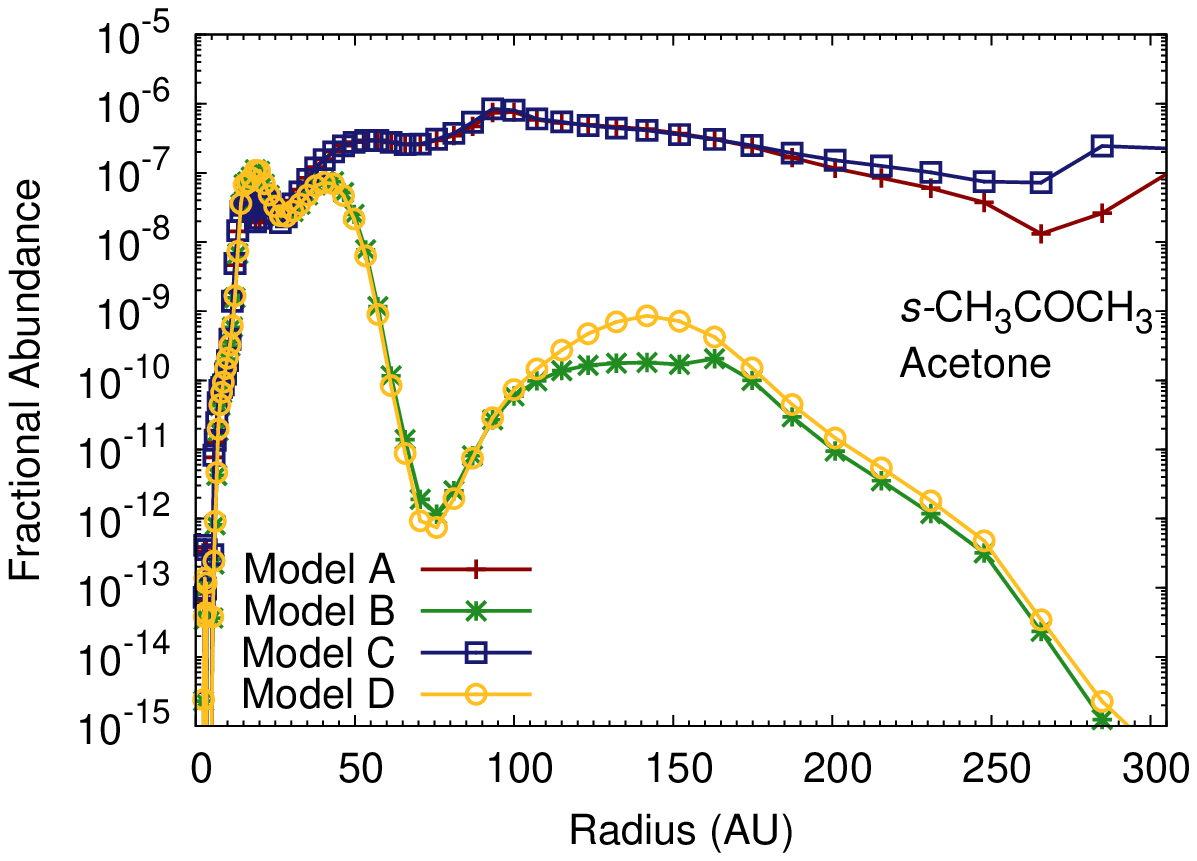}}
\subfigure{\includegraphics[width=0.33\textwidth]{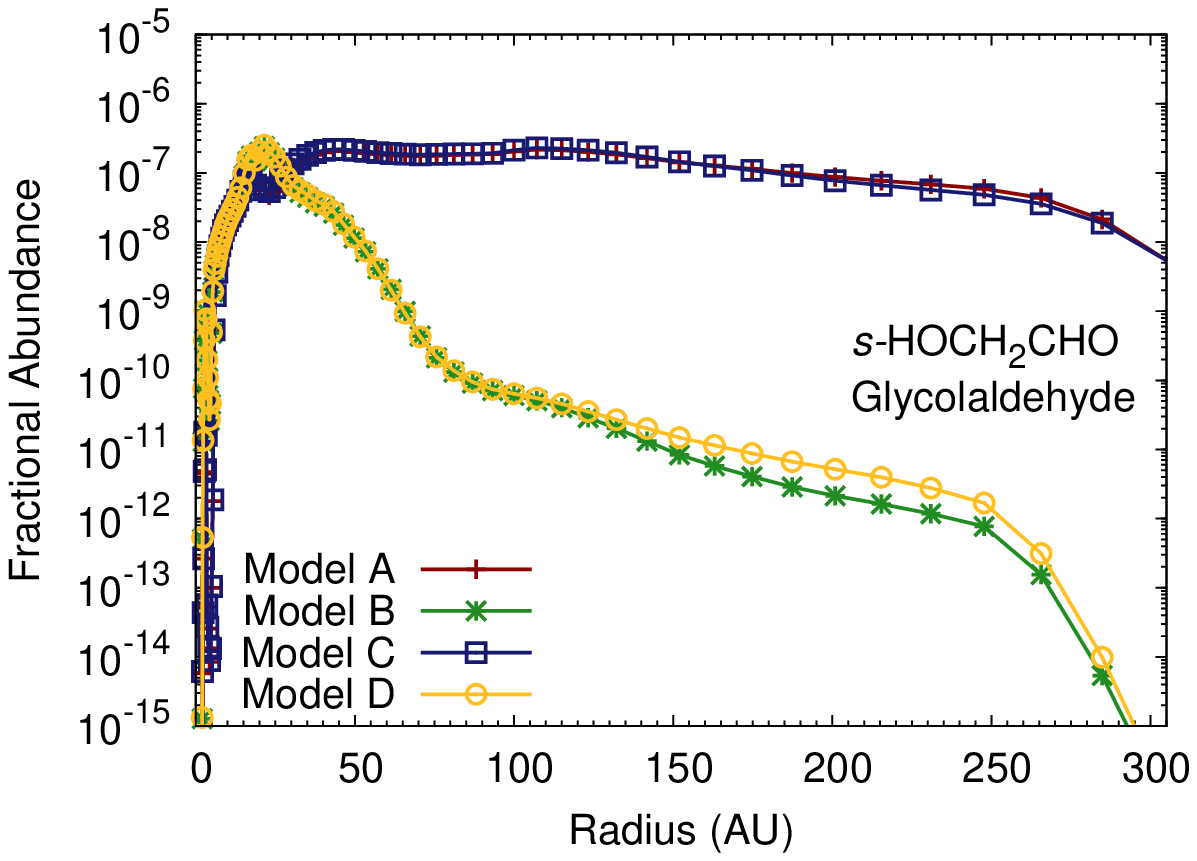}}
\subfigure{\includegraphics[width=0.33\textwidth]{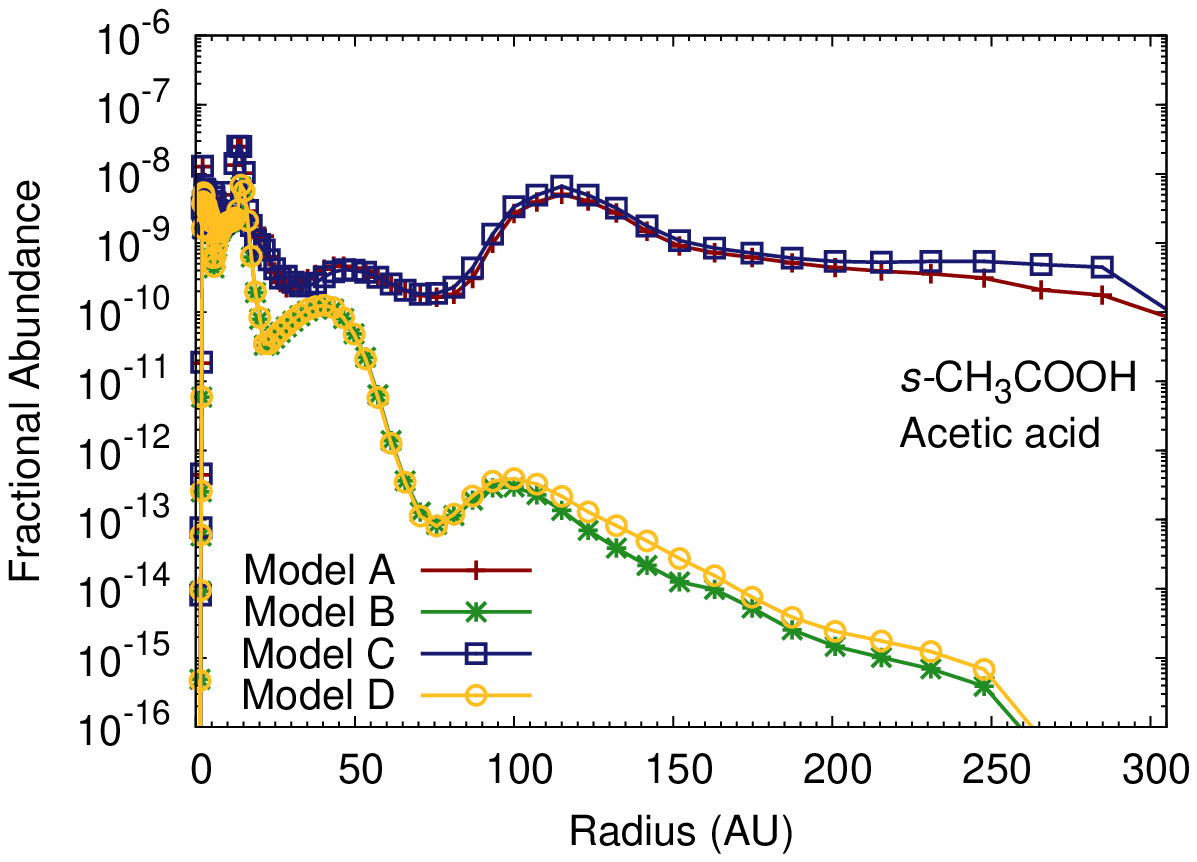}}
\caption{Same as Fig.~\ref{figurea1} for grain-surface species.}
\label{figurea2}
\end{figure*}

In Figures~\ref{figurea1} and \ref{figurea2} we present 
the fractional abundances of gas-phase and grain-surface COMs, 
respectively, relative to the H nuclei number density, as a function of disk radius, $R$.  
We show and discuss results for the disk midplane only\footnotemark[13].  
This is the region where grain-surface COMs form most efficiently 
in our fiducial disk model.  
In Sects.~\ref{diffusionbarrier} and \ref{reactivedesorption}, 
we discuss the effects and importance 
of the values adopted for the diffusion barrier and the probability for 
reactive desorption, respectively.  
   
\footnotetext[13]{The data used to plot Figs.~\ref{figurea1} and \ref{figurea2} are available upon request.}   
   
\begin{table}
\caption{Model parameters}
\centering
\begin{tabular}{ccc}
\hline\hline
Model & $E_{b}$/$E_{D}$ & P$_{rd}$ \\
\hline
A     & 0.3             & 0.01     \\
B     & 0.5             & 0.01     \\
C     & 0.3             & 0.10     \\
D     & 0.5             & 0.10     \\
\hline
\end{tabular}
\label{tablea1}
\end{table}

\subsection{Diffusion barrier}
\label{diffusionbarrier}

In Figs.~\ref{figurea1} and \ref{figurea2}, the red lines represent results 
from Model A (our fiducial model) in which we have adopted $E_{b}$/$E_{D}$~=~0.3 and 
the green lines represent results from Model~B  
in which we have used $E_{b}$/$E_{D}$~=~0.5.  
There are only minor differences (less than an order of magnitude) 
between the gas-phase and grain-surface abundances of  
\ce{H2CO}, \ce{CH3OH}, \ce{HC3N}, and \ce{CH3CCH} calculated 
using Model A and Model B.  
These are species which can form either in the gas phase 
or which depend on hydrogenation 
reactions for their formation.  

Lower abundances are calculated for Model B relative to Model A in the outer disk 
for those grain-surface species which require radical-radical association 
to enhance their abundance above that achieved in dark clouds. 
In Model B, the fractional abundances attained in the very outer disk midplane, 
$\approx$~300~AU, for most species are comparable to those achieved under dark cloud conditions 
(see Table~\ref{table1} and Fig.~\ref{figure4}).  
We see an enhancement in the fractional abundances of 
{\em s-}\ce{C2H5OH}, {\em s-}\ce{CH3OCH3}, and {\em s-}\ce{HCOOCH3} relative 
to their initial abundances.  
This is indicative that radical-radical grain-surface chemistry still operates 
in the outer disk midplane in Model~B, albeit significantly slower relative to Model A.  
Moving inwards along the midplane, the temperature and 
density both increase and there is a corresponding increase 
in the fractional abundances of most COMs.  
This increase is mirrored in the fractional abundances attained by the 
analogous gas-phase species.  
{\em s-}\ce{CH3CHO}, {\em s-}\ce{C2H5OH}, {\em s-}\ce{CH3OCH3}, and {\em s-}\ce{CH3COCH3} 
exhibit an interesting behaviour between $\approx$~50 and 150~AU.  
Within this region, the fractional abundances of all four species 
show a `dip' or minimum around 70~AU.  
The dust temperature within $\approx$~150~AU in the midplane is $\gtrsim$~22~K allowing thermal 
desorption of volatile molecules, for example, {\em s-}\ce{CO} ($E_D$~=~1150~K).  
The species showing this minimum all form via grain-surface reactions 
involving the relatively volatile methyl radical,  
{\em s-}\ce{CH3} ($E_D$~=~1180~K).  
In Model~B, the grain-surface reaction rates are not sufficiently fast to 
compete with the thermal desorption of {\em s-}\ce{CH3} until the 
dust temperature increases to a value which allows grain-surface thermal 
chemistry to operate efficiently.  

The mobility of grain-surface species is dependent upon 
$\exp(-E_{b}/T)$~=~$\exp(-\chi E_{D})$, where $\chi$~=~$(E_{b}/E_{D})/T$.  
In Model A, there is efficient mobility of grain-surface radicals and thus 
efficient grain-surface synthesis when $\chi$~$\gtrsim$~0.02.    
In Model B this value of $\chi$ (a measure of the degree of mobility) 
is attained when T~$\gtrsim$~28~K which is 
reached within $\approx$~70~AU in the disk midplane.  
In Model B, relatively high fractional abundances of grain-surface 
COMs are attained that are comparable with those from Model A. 
However, the radial range over which they reach their peak fractional abundance is 
restricted to regions where $T$~$\gtrsim$~28~K and where 
the temperature is also lower than the desorption temperature of each molecule.  
Results from Model A and Model B are similar within 
$\approx$~50~AU of the central star.  

\subsection{Reactive desorption}
\label{reactivedesorption}

In Figs.~\ref{figurea1} and \ref{figurea2}, the blue lines 
represent results from Model C in which we have adopted a higher probability for 
reactive desorption, $P_\mathrm{rd}$~=~0.1.  
The increased reactive desorption has little effect on the grain-surface abundances.  
However, in the outer disk midplane, there is around an order of magnitude 
enhancement in the gas-phase fractional abundances when using the higher probability.  
In the inner disk, thermal desorption is the most important mechanism for releasing grain mantle 
material back into the gas phase so that the results from all models converge at small radii.     

There is a similar effect seen when comparing results from Model B and Model D, 
in which the higher diffusion barrier, $E_b/E_D$~=~0.5, has been adopted.  
Results for Model B and Model D are represented by the green lines and yellow lines, 
respectively.  
Again, there is little change in the grain-surface species when the probability for reactive desorption 
is increased to 0.1.  
For the gas-phase abundances, in Model D there is the familiar `order-of-magnitude' 
enhancement when using $P_\mathrm{rd}$~=~0.1. 
Note that for the most `optimistic' model, Model C, the gas-phase COMs reach peak fractional abundances  
between 10$^{-13}$ and 10$^{-9}$ in the outer disk midplane ($R$~$\gtrsim$~10~AU).  
This enhancement in fractional abundance will increase the column densities of COMs; 
however, the main contribution to the COM gas-phase column density remains  
the photodesorbed material in the molecular layer.  

\clearpage

\onecolumn

\onltab{
\begin{longtab}
\begin{landscape}
\scriptsize
\centering
\renewcommand*{\arraystretch}{1.25}
\begin{longtable}{cccccccccccccccc}
\caption{\label{table5} Column densities (cm$^{-2}$) of gas-phase molecules as a function of radius.}\\
\hline\hline
Radius (AU) & \ce{H2CO} & \ce{CH3OH} & \ce{HCOOH} & \ce{HC3N} & \ce{CH3CN} & \ce{CH3CCH} & \ce{CH3CHO} & \ce{NH2CHO} & \ce{CH3NH2} & \ce{C2H5OH} & \ce{CH3OCH3} & \ce{HCOOCH3} & \ce{CH3COCH3}  & \ce{HOCH2CHO} & \ce{CH3COOH} \\ 
\hline
\endfirsthead
\caption{continued.}\\
\hline\hline
Radius (AU) & \ce{H2CO} & \ce{CH3OH} & \ce{HCOOH} & \ce{HC3N} & \ce{CH3CN} & \ce{CH3CCH} & \ce{CH3CHO} & \ce{NH2CHO} & \ce{CH3NH2} & \ce{C2H5OH} & \ce{CH3OCH3} & \ce{HCOOCH3} & \ce{CH3COCH3} & \ce{HOCH2CHO} & \ce{CH3COOH} \\ 
\hline
\endhead
\hline
\endfoot
\hline
1.07(0) & 7.8(12) & 1.9(16) & 7.4(14) & 8.0(16) & 8.2(16) & 2.0(18) & 5.9(14) & 6.6(15) & 1.2(17) & 2.3(13) & 1.4(15) & 8.0(15) & 2.5(09) & 4.8(07) & 6.6(10) \\
1.15(0) & 5.8(12) & 1.0(16) & 1.8(14) & 4.7(16) & 4.5(16) & 1.3(18) & 6.0(14) & 3.3(15) & 6.7(16) & 3.2(13) & 7.8(14) & 3.8(15) & 2.2(09) & 6.9(05) & 5.4(10) \\
1.23(0) & 6.7(12) & 2.4(14) & 6.0(12) & 1.0(16) & 8.8(15) & 4.5(17) & 5.5(14) & 1.1(15) & 3.3(15) & 3.6(13) & 1.2(13) & 4.7(13) & 1.4(08) & 1.2(05) & 4.1(09) \\
1.32(0) & 6.1(12) & 1.1(14) & 3.0(12) & 3.0(15) & 7.4(14) & 9.2(16) & 5.0(14) & 8.2(14) & 5.3(13) & 3.4(13) & 4.3(08) & 1.6(11) & 1.5(07) & 5.7(04) & 5.8(08) \\
1.42(0) & 7.3(12) & 1.5(14) & 1.4(12) & 1.1(15) & 1.7(14) & 3.7(16) & 2.0(14) & 7.0(14) & 3.3(13) & 1.3(13) & 6.1(08) & 9.7(10) & 8.2(06) & 9.7(04) & 4.6(08) \\
1.52(0) & 1.8(15) & 1.4(14) & 9.3(11) & 5.0(14) & 1.5(13) & 2.1(16) & 1.8(14) & 8.8(14) & 2.6(13) & 1.1(13) & 4.3(09) & 1.3(12) & 5.0(11) & 1.1(06) & 4.0(11) \\
1.63(0) & 3.2(15) & 8.2(13) & 2.8(12) & 2.8(14) & 1.6(13) & 1.9(16) & 1.8(14) & 2.1(14) & 2.0(13) & 8.2(12) & 2.3(09) & 9.7(11) & 1.7(12) & 1.3(07) & 3.2(12) \\
1.75(0) & 2.3(15) & 4.8(14) & 8.3(11) & 3.3(15) & 2.5(15) & 3.3(16) & 3.8(14) & 6.1(13) & 3.0(14) & 3.4(12) & 1.9(11) & 3.3(12) & 6.7(11) & 1.6(04) & 2.0(12) \\
1.87(0) & 1.5(15) & 4.4(14) & 2.1(11) & 3.5(15) & 2.3(15) & 4.0(16) & 4.1(14) & 1.7(12) & 1.4(14) & 3.2(11) & 4.0(11) & 4.4(12) & 9.6(10) & 2.0(04) & 2.2(12) \\
2.01(0) & 2.1(15) & 4.0(14) & 3.3(09) & 2.9(15) & 1.1(15) & 2.4(17) & 2.9(14) & 5.9(09) & 2.1(14) & 4.4(09) & 7.8(11) & 2.4(12) & 1.7(10) & 2.5(04) & 1.6(10) \\
2.15(0) & 1.5(15) & 4.1(12) & 1.8(09) & 1.0(15) & 8.3(14) & 1.5(17) & 9.3(13) & 2.2(07) & 4.4(13) & 1.2(07) & 6.2(11) & 1.5(11) & 3.1(09) & 3.8(04) & 3.1(07) \\
2.31(0) & 9.0(14) & 2.9(10) & 2.3(09) & 2.5(14) & 2.4(13) & 1.2(17) & 3.0(13) & 6.6(06) & 5.3(11) & 3.1(04) & 7.6(11) & 3.3(10) & 2.9(09) & 3.9(04) & 2.2(05) \\
2.48(0) & 1.3(14) & 6.0(08) & 2.7(09) & 2.3(13) & 5.4(12) & 1.2(16) & 1.1(13) & 1.6(06) & 1.1(10) & 1.6(04) & 1.2(11) & 9.9(08) & 4.6(08) & 6.2(03) & 2.3(05) \\
2.66(0) & 8.1(13) & 1.6(08) & 3.0(09) & 9.6(12) & 3.9(12) & 2.4(15) & 2.2(12) & 1.4(06) & 1.3(09) & 2.3(04) & 3.4(11) & 2.4(08) & 1.0(09) & 3.9(03) & 4.1(05) \\
2.85(0) & 8.4(13) & 1.1(08) & 3.0(09) & 7.0(12) & 3.8(12) & 6.0(14) & 5.6(11) & 1.7(06) & 2.3(08) & 3.0(04) & 1.2(12) & 6.2(07) & 1.2(09) & 3.0(03) & 7.6(05) \\
3.05(0) & 6.9(13) & 1.4(08) & 4.2(09) & 6.8(12) & 4.8(12) & 7.3(14) & 5.7(11) & 1.8(06) & 2.8(08) & 2.9(04) & 5.9(11) & 3.0(07) & 7.8(07) & 2.4(03) & 7.1(05) \\
3.28(0) & 1.5(14) & 2.4(08) & 8.6(09) & 8.7(12) & 7.3(12) & 8.9(14) & 9.2(11) & 3.9(06) & 4.9(08) & 4.0(04) & 1.4(12) & 5.7(08) & 7.4(09) & 1.9(04) & 8.6(05) \\
3.51(0) & 5.8(13) & 1.8(08) & 5.2(09) & 8.5(12) & 4.7(12) & 1.2(15) & 7.5(11) & 2.8(06) & 4.6(08) & 2.8(04) & 7.9(10) & 3.3(07) & 8.0(06) & 3.1(03) & 5.9(05) \\
3.77(0) & 5.6(13) & 2.3(08) & 7.7(09) & 8.6(12) & 5.5(12) & 1.3(15) & 8.2(11) & 3.3(06) & 5.5(08) & 2.9(04) & 3.1(10) & 3.6(07) & 4.5(06) & 3.5(03) & 5.6(05) \\
4.04(0) & 5.4(13) & 2.6(08) & 9.0(09) & 8.9(12) & 6.9(12) & 1.5(15) & 8.8(11) & 3.8(06) & 6.4(08) & 2.9(04) & 9.5(09) & 4.1(07) & 2.4(06) & 3.9(03) & 5.3(05) \\
4.33(0) & 5.4(13) & 2.8(08) & 1.0(10) & 8.8(12) & 6.1(12) & 1.7(15) & 9.6(11) & 4.2(06) & 7.0(08) & 3.2(04) & 3.6(09) & 4.7(07) & 1.5(06) & 4.4(03) & 5.1(05) \\
4.64(0) & 6.1(13) & 2.6(08) & 1.0(10) & 7.4(12) & 6.1(12) & 9.3(14) & 4.4(11) & 8.7(06) & 6.4(08) & 3.5(04) & 1.3(10) & 2.9(07) & 1.7(06) & 5.0(03) & 6.8(05) \\
4.98(0) & 7.2(13) & 6.2(08) & 9.4(09) & 5.2(12) & 5.2(12) & 1.5(14) & 6.7(10) & 2.4(07) & 5.4(08) & 1.4(06) & 3.2(11) & 9.6(06) & 8.2(06) & 4.4(03) & 2.2(06) \\
5.34(0) & 7.8(13) & 5.2(09) & 8.5(09) & 5.7(12) & 5.5(12) & 1.4(13) & 1.8(10) & 1.7(09) & 1.4(09) & 2.0(07) & 2.7(11) & 6.6(06) & 4.8(08) & 1.1(04) & 6.5(06) \\
5.72(0) & 6.3(13) & 1.1(10) & 9.6(09) & 4.5(12) & 6.0(12) & 1.4(12) & 8.1(09) & 3.5(08) & 2.7(09) & 1.9(08) & 7.3(10) & 6.7(07) & 3.7(09) & 7.6(04) & 1.2(07) \\
6.14(0) & 5.7(13) & 1.5(10) & 9.7(09) & 5.5(12) & 5.0(12) & 6.7(11) & 7.4(09) & 3.0(10) & 3.5(09) & 8.2(07) & 3.4(10) & 1.1(08) & 2.7(09) & 5.1(05) & 2.5(07) \\
6.58(0) & 3.3(13) & 6.2(09) & 1.1(10) & 4.2(12) & 6.0(12) & 4.5(11) & 6.2(09) & 6.5(09) & 1.6(09) & 2.9(07) & 1.5(10) & 9.9(07) & 1.7(09) & 1.4(06) & 2.8(07) \\
7.06(0) & 6.2(12) & 3.3(08) & 1.3(10) & 2.8(12) & 5.1(12) & 3.4(11) & 5.1(09) & 2.1(08) & 4.4(08) & 5.4(05) & 5.7(09) & 5.4(07) & 9.4(08) & 2.7(06) & 1.7(07) \\
7.57(0) & 3.6(12) & 3.5(08) & 1.5(10) & 2.6(12) & 5.8(12) & 3.2(11) & 4.8(09) & 2.4(08) & 5.1(08) & 7.0(05) & 2.3(09) & 2.6(07) & 4.6(08) & 3.3(06) & 1.4(07) \\
8.11(0) & 3.2(12) & 5.5(08) & 2.0(10) & 2.4(12) & 5.1(12) & 3.3(11) & 5.0(09) & 3.9(08) & 5.5(08) & 1.1(06) & 8.1(08) & 1.9(07) & 1.7(08) & 6.3(06) & 1.5(07) \\
8.67(0) & 3.1(12) & 5.4(08) & 3.8(10) & 2.3(12) & 5.6(12) & 3.5(11) & 6.3(09) & 6.7(08) & 5.9(08) & 1.7(06) & 2.2(08) & 2.2(07) & 4.7(07) & 1.2(07) & 1.7(07) \\
9.33(0) & 3.4(12) & 9.3(08) & 5.9(10) & 2.1(12) & 5.0(12) & 4.0(11) & 8.4(09) & 1.2(09) & 6.4(08) & 2.6(06) & 6.0(07) & 3.8(07) & 1.4(07) & 2.4(07) & 1.9(07) \\
1.00(1) & 3.7(12) & 1.0(09) & 8.1(10) & 2.0(12) & 5.5(12) & 4.5(11) & 1.1(10) & 1.7(09) & 7.2(08) & 3.8(06) & 2.2(07) & 5.8(07) & 5.8(06) & 4.2(07) & 2.1(07) \\
1.07(1) & 4.2(12) & 1.2(09) & 1.5(11) & 1.9(12) & 5.1(12) & 4.9(11) & 1.5(10) & 2.7(09) & 8.7(08) & 5.7(06) & 1.1(07) & 8.9(07) & 2.9(06) & 7.6(07) & 3.6(07) \\
1.15(1) & 4.8(12) & 1.4(10) & 3.1(11) & 1.8(12) & 5.3(12) & 5.5(11) & 2.7(10) & 4.2(09) & 1.0(09) & 9.6(06) & 1.1(07) & 2.2(08) & 2.1(06) & 1.3(08) & 6.3(07) \\
1.23(1) & 6.0(12) & 5.2(09) & 3.7(11) & 1.7(12) & 4.9(12) & 5.7(11) & 6.8(10) & 6.2(09) & 1.3(09) & 1.5(07) & 1.5(07) & 3.0(08) & 3.1(06) & 2.2(08) & 1.3(08) \\
1.32(1) & 9.0(12) & 2.6(10) & 4.4(11) & 1.5(12) & 5.0(12) & 6.1(11) & 7.8(10) & 9.1(09) & 1.6(09) & 2.7(07) & 3.0(07) & 1.2(09) & 8.2(06) & 4.6(08) & 2.5(08) \\
1.42(1) & 1.6(13) & 4.5(10) & 5.1(11) & 1.4(12) & 4.7(12) & 6.3(11) & 5.6(10) & 1.2(10) & 2.7(09) & 6.5(07) & 7.9(07) & 3.6(09) & 2.4(07) & 9.4(08) & 3.6(08) \\
1.52(1) & 2.0(13) & 2.4(10) & 5.6(11) & 1.3(12) & 4.8(12) & 5.8(11) & 3.3(10) & 1.3(10) & 4.2(09) & 2.0(08) & 1.4(08) & 6.4(09) & 5.5(07) & 1.1(10) & 6.1(08) \\
1.63(1) & 2.2(13) & 7.2(10) & 5.9(11) & 1.3(12) & 4.6(12) & 5.6(11) & 1.6(10) & 1.3(10) & 6.5(09) & 2.4(08) & 4.0(08) & 1.1(10) & 1.1(08) & 5.7(09) & 8.0(08) \\
1.75(1) & 3.0(13) & 9.0(10) & 6.5(11) & 1.2(12) & 4.4(12) & 5.6(11) & 1.1(10) & 1.4(10) & 1.0(10) & 1.0(09) & 1.0(09) & 4.1(10) & 1.9(08) & 1.6(10) & 9.9(08) \\
1.87(1) & 9.0(13) & 5.3(10) & 6.7(11) & 1.1(12) & 4.2(12) & 5.6(11) & 8.6(09) & 1.4(10) & 1.8(10) & 1.7(09) & 1.4(09) & 2.8(10) & 2.8(08) & 4.1(10) & 1.5(09) \\
2.01(1) & 2.2(14) & 1.4(11) & 6.6(11) & 1.0(12) & 4.1(12) & 5.4(11) & 1.8(10) & 1.3(10) & 2.8(10) & 4.5(09) & 7.5(09) & 5.5(10) & 3.9(08) & 1.5(10) & 4.0(09) \\
2.15(1) & 5.1(14) & 7.2(10) & 6.9(11) & 9.5(11) & 4.0(12) & 4.7(11) & 1.2(10) & 1.2(10) & 4.1(10) & 8.7(09) & 8.1(09) & 7.5(10) & 6.6(08) & 2.6(10) & 5.6(09) \\
2.31(1) & 7.3(14) & 8.2(10) & 8.1(11) & 8.8(11) & 3.9(12) & 4.3(11) & 1.8(10) & 1.4(10) & 5.4(10) & 1.3(10) & 1.6(10) & 4.5(10) & 8.2(08) & 2.4(10) & 9.3(09) \\
2.48(1) & 5.3(14) & 1.1(11) & 6.1(11) & 8.2(11) & 3.7(12) & 5.0(11) & 2.0(10) & 2.0(10) & 5.8(10) & 1.5(10) & 2.1(10) & 3.2(10) & 1.1(09) & 1.0(10) & 7.9(09) \\
2.66(1) & 2.9(14) & 1.5(11) & 7.5(11) & 7.8(11) & 3.5(12) & 6.1(11) & 1.7(10) & 2.8(10) & 6.2(10) & 2.1(10) & 2.2(10) & 5.0(10) & 1.4(09) & 1.4(10) & 8.4(09) \\
2.85(1) & 1.3(14) & 2.0(11) & 7.1(11) & 7.3(11) & 3.2(12) & 7.2(11) & 2.4(10) & 4.2(10) & 6.5(10) & 2.2(10) & 2.6(10) & 3.7(10) & 1.3(09) & 1.1(10) & 1.0(10) \\
3.05(1) & 5.1(13) & 2.2(11) & 7.5(11) & 6.9(11) & 2.9(12) & 8.2(11) & 2.0(10) & 5.6(10) & 6.6(10) & 1.9(10) & 2.2(10) & 4.6(10) & 1.5(09) & 1.2(10) & 6.5(09) \\
3.28(1) & 1.8(13) & 2.6(11) & 7.9(11) & 6.5(11) & 2.5(12) & 8.9(11) & 3.5(10) & 7.2(10) & 6.9(10) & 2.3(10) & 3.4(10) & 5.4(10) & 1.2(09) & 1.2(10) & 5.7(09) \\
3.51(1) & 5.4(12) & 3.1(11) & 8.7(11) & 6.1(11) & 2.1(12) & 9.6(11) & 3.7(10) & 8.1(10) & 7.2(10) & 1.9(10) & 2.7(10) & 6.7(10) & 1.6(09) & 1.4(10) & 3.8(09) \\
3.77(1) & 2.6(12) & 3.7(11) & 9.0(11) & 5.7(11) & 1.8(12) & 9.9(11) & 4.2(10) & 9.1(10) & 7.9(10) & 2.4(10) & 3.7(10) & 7.0(10) & 2.3(09) & 1.3(10) & 2.7(09) \\
4.04(1) & 2.1(12) & 4.3(11) & 9.9(11) & 5.4(11) & 1.5(12) & 1.0(12) & 6.1(10) & 9.5(10) & 8.7(10) & 2.6(10) & 4.0(10) & 8.3(10) & 3.1(09) & 1.5(10) & 3.5(09) \\
4.33(1) & 1.9(12) & 5.4(11) & 1.0(12) & 4.9(11) & 1.3(12) & 1.1(12) & 6.9(10) & 1.0(11) & 9.5(10) & 2.9(10) & 4.7(10) & 8.7(10) & 4.6(09) & 1.5(10) & 2.9(09) \\
4.64(1) & 1.8(12) & 6.5(11) & 1.2(12) & 4.5(11) & 1.2(12) & 1.1(12) & 7.4(10) & 1.0(11) & 1.0(11) & 2.8(10) & 4.1(10) & 9.3(10) & 6.6(09) & 1.6(10) & 3.3(09) \\
4.98(1) & 1.8(12) & 8.3(11) & 1.4(12) & 4.5(11) & 1.1(12) & 1.2(12) & 9.2(10) & 1.1(11) & 1.1(11) & 3.6(10) & 5.6(10) & 1.0(11) & 8.5(09) & 1.8(10) & 3.6(09) \\
5.34(1) & 1.7(12) & 9.9(11) & 1.7(12) & 4.2(11) & 9.9(11) & 1.2(12) & 1.1(11) & 1.1(11) & 1.1(11) & 3.7(10) & 5.5(10) & 1.1(11) & 1.1(10) & 1.9(10) & 3.6(09) \\
5.72(1) & 1.6(12) & 1.3(12) & 2.3(12) & 4.0(11) & 9.5(11) & 1.2(12) & 1.3(11) & 1.2(11) & 1.2(11) & 3.9(10) & 5.8(10) & 1.1(11) & 1.4(10) & 2.0(10) & 4.5(09) \\
6.14(1) & 1.6(12) & 1.7(12) & 2.9(12) & 3.6(11) & 9.2(11) & 1.3(12) & 1.4(11) & 1.3(11) & 1.3(11) & 4.1(10) & 6.2(10) & 1.1(11) & 1.7(10) & 2.0(10) & 5.0(09) \\
6.58(1) & 1.5(12) & 2.0(12) & 4.0(12) & 3.0(11) & 8.8(11) & 1.2(12) & 1.7(11) & 1.6(11) & 1.4(11) & 4.3(10) & 6.6(10) & 1.1(11) & 2.2(10) & 2.1(10) & 5.4(09) \\
7.06(1) & 1.4(12) & 2.5(12) & 5.4(12) & 2.4(11) & 8.5(11) & 1.1(12) & 2.2(11) & 1.9(11) & 1.6(11) & 5.0(10) & 7.4(10) & 1.1(11) & 3.0(10) & 2.4(10) & 6.4(09) \\
7.57(1) & 1.4(12) & 3.1(12) & 6.6(12) & 2.2(11) & 8.4(11) & 1.0(12) & 3.2(11) & 2.2(11) & 1.9(11) & 6.1(10) & 8.7(10) & 1.1(11) & 4.4(10) & 2.6(10) & 7.8(09) \\
8.11(1) & 1.4(12) & 3.6(12) & 7.5(12) & 2.2(11) & 8.1(11) & 9.7(11) & 4.6(11) & 2.5(11) & 2.3(11) & 7.6(10) & 1.0(11) & 1.1(11) & 6.1(10) & 2.9(10) & 8.8(09) \\
8.70(1) & 1.4(12) & 4.2(12) & 8.0(12) & 2.2(11) & 7.5(11) & 9.8(11) & 6.9(11) & 3.0(11) & 3.3(11) & 1.2(11) & 1.2(11) & 1.1(11) & 1.0(11) & 3.2(10) & 8.2(09) \\
9.33(1) & 1.5(12) & 5.0(12) & 8.5(12) & 2.2(11) & 7.1(11) & 1.0(12) & 7.7(11) & 3.8(11) & 5.8(11) & 1.4(11) & 1.3(11) & 1.1(11) & 1.6(11) & 3.7(10) & 7.2(09) \\
1.00(2) & 1.5(12) & 5.8(12) & 9.1(12) & 2.1(11) & 6.9(11) & 1.1(12) & 7.2(11) & 5.1(11) & 8.4(11) & 1.4(11) & 1.4(11) & 1.3(11) & 1.7(11) & 4.3(10) & 6.6(09) \\
1.07(2) & 1.5(12) & 6.8(12) & 9.8(12) & 2.0(11) & 6.7(11) & 1.3(12) & 6.6(11) & 6.8(11) & 9.7(11) & 1.3(11) & 1.4(11) & 1.4(11) & 1.4(11) & 4.8(10) & 6.0(09) \\
1.15(2) & 1.6(12) & 7.8(12) & 1.1(13) & 1.8(11) & 7.0(11) & 1.5(12) & 6.3(11) & 8.4(11) & 9.8(11) & 1.2(11) & 1.3(11) & 1.5(11) & 1.3(11) & 5.3(10) & 5.7(09) \\
1.23(2) & 1.6(12) & 8.5(12) & 1.5(13) & 8.4(10) & 5.1(11) & 1.6(12) & 6.1(11) & 9.7(11) & 8.1(11) & 1.1(11) & 1.3(11) & 1.6(11) & 1.1(11) & 5.5(10) & 5.4(09) \\
1.32(2) & 1.6(12) & 8.9(12) & 1.3(13) & 2.9(10) & 2.7(11) & 1.8(12) & 5.1(11) & 6.4(11) & 6.9(11) & 9.5(10) & 1.2(11) & 1.7(11) & 8.6(10) & 5.4(10) & 4.6(09) \\
1.42(2) & 1.8(12) & 9.7(12) & 1.4(13) & 2.5(10) & 2.7(11) & 1.9(12) & 4.8(11) & 6.1(11) & 5.0(11) & 8.5(10) & 1.2(11) & 1.7(11) & 6.9(10) & 5.3(10) & 4.5(09) \\
1.52(2) & 2.1(12) & 1.0(13) & 1.5(13) & 2.5(10) & 3.0(11) & 2.0(12) & 4.3(11) & 6.5(11) & 3.8(11) & 7.5(10) & 1.1(11) & 1.7(11) & 5.3(10) & 5.3(10) & 4.1(09) \\
1.63(2) & 2.4(12) & 1.1(13) & 1.7(13) & 2.7(10) & 3.6(11) & 2.1(12) & 3.9(11) & 7.3(11) & 3.6(11) & 7.1(10) & 1.0(11) & 1.8(11) & 4.2(10) & 5.4(10) & 3.8(09) \\
1.75(2) & 2.8(12) & 1.2(13) & 1.9(13) & 3.3(10) & 4.7(11) & 2.1(12) & 3.7(11) & 7.9(11) & 3.8(11) & 6.8(10) & 1.0(11) & 1.8(11) & 3.5(10) & 5.6(10) & 3.7(09) \\
1.87(2) & 3.4(12) & 1.3(13) & 2.0(13) & 4.5(10) & 6.2(11) & 2.0(12) & 3.5(11) & 8.8(11) & 4.2(11) & 6.7(10) & 1.0(11) & 1.9(11) & 2.5(10) & 5.8(10) & 3.6(09) \\
2.01(2) & 4.3(12) & 1.4(13) & 2.2(13) & 7.4(10) & 8.5(11) & 1.9(12) & 3.4(11) & 9.7(11) & 5.1(11) & 6.7(10) & 1.0(11) & 2.0(11) & 1.9(10) & 6.3(10) & 3.6(09) \\
2.15(2) & 5.6(12) & 1.5(13) & 2.3(13) & 1.4(11) & 1.1(12) & 1.9(12) & 3.4(11) & 1.0(12) & 6.5(11) & 6.5(10) & 9.2(10) & 2.2(11) & 1.2(10) & 7.2(10) & 3.3(09) \\
2.31(2) & 6.6(12) & 1.6(13) & 2.1(13) & 1.9(11) & 1.0(12) & 1.8(12) & 3.4(11) & 1.0(12) & 7.5(11) & 5.9(10) & 7.8(10) & 2.4(11) & 6.7(09) & 8.4(10) & 2.5(09) \\
2.48(2) & 7.2(12) & 1.7(13) & 1.6(13) & 1.7(11) & 7.3(11) & 1.8(12) & 3.4(11) & 9.2(11) & 7.9(11) & 6.1(10) & 7.6(10) & 2.7(11) & 4.2(09) & 9.8(10) & 1.9(09) \\
2.66(2) & 8.0(12) & 1.7(13) & 1.2(13) & 1.6(11) & 6.0(11) & 1.8(12) & 3.3(11) & 8.1(11) & 8.5(11) & 6.0(10) & 7.5(10) & 2.9(11) & 2.9(09) & 1.1(11) & 1.6(09) \\
2.85(2) & 8.2(12) & 1.7(13) & 9.6(12) & 1.3(11) & 4.8(11) & 1.8(12) & 3.3(11) & 7.4(11) & 9.4(11) & 6.2(10) & 7.7(10) & 3.2(11) & 2.6(09) & 1.2(11) & 1.3(09) \\
3.05(2) & 8.3(12) & 1.7(13) & 8.2(12) & 9.8(10) & 4.1(11) & 1.8(12) & 3.9(11) & 7.0(11) & 1.1(12) & 6.8(10) & 8.6(10) & 3.5(11) & 5.7(09) & 1.3(11) & 1.1(09) \\
\hline
\end{longtable}
\tablefoot{$a(b)$ represents $a\times10^b$}
\end{landscape}
\end{longtab}}

\onltab{
\begin{longtab}
\begin{landscape}
\scriptsize
\centering
\renewcommand*{\arraystretch}{1.25}
\begin{longtable}{cccccccccccccccc}
\caption{\label{table6} Column densities (cm$^{-2}$) of grain-surface (ice) molecules as a function of radius.}\\
\hline\hline
Radius (AU) & \ce{H2CO} & \ce{CH3OH} & \ce{HCOOH} & \ce{HC3N} & ce{CH3CN} & \ce{CH3CCH} & \ce{CH3CHO} & 
\ce{NH2CHO} & \ce{CH3NH2} & \ce{C2H5OH} & \ce{CH3OCH3} & \ce{HCOOCH3} & \ce{CH3COCH3}  & \ce{HOCH2CHO} & \ce{CH3COOH} \\ 
\hline
\endfirsthead
\caption{continued.}\\
\hline\hline
Radius (AU) & \ce{H2CO} & \ce{CH3OH} & \ce{HCOOH} & \ce{HC3N} & \ce{CH3CN} & \ce{CH3CCH} & \ce{CH3CHO} & 
\ce{NH2CHO} & \ce{CH3NH2} & \ce{C2H5OH} & \ce{CH3OCH3} & \ce{HCOOCH3} & \ce{CH3COCH3}  & \ce{HOCH2CHO} & \ce{CH3COOH} \\ 
\hline
\endhead
\hline
\endfoot
\hline
1.07(0) & 4.5(01) & 7.6(11) & 2.2(10) & 8.6(10) & 9.7(10) & 4.7(11) & 2.4(05) & 1.1(12) & 7.7(11) & 1.1(12) & 9.8(06) & 1.9(08) & 1.3(01) & 1.1(05) & 2.1(08) \\
1.15(0) & 1.9(01) & 8.9(11) & 1.7(10) & 6.5(10) & 8.3(10) & 4.5(11) & 2.9(05) & 9.0(11) & 9.4(11) & 1.2(12) & 1.1(07) & 1.8(08) & 1.3(01) & 2.6(04) & 2.2(08) \\
1.23(0) & 1.9(01) & 1.3(11) & 3.6(09) & 3.7(10) & 4.6(10) & 3.6(11) & 3.1(05) & 8.0(11) & 1.6(11) & 1.4(12) & 4.3(05) & 6.5(06) & 2.1(00) & 1.9(04) & 8.6(07) \\
1.32(0) & 1.7(01) & 8.0(10) & 2.2(09) & 1.4(10) & 4.9(09) & 1.1(11) & 2.7(05) & 6.6(11) & 4.6(09) & 1.4(12) & 2.1(01) & 4.4(04) & 0.4(00) & 1.6(04) & 2.7(07) \\
1.42(0) & 3.7(01) & 1.5(11) & 1.4(09) & 6.8(09) & 1.3(09) & 5.0(10) & 9.8(04) & 8.0(11) & 3.9(09) & 8.0(11) & 2.8(01) & 2.7(04) & 0.2(00) & 9.3(03) & 3.4(07) \\
1.52(0) & 9.5(04) & 1.0(13) & 7.5(10) & 1.0(11) & 1.3(09) & 8.4(11) & 8.0(05) & 7.9(13) & 1.8(11) & 7.6(13) & 4.6(03) & 8.7(06) & 2.0(05) & 5.3(06) & 4.0(12) \\
1.63(0) & 7.3(05) & 7.5(14) & 4.4(13) & 3.2(12) & 3.6(11) & 3.3(13) & 6.7(06) & 5.8(14) & 2.4(12) & 1.1(16) & 1.4(04) & 1.2(08) & 3.9(06) & 1.9(10) & 1.3(16) \\
1.75(0) & 4.4(06) & 6.2(18) & 3.0(13) & 5.6(16) & 1.5(17) & 3.4(16) & 7.9(08) & 4.3(14) & 8.7(17) & 2.3(16) & 1.7(09) & 1.6(11) & 2.5(07) & 6.8(11) & 1.2(17) \\
1.87(0) & 1.5(07) & 8.0(18) & 3.3(13) & 5.9(17) & 9.5(17) & 3.5(17) & 4.0(09) & 3.4(14) & 1.7(18) & 2.1(16) & 2.3(10) & 4.0(11) & 5.6(07) & 1.0(12) & 1.4(17) \\
2.01(0) & 9.5(07) & 1.0(19) & 1.1(13) & 2.9(18) & 1.3(18) & 3.5(19) & 4.9(09) & 2.0(14) & 2.4(18) & 1.5(16) & 2.6(11) & 1.7(12) & 2.3(08) & 2.8(12) & 6.8(16) \\
2.15(0) & 1.1(08) & 7.2(18) & 2.5(13) & 4.9(18) & 1.0(18) & 1.0(20) & 4.2(09) & 2.1(14) & 1.9(18) & 5.1(15) & 8.5(11) & 1.9(12) & 4.5(08) & 2.0(12) & 2.4(16) \\
2.31(0) & 1.5(08) & 1.1(19) & 1.0(14) & 7.7(18) & 1.2(18) & 2.4(20) & 1.5(10) & 5.0(14) & 3.2(18) & 6.2(15) & 5.2(12) & 4.8(12) & 2.9(09) & 4.0(12) & 3.4(16) \\
2.48(0) & 6.8(07) & 1.1(19) & 8.3(13) & 5.7(18) & 9.0(17) & 2.4(20) & 1.5(10) & 1.5(14) & 3.3(18) & 7.0(15) & 4.2(12) & 1.4(12) & 2.1(09) & 1.2(12) & 4.2(16) \\
2.66(0) & 1.1(08) & 1.1(19) & 2.3(14) & 4.6(18) & 7.6(17) & 2.3(20) & 8.4(09) & 4.4(14) & 3.4(18) & 8.7(15) & 8.2(13) & 3.3(12) & 3.2(10) & 3.5(12) & 6.9(16) \\
2.85(0) & 2.3(08) & 1.0(19) & 5.1(14) & 4.0(18) & 5.5(17) & 2.2(20) & 6.8(09) & 5.8(14) & 3.3(18) & 9.5(15) & 1.0(15) & 3.6(12) & 1.3(11) & 4.9(12) & 1.1(17) \\
3.05(0) & 1.3(08) & 9.0(18) & 3.5(14) & 3.6(18) & 5.0(17) & 1.9(20) & 4.7(09) & 1.1(14) & 2.7(18) & 6.9(15) & 3.6(14) & 7.7(11) & 6.0(09) & 8.4(11) & 7.7(16) \\
3.28(0) & 4.0(08) & 1.3(19) & 8.6(14) & 4.9(18) & 7.4(17) & 2.6(20) & 7.7(09) & 4.5(15) & 4.2(18) & 1.1(16) & 9.3(14) & 2.9(13) & 6.2(11) & 3.6(13) & 1.2(17) \\
3.51(0) & 5.2(07) & 7.1(18) & 1.8(14) & 3.0(18) & 4.2(17) & 1.6(20) & 2.8(09) & 2.3(13) & 1.9(18) & 3.8(15) & 2.4(13) & 2.0(11) & 2.8(08) & 1.5(11) & 3.6(16) \\
3.77(0) & 3.9(07) & 6.4(18) & 1.5(14) & 2.8(18) & 3.9(17) & 1.5(20) & 2.4(09) & 1.6(13) & 1.7(18) & 3.0(15) & 7.3(12) & 1.4(11) & 1.2(08) & 9.1(10) & 2.7(16) \\
4.04(0) & 2.8(07) & 5.6(18) & 1.2(14) & 2.6(18) & 3.4(17) & 1.3(20) & 2.0(09) & 1.1(13) & 1.4(18) & 2.2(15) & 1.7(12) & 9.6(10) & 4.9(07) & 5.6(10) & 1.9(16) \\
4.33(0) & 2.1(07) & 5.0(18) & 9.6(13) & 2.4(18) & 3.0(17) & 1.2(20) & 1.7(09) & 9.0(12) & 1.2(18) & 1.7(15) & 4.9(11) & 7.6(10) & 2.4(07) & 4.0(10) & 1.4(16) \\
4.64(0) & 2.5(07) & 4.5(18) & 1.3(14) & 2.1(18) & 2.6(17) & 1.1(20) & 1.1(09) & 1.2(13) & 1.1(18) & 1.5(15) & 3.0(12) & 1.0(11) & 4.3(07) & 5.4(10) & 1.4(16) \\
4.98(0) & 8.7(07) & 4.3(18) & 3.9(14) & 1.7(18) & 1.7(17) & 1.0(20) & 7.5(08) & 6.2(13) & 1.0(18) & 1.7(15) & 3.9(14) & 5.0(11) & 1.1(09) & 3.5(11) & 3.0(16) \\
5.34(0) & 2.7(08) & 4.0(18) & 1.6(15) & 1.5(18) & 8.7(16) & 8.4(19) & 9.8(08) & 6.4(14) & 9.8(17) & 2.2(15) & 2.1(15) & 4.2(12) & 4.7(11) & 4.3(12) & 4.6(16) \\
5.72(0) & 7.4(08) & 3.9(18) & 3.0(16) & 1.6(18) & 6.3(16) & 7.6(19) & 2.6(09) & 1.5(17) & 1.1(18) & 4.1(15) & 5.0(15) & 1.2(15) & 3.3(13) & 1.5(15) & 5.0(16) \\
6.14(0) & 7.6(08) & 3.6(18) & 1.2(17) & 1.5(18) & 5.8(16) & 6.9(19) & 4.5(09) & 5.9(17) & 1.1(18) & 5.1(15) & 6.3(15) & 6.2(15) & 6.0(13) & 6.3(15) & 4.2(16) \\
6.58(0) & 5.4(08) & 3.4(18) & 3.1(17) & 1.5(18) & 5.8(16) & 6.3(19) & 8.9(09) & 1.5(18) & 1.2(18) & 6.9(15) & 9.0(15) & 1.8(16) & 1.1(14) & 1.8(16) & 3.3(16) \\
7.06(0) & 3.1(08) & 3.2(18) & 5.8(17) & 1.8(18) & 6.4(16) & 5.8(19) & 2.1(10) & 2.7(18) & 1.3(18) & 9.8(15) & 1.4(16) & 3.6(16) & 2.1(14) & 3.5(16) & 2.6(16) \\
7.57(0) & 3.1(08) & 3.0(18) & 7.7(17) & 2.1(18) & 7.9(16) & 5.4(19) & 5.0(10) & 3.8(18) & 1.4(18) & 1.3(16) & 2.0(16) & 5.0(16) & 3.3(14) & 5.1(16) & 2.1(16) \\
8.11(0) & 5.1(08) & 2.8(18) & 9.1(17) & 2.1(18) & 1.0(17) & 5.2(19) & 1.4(11) & 4.7(18) & 1.4(18) & 1.6(16) & 2.3(16) & 6.2(16) & 3.8(14) & 6.6(16) & 1.7(16) \\
8.67(0) & 1.2(09) & 2.6(18) & 1.0(18) & 2.0(18) & 1.2(17) & 5.1(19) & 5.3(11) & 5.6(18) & 1.5(18) & 2.0(16) & 2.4(16) & 7.5(16) & 3.7(14) & 8.3(16) & 1.5(16) \\
9.33(0) & 3.1(09) & 2.5(18) & 1.1(18) & 1.9(18) & 1.3(17) & 5.1(19) & 2.0(12) & 6.6(18) & 1.7(18) & 2.5(16) & 2.8(16) & 8.9(16) & 4.6(14) & 1.0(17) & 1.5(16) \\
1.00(1) & 6.7(09) & 2.3(18) & 1.1(18) & 1.7(18) & 1.2(17) & 4.7(19) & 5.9(12) & 6.9(18) & 1.8(18) & 2.9(16) & 3.1(16) & 9.5(16) & 6.1(14) & 1.1(17) & 1.5(16) \\
1.07(1) & 1.7(10) & 2.1(18) & 1.1(18) & 1.5(18) & 1.1(17) & 4.5(19) & 2.1(13) & 7.2(18) & 2.0(18) & 3.4(16) & 4.0(16) & 1.1(17) & 1.0(15) & 1.3(17) & 1.7(16) \\
1.15(1) & 4.8(10) & 2.0(18) & 1.1(18) & 1.3(18) & 1.0(17) & 4.1(19) & 1.6(14) & 7.4(18) & 2.2(18) & 4.0(16) & 5.3(16) & 1.2(17) & 2.8(15) & 1.4(17) & 2.3(16) \\
1.23(1) & 1.6(11) & 1.9(18) & 1.1(18) & 1.0(18) & 6.9(16) & 3.8(19) & 1.7(15) & 7.6(18) & 2.6(18) & 5.0(16) & 7.6(16) & 1.5(17) & 9.9(15) & 1.6(17) & 5.5(16) \\
1.32(1) & 7.8(11) & 1.8(18) & 1.0(18) & 7.1(17) & 4.7(16) & 3.3(19) & 6.8(15) & 7.7(18) & 3.2(18) & 6.4(16) & 1.2(17) & 1.9(17) & 3.2(16) & 1.8(17) & 1.1(17) \\
1.42(1) & 5.0(12) & 1.8(18) & 1.0(18) & 5.1(17) & 4.5(16) & 2.8(19) & 1.9(16) & 7.8(18) & 4.3(18) & 8.5(16) & 2.0(17) & 2.8(17) & 6.9(16) & 2.3(17) & 1.2(17) \\
1.52(1) & 2.1(13) & 1.7(18) & 9.6(17) & 3.8(17) & 7.4(16) & 2.4(19) & 3.8(16) & 7.6(18) & 5.0(18) & 1.1(17) & 2.9(17) & 3.6(17) & 1.0(17) & 2.9(17) & 1.0(17) \\
1.63(1) & 8.1(13) & 1.6(18) & 8.6(17) & 2.9(17) & 2.0(17) & 2.1(19) & 7.4(16) & 7.2(18) & 5.8(18) & 1.5(17) & 3.8(17) & 4.0(17) & 1.2(17) & 3.7(17) & 5.9(16) \\
1.75(1) & 5.6(14) & 1.6(18) & 6.9(17) & 2.0(17) & 3.6(17) & 1.9(19) & 1.4(17) & 6.3(18) & 6.9(18) & 2.4(17) & 4.9(17) & 4.0(17) & 1.2(17) & 4.1(17) & 3.5(16) \\
1.87(1) & 1.0(16) & 1.7(18) & 5.2(17) & 1.3(17) & 5.4(17) & 1.8(19) & 2.6(17) & 4.8(18) & 9.0(18) & 4.0(17) & 6.5(17) & 3.2(17) & 8.9(16) & 3.6(17) & 2.3(16) \\
2.01(1) & 8.0(16) & 1.8(18) & 4.0(17) & 8.4(16) & 5.4(17) & 1.6(19) & 3.0(17) & 3.2(18) & 1.0(19) & 5.6(17) & 7.6(17) & 2.4(17) & 7.9(16) & 3.0(17) & 1.8(16) \\
2.15(1) & 6.7(17) & 2.1(18) & 3.1(17) & 4.7(16) & 4.8(17) & 1.4(19) & 3.8(17) & 2.0(18) & 1.1(19) & 7.7(17) & 8.7(17) & 1.7(17) & 8.9(16) & 2.5(17) & 1.5(16) \\
2.31(1) & 4.0(18) & 2.0(18) & 3.0(17) & 2.2(16) & 4.3(17) & 1.3(19) & 7.4(17) & 2.3(18) & 1.1(19) & 6.8(17) & 8.6(17) & 2.0(17) & 1.0(17) & 2.2(17) & 1.3(16) \\
2.48(1) & 6.9(18) & 1.5(18) & 2.8(17) & 1.1(16) & 3.4(17) & 1.3(19) & 7.2(17) & 3.6(18) & 8.2(18) & 4.4(17) & 7.2(17) & 2.8(17) & 8.5(16) & 1.9(17) & 1.2(16) \\
2.66(1) & 7.8(18) & 1.2(18) & 2.5(17) & 5.7(15) & 2.7(17) & 1.2(19) & 6.0(17) & 4.6(18) & 6.2(18) & 3.0(17) & 6.0(17) & 3.7(17) & 6.2(16) & 1.9(17) & 9.0(15) \\
2.85(1) & 7.5(18) & 9.9(17) & 2.4(17) & 2.6(15) & 2.3(17) & 1.1(19) & 5.9(17) & 5.1(18) & 4.9(18) & 2.2(17) & 5.2(17) & 4.4(17) & 4.8(16) & 2.0(17) & 8.6(15) \\
3.05(1) & 6.4(18) & 8.4(17) & 2.4(17) & 1.3(15) & 2.1(17) & 1.0(19) & 6.2(17) & 5.1(18) & 4.1(18) & 1.8(17) & 4.7(17) & 4.8(17) & 5.1(16) & 2.1(17) & 6.5(15) \\
3.28(1) & 5.1(18) & 7.3(17) & 2.3(17) & 6.7(14) & 2.0(17) & 9.4(18) & 6.5(17) & 4.7(18) & 3.6(18) & 1.7(17) & 4.4(17) & 4.9(17) & 6.3(16) & 2.3(17) & 3.5(15) \\
3.51(1) & 3.9(18) & 6.6(17) & 2.1(17) & 3.4(14) & 1.8(17) & 8.7(18) & 6.7(17) & 4.3(18) & 3.2(18) & 1.7(17) & 4.2(17) & 5.1(17) & 8.8(16) & 2.4(17) & 2.7(15) \\
3.77(1) & 2.8(18) & 6.2(17) & 1.9(17) & 1.6(14) & 1.7(17) & 8.0(18) & 6.6(17) & 3.8(18) & 2.8(18) & 1.7(17) & 4.0(17) & 5.1(17) & 1.2(17) & 2.6(17) & 1.7(15) \\
4.04(1) & 2.1(18) & 5.9(17) & 1.8(17) & 1.5(14) & 1.5(17) & 7.2(18) & 6.1(17) & 3.2(18) & 2.5(18) & 1.6(17) & 3.8(17) & 4.9(17) & 1.5(17) & 2.5(17) & 1.7(15) \\
4.33(1) & 1.4(18) & 5.8(17) & 1.6(17) & 8.0(13) & 1.3(17) & 6.6(18) & 5.3(17) & 2.7(18) & 2.2(18) & 1.6(17) & 3.5(17) & 4.7(17) & 1.8(17) & 2.5(17) & 1.4(15) \\
4.64(1) & 9.8(17) & 5.9(17) & 1.4(17) & 6.9(13) & 1.2(17) & 6.0(18) & 4.4(17) & 2.1(18) & 1.9(18) & 1.5(17) & 3.3(17) & 4.4(17) & 2.2(17) & 2.3(17) & 1.3(15) \\
4.98(1) & 7.4(17) & 6.1(17) & 1.3(17) & 6.0(13) & 1.0(17) & 5.5(18) & 3.7(17) & 1.6(18) & 1.7(18) & 1.4(17) & 3.1(17) & 4.1(17) & 2.4(17) & 2.1(17) & 1.3(15) \\
5.34(1) & 5.8(17) & 6.3(17) & 1.2(17) & 5.3(13) & 9.1(16) & 5.1(18) & 3.1(17) & 1.3(18) & 1.5(18) & 1.3(17) & 2.8(17) & 3.7(17) & 2.4(17) & 1.9(17) & 1.1(15) \\
5.72(1) & 4.7(17) & 6.6(17) & 1.1(17) & 3.8(13) & 7.7(16) & 4.7(18) & 2.6(17) & 9.5(17) & 1.2(18) & 1.1(17) & 2.6(17) & 3.3(17) & 2.4(17) & 1.6(17) & 1.1(15) \\
6.14(1) & 4.1(17) & 7.0(17) & 1.0(17) & 3.0(13) & 6.4(16) & 4.3(18) & 2.2(17) & 7.1(17) & 1.0(18) & 1.0(17) & 2.3(17) & 3.0(17) & 2.3(17) & 1.4(17) & 1.1(15) \\
6.58(1) & 3.6(17) & 7.5(17) & 9.5(16) & 2.0(13) & 5.2(16) & 3.9(18) & 1.9(17) & 5.4(17) & 8.3(17) & 9.1(16) & 2.0(17) & 2.7(17) & 2.1(17) & 1.3(17) & 9.0(14) \\
7.06(1) & 3.5(17) & 7.9(17) & 8.6(16) & 1.4(13) & 4.2(16) & 3.5(18) & 1.8(17) & 4.1(17) & 6.8(17) & 8.2(16) & 1.8(17) & 2.4(17) & 1.8(17) & 1.1(17) & 7.0(14) \\
7.57(1) & 3.5(17) & 8.2(17) & 7.8(16) & 1.2(13) & 3.4(16) & 2.9(18) & 1.7(17) & 3.2(17) & 5.6(17) & 7.5(16) & 1.6(17) & 2.2(17) & 1.7(17) & 1.0(17) & 5.1(14) \\
8.11(1) & 3.4(17) & 8.2(17) & 7.4(16) & 1.0(13) & 2.9(16) & 2.3(18) & 1.8(17) & 2.6(17) & 4.9(17) & 7.0(16) & 1.5(17) & 1.9(17) & 1.7(17) & 9.2(16) & 4.2(14) \\
8.70(1) & 3.5(17) & 8.6(17) & 7.3(16) & 9.3(12) & 2.7(16) & 1.7(18) & 2.5(17) & 2.1(17) & 5.0(17) & 8.0(16) & 1.4(17) & 1.8(17) & 1.8(17) & 8.7(16) & 3.7(14) \\
9.33(1) & 3.5(17) & 1.0(18) & 8.2(16) & 8.3(12) & 2.8(16) & 1.3(18) & 4.2(17) & 1.7(17) & 8.4(17) & 1.3(17) & 1.6(17) & 1.5(17) & 2.4(17) & 8.0(16) & 4.5(14) \\
1.00(2) & 3.4(17) & 1.1(18) & 1.1(17) & 8.2(12) & 2.7(16) & 1.1(18) & 4.1(17) & 2.0(17) & 1.3(18) & 1.4(17) & 1.6(17) & 1.3(17) & 2.6(17) & 7.6(16) & 7.1(14) \\
1.07(2) & 3.1(17) & 1.1(18) & 1.5(17) & 8.2(12) & 2.5(16) & 9.5(17) & 2.9(17) & 3.0(17) & 1.5(18) & 1.2(17) & 1.5(17) & 1.2(17) & 2.2(17) & 7.2(16) & 9.7(14) \\
1.15(2) & 2.9(17) & 1.0(18) & 1.9(17) & 7.4(12) & 2.1(16) & 8.5(17) & 2.2(17) & 4.1(17) & 1.4(18) & 9.0(16) & 1.2(17) & 1.1(17) & 1.8(17) & 6.8(16) & 1.0(15) \\
1.23(2) & 2.9(17) & 9.7(17) & 1.9(17) & 5.7(12) & 1.8(16) & 7.7(17) & 2.0(17) & 3.6(17) & 1.1(18) & 7.2(16) & 1.0(17) & 9.6(16) & 1.6(17) & 6.1(16) & 1.1(15) \\
1.32(2) & 2.9(17) & 9.1(17) & 1.8(17) & 5.1(12) & 1.6(16) & 7.0(17) & 1.7(17) & 2.8(17) & 8.5(17) & 5.7(16) & 8.6(16) & 8.4(16) & 1.3(17) & 5.3(16) & 9.2(14) \\
1.42(2) & 2.9(17) & 8.6(17) & 1.5(17) & 4.7(12) & 1.3(16) & 6.4(17) & 1.5(17) & 1.9(17) & 6.6(17) & 4.6(16) & 7.2(16) & 7.1(16) & 1.1(17) & 4.5(16) & 7.0(14) \\
1.52(2) & 2.9(17) & 8.5(17) & 1.3(17) & 4.4(12) & 1.1(16) & 5.8(17) & 1.3(17) & 1.1(17) & 4.9(17) & 3.8(16) & 6.2(16) & 5.9(16) & 9.4(16) & 3.7(16) & 5.5(14) \\
1.63(2) & 2.8(17) & 8.4(17) & 1.1(17) & 4.2(12) & 9.7(15) & 5.3(17) & 1.1(17) & 7.1(16) & 3.7(17) & 3.1(16) & 5.3(16) & 5.0(16) & 7.6(16) & 3.0(16) & 4.4(14) \\
1.75(2) & 2.8(17) & 8.5(17) & 1.0(17) & 4.0(12) & 8.2(15) & 4.8(17) & 9.4(16) & 4.8(16) & 2.7(17) & 2.5(16) & 4.4(16) & 4.1(16) & 5.8(16) & 2.5(16) & 3.7(14) \\
1.87(2) & 2.7(17) & 8.8(17) & 8.9(16) & 3.8(12) & 6.9(15) & 4.4(17) & 7.6(16) & 3.4(16) & 1.9(17) & 2.0(16) & 3.5(16) & 3.4(16) & 3.9(16) & 2.0(16) & 3.2(14) \\
2.01(2) & 2.7(17) & 8.7(17) & 8.0(16) & 3.7(12) & 5.8(15) & 4.0(17) & 6.3(16) & 2.6(16) & 1.4(17) & 1.6(16) & 2.8(16) & 2.8(16) & 2.6(16) & 1.6(16) & 2.7(14) \\
2.15(2) & 2.7(17) & 8.5(17) & 7.4(16) & 3.5(12) & 4.6(15) & 3.6(17) & 5.2(16) & 2.0(16) & 1.0(17) & 1.2(16) & 2.2(16) & 2.3(16) & 1.7(16) & 1.3(16) & 2.1(14) \\
2.31(2) & 2.7(17) & 8.1(17) & 6.9(16) & 3.6(12) & 3.4(15) & 3.2(17) & 4.4(16) & 1.6(16) & 7.7(16) & 9.3(15) & 1.6(16) & 1.8(16) & 1.1(16) & 1.1(16) & 1.4(14) \\
2.48(2) & 2.9(17) & 7.7(17) & 6.1(16) & 4.0(12) & 2.5(15) & 2.9(17) & 3.7(16) & 1.4(16) & 5.5(16) & 6.9(15) & 1.2(16) & 1.4(16) & 6.2(15) & 8.3(15) & 8.9(13) \\
2.66(2) & 3.9(17) & 8.0(17) & 5.7(16) & 4.5(12) & 2.1(15) & 2.7(17) & 3.3(16) & 1.2(16) & 3.8(16) & 5.0(15) & 8.5(15) & 1.0(16) & 3.1(15) & 6.4(15) & 6.3(13) \\
2.85(2) & 5.2(17) & 8.4(17) & 4.6(16) & 5.0(12) & 1.9(15) & 2.6(17) & 4.9(16) & 8.9(15) & 3.3(16) & 5.4(15) & 9.1(15) & 7.0(15) & 2.6(15) & 4.4(15) & 4.6(13) \\
3.05(2) & 6.0(17) & 8.8(17) & 3.3(16) & 5.5(12) & 2.0(15) & 2.5(17) & 9.4(16) & 6.7(15) & 4.0(16) & 9.5(15) & 1.5(16) & 4.2(15) & 5.7(15) & 2.7(15) & 3.2(13) \\
\hline
\end{longtable}
\tablefoot{$a(b)$ represents $a\times10^b$}
\end{landscape}
\end{longtab}}

\end{document}